\documentclass[prd,aps,showpacs,preprintnumbers,amsmath,amssymb,superscriptaddress,twocolumn,floatfix]{revtex4}
\usepackage{alltt}
\usepackage{epsfig}
\usepackage{calc,rotating,xspace}
\usepackage{pifont}

\newcommand{\met}{\mbox{${\hbox{$p$\kern-0.5em\lower+.2ex\hbox{/}}}_T\:$}}
\newcommand{\vecmet}{\mbox{${\hbox{$\vec{p}$\kern-0.5em\lower+.4ex\hbox{/}}}_T\:$}}

\begin{document}

\title{\boldmath{First Run II Measurement of the $W$ Boson Mass}}

\affiliation{Institute of Physics, Academia Sinica, Taipei, Taiwan 11529, Republic of China} 
\affiliation{Argonne National Laboratory, Argonne, Illinois 60439} 
\affiliation{Institut de Fisica d'Altes Energies, Universitat Autonoma de Barcelona, E-08193, Bellaterra (Barcelona), Spain} 
\affiliation{Baylor University, Waco, Texas  76798} 
\affiliation{Istituto Nazionale di Fisica Nucleare, University of Bologna, I-40127 Bologna, Italy} 
\affiliation{Brandeis University, Waltham, Massachusetts 02254} 
\affiliation{University of California, Davis, Davis, California  95616} 
\affiliation{University of California, Los Angeles, Los Angeles, California  90024} 
\affiliation{University of California, San Diego, La Jolla, California  92093} 
\affiliation{University of California, Santa Barbara, Santa Barbara, California 93106} 
\affiliation{Instituto de Fisica de Cantabria, CSIC-University of Cantabria, 39005 Santander, Spain} 
\affiliation{Carnegie Mellon University, Pittsburgh, PA  15213} 
\affiliation{Enrico Fermi Institute, University of Chicago, Chicago, Illinois 60637} 
\affiliation{Comenius University, 842 48 Bratislava, Slovakia; Institute of Experimental Physics, 040 01 Kosice, Slovakia} 
\affiliation{Joint Institute for Nuclear Research, RU-141980 Dubna, Russia} 
\affiliation{Duke University, Durham, North Carolina  27708} 
\affiliation{Fermi National Accelerator Laboratory, Batavia, Illinois 60510} 
\affiliation{University of Florida, Gainesville, Florida  32611} 
\affiliation{Laboratori Nazionali di Frascati, Istituto Nazionale di Fisica Nucleare, I-00044 Frascati, Italy} 
\affiliation{University of Geneva, CH-1211 Geneva 4, Switzerland} 
\affiliation{Glasgow University, Glasgow G12 8QQ, United Kingdom} 
\affiliation{Harvard University, Cambridge, Massachusetts 02138} 
\affiliation{Division of High Energy Physics, Department of Physics, University of Helsinki and Helsinki Institute of Physics, FIN-00014, Helsinki, Finland} 
\affiliation{University of Illinois, Urbana, Illinois 61801} 
\affiliation{The Johns Hopkins University, Baltimore, Maryland 21218} 
\affiliation{Institut f\"{u}r Experimentelle Kernphysik, Universit\"{a}t Karlsruhe, 76128 Karlsruhe, Germany} 
\affiliation{Center for High Energy Physics: Kyungpook National University, Taegu 702-701, Korea; Seoul National University, Seoul 151-742, Korea; SungKyunKwan University, Suwon 440-746, Korea; Korea Institute of Science and Technology Information, Daejeon, 305-806, Korea; Chonnam National University, Gwangju, 500-757, Korea} 
\affiliation{Ernest Orlando Lawrence Berkeley National Laboratory, Berkeley, California 94720} 
\affiliation{University of Liverpool, Liverpool L69 7ZE, United Kingdom} 
\affiliation{University College London, London WC1E 6BT, United Kingdom} 
\affiliation{Centro de Investigaciones Energeticas Medioambientales y Tecnologicas, E-28040 Madrid, Spain} 
\affiliation{Massachusetts Institute of Technology, Cambridge, Massachusetts  02139} 
\affiliation{Institute of Particle Physics: McGill University, Montr\'{e}al, Canada H3A~2T8; and University of Toronto, Toronto, Canada M5S~1A7} 
\affiliation{University of Michigan, Ann Arbor, Michigan 48109} 
\affiliation{Michigan State University, East Lansing, Michigan  48824} 
\affiliation{University of New Mexico, Albuquerque, New Mexico 87131} 
\affiliation{Northwestern University, Evanston, Illinois  60208} 
\affiliation{The Ohio State University, Columbus, Ohio  43210} 
\affiliation{Okayama University, Okayama 700-8530, Japan} 
\affiliation{Osaka City University, Osaka 588, Japan} 
\affiliation{University of Oxford, Oxford OX1 3RH, United Kingdom} 
\affiliation{University of Padova, Istituto Nazionale di Fisica Nucleare, Sezione di Padova-Trento, I-35131 Padova, Italy} 
\affiliation{LPNHE, Universite Pierre et Marie Curie/IN2P3-CNRS, UMR7585, Paris, F-75252 France} 
\affiliation{University of Pennsylvania, Philadelphia, Pennsylvania 19104} 
\affiliation{Istituto Nazionale di Fisica Nucleare Pisa, Universities of Pisa, Siena and Scuola Normale Superiore, I-56127 Pisa, Italy} 
\affiliation{University of Pittsburgh, Pittsburgh, Pennsylvania 15260} 
\affiliation{Purdue University, West Lafayette, Indiana 47907} 
\affiliation{University of Rochester, Rochester, New York 14627} 
\affiliation{The Rockefeller University, New York, New York 10021} 
\affiliation{Istituto Nazionale di Fisica Nucleare, Sezione di Roma 1, University of Rome ``La Sapienza," I-00185 Roma, Italy} 
\affiliation{Rutgers University, Piscataway, New Jersey 08855} 
\affiliation{Texas A\&M University, College Station, Texas 77843} 
\affiliation{Istituto Nazionale di Fisica Nucleare, University of Trieste/\ Udine, Italy} 
\affiliation{University of Tsukuba, Tsukuba, Ibaraki 305, Japan} 
\affiliation{Tufts University, Medford, Massachusetts 02155} 
\affiliation{Waseda University, Tokyo 169, Japan} 
\affiliation{Wayne State University, Detroit, Michigan  48201} 
\affiliation{University of Wisconsin, Madison, Wisconsin 53706} 
\affiliation{Yale University, New Haven, Connecticut 06520} 
\author{T.~Aaltonen}
\affiliation{Division of High Energy Physics, Department of Physics, University of Helsinki and Helsinki Institute of Physics, FIN-00014, Helsinki, Finland}
\author{A.~Abulencia}
\affiliation{University of Illinois, Urbana, Illinois 61801}
\author{J.~Adelman}
\affiliation{Enrico Fermi Institute, University of Chicago, Chicago, Illinois 60637}
\author{T.~Akimoto}
\affiliation{University of Tsukuba, Tsukuba, Ibaraki 305, Japan}
\author{M.G.~Albrow}
\affiliation{Fermi National Accelerator Laboratory, Batavia, Illinois 60510}
\author{B.~\'{A}lvarez~Gonz\'{a}lez}
\affiliation{Instituto de Fisica de Cantabria, CSIC-University of Cantabria, 39005 Santander, Spain}
\author{S.~Amerio}
\affiliation{University of Padova, Istituto Nazionale di Fisica Nucleare, Sezione di Padova-Trento, I-35131 Padova, Italy}
\author{D.~Amidei}
\affiliation{University of Michigan, Ann Arbor, Michigan 48109}
\author{A.~Anastassov}
\affiliation{Rutgers University, Piscataway, New Jersey 08855}
\author{A.~Annovi}
\affiliation{Laboratori Nazionali di Frascati, Istituto Nazionale di Fisica Nucleare, I-00044 Frascati, Italy}
% ...
\collaboration{CDF Collaboration\footnote{With visitors from $^a$University of Athens, 15784 Athens, Greece, 
$^b$Chinese Academy of Sciences, Beijing 100864, China, 
$^c$University of Bristol, Bristol BS8 1TL, United Kingdom, 
$^d$University Libre de Bruxelles, B-1050 Brussels, Belgium, 
$^e$University of California, Irvine, Irvine, CA  92697, 
$^f$University of California Santa Cruz, Santa Cruz, CA  95064, 
$^g$Cornell University, Ithaca, NY  14853, 
$^h$University of Cyprus, Nicosia CY-1678, Cyprus, 
$^i$University College Dublin, Dublin 4, Ireland, 
$^j$University of Edinburgh, Edinburgh EH9 3JZ, United Kingdom, 
$^k$University of Heidelberg, D-69120 Heidelberg, Germany, 
$^l$Universidad Iberoamericana, Mexico D.F., Mexico, 
$^m$University of Manchester, Manchester M13 9PL, England, 
$^n$Nagasaki Institute of Applied Science, Nagasaki, Japan, 
$^o$University de Oviedo, E-33007 Oviedo, Spain, 
$^p$Queen Mary's College, University of London, London, E1 4NS, England, 
$^q$Texas Tech University, Lubbock, TX  79409, 
$^r$IFIC(CSIC-Universitat de Valencia), 46071 Valencia, Spain, 
}}
\noaffiliation


\begin{abstract}
We describe a measurement of the $W$ boson mass $m_W$ using 200 pb$^{-1}$ 
of $\sqrt{s}$=1.96 TeV $p\bar{p}$ collision data taken with the CDF II 
detector.  With a sample of 63,964 $W\rightarrow e\nu$ candidates and 
51,128 $W\rightarrow \mu\nu$ candidates, we measure 
$m_W = [80.413 \pm 0.034 ({\rm stat}) \pm 0.034 ({\rm sys}) = 80.413 \pm 0.048]$ 
GeV/$c^2$.  This is the single most precise $m_W$ measurement to date.  When 
combined with other measured electroweak parameters, this result further 
constrains the properties of new unobserved particles coupling to $W$ and $Z$ 
bosons.
\end{abstract}

\maketitle


\section{Introduction}
\label{sec:introduction}

The discovery of the $W$ and $Z$ bosons in 1983 \cite{WZdiscovery} 
confirmed a central prediction of the unified model of electromagnetic
and weak interactions \cite{GWS}.  Initial $W$ and $Z$ boson mass measurements
verified the tree-level predictions of the theory, with subsequent measurements
probing the predicted ${\cal O}(3$~GeV/$c^2)$ \cite{corrections, sirlin} 
radiative corrections to the masses.  The current knowledge of these masses 
and other electroweak parameters constrains additional radiative corrections 
from unobserved particles such as the Higgs boson or supersymmetric 
particles.  These constraints are however limited by the precision of the 
$W$ boson mass $m_W$, making improved measurements of $m_W$ a high priority 
in probing the masses and electroweak couplings of new hypothetical particles.  
We describe in this article the single most precise $m_W$ measurement 
\cite{prlreference} to date.
\par
\begin{figure}[!htbp]
\begin{center}
\epsfysize = 3.5cm
\epsffile{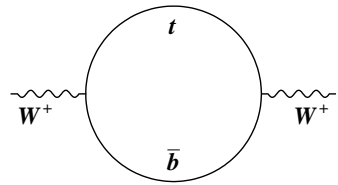}
\caption{The one-loop contribution to the $W$ boson mass from top and bottom 
quarks. }
\label{fig:toploop}
\end{center}
\end{figure}
\begin{figure}[!htbp]
\begin{center}
\epsfysize = 3.cm
\epsffile{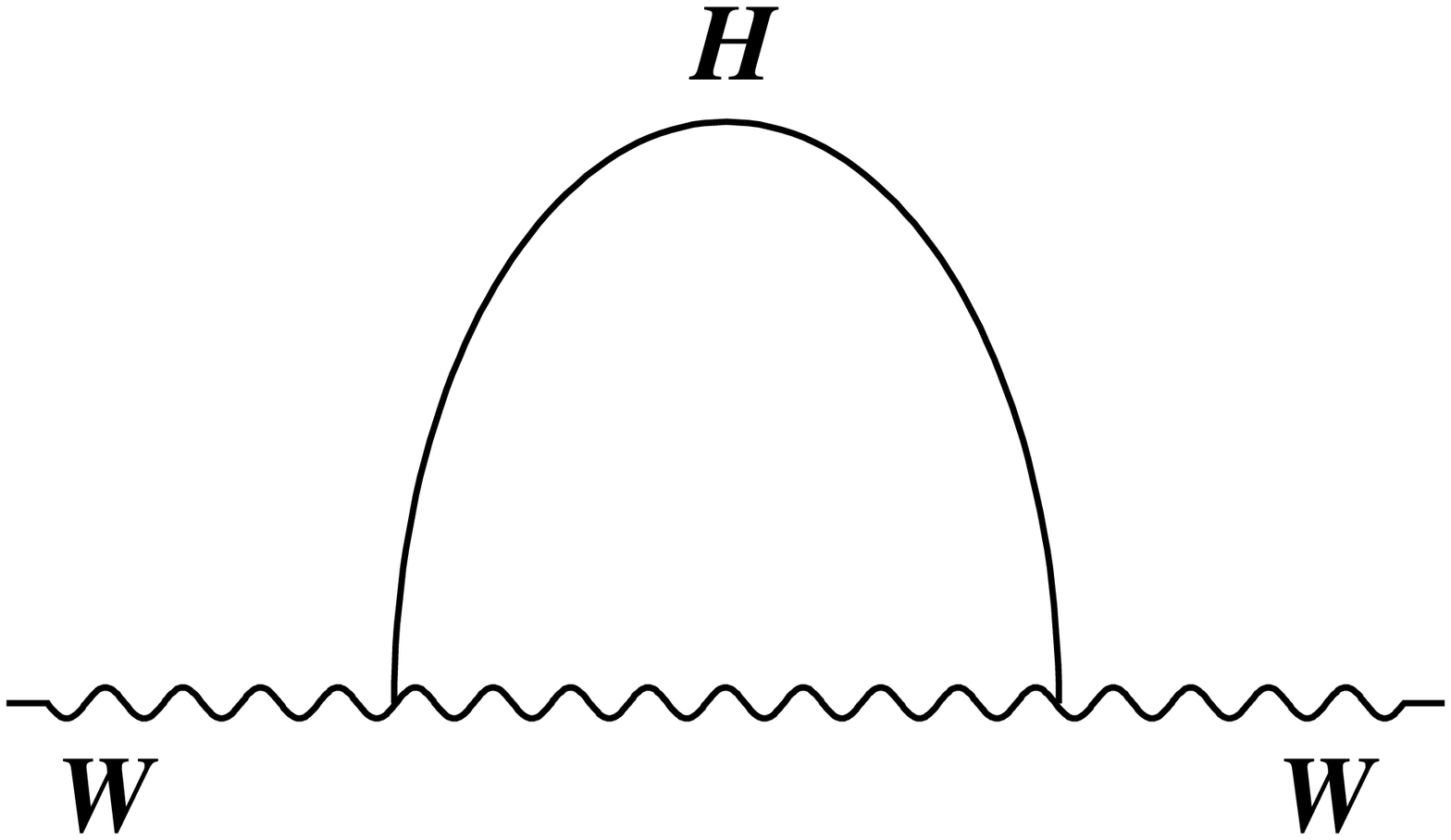}
\hskip -0.05in
\epsfysize = 3.cm
\epsffile{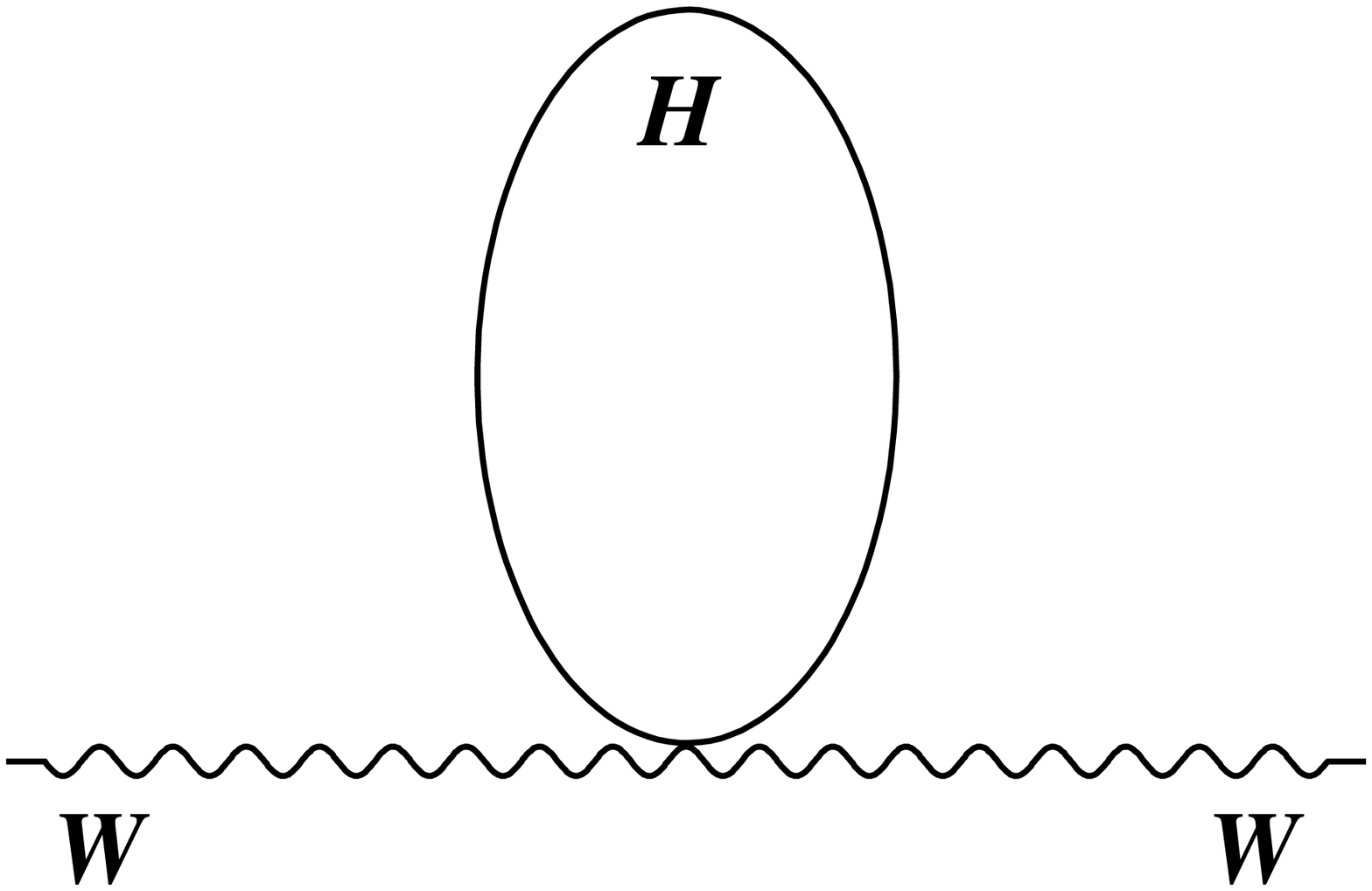}
\caption{Higgs one-loop contributions to the $W$ boson mass. }
\label{fig:hloop}
\end{center}
\end{figure}
\begin{figure}[!htbp]
\begin{center}
\epsfysize = 3.cm
\epsffile{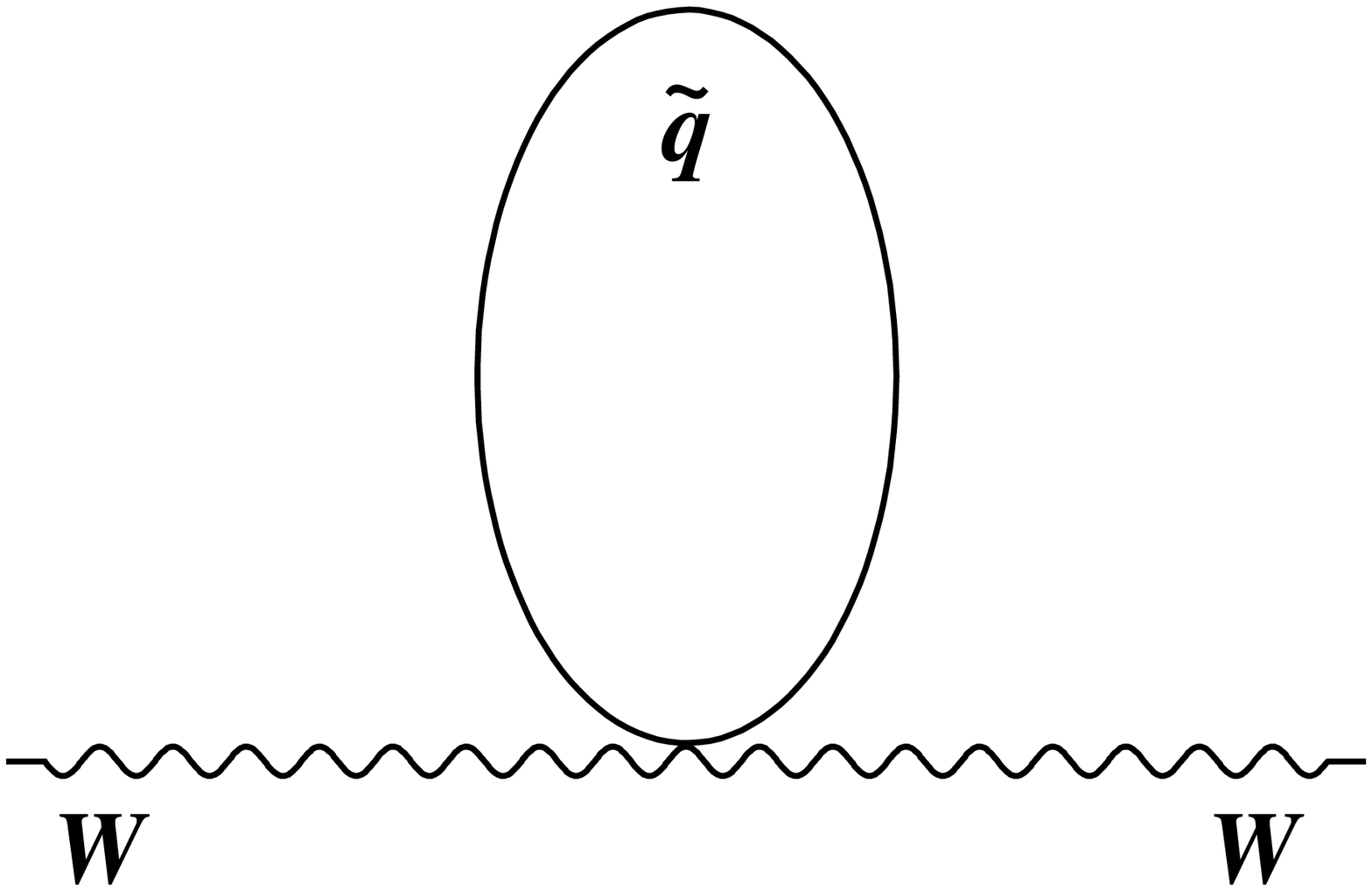}
\hskip -0.05in
\epsfysize = 3.cm
\epsffile{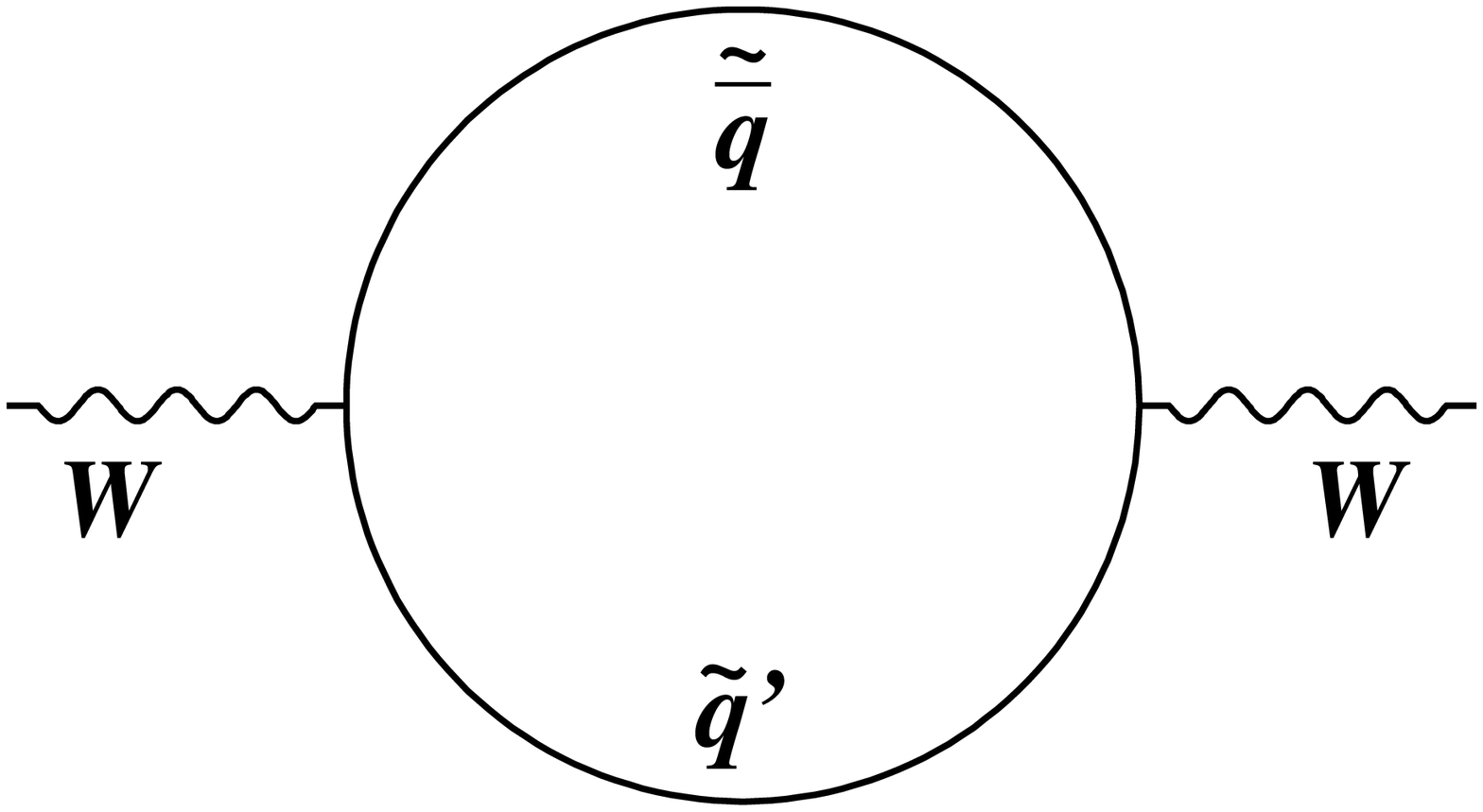}
\caption{One-loop squark contributions to the $W$ boson mass. }
\label{fig:susyloop}
\end{center}
\end{figure}

The $W$ boson mass can be written in terms of other precisely measured 
parameters in the ``on-shell'' scheme as \cite{sirlin}:
\begin{equation}
m_W^2 =\frac{\hbar^3}{c}\frac{\pi\alpha_{EM}}{\sqrt{2}G_F(1-m_W^2/m_Z^2)(1-\Delta r)},
\end{equation}

\noindent
where $\alpha_{EM}$ is the electromagnetic coupling at the renormalization energy 
scale $Q = m_Z c^2$, $G_F$ is the Fermi weak coupling extracted from the muon 
lifetime, $m_Z$ is the $Z$ boson mass, and $\Delta r$ includes all radiative 
corrections.   Fermionic loop corrections increase the $W$ boson mass by terms 
proportional to $\ln(m_Z/m_f)$ for $m_f \ll m_Z$ \cite{sirlin}, while the loop 
containing top and bottom quarks (Fig.~\ref{fig:toploop}) increases $m_W$ 
according to \cite{wtb}:
\begin{align}
\Delta r_{tb} & =  \frac{c}{\hbar^3}\frac{-3G_F m_W^2}{8\sqrt{2}\pi^2(m_Z^2 - m_W^2)} 
\times \notag \\ 
 & \; \; \; \; \; \; \; \; \; \; \; \; \; \left[m_t^2 + m_b^2 - 
  \frac{2m_t^2m_b^2}{m_t^2-m_b^2}\ln(m_t^2/m_b^2)\right],
\end{align}

\noindent
where the second and third terms can be neglected since $m_t \gg m_b$. 
Higgs loops (Fig.~\ref{fig:hloop}) decrease $m_W$ with a contribution 
proportional to the logarithm of the Higgs mass ($m_H$).  Contributions from 
possible supersymmetric particles are dominated by squark loops 
(Fig.~\ref{fig:susyloop}) and tend to increase $m_W$.  Generally, the lighter the 
squark masses and the larger the squark weak doublet mass splitting, the 
larger the contribution to $m_W$.  The total radiative correction from 
supersymmetric particles can be as large as several hundred MeV/$c^2$ 
\cite{susyloop}.
\par
Table~\ref{tbl:dmwdpar} \cite{dmdpar} shows the change in $m_W$ for $+1\sigma$ 
changes in the measured standard model input parameters and the effect of 
doubling $m_H$ from 100 GeV/$c^2$ to 200 GeV/$c^2$.  In addition to the 
listed parameters, a variation of $\pm 1.7$ MeV/$c^2$ on the predicted $m_W$ arises 
from two-loop sensitivity to $\alpha_s$, e.g. via gluon exchange in the quark loop in 
Fig.~\ref{fig:toploop}.  Theoretical corrections beyond second order, which have yet 
to be calculated, are estimated to affect the $m_W$ prediction by $\pm 4$ MeV/$c^2$ 
\cite{dmdpar}.  

\begin{table}[ht]
\begin{center}
\begin{tabular}{lc}
\hline  
\hline  
Parameter Shift                                                & $m_W$ Shift \\
                                                               & (MeV/$c^2$) \\
\hline  
$\Delta \ln m_H = + 0.693$                                     &  -41.3 \\
$\Delta m_t = + 1.8$ GeV/$c^2$ \cite{top}                      & 11.0   \\
$\Delta \alpha_{EM}(Q = m_Z c^2) = + 0.00035$ \cite{alpha} & -6.2   \\
$\Delta m_Z = + 2.1$ MeV/$c^2$ \cite{pdg}                      & 2.6    \\
\hline  
\hline  
\end{tabular}
\end{center}
\vskip 0.1in
\caption{The effect on $m_W$ of $+1\sigma$ increases of the input parameters 
dominating the uncertainty on the $m_W$ prediction.  Since the Higgs boson has 
not been observed, we show the effect of doubling the Higgs boson mass 
from 100 GeV/$c^2$ to 200 GeV/$c^2$ \cite{dmdpar}. } 
\label{tbl:dmwdpar}
\end{table}

The uncertainties on the $m_W$ prediction can be compared to the 29 MeV/$c^2$ 
uncertainty on the world average from direct $m_W$ measurements 
(Table~\ref{tbl:wmeasurements}), which include results from four experiments, 
ALEPH \cite{ALEPH}, DELPHI \cite{DELPHI}, L3 \cite{L3}, and OPAL \cite{OPAL},
studying $\sqrt{s} = 161-209$ GeV $e^+ e^-$ collisions at the Large Electron 
Positron collider (LEP), and from two experiments, CDF \cite{CDF} and 
D\O\ \cite{DZERO, DZEROEC}, studying $\sqrt{s} = 1.8$ TeV $p\bar{p}$ collisions 
in Run~I of the Fermilab Tevatron.  The current experimental $m_W$ uncertainty is 
a factor of two larger than the uncertainty from radiative corrections, excluding 
the Higgs contribution (Table \ref{tbl:dmwdpar}).  The Higgs mass constraint 
extracted from the $W$ boson mass is thus limited by the direct $m_W$ measurement.  
The precise $m_W$ measurement described in this article has a significant impact 
on the world average $m_W$.

\begin{table}[h]
\begin{center}
\begin{tabular}{l c}
\hline
\hline
Experiment                        & $m_W$ (GeV/$c^2$)   \\ 
\hline
ALEPH~\cite{ALEPH}                & $80.440 \pm 0.051$ \\
OPAL~\cite{OPAL}                  & $80.416 \pm 0.053$ \\
L3~\cite{L3}                      & $80.270 \pm 0.055$ \\
DELPHI~\cite{DELPHI}              & $80.336 \pm 0.067$ \\
CDF Run I \cite{CDF}              & $80.433 \pm 0.079$ \\
D\O\ Run I \cite{DZERO, DZEROEC}  & $80.483 \pm 0.084$ \\
\hline
LEP Average \cite{LEP}            & $80.376 \pm 0.033$ \\
Tevatron Run I Average \cite{TEV} & $80.456 \pm 0.059$ \\
\hline
World Average                     & $80.392 \pm 0.029$ \\
\hline
\hline
\end{tabular}
\end{center}
\caption{Direct measurements of the $W$ boson mass, the preliminary combined
LEP average, the combined Tevatron Run I average, and the preliminary 
world average.}
\label{tbl:wmeasurements}
\end{table}

\section{Overview}
\label{sec:overview}

A measurement of $m_W$ at a $p\bar{p}$ collider \cite{theses} is complementary 
to that at an $e^+ e^-$ collider.  Individual $u$ ($d$) quarks inside the proton 
can interact with $\bar{d}$ ($\bar{u}$) quarks inside the anti-proton (or vice 
versa), allowing single $W^+$ ($W^-$) boson production, which is not possible at 
an $e^+ e^-$ collider.  In addition, $p\bar{p}$ colliders have higher center of 
mass energies and $W$ boson production cross sections.  This provides high 
statistics for the leptonic decays of the $W$ boson, which are studied exclusively 
because of the overwhelming hadronic-jet background in the quark decay channels.  
The leptonic decays of singly produced $Z$ bosons provide important control samples, 
since both leptons from $Z$ boson decay are well measured.  The production and decay 
uncertainties on the measurement of $m_W$ from $p\bar{p}$ and $e^+ e^-$ collider 
data are almost completely independent \cite{correlations}.
\par
We present in this Section an overview of $W$ and $Z$ boson production at the 
Tevatron, a description of the coordinate definitions and symbol conventions 
used for this measurement, and a broad discussion of our $m_W$ measurement 
strategy.

\subsection{$W$ and $Z$ Boson Production and Decay}

\begin{figure}[!htbp]
\begin{center}
\epsfysize = 4.cm
\epsffile{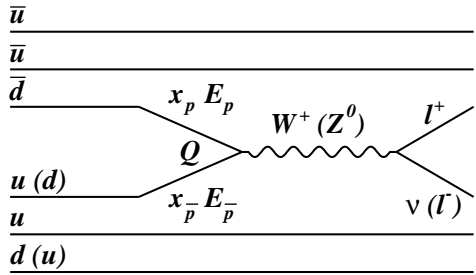}
\caption{Leading-order annihilation of a quark and antiquark inside the 
proton and antiproton, respectively, producing a $W^+$ or $Z^0$ boson.  
The quark (antiquark) has energy $x_p E_p$ ($x_{\bar{p}} E_{\bar{p}}$), 
where $E_p$ ($E_{\bar{p}}$) represents the total proton (antiproton) 
energy.  The production occurs at a partonic center-of-mass energy $Q$.  
The $u\bar{u}\rightarrow Z^0$ and $d\bar{u}\rightarrow W^-$ processes
are similar. }
\label{fig:wzprod}
\end{center}
\end{figure}

$W$ and $Z$ bosons are produced in $\sqrt{s} = 1.96$ TeV $p\bar{p}$
collisions primarily through $s-$channel annihilation of valence
$u$ and/or $d$ quarks (Fig. \ref{fig:wzprod}), with a smaller {\cal O}(20\%) 
contribution from sea quarks.  The quark (antiquark) has a fraction $x_p$ 
($x_{\bar{p}}$) of the proton's (antiproton's) total momentum, producing a 
$W$ or $Z$ boson at center of mass energy $\sqrt{\hat{s}} \equiv Q$ equal 
to its mass times $c^2$.  The rate of production can be predicted from two 
components: (1) the momentum fraction distributions of the quarks, $f_q(x,Q^2)$, 
which are determined from fits to world data \cite{MRST, CTEQ}; and (2) a 
perturbative calculation of the $q\bar{q'} \rightarrow W$ or $Z$ boson process 
\cite{WZxsec}.  

$W$ and $Z$ bosons can decay to lepton or quark pairs.  Decays to quark pairs 
are not observable given the large direct $q\bar{q'}$ background, and decays 
to $\tau \rightarrow \nu_{\tau} +$hadrons are not as precisely measured as 
boson decays to electrons or muons.  For these reasons we restrict ourselves 
to the direct electronic and muonic decays ($W\rightarrow e\nu$, 
$W\rightarrow \mu\nu$, $Z\rightarrow ee$, and $Z\rightarrow \mu\mu$), with 
the corresponding decays to $\tau \rightarrow$~leptons considered as 
backgrounds to these processes (Section~\ref{sec:background}).  The branching 
ratio for each leptonic decay $W\rightarrow l\nu$ ($Z\rightarrow ll$) is 
$\approx$11\% (3.3\%), and the measured cross section times branching ratio 
is $(2749 \pm 174)$ pb [($254.9 \pm 16.2$) pb] \cite{wzprd}.  

\subsection{Conventions}

We use both Cartesian and cylindrical coordinate systems, in which 
$+z$ points in the direction of the proton beam (east) and the origin
is at the center of the detector.  In the right-handed Cartesian 
coordinate system, $+x$ points north (outward from the ring) and $+y$ 
points upwards; in the cylindrical system, $\phi$ is the azimuthal angle 
and $r$ is the radius from the center of the detector in the $x-y$ plane.  
The rapidity $y = -\frac{1}{2}\ln[(E - p_z c)/(E + p_z c)]$ is additive 
under Lorentz boosts along the $z$ axis.  For massless particles, this 
quantity is equal to the pseudorapidity $\eta = -\ln[\tan(\theta/2)]$, 
where $\theta$ is the polar angle with respect to the $z$ axis.  All 
angles are quoted in radians unless otherwise indicated.  
\par
\begin{figure}[!tp]
\begin{center}
\epsfysize = 4.cm
\epsffile{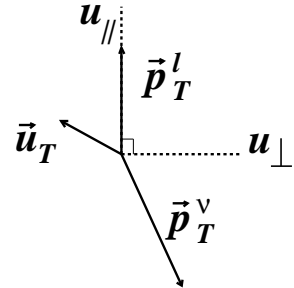}
\caption{A $W$ boson event, with the recoil hadron momentum ($\vec{u}_T$) 
separated into axes parallel ($u_{||}$) and perpendicular ($u_{\perp}$) to 
the charged lepton. }
\label{fig:recoilw}
\end{center}
\end{figure}

Because the interacting quarks' longitudinal momenta $p_z$ are not 
known for each event, we generally work with momenta transverse to the 
beam line.  The interacting protons and antiprotons have no net 
transverse momentum.  Electron energy (muon momentum) measured using the 
calorimeter (tracker) is denoted as $E$ ($\vec{p}$), and the corresponding 
transverse momenta $\vec{p}_T$ are derived using the measured track direction 
and neglecting particle masses.  The event calorimetric $\vec{p}_T$, 
excluding the lepton(s), is calculated assuming massless particles using 
calorimeter tower energies (Section~\ref{sec:calorimeter}) and the lepton 
production vertex, and provides a measurement of the recoil momentum vector 
$\vec{u}_T$.  The component of recoil projected along the lepton direction is 
denoted $u_{||}$ and the orthogonal component is $u_{\perp}$ 
(Fig.~\ref{fig:recoilw}).  The transverse momentum imbalance in a $W$ boson 
event is a measure of the neutrino transverse momentum $\vec{p}^{~\nu}_T$ and 
is given by \vecmet~$= -(\vec{p}^{~l}_T + \vec{u}_T)$, where $\vec{p}^{~l}_T$ 
is the measured charged lepton transverse momentum.  
\par
When electromagnetic charge is not indicated, both charges are considered.
We use units where $\hbar = c \equiv 1$ for the remainder of this paper.  

\subsection{Measurement Strategy}

The measurement of the final state from $W\rightarrow l\nu$ decays involves 
a measurement of $\vec{p}_T^{~l}$ and the total recoil $\vec{u}_T$.  The 
neutrino escapes detection and the unknown initial partonic $p_z$ precludes 
the use of $p_z$ conservation in the measurement.  The boson invariant mass
is thus not reconstructable; rather, the 2-dimensional ``transverse mass''
$m_T$ is used in the $m_W$ fit:
\begin{equation}
m_T = \sqrt{2 p_T^{~l} \met (1 - \cos\Delta\phi)},
\end{equation}

\noindent
where $\Delta\phi$ is the angle in the transverse plane between the leptons, 
whose masses are negligible.  The fit to the $m_T$ distribution provides the 
statistically most precise measurement of $m_W$. 
\par
The charged lepton, which can be measured precisely, carries most of the
observable mass information in the event.  We calibrate the muon momentum 
using high statistics samples of the meson decays $J/\psi\rightarrow \mu\mu$ 
and $\Upsilon\rightarrow\mu\mu$, which are fully reconstructable and have 
well known masses.  This results in a precise track momentum calibration, which 
we transfer to the calorimeter with a fit to the ratio of calorimeter energy to 
track momentum ($E/p$) of electrons from $W$ boson decays.   The accuracy of 
these calibrations is demonstrated by applying them to measurements of the $Z$ 
boson mass in the muon and electron decay channels.  We then incorporate the 
known $Z$ boson mass as an additional calibration constraint.
\par
The other directly measurable quantity needed for the calculation of $m_T$ 
is the recoil transverse momentum $\vec{u}_T$.  Since the $W$ and $Z$ bosons 
are produced at a similar $Q^2$, they have similar recoil distributions.  We 
use the leptons from the $Z$ boson decay to measure the $p_T$ of the $Z$ 
boson.  We then calibrate our model of $\vec{u}_T$ by measuring the balance
between the recoil and $Z$ boson $\vec{p}_T$.  The $Z$ boson statistics are 
sufficient to perform a recoil calibration to 1\% accuracy, which leads to 
a systematic uncertainty commensurate with other uncertainties on $m_W$.
\par
To accurately model the shape of the $m_T$ distribution, we use a fast 
Monte Carlo simulation of the $p\bar{p} \rightarrow W \rightarrow l\nu$
process including the recoil and the detector response.  The custom fast 
simulation allows flexibility in parametrizing the detector response and 
in separating the effects of the detector model components.  We use a 
binned likelihood to fit the measured $m_T$ distributions to templates 
(Section~\ref{sec:templates}) generated from the fast simulation, with $m_W$ 
as the free parameter.  All $m_W$ and lepton energy scale fits are performed 
with this procedure.
\par
Though less statistically precise, the $p^l_T$ and \met distributions provide 
additional information on the $W$ boson mass and are used as important
tests of consistency.  We separately fit these distributions for $m_W$ and 
combine all fits in our final result.
\par
During the measurement process, all $W$ boson mass fits were offset by a single 
unknown random number chosen from a flat distribution in the range [-100,100] 
MeV.  The fit result was thus blinded to the authors until the analysis was 
complete \cite{blinding}.  The final measured $m_W$ and its uncertainty have not 
changed since the random offset was removed from the fit results.
\par
We give a brief overview of the template likelihood fitting procedure in 
Section~\ref{sec:templates}.  Section~\ref{sec:detector} describes the detector 
and the fast detector simulation used in the analysis.  The $W$ boson measurement 
samples are defined in Section~\ref{sec:wsample}.  We describe the precision 
measurements of muons and electrons in Sections~\ref{sec:muons} and \ref{sec:electrons}, 
respectively.  These sections include event selection, calibration, and resolution 
studies from the dilepton and $W$ boson data samples.  Measurement of the recoil 
response and resolution is presented in Section~\ref{sec:recoil}.  The backgrounds 
to the $W$ boson sample are discussed in Section~\ref{sec:background}.  Theoretical 
aspects of $W$ and $Z$ boson production and decay, including constraints from the 
current data sample, are described in Section~\ref{sec:production}.  We present the 
$W$ boson mass fits and cross-checks in Section~\ref{sec:fits}.  Finally, in 
Section~\ref{sec:summary} we show the result of combining our measurement with 
previous measurements, and the corresponding implications on the predicted standard 
model Higgs boson mass.

\subsection{Template Likelihood Fits}
\label{sec:templates}

All the fits involving mass measurements and the energy scale 
(Sections~\ref{sec:muons}, \ref{sec:electrons}, and \ref{sec:fits}) are 
performed with a template binned likelihood fitting procedure.  A given 
distribution to be fit is generated as a discrete function of the fit 
parameter, using the fast simulation.  These simulated distributions are 
referred to as ``templates.''  For each value of the fit parameter, the 
simulated distribution is compared to the data distribution and the 
logarithm of a binned likelihood is calculated.  The binned likelihood is 
the Poisson probability for each bin to contain the $n_i$ observed data 
events given $m_i$ expected events, multiplied over the $N$ bins in the 
fit range:
\begin{equation}
{\cal{L}} = \prod_{i=1}^N \frac{e^{-m_i} n_i^{m_i}}{n_i!}.
\end{equation}

\noindent
We calculate the logarithm of the likelihood using the approximation 
$\ln n! \approx (n + 1/2) \ln (n + 1) - n$:
\begin{equation}
\ln {\cal {L}} \approx \sum_{i=1}^N [n_i \ln m_i - m_i - (n_i + 1/2) \ln (n_i + 1) + n_i].
\end{equation}

\noindent
The best-fit value of the parameter maximizes the likelihood (or equivalently 
minimizes $-\ln{\cal L}$), and the $\pm 1\sigma$ values are those that increase 
$-\ln{\cal L}$ by 1/2.  The approximation for $\ln n!$ only affects the shape 
of the likelihood about the minimum and not the position of the minimum.  The 
procedure is validated by fitting simulated data (``pseudoexperiments'') and no 
bias is found.  We symmetrize the uncertainty by taking half the difference between 
the $+1\sigma$ and $-1\sigma$ values.  For the $E/p$ fits in the $W$ boson sample, 
we reduce the effect of finite template statistics by fitting $-\ln{\cal L}$ to a 
parabola, and extracting the best-fit value and the uncertainty from this parabola.

\section{Detector and Model}
\label{sec:detector}

The CDF II detector \cite{wzprd, jpsi} is well suited for the $m_W$ 
measurement.  Its high-resolution tracker and calorimeter measure 
individual charged lepton momenta from $W$ and $Z$ boson decays with 
a resolution of $\approx 2$\%.  It has similar acceptance and resolution 
for central electrons and muons, giving the two channels similar weight 
in a combined mass measurement.

\subsection{Detector Components}
\label{sec:components}

\begin{figure*}[!tp]
\begin{center}
\epsfysize = 11.cm
\epsffile{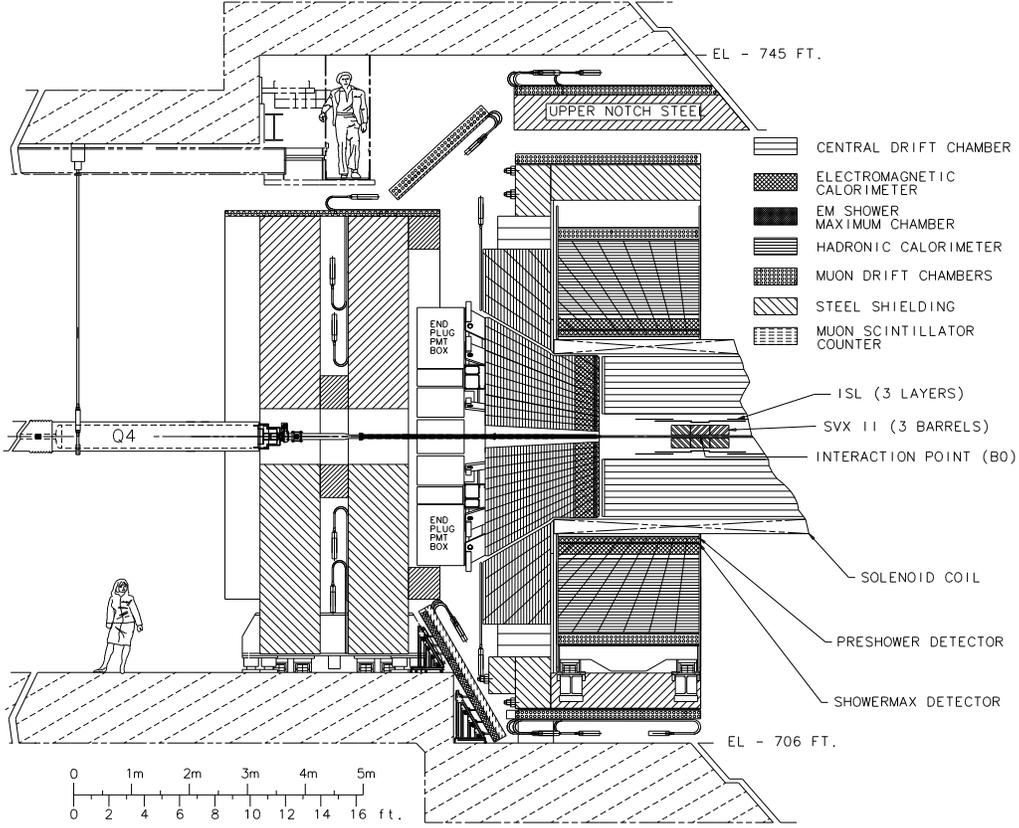}
\caption{A cut-away view of a section of the CDF detector.  The 
slice is along the $y$-axis at $x = 0$ cm. }
\label{fig:detector}
\end{center}
\end{figure*}

The CDF II detector (Fig. \ref{fig:detector}) is a multi-purpose detector 
consisting of:  an inner silicon tracker designed to measure the 
production vertex of charged particles with high precision; an outer 
tracking drift chamber to measure charged particle momenta; a solenoid 
to provide a uniform 1.4 T magnetic field inside the trackers; 
electromagnetic calorimeters to contain and measure electron and photon 
showers; hadronic calorimeters for hadron energy measurements; and a 
muon system to detect muons escaping the calorimeters.  The detector 
information is read out on-line and saved for later analysis when event 
topologies consistent with a particular physics process (or class of 
processes) are selected.  The read-out decision is made with a fast 
three-level trigger system that has high efficiency for selecting the 
$W$ and $Z$ bosons to be used in the offline analysis.

\subsubsection{Tracking System}

\begin{figure}[!tp]
\begin{center}
\epsfysize = 6.25cm
\epsffile{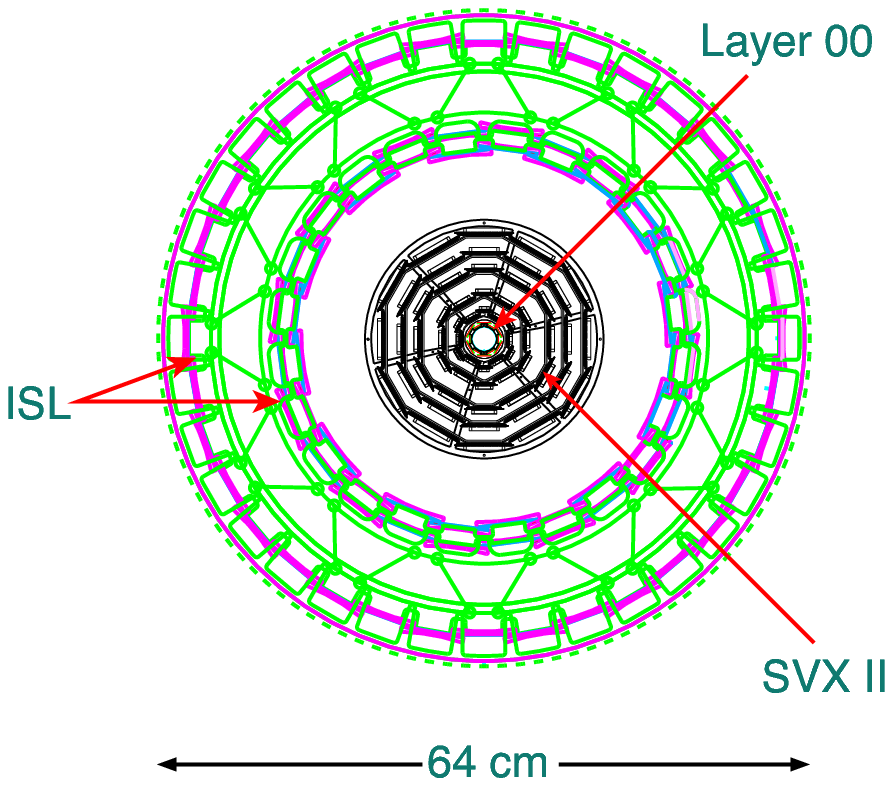}
\caption{End view of the silicon detector.  The innermost layer (Layer 00)
is attached to the beam pipe, and is surrounded by five concentric layers
of silicon wafers (SVX II).  The outermost layers are the intermediate 
silicon layers (ISL), which sit just inside the outer tracking chamber. }
\label{fig:silicon}
\end{center}
\end{figure}

The silicon tracker (Fig.~\ref{fig:silicon}) 
consists of three separate detectors: Layer 00, SVX II, and ISL.  
Layer 00 is a single layer of 300 $\mu$m thick sensors attached
to the beam pipe at a radius of 1.3 cm.  Five additional layers of
sensors at radii ranging from 2.5 cm to 10.6 cm comprise
SVX II.  Surrounding these sensors are port cards, which transport 
deposited charge information from the silicon wafers to the readout 
system.  The intermediate silicon layers (ISL) are located at radii 
of 20.2 cm and 29.1 cm.  The SVX II is segmented longitudinally into 
three barrels in the region \mbox{$|z| < 45$ cm}.  This covers the 
$p\bar{p}$ interaction region, which is well approximated by a Gaussian 
distribution with $\sigma_z \approx 30$ cm.  We do not  use the 
silicon measurements in this analysis, though we model the tracker's 
effects on leptons and photons (Section~\ref{sec:model}).

An open-cell drift chamber, the central outer tracker (COT) \cite{COT}, 
surrounds the silicon tracker and covers the region \mbox{$|z| < 155$~cm} 
($|\eta| \lesssim 1$) and \mbox{40~cm~$< r < 137$~cm}.  The COT consists of 
eight concentric `superlayers,' separated azimuthally into cells.  Each cell 
contains 12 sense wires to measure the ionization produced by a charged 
particle in the ambient argon-ethane gas mixture.  The superlayers alternate 
between a purely axial configuration, with sense wires parallel to the beam 
line, and a small-angle stereo configuration, with sense wires at a 2$^{\circ}$ 
angle relative to the $z$ axis.  
\par
\begin{figure*}[htbp]
\begin{center}
\epsfysize =8.cm
\epsffile{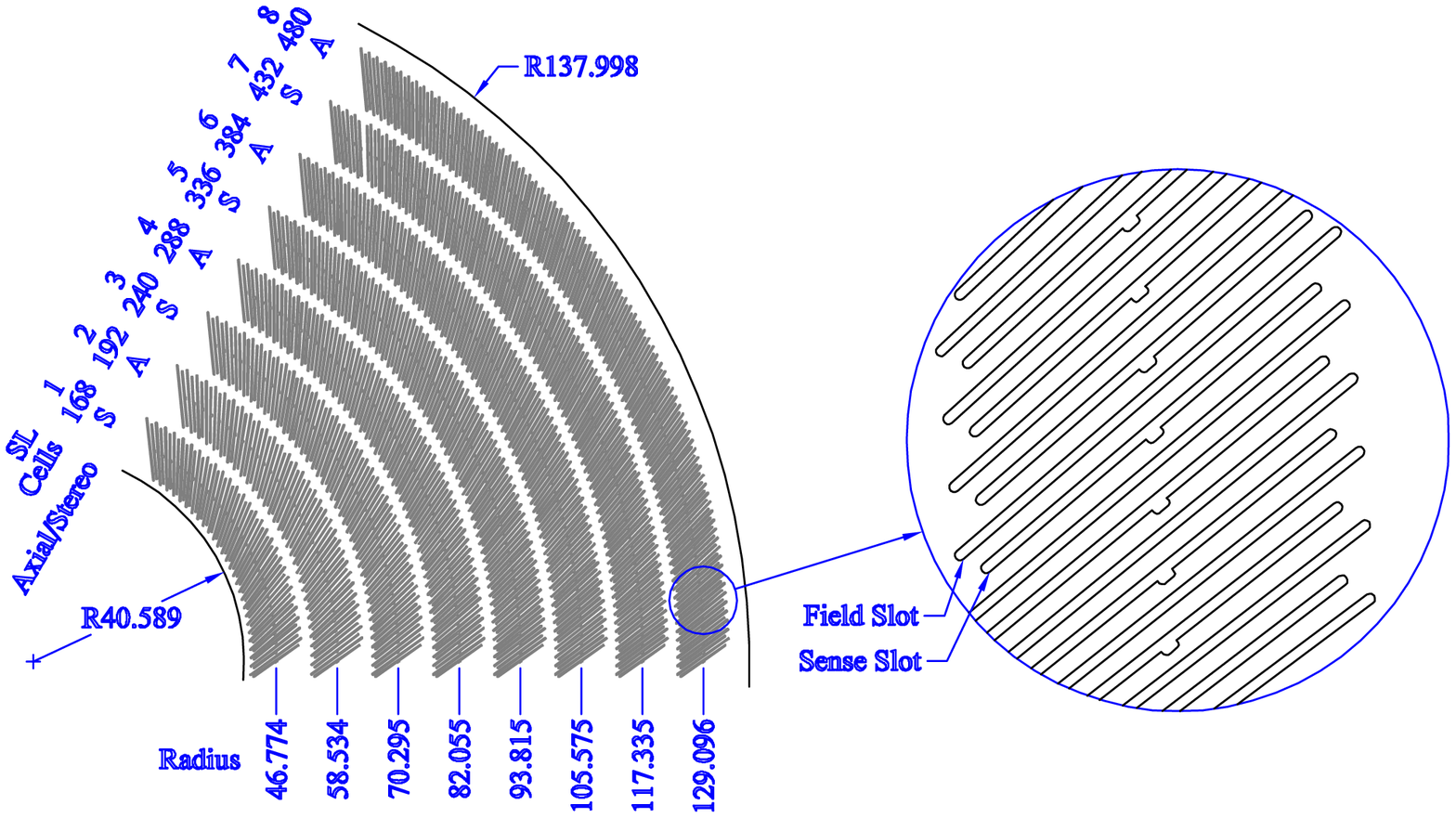}
\caption{End view of a section of a central outer tracker (COT) endplate.  
The COT consists of eight concentric `superlayers,' separated azimuthally into 
cells, each containing 12 sense wires and sandwiched by field sheets.  The 
endplates contain precision-machined slots where each cell's sense wires and 
field sheets are held under tension.  The radius at the center of each 
superlayer is shown in cm. }
\label{fig:COT}
\end{center}
\end{figure*}

The sense wires are strung from end to end in $z$ and held under tension at 
each aluminum endplate (Fig.~\ref{fig:COT}).  The wires are azimuthally 
sandwiched by field sheets, which provide a 1.9 kV/cm electric field.  All 
cells are rotated at a $35^{\circ}$ angle relative to a radial line, such that 
the ionized electrons travel approximately azimuthally to the wire under the 
combined influence of the local electric field and the global magnetic field 
from the solenoid.

Within a given cell the sense wires are slightly off-center relative to the 
field sheets.  In addition, the sense wires and field sheets sag under the 
influence of gravity, with the field sheets sagging more due to their larger 
masses.  These effects cause a small electrostatic deflection of the sense 
wires toward a particular field sheet.  To prevent the relative deflection of 
sense wires within a cell, a support rod connects the sense wires at the 
center of the detector.  The support rod results in a small ($\approx 2$~mm) 
region at $z=0$~cm where charged particles are not measured.
\par
Between the solenoid and the COT is a time-of-flight system (TOF) 
consisting of scintillator bars that precisely measure the time of incidence 
of charged particles.  From this measurement and the tracker information, a 
particle's velocity and mass can be inferred.  The TOF is not utilized in this 
analysis.

\subsubsection{Calorimeter System}
\label{sec:calorimeter}

The CDF calorimeter is segmented radially into electromagnetic
and hadronic sections.  The central calorimeter covers $|\eta| < 1.1$ 
and is split at the center into two separate barrels covering $+\eta$ 
and $-\eta$.  Each barrel consists of 24 azimuthal ``wedges'' of size 
$0.26$ radians ($15^{\circ}$) with ten projective towers of size 
\mbox{$\Delta \eta \approx 0.11$}.  To allow a pathway for the solenoid 
cryogenic tubes, a two-tower region is removed, corresponding to 
\mbox{$0.77 < \eta < 1.0$}, \mbox{$75^{\circ} < \phi < 90^{\circ}$}, and 
$z > 193$ cm.  The forward calorimeter covers \mbox{$1.1 < |\eta| < 3.6$}, 
filling the forward gaps with a plug shape (Fig.~\ref{fig:detector}).  
\par
The central electromagnetic calorimeter (CEM) \cite{CEM} has a thickness 
of 18 radiation lengths, consisting of 31 radial scintillator layers 
interleaved with 30 layers of lead-aluminum plates.  At a radius of 184 cm 
electromagnetic showers have traversed about six radiation lengths (including 
the solenoidal coil) and have their maximum energy deposition.  At this radius 
finely segmented strip and wire chambers (CES) measure the energy deposition 
with a position resolution of $\approx 2$ mm.  
\par
The local shower position in the azimuthal direction in the tower
is denoted as CES $x$, which ranges from -24.1 cm to 24.1 cm.  The 
wire chambers extend only to $|x| \leq 22.5$ cm, and for $|x| > 23.1$ cm
no energy measurements are made.  In this region wavelength shifters 
read out the light from the scintillator, and steel and foam separate the 
towers.  Light follows a waveguide to a phototube positioned at the 
back end of the hadronic calorimeter.  
\par
Parallel to the beam line, the position at shower maximum is denoted
CES $z$.  The strip chambers extend from 6-239 cm in $|z|$, and there
is no scintillator for $|z| < 4.2$ cm, where the two calorimeter barrels 
meet.
\par
The central hadronic calorimeter \cite{HAD} is separated into a central 
region (CHA, $|\eta| < 0.6$) with 32 longitudinal layers of scintillator 
sandwiched with steel and a forward ``wall'' calorimeter (WHA, 
$0.6 < |\eta| < 1.1$) with 15 such layers.  These calorimeters have thicknesses
of $\approx 4.5$ interaction lengths.
\par
The plug calorimeter \cite{plug} has a comparable design to the central
calorimeter with scintillator-lead electromagnetic calorimeters and 
scintillator-steel hadronic calorimeter compartments.  The $\phi$ 
segmentation is $0.13$ radians up to $|\eta| = 2.1$, and then broadens to 
$0.26$ radians.  The two furthest forward plug towers cover the $|\eta|$ 
regions $2.6-3.0$ and $3.0-3.6$, while the remaining towers have a size 
$\Delta \eta = 0.1$.

\subsubsection{Muon Detectors}

The muon systems relevant for the $W$ mass measurement cover the region 
$|\eta| \leq 1$.  The central muon detector (CMU) and the central muon 
upgrade (CMP) cover $|\eta| \leq 0.6$, while the central muon extension 
(CMX) covers $0.6 < |\eta| \leq 1$. 
\par
The CMU detector \cite{CMU} is located at the outer edge of the CHA, 
347 cm from the $z$ axis.  The CMU is segmented into $15^{\circ}$ azimuthal 
wedges containing four layers of proportional drift chambers that cover 
$12.6^{\circ}$.  The maximum drift time within a chamber is 800 ns, about 
twice as long as the 396 ns spacing between $p\bar{p}$ crossings.  CMU 
information must therefore be combined with reconstructed COT particle 
tracks to determine the appropriate $p\bar{p}$ crossing.  
\par
Because the total thickness of the central calorimeter is about five  
interaction lengths, approximately 0.5\% of high-momentum pions reach the 
CMU.  To reduce this background, the CMP detector is located behind an 
additional 60 cm of steel.  The CMP has a similar construction to the CMU, 
with the exception that wider drift chambers are used to cover the same 
solid angle, resulting in a maximum drift time of 1.8~$\mu$s rather than 
800 ns.  
\par
The CMX detector \cite{cdftdr} consists of eight drift chamber layers 
beyond both the calorimeter and the steel detector support structure 
($6-10$ interaction lengths).  The CMX $\phi$ regions used in this analysis 
are $-45^{\circ} < \phi < 75^{\circ}$ and $105^{\circ} < \phi < 225^{\circ}$.  
New detectors for Run II cover much of the remaining $\phi$ region, but 
were not fully commissioned for the data-taking period of this analysis.  
Scintillator detectors (CSX) at the inner and outer surfaces of the CMX 
provide timing information to the trigger to separate collision particles 
from other sources such as beam halo or cosmic rays.

\subsubsection{Trigger System}
\label{sec:trigger}

The trigger consists of three stages with progressively greater 
sophistication of event reconstruction.  The first stage is hardware-based, 
the second a mix of hardware and software, and the third a farm of 
processors performing full event reconstruction.
\par
The first trigger stage, level 1, includes tracker, calorimeter, and muon 
reconstruction.  The charged particle track reconstruction is performed 
with the extremely fast tracker (XFT) \cite{XFT} based on the four axial COT 
superlayers.  A track segment is reconstructed in a given superlayer if at 
least 11 of the 12 sense wires \cite{XFTcaveat} in a wide road have charge 
deposition above a given threshold (``hits'').   The list of segments from 
the full tracker is compared to predefined groups of segments expected
from charged particles above a given momentum threshold.  When matches are 
found, track candidates are created and passed to the track extrapolator 
(XTRP) \cite{XTRP}.  The XTRP determines the expected $\phi$ positions of 
the tracks in the calorimeter and muon detectors, for the purpose of forming 
electron and muon candidates.
\par
The calorimeter reconstruction at level 1 defines separate electromagnetic 
and hadronic ``trigger towers'' as tower pairs adjacent in $\eta$.  The tower 
$p_T$ is calculated assuming a collision vertex $z = 0$ and an electron 
candidate is formed if the ratio of hadronic to electromagnetic energy (Had/EM) 
in a trigger tower is less than 1/8.  The high-momentum electron trigger used 
in this analysis requires a level 1 trigger tower with electromagnetic $p_T > 8$ 
GeV matched to a track with $p_T > 8$ GeV, and drops the Had/EM requirement 
for electromagnetic $p_T > 14$ GeV.
\par
Level 1 muon reconstruction includes a $p_T$ estimate within the CMU and CMX
chambers from the relative timing of the hits in different layers.  The CMU
track segments are combined with reconstructed CMP track segments to create
``CMUP'' muon candidates.  For the majority of the data CMX candidates also 
require local CSX hits consistent with particles originating from the collision.  
For our $W$ and $Z$ boson samples we use a muon trigger that requires CMU or 
CMX $p_T > 6$ GeV matched to an XFT track with $p_T > 4$ GeV (CMUP) or $p_T > 8$ 
GeV (CMX). 
\par
The level 2 calorimeter reconstruction uses a more sophisticated clustering
algorithm for electromagnetic objects.  This improves energy measurement 
resolution and allows a higher threshold ($p_T > 16$ GeV) to be applied.  To 
reduce rates, the XFT track requirement for CMUP candidates was raised to 
$p_T > 8$ GeV for most of the data-taking period.
\par
At level 3, approximately 300 dual processor computers allow full track pattern 
recognition, muon reconstruction, and calorimeter clustering.  Variables used 
to select electrons at level 3 are the lateral shower profile, $L_{shr}$ 
(Section~\ref{sec:wesample}), and the distance between CES $z$ and the 
$z$-position of the track extrapolated to the CES ($\Delta z$).  The $L_{shr}$ 
variable quantifies the difference between the measured energies of towers 
neighboring the electron in $\eta$ and the expected energies determined from 
electron test beam data.  The trigger requirements of $L_{shr} < 0.4$ and 
$|\Delta z| < 8$ cm are $\approx$100\% efficient for electrons from $W$ and $Z$ 
boson decays.  The high-momentum electron trigger also requires electromagnetic 
$p_T > 18$ GeV and track $p_T > 9$ GeV.  For efficiency studies we use a separate 
trigger that requires electromagnetic $p_T > 25$ GeV and \met$\! \! \! ^{L3} > 25$ 
GeV, but has no quality requirements at level 3 and no trigger track requirements.  
At level 3, \vecmet$\! \! \! ^{L3}$ is defined as the negative of the vector sum of 
the transverse momenta in all calorimeter towers.  The high-momentum muon trigger 
requires a COT track with $p_T > 18$ GeV matched to a CMUP or CMX track segment.

\subsubsection{Luminosity Detector}
\label{sec:clc}
The small-angle Cherenkov luminosity counters (CLC) \cite{CLC}
are used to measure the instantaneous and integrated luminosity of our 
data samples.  The CLC consists of two modules installed around the 
beampipe at each end of the detector, providing coverage in the regions 
$3.6 < |\eta| < 4.6$.  Each module contains 48 conical gas Cherenkov 
counters pointing to the collision region.  Signals in both CLC modules 
coinciding in time with the bunch crossing are used to measure the 
instantaneous luminosity and to trigger collision events.  Events 
collected with this trigger, known as ``minimum bias'' events, are used 
to study the detector response to generic inelastic $p\bar{p}$ collisions 
(Section \ref{sec:recoil}).

\subsection{Detector Model}
\label{sec:model}

We use a parametrized model of the detector response to 
electrons, muons, and the hadronic recoil.  The model is 
incorporated into a custom fast simulation that includes 
lepton and recoil reconstruction, event selection, and fit 
template generation.  The simulation provides both flexibility 
in determining the effects of various inputs, and computing 
speed to allow frequent high-statistics studies.  A sample 
of {\cal{O}}($10^7$) events can be generated using a 
single-processor machine in one day.  This is several orders 
of magnitude more than the {\cal{O}}($10^3$) events that can 
be produced with the standard {\sc geant}-based CDF simulation
\cite{GEANT} \cite{CDFSIM}.
\par
We describe in this section the simulation of electrons and 
muons.  Fits to the data that determine the values of some of 
the model parameters are described in 
Secs.~\ref{sec:muons}~and~\ref{sec:electrons}.  The detector 
model of hadronic recoil response and resolution is discussed 
in Sec.~\ref{sec:recoil}.
\par
The model components common to muons and electrons are: ionization 
energy loss and multiple scattering in the beam pipe and tracker 
volume; parametrized track hit resolutions and efficiencies; and 
track reconstruction.  We describe these components in the muon 
simulation overview, and then discuss the electron- and 
photon-specific simulation.

\subsubsection{Muon Simulation}
\label{sec:muonsim}

Muon and electron tracks are reconstructed using only COT hit 
and beam position information (Section \ref{sec:wsample}).  Thus, 
the simulation of the silicon detector consists entirely of energy 
loss and multiple scattering.  In the COT, hit resolutions and 
efficiencies are additionally simulated, and track reconstruction 
is performed.  The total measured muon EM calorimeter energy is 
simulated by combining the minimum-ionizing energy deposition with 
energy from final-state photon radiation (Section~\ref{sec:fsr}) and 
the recoil and underlying event \cite{ue}.  Finally, the detector 
fiduciality of muons is calculated using a map of the muon detector 
geometry as a function of $\eta$ and $\phi$.  The map is extracted 
from a full {\sc geant}-based simulation of the CDF II detector 
\cite{GEANT, CDFSIM}.

\subsubsection*{Ionization Energy Loss}
\label{sec:ionization}

The differential ionization energy loss of muons and electrons in the 
tracking system is simulated according to the Bethe-Bloch equation 
\cite{pdg}:
\begin{equation}
-\frac{dE}{dx} = \frac{K Z}{A\beta^2}
  \left[\frac{1}{2}\ln\frac{2m_e \beta^2T_{max}}{(1 - \beta^2) I^2} - 
  \beta^2 - \frac{\delta}{2}\right],
\end{equation}

\noindent
where $K = 4\pi N_A r_e^2 m_e$, $N_A$ is Avogadro's number, $r_e$ is 
the classical electron radius, $Z (A)$ is the atomic (mass) number, 
$\beta$ is the particle velocity, $I$ is the mean excitation energy, 
$T_{max}$ is the maximum kinematic energy that can be transferred to 
a free electron in a single collision, and $\delta$ is the 
material-dependent density effect as a function of $\beta$ \cite{pdg}.  
When calculating the effect of $\delta$, we take the material to be 
silicon throughout.
\par
To calculate muon energy loss in the material upstream of the COT 
($r < 40$ cm), we use a three-dimensional lookup table of the material 
properties of the beam pipe, the silicon detector, and the wall of 
the alumnium can at the inner radius of the COT.  The lookup table 
determines the appropriate $Z/A$ and $I$ values, along with the 
radiation length $X_0$ (Appendix \ref{app:egammaint}), for each of 
32 radial layers.  Except for the inner and outer layers, the map 
is finely segmented longitudinally and in azimuth to capture the 
material variation in the silicon detector \cite{silimap}.  Inside 
the COT fiducial volume we calculate the energy loss between each 
of the 96 radial sense wires.
\par
The energy loss model is tuned using the data.  We apply a global 
correction factor of 0.94 to the calculated energy loss in the 
material upstream of the COT in order to obtain a 
$J/\psi \rightarrow \mu\mu$ mass measurement that is independent of 
the mean inverse momentum of the decay muons (Section~\ref{sec:jpsipscale}).

\subsubsection*{Multiple Coulomb Scattering}

Multiple Coulomb scattering in the beampipe, silicon detector, and 
COT affects the resolution of the reconstructed track parameters 
for low-momentum tracks.  We model the scattering using a Gaussian 
distribution for 98\% of the scatters \cite{ms} with an angular 
resolution $\sigma_{\vartheta}$ defined by
\begin{equation}
\sigma_{\vartheta} = \frac{13.6~{\rm MeV}}{\beta p}\sqrt{x/X_0}, 
\end{equation}

\noindent
where $x$ is the thickness of the layer and $X_0$ is the layer's 
radiation length (Section \ref{sec:elesim}).  Simulation of multiple 
scattering is implemented for each radial layer of the three-dimensional 
lookup table and between each COT layer.
\par
Based on the results of low-energy muon scattering data \cite{mstail}, 
we model the non-Gaussian wide-angle scatters by increasing 
$\sigma_{\theta}$ by a factor of 3.8 for 2\% of the scatters.

\subsubsection*{COT Simulation and Reconstruction}

The charged track measurement is modeled with a full hit-level 
simulation of the charge deposition in the COT and a helical track 
fit.  The parameter resolution of reconstructed tracks is affected 
by the individual hit resolution, and by the distribution of the 
number of hits ($N_{hit}$) used in the fit \cite{trackresolution}.  
\par
We tune the COT hit resolution using the width of the 
$\Upsilon\rightarrow \mu\mu$ mass distribution reconstructed with 
non-beam-constrained tracks.  The tuned value of $[150 \pm 3 ({\rm stat})]$ 
$\mu$m is consistent with the 149 $\mu$m RMS of the observed hit residual 
distribution for the muon tracks in $Z \rightarrow \mu \mu$ data.  We 
use a 150 $\mu$m hit resolution for the simulation of the $\Upsilon$, 
$W$, and $Z$ bosons.
\par
We use a dual-resolution model to describe the narrower mass peak in the
high-statistics $J/\psi \rightarrow \mu\mu$ sample, where the muons 
generally have lower momenta than the other samples.  The $J/\psi$ mass 
peak width is particularly sensitive to multiple scattering and relative 
energy loss, and our hit-resolution model compensates for any mismodeling 
that affects the peak width.  We find that a single-hit resolution of 155 
$\mu$m applied to 70\% of the tracks and 175 $\mu$m applied to the remaining 
30\% adequately describes the width and lineshape of the 
$J/\psi \rightarrow \mu\mu$ mass peak.
\par
To describe the $N_{hit}$ distribution, we use a dual-hit-efficiency model, 
the larger one applied to the majority of the tracks.  The lower efficiency 
accounts for events with high COT occupancy, where fewer hits are attached 
to reconstructed tracks.  The two parameters are tuned to match the mean 
and RMS of the data $N_{hit}$ distributions.  We independently tune these 
parameters for the $J/\psi$ sample, the $\Upsilon$ sample, and the $W$ and 
$Z$ boson samples.
\par
COT hit positions from a charged track are used to reconstruct
a helix with a $\chi^2$-minimization procedure.  The axial helix 
parameters \cite{ionim} are the impact parameter with respect 
to the nominal beam position, $d_0$, the azimuthal angle at the
closest approach to the beam, $\phi_0$, and the curvature of 
the track, $c$, defined to be ($2R$)$^{-1}$, where $R$ is the radius 
of curvature.  The stereo helix parameters are the longitudinal 
position at the closest approach to the beam, $z_0$, and the cotangent 
of the polar angle, $\cot \theta$.  
\par
When optimizing resolution of lepton tracks from prompt resonance decays, 
we constrain the helix to originate from the location of the beam.  The 
transverse size of the beam is $\approx 30~\mu$m at $z = 0$ cm and increases 
to $50 - 60~\mu$m at $|z| = 40$ cm \cite{ttxsec}.  For simplicity we assume 
an average beam size of $[39 \pm 3 ({\rm stat})]~\mu$m, which is determined 
from a fit to the width of the $Z\rightarrow \mu\mu$ mass peak.  The beam 
constraint improves the intrinsic fractional momentum resolution by about a 
factor of three, to $\delta p_T/p_T \approx 0.0005~p_T$/GeV.  
\par
We perform a track fit on our simulated hits in the same manner as the 
data.  The hits are first fit to a helix without a beam constraint; hits 
with large residuals ($> 600~\mu$m) are dropped from the track (in order to 
remove spurious hits added in data pattern recognition); and the track is 
fit again with an optional beam constraint.  This option is applied to 
prompt lepton tracks from $W$ and $Z$ boson decays, but not to tracks from 
$J/\psi$ decays, approximately 20\% of which are not prompt.  The prompt muons 
from $\Upsilon$ decays are fit twice, both with and without the beam constraint, 
as a consistency check.

\subsubsection*{Calorimeter Response}

Muons deposit ionization energy in the calorimeter.  We 
simulate a muon's EM energy deposition using a distribution 
taken from cosmic ray muons passing through the center of 
the detector, in events with no other track activity.  An 
additional contribution comes from energy flow into the 
calorimeter from the underlying event \cite{ue}.  We model 
this energy using a distribution taken from $W \rightarrow \mu\nu$ 
data events, using towers separated in azimuth from the muon.  
\par
Muons with a CES $z$ position within 1.58 cm of a tower 
boundary typically deposit energy in two calorimeter towers.
We use this criterion in the simulation to apply the underlying 
event and final-state photon radiation (Section \ref{sec:fsr}) 
contributions for one or two towers.  The simulated underlying 
event energy includes its dependence on $u_{||}$ and $u_{\perp}$
(Fig.~\ref{fig:recoilw}), and on the tower $\eta$ position of 
the muon when it crosses the CES (Section~\ref{sec:leptonremoval}).

\subsubsection*{Detector Fiduciality}

The CMUP and CMX muon systems do not have complete azimuthal 
or polar angle coverage.  We create an $\eta-\phi$ map of each 
muon detector's coverage using muons simulated \cite{CDFSIM} 
with a detector geometry based on {\sc geant} \cite{GEANT}. 
We use the map in the fast simulation to determine the fiduciality 
of a muon at a given $\eta-\phi$ position.
\par
We incorporate the relative efficiency of the CMUP to CMX triggers 
in the fast simulation by matching the ratio of CMUP to CMX events 
in the $W\rightarrow \mu\nu$ data (Section \ref{sec:wmusample}).

\subsubsection{Electron and Photon Simulation}
\label{sec:elesim}

The dominant calibration of the calorimeter energy measurement $E$ 
of electrons uses their track momenta $p$ and a fit to the peak of 
the $E/p$ distribution.  An additional calibration results from a 
mass fit to the $Z$ boson resonance and reduces the calibration 
uncertainty by 20\% relative to the $E/p$ calibration alone.  
\par
The $E/p$ method relies on an accurate modeling of radiative effects 
that reduce the track momentum measured in the COT.  A given electron 
loses $\approx 20$\% of its energy through bremsstrahlung radiation 
in the silicon detector, and this process has the most significant 
impact on the $E/p$ calibration.  The total amount of silicon detector
material is tuned with data using highly radiative electrons 
(Section~\ref{sec:eop}).  We additionally model processes that affect 
the shape of the $E/p$ distribution:  photon conversion in the tracker; 
energy loss in the solenoid and the time-of-flight system; 
electromagnetic calorimeter response and resolution; and energy loss 
into the hadronic calorimeter.  The models of ionization energy loss 
and multiple scattering in the tracker, as well as the COT track 
simulation and reconstruction, are the same as for muons 
(Section~\ref{sec:muonsim}).

\subsubsection*{Bremsstrahlung}

\begin{figure}[!tp]
\begin{center}
\epsfysize = 5.5cm
\epsffile{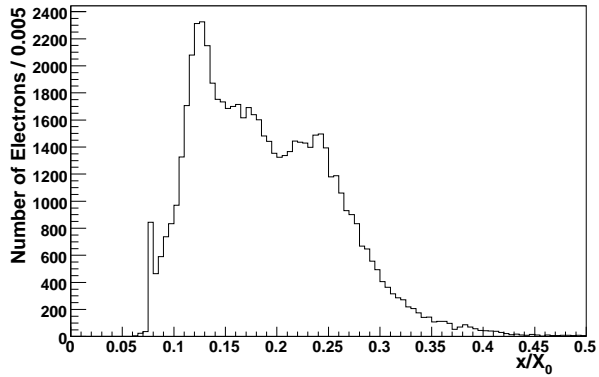}
\caption{The distribution of material upstream of the COT traversed by 
reconstructed electron trajectories in $W\rightarrow e\nu$ data events, in 
units of radiation lengths.  The peaks at $\approx 0.08$, $\approx 0.13$, and 
$\approx 0.24$ correspond respectively to trajectories outside the silicon 
detector ($|z| > 45$ cm), within the silicon detector, and crossing silicon 
barrels ($|z| \approx 15, 45$ cm).  The mean of the distribution is 19\%. }
\label{fig:matx0}
\end{center}
\end{figure}

The differential cross section for an electron of energy $E_e$ to radiate 
a photon of energy $E_{\gamma}$ is given by the screened Bethe-Heitler 
equation \cite{tsai} over most of the $y \equiv E_{\gamma}/E_{e}$ spectrum.  
In terms of the material's radiation length $X_0$, the differential cross 
section for bremsstrahlung radiation is:
\begin{equation}
\label{eq:pdgbrem}
\frac{d\sigma }{dy} = \frac{A}{N_A X_0\rho} \left[\left(\frac{4}{3} + C\right)
\left(\frac{1}{y}-1\right) + y\right], 
\end{equation}

\noindent
where $C$ is a small material-dependent correction (Appendix 
\ref{app:egammaint}).  Figure \ref{fig:matx0} shows the integrated thickness 
of material upstream of the COT, in terms of radiation lengths, traversed by 
the reconstructed electron tracks in $W\rightarrow e\nu$ data.  The number of 
photons emitted per layer is given by:
\begin{equation}
N_{\gamma} = \frac{x}{X_0}\left[\left(\frac{4}{3} + C\right)(y_0 - \ln y_0 - 1) + 
            \frac{1}{2}\left(1-y_0\right)^2\right],
\end{equation}

\noindent
where $x$ is the thickness of the layer and $y_0$ is a lower threshold 
introduced to avoid infrared divergences.  We use $y_0 = 10^{-4}$ \cite{ymin} 
and determine $C = 0.0253$ using the silicon atomic number $Z=14$.

For each layer of the silicon or COT material, we use a Poisson distribution 
with mean $N_{\gamma}$ to determine the number of photons radiated in that 
layer.  For each radiated photon, we calculate $y$ from the spectrum in 
Eq.~(\ref{eq:pdgbrem}).  To correct for inaccuracies of the screened 
Bethe-Heitler equation at the ends of the $y$ spectrum, we apply a suppression 
factor if $y \leq 0.005$ or $y \geq 0.8$.   
\par
For radiation of high-momentum photons ($y \gtrsim 0.8$), the approximation of 
complete screening of the nuclear electromagnetic field by the atomic electrons 
breaks down.  In this region, the full Bethe-Heitler equation for incomplete 
screening \cite{tsai} must be used.  We implement this correction by removing 
generated photons in the high-$y$ region such that we match the reduced cross 
section from incomplete screening. 
\par
Two effects reduce the cross section for low-momentum photon radiation 
\cite{expsuppression}: multiple scattering and Compton scattering.  Multiple 
Coulomb scattering suppresses long-distance interactions, and the resulting $LPM$ 
suppression \cite{migdal} in low-momentum radiation can be expressed in terms of 
the Bethe-Heitler cross section \cite{klein}:
\begin{equation}
\label{eq:migdal}
S_{LPM} \equiv \frac{d\sigma_{LPM}/dy}{d\sigma_{BH}/dy} = 
\sqrt{\frac{E_{LPM}}{E_e}\frac{y}{(1-y)}},
\end{equation}

\noindent
where $E_{LPM}$ depends on the material.  We use $E_{LPM} = 72$ TeV, appropriate 
for silicon, and apply the suppression when $S_{LPM} < 1$.
\par
Radiated photons scatter off the atomic electrons, and destructive interference 
of low-momentum photons suppresses this radiation \cite{dielectric}.  The 
suppression factor is:
\begin{equation}
S_{Compton} = \frac{y^2}{y^2 + E_p^2/E_e^2},
\end{equation}

\noindent
where $E_p = \gamma \omega_p$ is 2.4 MeV for a 40 GeV electron in silicon, using
the silicon plasma frequency $\omega_p$, and $\gamma$ is the Lorentz factor.
\par
In any given simulated event, the product of $S_{LPM}$ and $S_{Compton}$ provides 
the probability that a photon generated from the screened Bethe-Heitler equation 
with $y \leq 0.005$ survives the low-momentum suppression.  For a 40 GeV electron
radiating a 20 MeV (8 MeV) photon, the suppression factors are 
$S_{LPM} = 0.95~(0.60)$ and $S_{Compton} = 0.99~(0.92)$.  Our simulated $y$ 
spectrum from $W$ boson decay electrons reproduces the spectrum obtained by a 
{\sc geant} \cite{GEANT} simulation.

\subsubsection*{Photon Conversion}

Photons can convert to an electron-positron pair by interacting with the 
tracker material.  The differential cross section for a photon of energy 
$E_{\gamma} \gtrsim 1$ GeV to convert into an electron with energy $E_e$ 
is given by the screened Bethe-Heitler equation \cite{tsai}:
\begin{equation}
\frac{d\sigma}{dy} = \frac{A}{N_A X_0 \rho}\left[1 - (4/3 + C)y(1-y)\right] ,
\label{eq:pp}
\end{equation}

\noindent
where $y = E_e/E_{\gamma}$.  Integrating over $y$ and multiplying by 
$\rho x N_A/A$ gives the total cross section, from which we obtain the 
following conversion probability at high photon energy:
\begin{equation}
\label{eq:convprob}
P_{\gamma \rightarrow e^+e^-}(E_{\gamma} \rightarrow \infty) = 1 - e^{-(7/9 - C/6) x/X_0}.
\end{equation}

\begin{figure}[!tp]
\begin{center}
\epsfysize = 6.cm
\epsffile{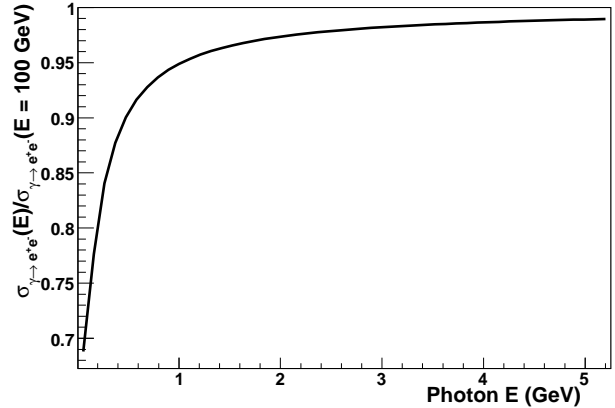}
\caption{The ratio of the photon conversion cross section at a given photon energy
to the cross section at $E_{\gamma} = 100$ GeV \cite{hubble}.  We use this function 
to scale down the cross section obtained from the Bethe-Heitler equation 
[Eq.~(\ref{eq:convprob})] \cite{tsai}. }
\label{fig:convfit}
\end{center}
\end{figure}

\noindent
We parametrize the cross section as a function of photon energy using the 
tables for photon cross sections in silicon given in \cite{hubble}.  We apply 
the ratio shown in Fig. \ref{fig:convfit} to the high-energy cross section 
when calculating the conversion probability.

For each radiated photon upstream of the COT, we integrate the material 
between the radiation point and the COT inner can.  If the photon converts, 
we take the conversion point to be halfway between the radiation point and
the inner can.  If the photon does not convert before the COT, we integrate 
the material in the COT and take a converting photon to convert halfway 
through the COT.  
\par
We use the conversion electron momentum spectrum from Eq.~(\ref{eq:pp}), 
ignoring the small effect of the $C$ term on the shape.  If a radiated photon 
has high momentum, a conversion electron's measured momentum can be larger 
than that of the electron from the $W$ boson decay.  To mimic the offline 
reconstruction, we assign the track from the highest momentum electron to 
the electron cluster.

\subsubsection*{Compton Scattering}

The cross section for a low-momentum photon to scatter off an electron 
is similar to that of conversion into an $e^+ e^-$ pair.  The differential 
cross section with respect to the photon fractional energy loss $y$ can be 
approximated as (Appendix~\ref{app:egammaint}):
\begin{equation}
\frac{d\sigma}{dy} \propto 1/y + y.\\
\end{equation}

\noindent
Using a lower bound of $y=0.001$, this spectrum approximates the Compton 
energy loss distribution for photons radiated from electrons from $W$ boson 
decays.  
\par
\begin{figure}[!tp]
\begin{center}
\epsfysize = 6.cm
\epsffile{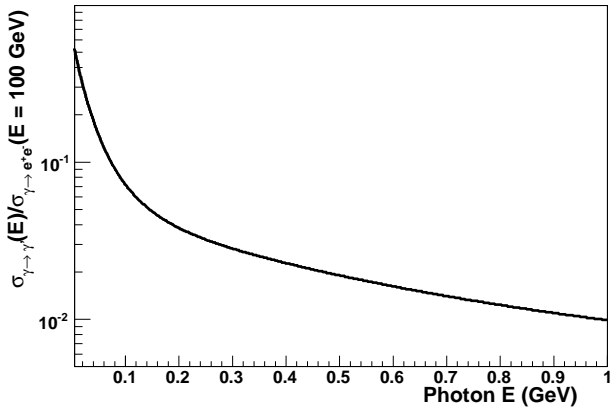}
\caption{The ratio of the Compton scattering cross section at a given 
photon energy to the pair-production cross section at $E_{\gamma} = 100$ 
GeV \cite{hubble}.  This ratio is applicable for photons traversing silicon. }
\label{fig:compton}
\end{center}
\end{figure}

We calculate the total cross section in terms of the pair production cross 
section using the tables for photon interactions in silicon in \cite{hubble}.  
The ratio of cross sections as a function of energy is parametrized as 
(Fig.~\ref{fig:compton}):
\begin{equation}
R_{Com} \equiv \frac{\sigma_{\gamma\rightarrow\gamma'}}{\sigma_{\gamma \rightarrow e^+e^-}
   (E_{\gamma}\rightarrow \infty)} = e^{F(E_{\gamma})},
\end{equation}

\noindent
where $F(E_{\gamma}) = 2.35e^{-1.16E_{\gamma}} + 2.42e^{-15.8E_{\gamma}} - 
5.21 - 0.151E_{\gamma}$, with $E_{\gamma}$ in GeV, and $\gamma$ and $\gamma'$ 
are the initial- and final-state photons, respectively.  We thus use the following 
Compton scattering probability per layer:
\begin{equation}
P_{\gamma \rightarrow \gamma'} = R_{Com}(7/9 - C/6) x/X_0.
\end{equation}

\subsubsection*{Energy Loss in Solenoid}

After exiting the tracker electrons and photons travel
through the time-of-flight (TOF) system and the solenoid.  
These systems have thicknesses of $\approx 10\%$ and 
$\approx 85\%$ of a radiation length, respectively.  With 
this much material it becomes prohibitive to model 
individual radiative processes, and we instead use a 
parametrized energy-loss model determined from a {\sc geant} 
simulation \cite{GEANT}.  The energy loss is defined as the 
difference in energy of a single particle entering the TOF 
and the total energy of particles exiting the solenoid.  
\par
\begin{figure}[!tp]
\begin{center}
\epsfysize = 6.0cm
\epsffile{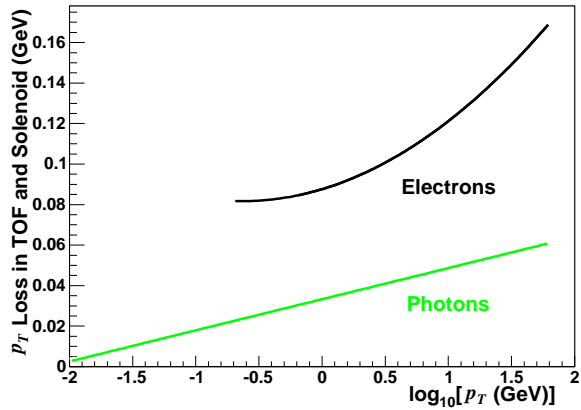}
\caption{The mean $p_T$ loss as a function of $\log_{10}(p_T/{\rm GeV})$ 
for electrons with $p_T > 400$ MeV and photons traversing the 
time-of-flight system and solenoid.}
\label{fig:solenoid}
\end{center}
\end{figure}

Figure \ref{fig:solenoid} shows the mean energy loss as a 
function of $\log_{10}(p_T/{\rm GeV})$ of the incoming 
particle for both photons and electrons.  Electrons lose more 
energy than photons due to their ionization of the material.  
Since electrons with $p_T \lesssim 400$ MeV curve back to 
the center of the detector before exiting the solenoid, we
do not parametrize energy loss in this energy region.  
\par
The energy loss distribution at a given particle $p_T$ is 
reasonably described by an exponential.  We use this distribution, 
with a mean determined by Fig.~\ref{fig:solenoid}, to model the 
energy loss of a given particle passing through the TOF and 
solenoid.

\subsubsection*{Calorimeter Response and Fiduciality}
\label{sec:ecal} 

The calorimeter simulation models the response of 
the electromagnetic calorimeter as a function of
each particle's energy and position, and the fraction 
of shower energy leaking into the hadronic calorimeter. 
\par
\begin{figure}[!t]
\begin{center}
\epsfysize = 6.0cm
\includegraphics*[width=8.5cm]{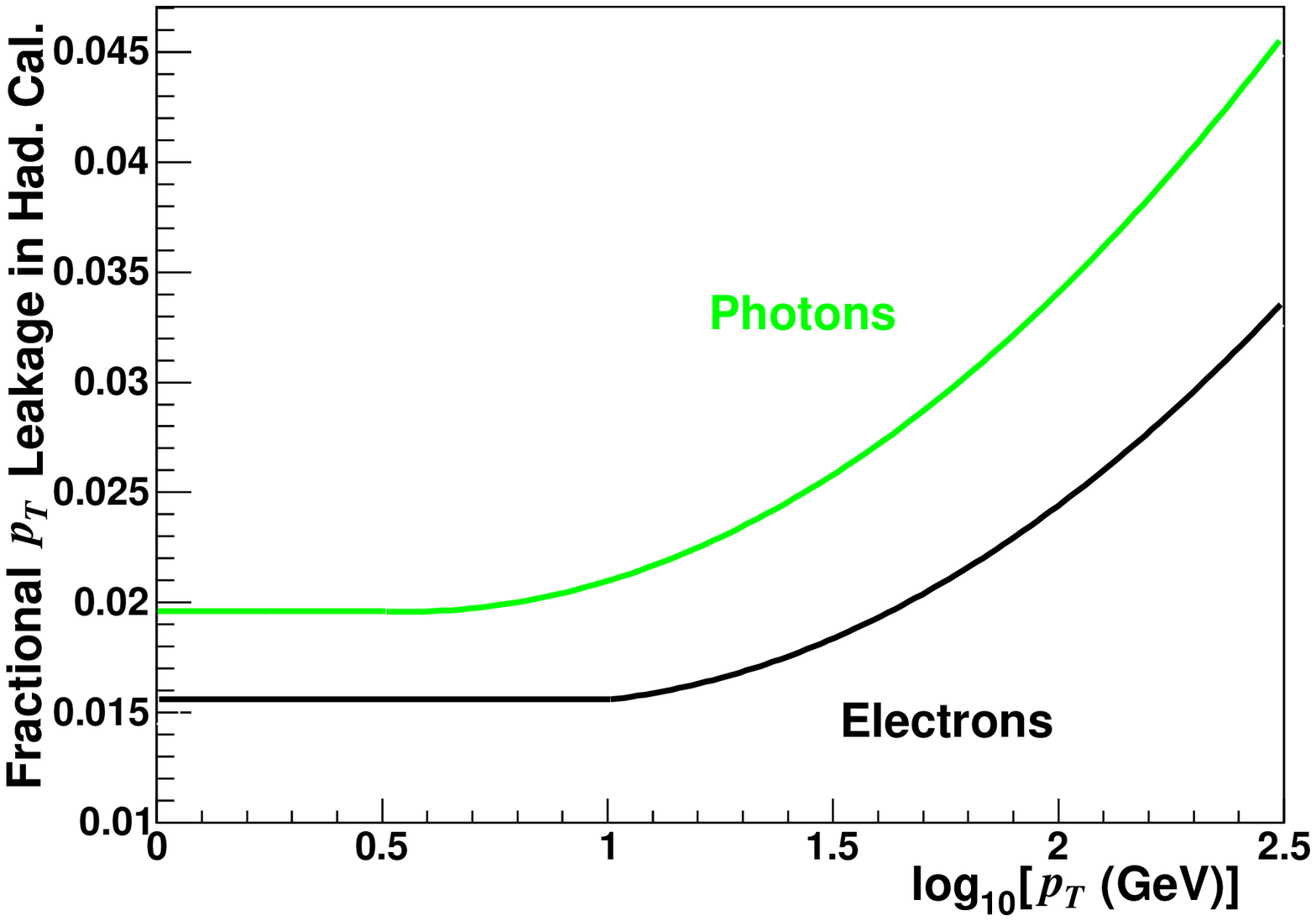}
\includegraphics*[width=8.5cm]{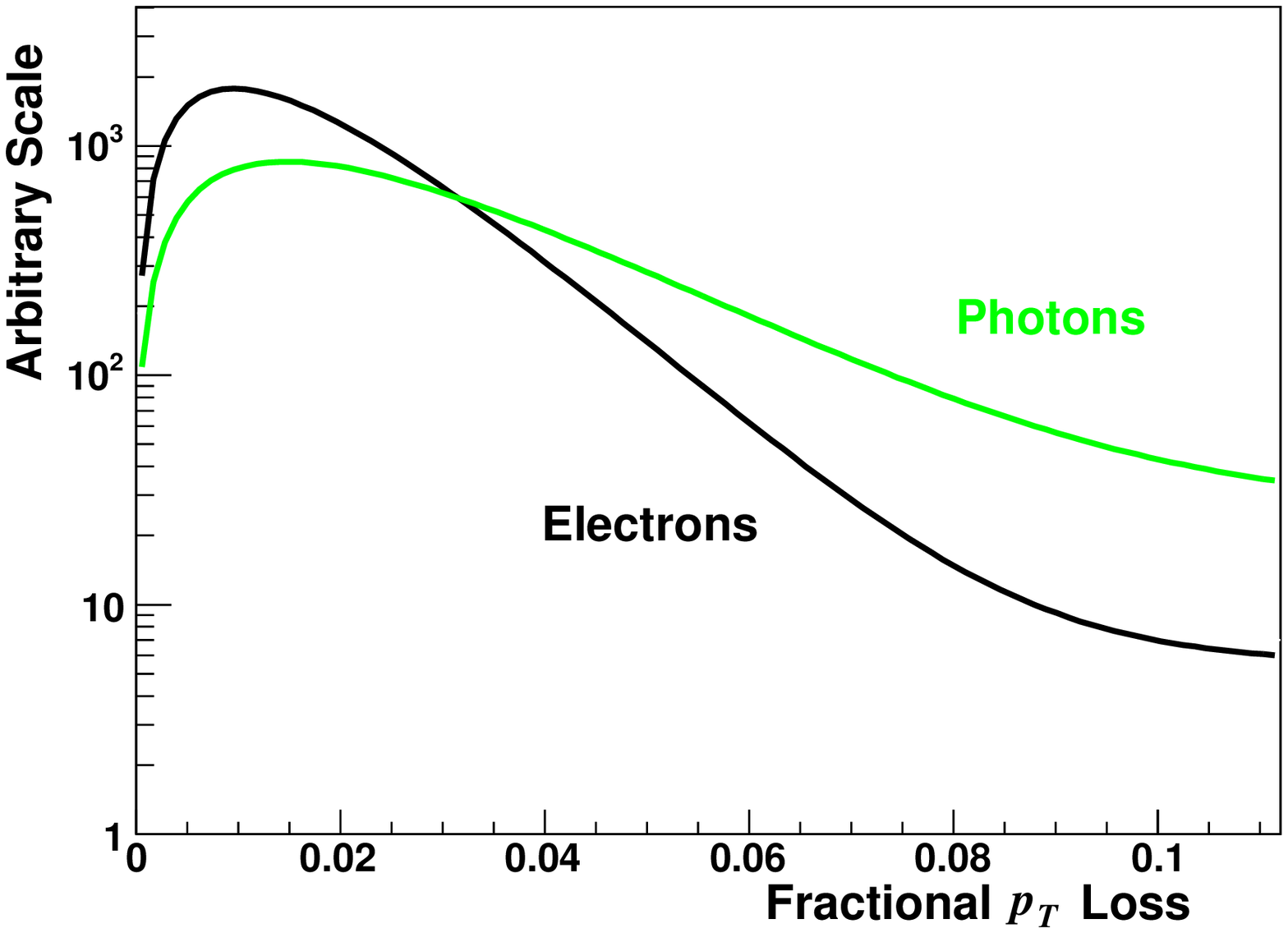}
\caption{The photon and electron $p_T$ leakage into the hadronic 
calorimeter.  Top: The mean $p_T$ leakage as a function of 
$\log_{10}(p_T/{\rm GeV})$.  Bottom:  The distributions of $p_T$ leakage
for high-$p_T$ ($> 10$ GeV) photons and electrons. }
\label{fig:leakage}
\end{center}
\end{figure}

The electromagnetic calorimeter response, or the 
average measured energy divided by the true particle 
energy entering the calorimeter, can depend on each 
particle's energy.  Possible sources of this dependence 
are variations in light yield as a function of 
calorimeter depth, attenuation in the light guide from 
the scintillator to the phototube, or leakage of showering 
particles into the hadronic calorimeter.  The mean 
fractional energy leakage into the hadronic calorimeter 
for particles exiting the tracker, determined using 
the {\sc geant} calorimeter simulation, is shown as a 
function of $\log_{10}(p_T/{\rm GeV})$ in Fig.~\ref{fig:leakage}.  
\par
For a low-$p_T$ particle exiting the tracker, the 
distribution of energy loss into the hadronic calorimeter 
is adequately described by an exponential.  For high-$p_T$ 
particles ($\gtrsim 10$ GeV), the distribution has a peak 
at non-zero values of energy loss.  In this energy region 
we model the hadronic energy loss fluctuations with the 
distributions shown in Fig. \ref{fig:leakage}.  Because a 
non-negligible fraction of electrons lose a significant 
amount of energy ($5-10$\%) in the hadronic calorimeter, it 
is important to model the energy loss spectrum in addition 
to the mean hadronic energy loss.

To correct for any unaccounted dependence of the response on 
incoming particle energy, we use an empirical model of response 
that increases linearly with particle $p_T$:
\begin{equation}
\label{eq:nonlinearity}
 R_{EM}(p_T) =  S_E [ 1 + \xi (p_T / {\rm GeV} - 39) ].
\end{equation}

\noindent
We determine the slope parameter $\xi = [6 \pm 7 ({\rm stat})] \times 10^{-5}$ 
using fits to the electron $E/p$ distribution as a function of $p_T$ 
in $W\rightarrow e\nu$ and $Z\rightarrow ee$ events 
(Section~\ref{sec:electrons}).  The inclusive $E/p$ distribution from 
$W \rightarrow e\nu$ events is used to calibrate the absolute response 
$S_E$.  Since electrons in this sample have a mean $p_T$ of 39 GeV, 
the fitted values for $S_E$ and $\xi$ are uncorrelated.  The parameter 
$\xi$ describes the ``non-linearity'' of the calorimeter response.
\par
Light attenuation in the scintillator results in non-uniform response as 
a function of distance from the wavelength-shifting light guides.  The 
attenuation function was measured using test beam data at construction, 
and aging effects are measured $in situ$ using electrons from $W$ boson 
decays.  The function is parametrized as a quadratic function of the 
CES~$x$ position within a tower and corresponds to a reduction in 
response of $\approx 10$\% at the edge of the tower.  We simulate the 
light attenuation by reducing the energy deposited by each particle 
according to this function, evaluated at the particle's CES $x$ position.
\par
To improve measurement resolution in data, we correct for 
attenuation effects by applying the inverse of the quadratic 
attenuation function to the measured EM energy.  We match this 
procedure in the simulation.
\par
The EM calorimeter response drops rapidly as a particle crosses the 
edge of the scintillator and into the dead region between towers 
\cite{CEM}.  We take the calorimeter to have zero response for any 
particle with $|$CES~$x| > 23.1$ cm or $|$CES~$z| < 4.2$ cm.  For the 
$m_W$ measurement we only use high-energy electrons far from the 
dead regions (Section~\ref{sec:wesample}).
\par
We apply the following smearing to the calorimeter cluster energy:
\begin{equation}
\sigma_E/E = \sqrt{0.135^2/p_T + \kappa^2},
\end{equation}

\noindent
where the constant term $\kappa$ is determined to be 
$[0.89 \pm 0.06 ({\rm stat}) \pm 0.13 ({\rm sys})]$\% from a fit to the 
width of the electron $E/p$ peak in $W$ boson decays \cite{eop}.  We find 
further energy smearing is necessary to model the multi-particle energy 
clusters populating the high $E/p$ region.  When a simulated $W$ or $Z$ 
decay electron radiates in the tracker, we apply an additional fractional
resolution of $\kappa_{\gamma} = [8.3 \pm 2.2 ({\rm stat})]$\% to each 
radiated particle.  This smearing contributes $\approx 1.3\%$ to the effective 
constant term, and is determined from a fit to the width of the $Z$ boson 
mass peak reconstructed from radiative electrons ($E/p > 1.06$).  
\par
The final contribution to the electron cluster energy comes from the 
underlying event \cite{ue} and additional $p\bar{p}$ interactions.  As 
with muons, we measure this energy distribution in $W$ boson data as a 
function of $u_{||}$, $u_{\perp}$, and the electron tower $\eta$ 
(Section~\ref{sec:leptonremoval}).  These measurements are incorporated 
in the simulation.

\section{$W$ Boson Selection}
\label{sec:wsample}

The $W$ boson samples are collected with triggers requiring
at least one central ($|\eta| \lesssim 1$) lepton candidate 
in the event.  A narrow kinematic region is defined for $W$ 
boson selection:  $30$ GeV $<$ lepton $p_T < 55$ GeV; $30$ 
GeV $<$ \met \! \!$< 55$ GeV; $60$ GeV $< m_T(l,$\met$) < 100$ 
GeV; and $u_T < 15$ GeV.  This selection results in low 
background while retaining events with precise $m_W$ 
information.  Additional background rejection is achieved 
through event selection targeting the removal of $Z$ boson 
decays to leptons.  To minimize bias, lepton selection 
criteria are required to have high efficiency or to be 
explicitly modeled by our fast simulation.

\subsection{$W \rightarrow \mu\nu$ Selection}
\label{sec:wmusample}

Muons are identified based on their reconstructed COT track 
quality and production vertex, minimum ionizing energy deposited
in the calorimeter, and the consistency of the track segments 
reconstructed in the muon chambers with the COT tracks. 
\par
All charged lepton candidates from $W$ and $Z$ boson decay are 
required to have fully-fiducial central (\mbox{$|z_0| < 60$~cm}) COT 
tracks with at least 5 hits on each of $\geq 3$ axial superlayers 
and $\geq 3$ small-angle stereo superlayers.  For muon candidates 
we remove background from decays of long-lived hadrons to muons 
(``decays in flight'') by requiring the track impact parameter to 
be small ($|d_0| < 1$ mm) and the track fit quality to be good 
($\chi^2/$dof $< 3$).  After this initial selection, the COT track 
parameters are updated with an additional constraint to the transverse 
position of the beam, which has a size of $\approx 30~\mu$m in the 
luminous region.  The beam constraint results in a factor of $\approx 3$ 
improvement in momentum resolution for muons from $W$ boson decays.
\par
Each muon candidate's COT track is extrapolated to the 
calorimeter and its energy deposition in the electromagnetic
and hadronic calorimeters is separately measured.  Muons near 
a tower edge in the $z$ direction cross two calorimeter towers, and 
those tower energies are combined to determine the muon's total 
energy deposition.  We require the muon's electromagnetic energy 
deposition $E_{EM}$ to be less than 2 GeV and its hadronic energy 
deposition $E_{Had}$ to be less than 6 GeV \cite{wzprd}.
\par
All $W$ muon candidates must have a track segment in either the 
CMU and CMP detectors, or the CMX detector.  COT tracks extrapolated
to these detectors must have $r-\phi$ positions that match to within
3, 5, or 6 cm of the CMU, CMP, or CMX track segment positions, 
respectively.
\par
The $Z/\gamma^* \rightarrow \mu\mu$ process presents a significant 
background to the $W\rightarrow \mu\nu$ sample.  We reduce this 
background by removing events with a second opposite-charge muon 
candidate passing the above selection, or passing the following 
looser set of criteria: an opposite-charge track with $p_T > 10$ GeV, 
$|d_0| < 1$ mm, $\geq 2$ axial superlayers with $\geq 5$ hits, and 
$\geq 2$ (1) small-angle stereo superlayers with $\geq 5$ hits for 
tracks fully (partially) fiducial to the COT; $E_{EM} < 2$ GeV and 
$E_{Had} < 6$ GeV; and calorimeter isolation $< 0.1$.  Calorimeter 
isolation is defined as the calorimeter $p_T$ in an $\eta-\phi$ cone 
of radius 0.4 surrounding the muon calorimeter towers, divided by the 
muon track $p_T$.  For events with one identified $W$ decay muon and a 
second muon candidate passing the looser criteria, the identified $W$ 
decay muon must also have isolation $< 0.1$ for the event to be rejected 
from the $W$ boson sample.  The full $W$ boson sample, after kinematic 
selection and $Z$ boson rejection, contains 51,128 events in 
$(190.8 \pm 11.1)$ pb$^{-1}$ of data.
\par
\begin{figure}[!tp]
\begin{center}
\epsfysize = 6.cm
\epsffile{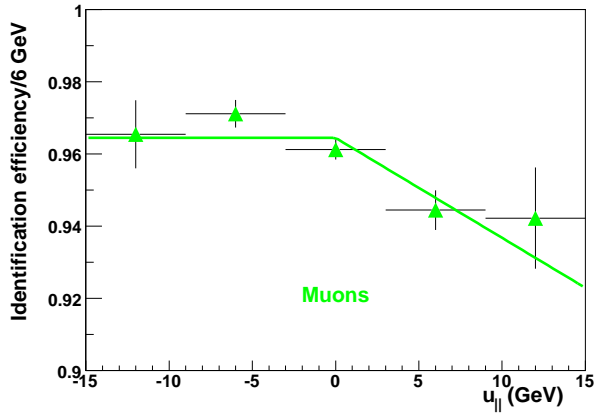}
\caption{The muon identification efficiency as a function of the recoil 
component in the direction of the muon ($u_{||}$).  }
\label{fig:uparmu}
\end{center}
\end{figure}

The identification efficiency of muons has a small dependence on the recoil
in $W\rightarrow \mu\nu$ and $Z\rightarrow \mu\mu$ events, due primarily to 
the track $\chi^2$ and $d_0$ requirements.  We measure this dependence using 
$Z\rightarrow \mu\mu$ events, selected with one muon passing the $W$ muon 
candidate criteria and a second ``probe'' muon identified as a track with 
$p_T > 30$ GeV.  The two muons must have opposite charge and reconstruct to 
an invariant mass in the $81-101$ GeV range.  The fraction of probe muons 
passing the additional $W$ muon candidate selection criteria is shown in 
Fig.~\ref{fig:uparmu} as a function of net recoil energy along the muon 
direction ($u_{||}$).  The observed dependence is parametrized as:
\begin{equation}
\label{eq:upareff}
\epsilon = a[1 + b(u_{||} + |u_{||}|)],
\end{equation}

\noindent
where $a$ is a normalization factor that does not affect the $m_W$ measurement and 
$b = [-1.32 \pm 0.40 ({\rm stat})] \times 10^{-3}$.  We vary $b$ by $\pm 3\sigma$ 
in simulated data and fit for $m_W$.  Assuming a linear variation of $m_W$ with $b$, 
we derive uncertainties of $\delta m_W = 1, 6,$ and 13 MeV for the $m_T$, $p_T$, 
and \met fits, respectively.

\subsection{$W \rightarrow e\nu$ Selection}
\label{sec:wesample}

Electron identification uses information from the COT track quality 
and production vertex, the matching of the track to calorimeter 
energy and position, and the longitudinal and lateral calorimeter 
energy profiles.
\par
An electron candidate's COT track has the same fiduciality and hit 
usage requirements as a muon candidate track, and utilizes the same 
beam-constrained track fit.  The track is required to have $p_T > 18$ 
GeV, a kinematic region where the trigger track-finding efficiency 
has no $p_T$ dependence.
\par
The clustering of showers in the CES produces an energy-weighted
position at the electron shower maximum.  We require the CES 
cluster to be well separated from the edges of the towers, 
$|{\rm CES}~x| < 18$ cm and \mbox{$|{\rm CES}~z| > 9$~cm}.  The cluster 
$z$ position is compared to the extrapolated track $z$ position, 
and the difference is required to be less than 5 cm, consistent 
with the trigger requirement.  The ratio of the measured calorimeter 
energy to the track momentum, $E/p$, must be less than 2.
\par
Electrons are differentiated from hadrons by their high fraction of 
energy deposited in the electromagnetic calorimeter.  The electron's 
EM energy is measured in two neighboring towers in $\eta$, while the 
energy collected in the hadronic calorimeter is measured in three 
towers.  The ratio, $E_{Had}/E_{EM}$, is required to be less than 
0.1.  Only the EM calorimeter measurement is used to determine the 
electron's $p_T$.
\par
An electron shower will typically be confined to a single tower, with 
a small amount of energy flowing into the nearest tower in $\eta$.  We 
define an error-weighted difference between the observed and expected 
energies in the two towers neighboring the electron in the $\eta$ 
direction \cite{wzrun1}:
\begin{equation}
L_{shr} = 0.14 \sum_i \frac{E_i^{adj} - E_i^{exp}}{\sqrt{0.14^2E_i^{adj} + 
(\Delta E_i^{exp})^2}},
\end{equation}

\noindent
where $E_i^{adj}$ is the energy in a neighboring tower, $E_i^{exp}$ is 
the expected energy contribution to that tower, $\Delta E_i^{exp}$ is 
the RMS of the expected energy, energies are measured in GeV, and the 
sum is over the two neighboring towers.  We require $L_{shr} < 0.3$, 
consistent with the trigger criterion (Section~\ref{sec:trigger}).
\par
The $Z\rightarrow ee$ background is highly suppressed by the 
$u_T < 15$ GeV requirement for the $W$ boson sample.  Residual 
background results from electrons passing through dead calorimeter 
regions, which reduces $u_T$ and increases \met.  We remove events 
from the $W$ sample if a track with $p_T > 20$ GeV and 
$|d_0| < 0.3$ cm extrapolates to a calorimeter region with reduced 
response ($|{\rm CES}~x| > 22$ cm or $|{\rm CES}~z| < 6$ cm), and the 
track's calorimeter isolation is $< 0.1$ (Section \ref{sec:wmusample}).  
The full $W \rightarrow e\nu$ selection results in a sample of 63,964 
candidate events in $(218.1 \pm 12.6)$ pb$^{-1}$ of integrated luminosity.
\par
\begin{figure}[!tp]
\begin{center}
\epsfysize = 6.cm
\epsffile{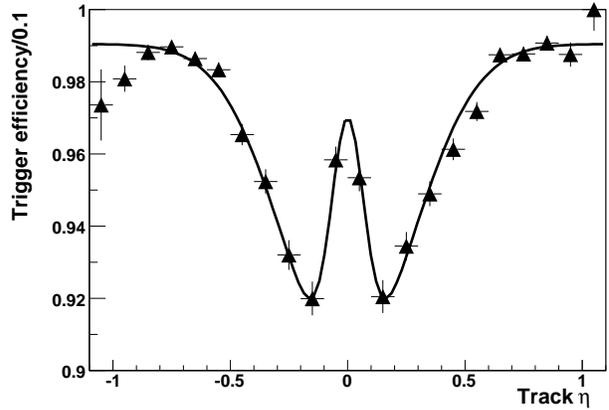}
\caption{The electron track trigger efficiency as a function of 
track $\eta$, for electrons identified in the calorimeter.  The 
solid line shows the double-Gaussian parametrization of the data. }
\label{fig:effvseta}
\end{center}
\end{figure}

The track selection in the single-electron trigger (Section 
\ref{sec:trigger}) results in an $\eta$-dependent trigger efficiency 
for reconstructed electrons (Fig. \ref{fig:effvseta}).  We study this 
efficiency using $W$ events selected with a trigger where the track 
requirements are replaced by a \met threshold.  The efficiency 
decreases as $|\eta|$ decreases because the reduced path length 
reduces the ionization charge collected by each wire, thus reducing 
the single hit efficiency.  There is an additional decrease in efficiency 
due to the dead region at $|z| \lesssim 2$ mm.  Electrons crossing this 
region at track $|\eta| = 0$ are not included in the efficiency plot, 
since we only measure electrons with \mbox{$|{\rm CES}~z| > 9$~cm}.  Thus, 
at $|\eta| = 0$ there is no inefficiency due to the dead COT region, and 
the measured efficiency increases.

\begin{figure}[!tp]
\begin{center}
\epsfysize = 6.cm
\epsffile{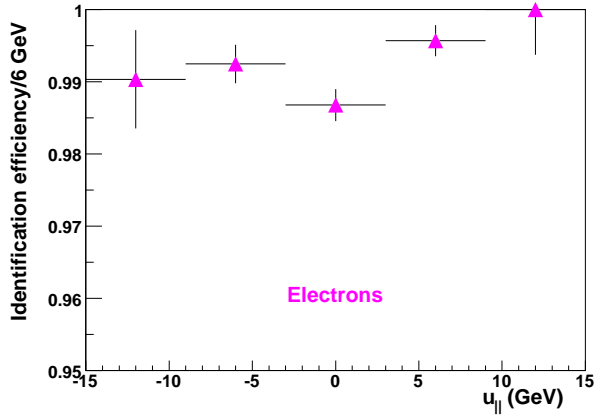}
\caption{The electron identification efficiency measured in $Z\rightarrow ee$ 
data as a function of the recoil component in the direction of the electron 
($u_{||}$).  Background is subtracted using the number of like-charge lepton
events observed at a given $u_{||}$.  The $E/p <2$ requirement is not included 
in this efficiency measurement. } 
\label{fig:uparele}
\end{center}
\end{figure}

We measure the $u_{||}$ dependence of the electron identification efficiency 
(Fig. \ref{fig:uparele}) using $Z\rightarrow ee$ events, selected with one 
electron passing the $W$ electron candidate criteria and a second ``probe'' 
electron identified as an EM energy cluster with $p_T > 30$ GeV, an associated 
track with $p_T > 18$ GeV, and $E/p < 2$.  Since the probe electron definition 
includes an $E/p$ requirement, this cut is not included in the efficiency 
measurement.  We instead study the unbiased $E/p < 2$ efficiency by 
recalculating $E$ and $u_{||}$ for towers separated in $\phi$ from the 
identified electron in $W \rightarrow e\nu$ events, and find no significant 
$u_{||}$ dependence in this efficiency.  In the simulation we use 
$b = 0 \pm 0.54 \times 10^{-3}$, obtained by fitting the measured efficiencies 
to the function in Eqn.~\ref{eq:upareff}.

We vary $b$ by $\pm 3\sigma$ in pseudoexperiments and assume linear variation 
of $m_W$ with $b$ to derive uncertainties of $\delta m_W = 3, 5,$ and 16 MeV for 
the $m_T$, $p_T$, and \met fits, respectively.  Since $b$ is measured with different
data samples for the electron and muon channels, there is no correlation between
the corresponding systematic uncertainties.

\section{Track Momentum Measurement}
\label{sec:muons}

Muon momenta are determined from helical fits to tracks 
reconstructed using COT information.  The momentum resolution 
of prompt muons is improved by constraining the helix to 
originate from the transverse beam position.  A given muon's 
transverse momentum is determined by the Lorentz equation,
\begin{eqnarray}
\begin{array}{cll}
\label{eq:ptvsc}
mv^2/R & = & e v B, \\
p_T & = & e B / (2|c|),
\end{array}
\end{eqnarray}

\noindent
where $B$ is the magnetic field, $R$ is the radius of curvature, 
$c \equiv q/(2R)$ is the curvature of the helix, and $q$ is the 
muon charge.  The {\it a priori} momentum scale is determined 
by the measurements of the magnetic field and the radius of the 
tracker.  At CDF, $eB/2 = 2.11593 \times 10^{-3}$ GeV/cm, where 
$B$ is measured using an NMR probe at a COT endplate.  Measurements 
of the local field nonuniformities and tracker geometry were 
performed during construction and installation and are used to 
determine the positions of individual track hits.  We find these 
measurements provide an {\it a priori} momentum scale accuracy 
of $\approx 0.15$\%.
\par
We refine the momentum scale calibration with data.  Using 
reconstructed cosmic ray muon tracks, we align the relative 
positions of the tracker wires.  Track-level corrections
derived from $W\rightarrow e\nu$ data reduce relative curvature 
bias between positive and negative particles.  Finally, we perform 
an absolute calibration of the momentum scale using high-statistics 
data samples of $J/\psi$, $\Upsilon$, and $Z$ boson decays to muons.  
The final calibration is applied as a relative momentum correction 
$\Delta p/p$ to the $W$ boson data and has an accuracy of $\approx 0.02$\%.

\subsection{COT Alignment}
\label{sec:alignment}

The COT contains 30,240 sense wires for measuring the positions 
of charged particles passing through the detector.  The position 
measurements rely on an accurate knowledge of the wire positions 
throughout the chamber.  We determine these positions using a 
combination of alignment survey, computer modeling, and cosmic-ray 
muon data.  Any remaining biases in track parameter measurements 
are studied with $J/\psi\rightarrow\mu\mu$ and $W\rightarrow e\nu$ 
data, from which final track-level corrections are derived.
\par
After construction of the COT endplates, the position of each 
12-wire cell was measured with an accuracy of $\pm 13~\mu$m 
using a coordinate measuring machine.  The effect of the load of 
the wire plane and field sheets was modeled with a finite element 
analysis (FEA) and found to cause an endplate bend towards 
$z = 0$~cm, with the maximum bend of $\approx 6$ mm in the fifth 
superlayer~\cite{COT}.  An equivalent load was applied to the 
detector and further measurements found the FEA to be accurate to 
within $\approx 20\%$.  The FEA results were scaled to match the 
measurements, and the positions determined from the FEA were set 
as the directly-determined cell positions.
\par
While each cell position determines the average positions of its 
12 sense wires within the chamber, several effects create a 
non-linear wire shape as a function of $z$.  Gravity has the 
most significant effect, causing each wire to sag 
$\approx 260~\mu$m in $y$ at $z=0$ cm.  Electrostatic deflection 
towards the nearest field sheet occurs for cells where the sense 
wire is not centered between the field sheets.  By construction, 
the wires are slightly offset within a cell; in addition, the 
gravitational sag of the field sheets is larger than that of the 
sense wires, resulting in an electrostatic deflection that 
partially counteracts the sag of the sense wires.  Combined, 
the electrostatic effects cause a $\phi$-dependent wire shift that 
has a maximum of $74~\mu$m at $\phi = 145^o$ and $z = 0$ cm.  The 
gravitational and electrostatic effects were combined to 
determine the best {\it a priori} estimate of the wire shapes.
\par
\begin{figure}[!tp]
\begin{center}
\epsfysize = 6.25cm
\epsffile{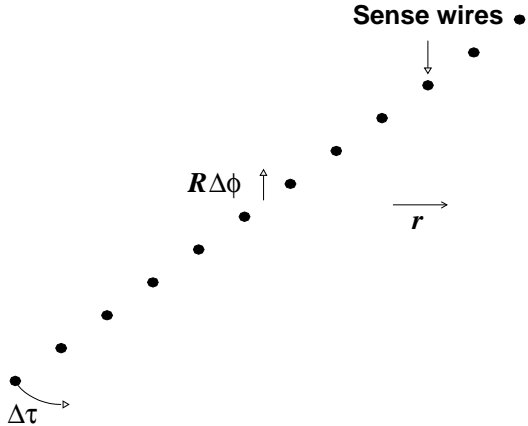}
\caption{The definitions of the local tilt ($\Delta \tau$) and azimuthal 
shift ($R\Delta\phi$) alignment corrections applied to each COT cell. }
\label{fig:alignment}
\end{center}
\end{figure}
\begin{figure}[!htbp]
\begin{center}
\includegraphics*[width=8.5cm]{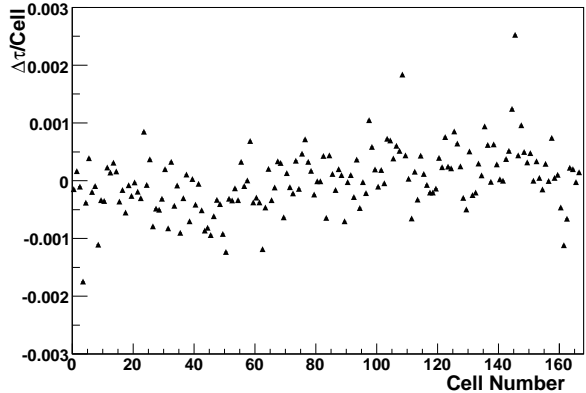}
\includegraphics*[width=8.5cm]{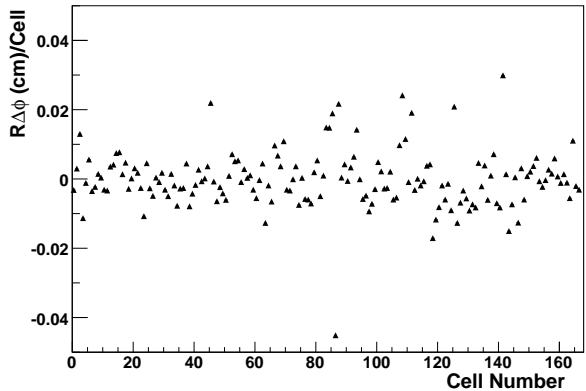}
\caption{The local tilt (top) and azimuthal shift (bottom) alignment 
corrections applied to each cell of the inner superlayer of the west 
endplate.  Not shown are a global 0.0021 tilt correction and a small global 
rotation and shift of the COT that does not affect track measurements. }
\label{fig:corrections}
\end{center}
\end{figure}

Starting from the predicted cell and wire positions, we develop
{\it in situ} corrections based on cosmic-ray muon data taken
during $p\bar{p}$ crossings with the single muon trigger.  The 
data are selected by requiring exactly two reconstructed tracks
in the event, eliminating effects from overlapping hits from 
collision-induced particles.  Since the two tracks on opposite 
sides of the COT result from a single cosmic-ray muon, we refit 
both tracks to a single helix and determine hit residuals with 
respect to this helix \cite{cosmic}.  For each cell, we use the 
residuals to determine a tilt correction about its center, and a 
shift correction along the global azimuth (Fig. \ref{fig:alignment}).  
We show the tilt and shift corrections for the inner superlayer of 
the west endplate in Fig. \ref{fig:corrections}, after removing 
global corrections.  We apply these corrections to each cell of 
each superlayer in each endplate.  In addition, we measure a 
relative east-west shift and include it in each cell's correction.

\begin{figure}[!tp]
\begin{center}
\epsfysize = 6.cm
\epsffile{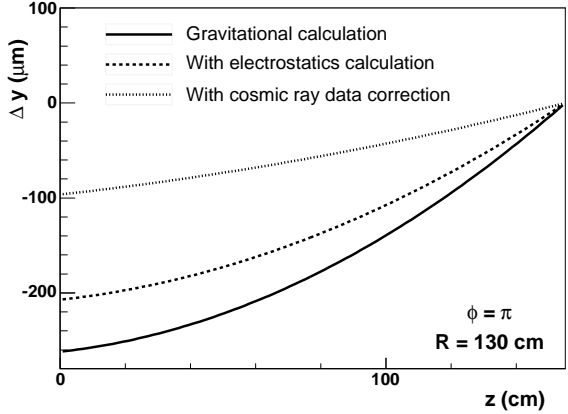}
\caption{The net wire shift in $y$ as a function of $z$ from gravitational 
sag only (solid), including electrostatic effects (dashed), and including 
data-based corrections from Eq. (\ref{eq:datacor}) (dotted).  The shift is 
shown at $\phi = \pi$ and $R = 130$ cm. }
\label{fig:zcorrections}
\end{center}
\end{figure}

We combine the cell-based corrections with wire-based corrections for 
the shapes of the wires between the endplates.  We measure these 
corrections as functions of $z$ and radius $R$ using the differences 
in the measured $d_0$ and curvature parameters for the helix fits on 
opposite sides of the COT for a cosmic ray muon.  The corrections are 
applied as additional offsets $\Delta \xi$ of the wires at $z = 0$ cm, 
with a parabolic wire shape as a function of $z$.  The corrections 
include a radial dependence,
\begin{equation}
\label{eq:datacor}
\Delta \xi = -160 + 380 (R/140) - 380 (R/140)^2,
\end{equation}

\noindent
where $R$ is measured in cm and $\Delta \xi$ in $\mu$m.  Figure 
\ref{fig:zcorrections} shows the gravitational and electrostatic
shifts of a wire as a function of $z$ at $\phi = \pi$, as well as the
data-based correction at $R = 130$ cm (the outer superlayer).  

The cell- and wire-based corrections are implemented for the track-finding
and fitting stage, and reduce the measured hit resolution for high-momentum 
muons from $\approx$180 $\mu$m to $\approx$140 $\mu$m.  Final track-based 
corrections are applied to the measured track curvature, which is inversely 
related to the transverse momentum [Eq. (\ref{eq:ptvsc})].  Expanding the 
measured curvature $c$ as a function of the true curvature $c_t$ in a Taylor 
series around zero,
\begin{equation}
\label{eq:taylor}
c = \epsilon_1 + (1 + \epsilon_2) c_t + \epsilon_3 c_t^2 + \epsilon_4 c_t^3 + ...,
\end{equation}

\noindent
the terms even in $c_t$ cause biases in positive tracks relative to negative 
tracks, which tend to cancel when the two are averaged.  The term linear in 
$c_t$ scales the true curvature and is determined by the momentum calibration.  
The $\epsilon_4 c_t^3$ term is the first to directly affect mass measurements 
and is suppressed by the $c_t^3$ factor at low curvature (high momentum).
\par
\begin{figure}[!tp]
\begin{center}
\includegraphics*[width=8.5cm]{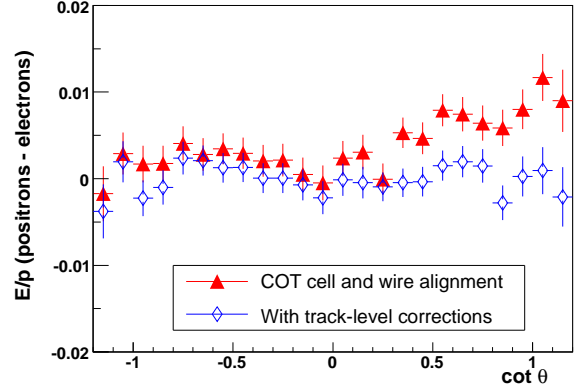}
\includegraphics*[width=8.5cm]{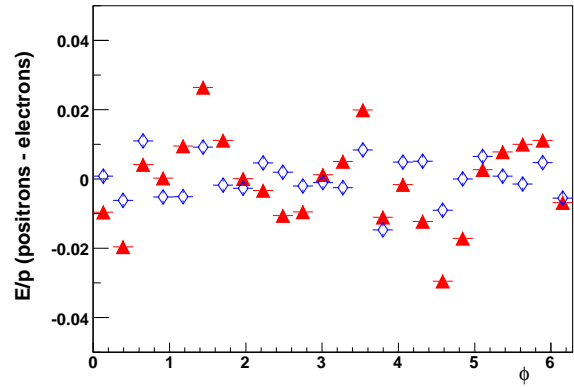}
\caption{The difference between $e^+$ and $e^-~E/p$ as a function of $\cot \theta$ 
(top) and $\phi$ (bottom) before (triangles) and after (diamonds) track-level 
corrections. }
\label{fig:eopalign}
\end{center}
\end{figure}

Corrections for high-momentum tracks from $W$ and $Z$ decay particles are
determined using the difference in $E/p$ for $e^+$ and $e^-$ from $W$ decays, 
which should be zero in the absence of misalignments.  This difference can be 
used to constrain $\epsilon_1$, the first term in the Taylor expansion.  Figure 
\ref{fig:eopalign} shows the differences in $E/p$ as functions of $\cot \theta$ 
and $\phi$, before and after corrections of the following form:
\begin{eqnarray}  
\label{eq:eopalign}
\begin{array}{lll}
\delta c & = & a_0 + a_1 \cot\theta + a_2 \cot^2\theta + \\
         &   & b_1 \sin(\phi + 0.1) + b_3 \sin(3\phi + 0.5). \\
\end{array}
\end{eqnarray}

\noindent
The terms can be interpreted as arising from the following physical effects:  a 
relative rotation of the outer edge to the inner edge of each endplate ($a_0$); 
a relative rotation of the east and west endplates ($a_1 \cot\theta$); and a 
mismeasurement of the beam position ($b_1 \sin(\phi + 0.1)$).  The measured 
values of the parameters $a_0, a_1, a_2, b_1,$ and $b_3$, are shown in 
Table~\ref{tbl:eoppars}.  
\par
Varying $a_1$ by $\pm 3\sigma$ in pseudoexperiments and assuming linear variation
of the momentum scale with $a_1$, we find the $a_1$ uncertainty results in a relative 
momentum scale uncertainty of $\pm 0.07 \times 10^{-3}$ for $W$ and $Z$ boson mass 
measurements.  The other parameter uncertainties, as well as residual higher-order 
terms, have a negligible impact on the momentum scale for the $m_W$ measurement.  

\begin{table}[ht]
\begin{center}
\begin{tabular}{cc}
\hline  
\hline  
Parameter & Value ($\times 10^{-7}$ cm$^{-1}$) \\
\hline  
$a_0$     & $-0.66 \pm 0.17$ \\
$a_1$     & $-1.6 \pm 0.3$ \\
$a_2$	  & $-2.1 \pm 0.5$ \\
$b_1$	  & $-2.1 \pm 0.2$ \\
$b_3$	  & $5.7 \pm 1.7$ \\
\hline  
\hline  
\end{tabular}
\end{center}
\vskip 0.1in
\caption{The parameters used to correct the track curvature of electrons and 
muons from $W$ and $Z$ boson decays.  The values and statistical uncertainties are 
determined from fits to the $E/p$ difference between positrons and electrons. }
\label{tbl:eoppars}
\end{table}

\subsection{$J/\psi\rightarrow \mu\mu$ Calibration}
\label{sec:jpsi}

With a measured $\sigma \times BR$ of $16.3^{+1.4}_{-1.3}$ nb 
\cite{jpsi}, $J/\psi$ mesons are the Tevatron's most prolific 
source of resonant decays to muon pairs.  In addition to its 
high statistics, the $J/\psi$'s precisely known mass 
($m_{J/\psi} = 3096.88 \pm 0.04$ MeV \cite{pdg2002}) and narrow 
width ($\Gamma_{J/\psi} = 0.0934 \pm 0.0021$ MeV \cite{pdg}) 
make it a key component of the track momentum calibration.  We 
perform measurements of the $J/\psi$ mass as a function of mean 
inverse muon $p_T$ to determine a momentum scale correction and 
extrapolate to the high-$p_T$ region relevant for $W$ and $Z$ 
boson decays.

\subsubsection{Data Sample}
\label{sec:jpsidata}

The $J/\psi$ data sample is collected with a Level 1 
trigger requiring one $p_T > 1.5$ GeV XFT track with a 
matching CMU track segment, and a second $p_T > 1.5$ (2) 
GeV XFT track with a matching CMU (CMX) segment.  At 
Level 3, the two corresponding COT tracks must have 
opposite charge and consistent $z$ vertex positions 
($|\Delta z_0| < 5$ cm), and must form an invariant mass 
between 2.7 and 4 GeV.  The resolution on the invariant 
mass measurement degrades at high track momentum, so to 
avoid trigger bias the mass range is extended to 
2 GeV $< m_{\mu \mu} < 5$ GeV when the $p_T$ of the muon 
pair $p_T^{\mu\mu}$ is greater than 9 GeV.
\par
Candidate events are selected offline by requiring two COT 
tracks, each with $p_T > 2$ GeV, $|d_0| < 0.3$ cm,  and 
$\geq 7$ hits on each of the eight superlayers.  The tracks 
must originate from a common vertex ($|\Delta z_0| < 3$ cm) 
and form an invariant mass in the range $(2.95, 3.21)$ GeV. 
\par
A significant fraction ($\approx 20\%$) of the $J/\psi$ 
mesons in our data sample result from decays of $B$ hadrons, 
which have an average proper decay length of $\approx 0.5$ mm.  
The muons from the $J/\psi$ decay can thus originate outside 
the beam radius.  Therefore, no beam constraint is applied in 
the COT track fit of muon candidates from $J/\psi$ decays.  
\par
The total sample consists of 606,701 $J/\psi$ candidates in 
$(194.1 \pm 11.3)$ pb$^{-1}$ of integrated luminosity.

\subsubsection{Monte Carlo Generation}

We use {\sc pythia} \cite{pythia} to generate $J/\psi \rightarrow \mu\mu$ 
events, from which templates are constructed to fit the data for the 
momentum scale.  The shape of the $m_{\mu\mu}$ distribution from $J/\psi$ 
decays is dominated by the $p_T$-dependent detector resolution.  We therefore 
model the $p_T^{J/\psi}$ distribution as well as the $p_T$ and relative 
$p_T$ of the muons in a $J/\psi$ decay.  To obtain an adequate model, 
we empirically tune the generated $J/\psi$ kinematics to describe the 
relevant data distributions for the $J/\psi$ mass fits.
\par
\begin{figure}[!htbp]
\begin{center}
\includegraphics*[width=8.5cm]{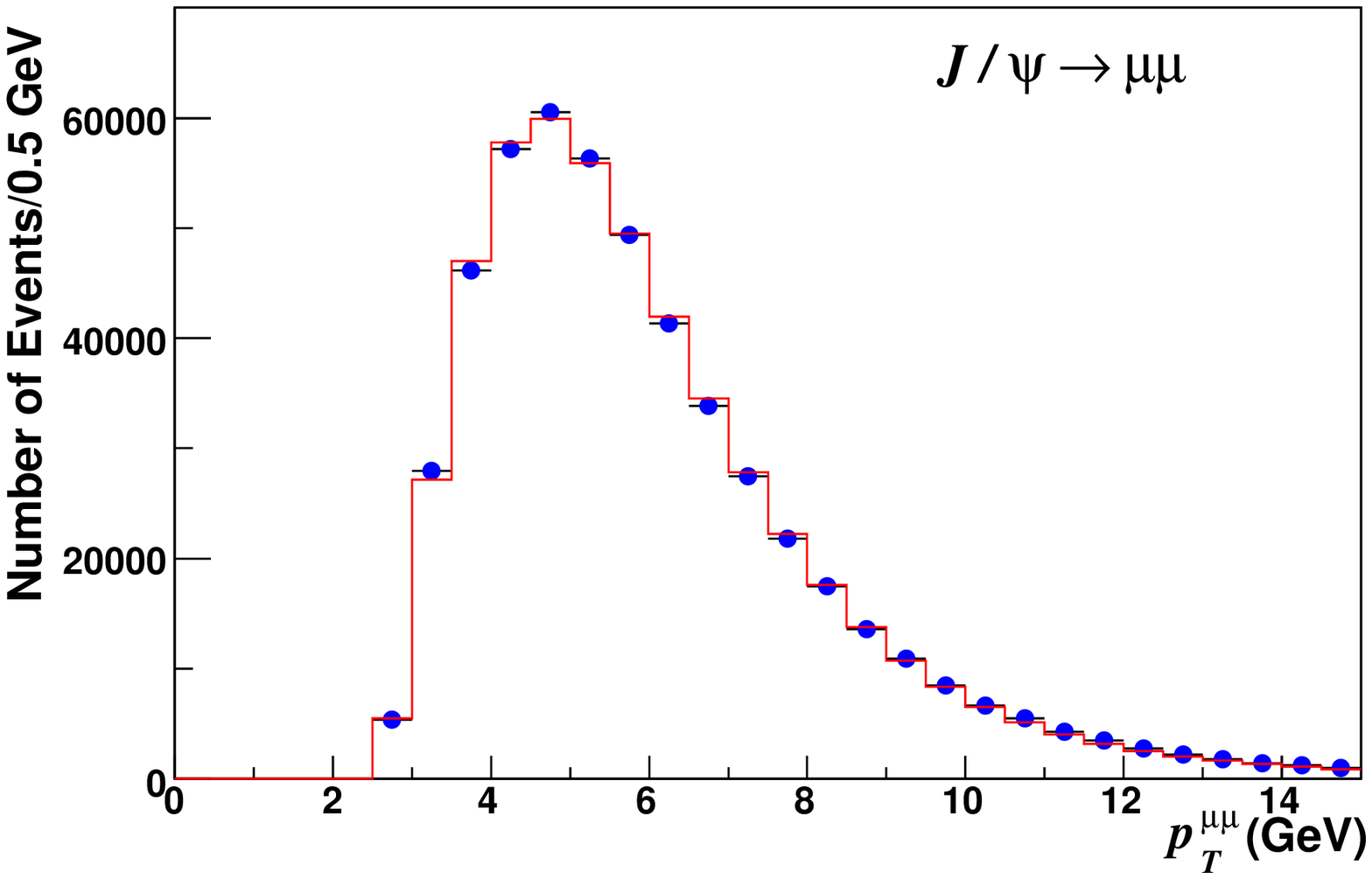}
\includegraphics*[width=8.5cm]{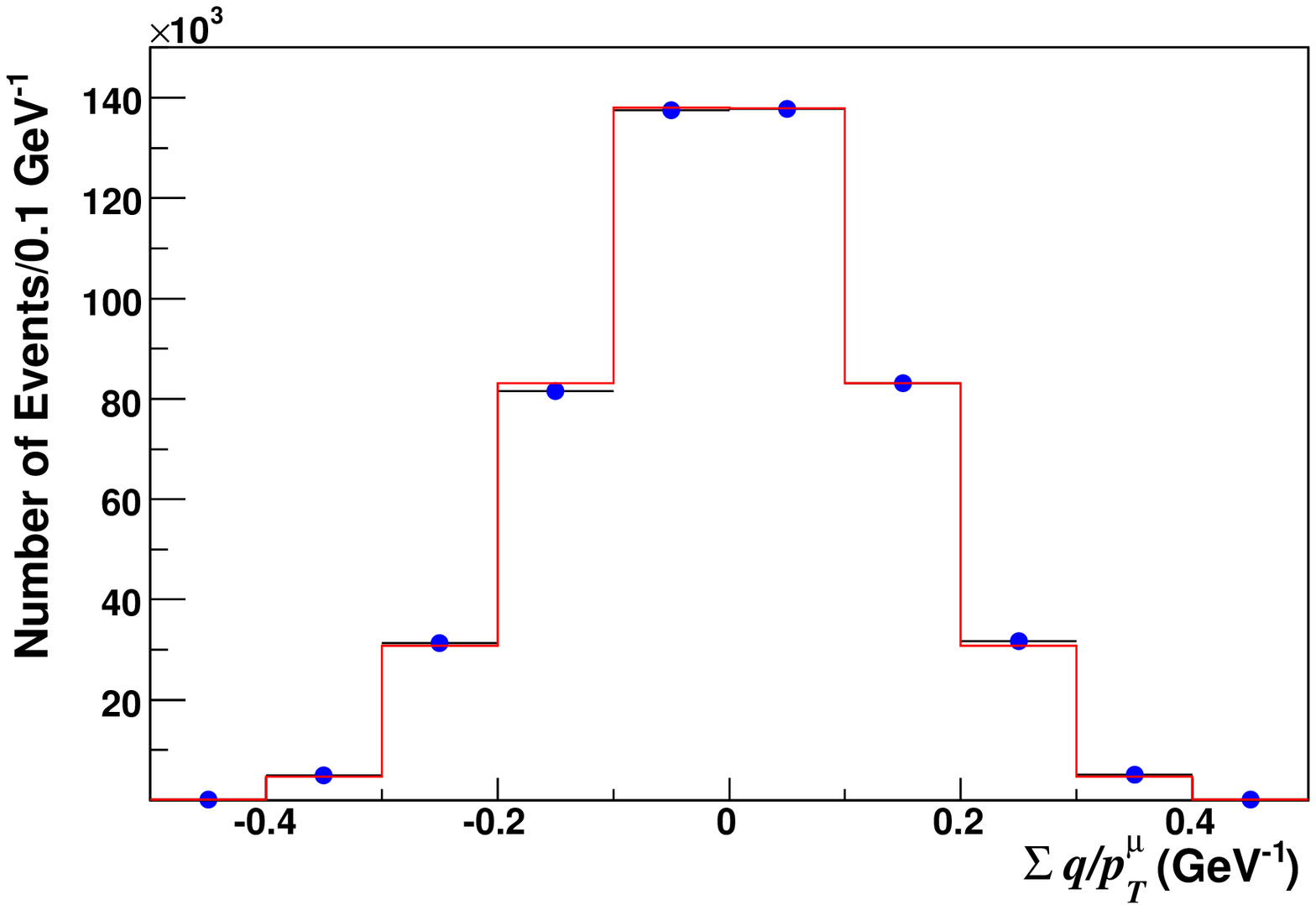}
\caption{The $J/\psi \rightarrow \mu\mu$ data (points) and tuned 
simulation (histogram) distributions of $p_T^{\mu\mu}$ (top) and 
$\sum q/p_T^{\mu}$ (bottom).  The $\sum q/p_T^{\mu}$ is equal to 
the sum of the track curvatures of muons from a $J/\psi$ decay, 
divided by $2.11593 \times 10^{-3}$. }
\label{fig:jpsiptsumcurv}
\end{center}
\end{figure}

To tune the $p_T^{J/\psi}$ distribution, we boost the $J/\psi$ momentum 
by changing its rapidity ($y_{J/\psi}$) along its direction of motion 
$\hat{p}_{J/\psi}$.  In 50\% of the generated events we multiply 
$y_{J/\psi}$ by 1.215, and in the other 50\% we multiply it by 1.535.  
The decay angle $\theta^*$ in the $J/\psi$ rest frame relative to 
$\hat{p}_{J/\psi}$ is tuned by multiplying $\cot{\theta^*}$ by 1.3.  
After tuning, the simulation matches the relevant background-corrected 
data distributions, as shown in Fig. \ref{fig:jpsiptsumcurv}.

The {\sc pythia} event generator does not include energy loss due to 
final-state photon radiation from the muons in $J/\psi$ decays.  To 
simulate this effect, we scale each muon's momentum by a factor 
$x$ determined from the following leading-log probability 
distribution for soft photon radiation \cite{pythia, fsrapprox}:
\begin{equation}
\label{eq:fsrpdf1}
f(x) = \beta (1 - x)^{\beta - 1},
\end{equation}
\noindent 
with
\begin{equation}
\label{eq:fsrpdf2}
\beta = \frac{\alpha_{EM}}{\pi}[\ln(Q^2/m_{\mu}^2) - 1]
\end{equation}
\noindent
and $Q^2 = m^2_{J/\psi}$.

\subsubsection{Momentum Scale Measurement}
\label{sec:jpsipscale}

The momentum scale is calibrated using $J/\psi$ decays
by fitting the dimuon mass as a function of mean inverse
$p_T$ of the two muons, and then extrapolating to high 
$p_T$ ($\left\langle p_T^{-1} \right\rangle \approx 0$ 
GeV$^{-1}$).  This procedure results in a track momentum 
calibration accuracy of 0.025\%. 
\par
The momentum scale calibration requires an accurate modeling 
of the muon ionization energy loss in the tracker.  Each 
muon passing through the silicon and COT detectors loses on 
average 9 MeV at normal incidence.  The combined effect on 
the reconstructed $m_{\mu\mu}$ is about 0.6\% of $m_{J/\psi}$, 
a factor of $\approx 20$ larger than our total uncertainty.  
Since the ionization energy loss $E_I$ varies only 
logarithmically with $p_T$ (Section \ref{sec:ionization}), the 
relative effect on the reconstructed mass is:
\begin{equation}
\frac{\Delta m}{m} = \frac{E_I^{\mu^+}}{2 p^{\mu^+}_T} + 
              \frac{E_I^{\mu^-}}{2 p^{\mu^-}_T}
              \approx E_I \left\langle p_T^{-1} \right\rangle.
\end{equation}

\begin{figure}[!tp]
\begin{center}
\epsfysize = 6.cm
\epsffile{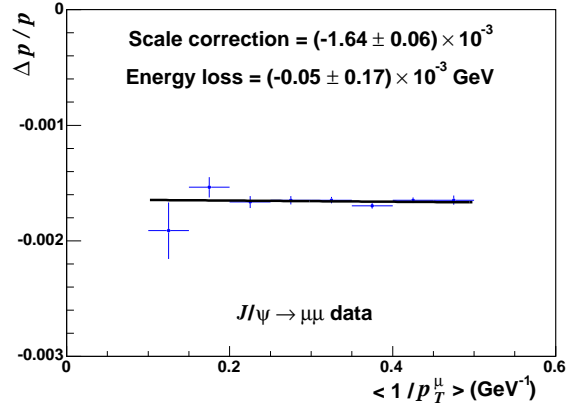}
\caption{The fractional momentum correction for data as a function 
of the mean inverse momentum of the muons from $J/\psi$ decays.  In 
a linear fit, the intercept corresponds to the scale correction 
relevant for $W$ and $Z$ boson decays, and the slope corresponds to 
the remaining unmodeled ionization energy loss after material tuning. 
The uncertainties are statistical only.}
\label{fig:jpsiinvpt}
\end{center}
\end{figure}

\noindent
Thus, in a linear fit of $\Delta m/m$ as a function of mean 
inverse $p_T$, a non-zero slope approximately corresponds to 
$E_I$.  Since we model the ionization energy loss based 
on the known detector material, this slope should be zero.  We 
however find that we need to scale down the ionization energy 
loss from the detector parametrization (Section~\ref{sec:ionization}) 
by 6\% to achieve a zero slope.  We show the result of this tuning 
in Fig.~\ref{fig:jpsiinvpt}, replacing $\Delta m/m$ on the $y-$axis 
with the relative momentum correction $\Delta p/p$ to be applied to 
the data in order to measure $m_{J/\psi} = 3096.88$ MeV.  The tuning 
is based on a $\left\langle p_T^{-1} \right\rangle$ region of 
$(0.1, 0.5)$~GeV$^{-1}$, divided into eight bins.  We find a scale 
correction of $\Delta p/p = [-1.64 \pm 0.06({\rm stat})] \times 10^{-3}$ 
from a linear fit to $\Delta p/p$ as a function of 
$\left\langle p_T^{-1} \right\rangle$.

\begin{figure}[!tp]
\begin{center}
\includegraphics*[width=8.5cm]{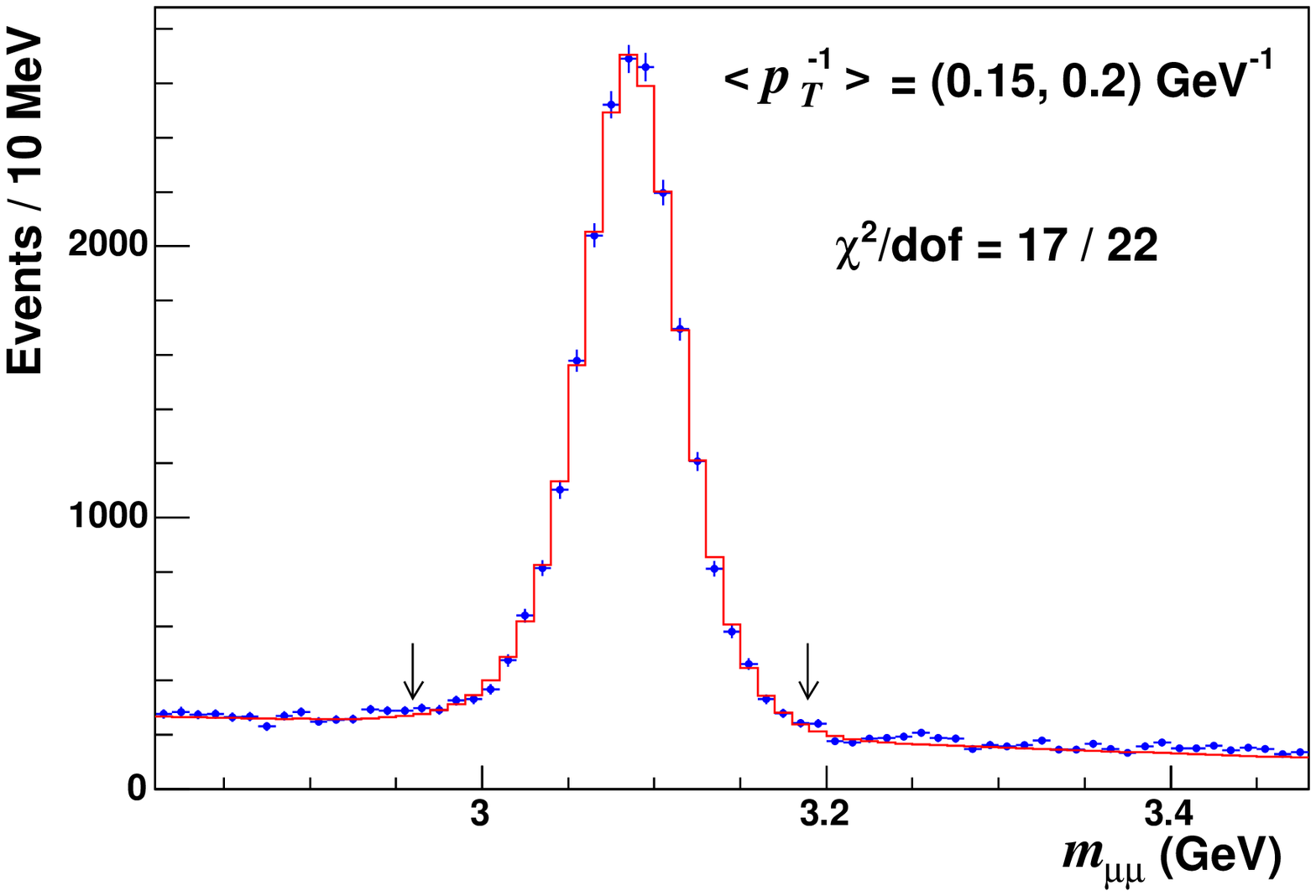}
\includegraphics*[width=8.5cm]{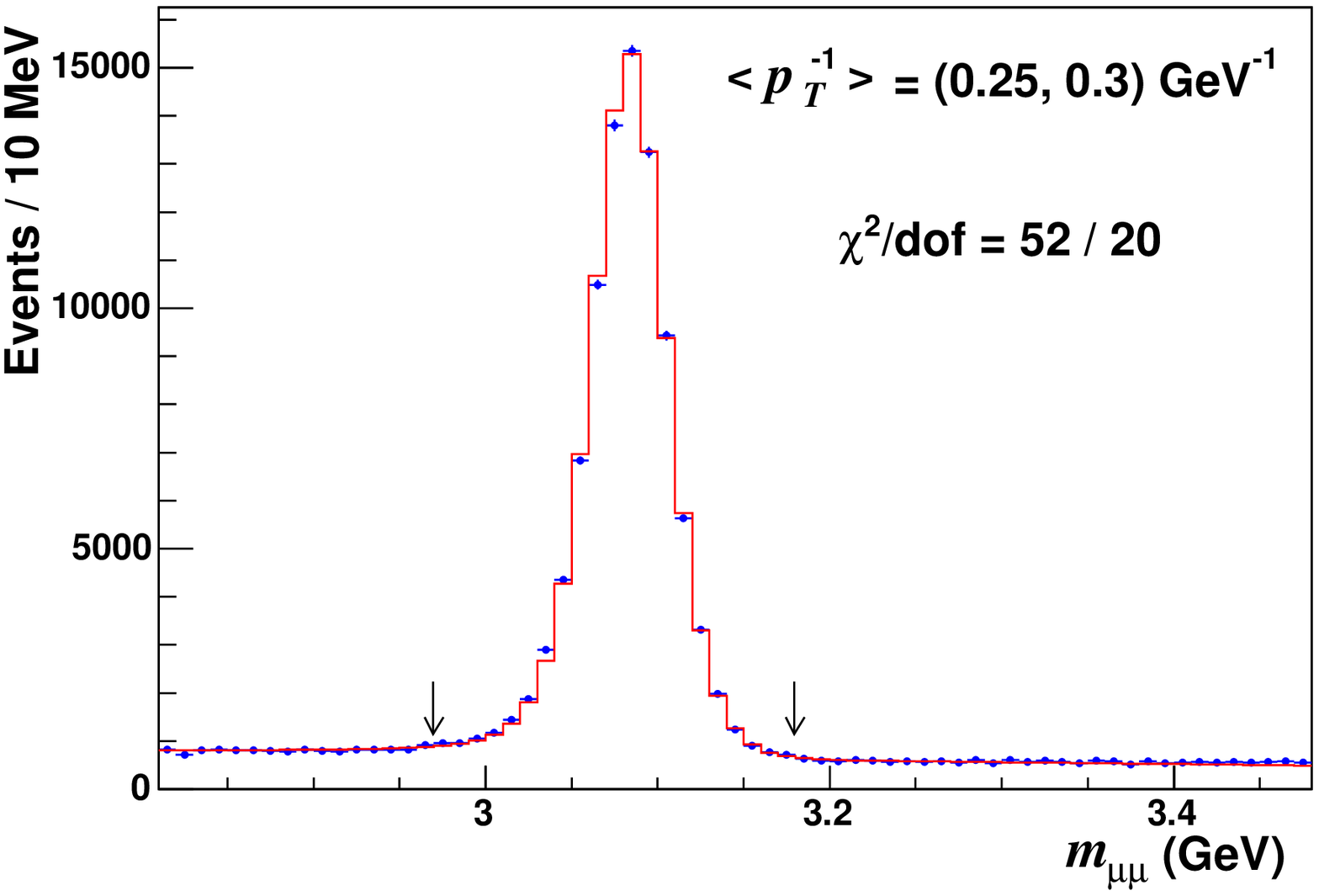}
\caption{The $m_{\mu\mu}$ fits to data (points) with 
$\left\langle p_T^{-1}\right\rangle  = (0.15,0.2)$ GeV$^{-1}$ (top) 
and $\left\langle p_T^{-1} \right\rangle = (0.25,0.3)$ GeV$^{-1}$ 
(bottom).  The best fit to the $m_{\mu\mu} = (3.08 \pm 0.12)$ GeV (top) 
and $m_{\mu\mu} = (3.08 \pm 0.11)$ GeV (bottom) regions correspond to 
momentum scale corrections of $(-1.54 \pm 0.09) \times 10^{-3}$ (top) 
and $(-1.65 \pm 0.04) \times 10^{-3}$ (bottom).  The arrows indicate the 
fit regions and the uncertainties are statistical only.  The fit $\chi^2$ 
can be improved by adjusting the final-state radiation model, and this 
effect is incorporated into the systematic uncertainty 
(Section~\ref{sec:jpsiuncertainty}). }
\label{fig:jpsimassfit}
\end{center}
\end{figure}

Each $\Delta p/p$ value in Fig. \ref{fig:jpsiinvpt} is extracted 
via a binned likelihood fit to the $m_{\mu\mu}$ distribution for 
each $\left\langle p_T^{-1} \right\rangle$ bin.  Since the mass 
resolution varies significantly with $\left\langle p_T^{-1} \right\rangle$, 
the fit ranges are adjusted from $3.08 \pm 0.13$ GeV for 
$\left\langle p_T^{-1} \right\rangle = (0.1,0.15)$ GeV$^{-1}$ to 
$3.08 \pm 0.08$ GeV for $\left\langle p_T^{-1} \right\rangle = (0.45,0.5)$ 
GeV$^{-1}$.  The background is modeled as a linear function of 
$m_{\mu\mu}$, with normalization and slope determined from upper and
lower sideband regions whose combined width is equal to that of the 
mass fit window.  The results of the fits in the 
$\left\langle p_T^{-1} \right\rangle = (0.15,0.2)$ GeV$^{-1}$ and 
$\left\langle p_T^{-1} \right\rangle = (0.25,0.3)$ GeV$^{-1}$ ranges 
are shown in Fig. \ref{fig:jpsimassfit}.  

The $J/\psi$ momentum calibration includes corrections to the curvature $c$ 
derived from the measured dimuon mass as a function of $\Delta \cot \theta$ 
between the positive and negative muons from the $J/\psi$ decay.  Biases linear 
in $\Delta \cot \theta$ are removed with a curvature correction linear in 
$\cot \theta$:
\begin{equation}
\label{eq:jpsicurvcor}
\delta c = [(-7 \pm 1) \times 10^{-7}~{\rm cm}^{-1}] \cot \theta,
\end{equation}

\noindent
where the uncertainty is statistical only.  Biases quadratic in $\Delta \cot \theta$ 
are removed with the following correction to the absolute length scale of the COT 
along the $z$ axis (statistical uncertainty only):
\begin{equation}
\label{eq:zscale}
\delta \cot \theta = [(-3.75 \pm 1.00) \times 10^{-4}] \cot\theta.
\end{equation}

The $\cot \theta$-dependent correction to the curvature [Eq.~(\ref{eq:jpsicurvcor})] 
is larger than the correction derived from $E/p$ in $W\rightarrow e\nu$ data 
($a_1$ in Table \ref{tbl:eoppars}).  Muons from $J/\psi$ decay have a broader 
curvature range and thus a greater dependence on misalignments affecting higher order 
terms in curvature.  Since we derive a curvature correction averaged over all of the 
terms in Eq.~(\ref{eq:taylor}), the $J/\psi$ correction can be larger than the 
correction for electrons and muons from $W$ and $Z$ boson decays.

\subsubsection{Momentum Scale Uncertainties}
\label{sec:jpsiuncertainty}

Systematic uncertainties on the momentum scale correction extracted from
$J/\psi \rightarrow \mu\mu$ decays (Table \ref{tbl:psiupssys}) are dominated
by the incompleteness of the QED and energy loss models.  At low muon $p_T$ 
(high $\left\langle p_T^{-1} \right\rangle$), the mass fits become increasingly
sensitive to QED and energy loss modeling because of the better mass 
resolution and higher statistics.  Since we only model the mean ionization 
energy loss, our modeling of the mass region below the peak is imperfect.  
Additionally, our neglect of higher-order QED corrections affects the modeling 
of this region.  We study possible bias from our incomplete model by changing 
the $Q^2$ value in the photon radiation probability function [Eq.~(\ref{eq:fsrpdf1})] 
such that the $\chi^2$ of the inclusive $m_{\mu\mu}$ fit is minimized.  We find 
that this change affects $\Delta p/p$ by $0.2 \times 10^{-3}$.
\par
If there is a relative tilt between the solenoid and the tracker axes, the extracted 
momentum scale correction will have a linear dependence on $\Delta \cot\theta$.  In 
addition, incomplete corrections of the magnetic field nonuniformities near the ends 
of the solenoid can cause a quadratic $\cot\theta$ variation.  We study the 
$\cot\theta$ dependence of $\Delta p/p$ using $J/\psi$ decays where both muons are 
measured in the same $\cot\theta$ region ($|\Delta\cot\theta (\mu\mu)| < 0.1$).  We 
find that if we correct for the observed quadratic dependence, the extracted 
$\Delta p/p$ changes by $0.1 \times 10^{-3}$.
\par
The uncertainty on the material correction propagates to a momentum scale uncertainty 
of $0.06 \times 10^{-3}$ when extrapolated to high momentum, as shown in 
Fig.~\ref{fig:jpsiinvpt}.  An additional statistical uncertainty of 
$0.01 \times 10^{-3}$ on the scale is determined by fixing the material correction 
and fitting for the scale.
\par
The statistical uncertainties on the $J/\psi$ alignment corrections 
[Eq.~(\ref{eq:jpsicurvcor})~and~(\ref{eq:zscale})] have a $0.05 \times 10^{-3}$ 
effect on $\Delta p/p$.  We test our model of the $m_{\mu\mu}$ lineshape by changing 
the fit range by $\pm 20\%$, and find a $\pm 0.05 \times 10^{-3}$ change in 
$\Delta p/p$.
\par
We apply the same $p_T$ thresholds offline as in the trigger for muons with CMU 
segments.  Since we do not model a $p_T$-dependent trigger efficiency, any 
inefficiency could cause a bias in the reconstructed $m_{\mu\mu}$.  We investigate 
this possibility by varying the offline $p_T$ thresholds by $\pm 5\%$, and find a 
$\Delta p/p$ variation of $\pm 0.04 \times 10^{-3}$.
\par
The quality of the fit is highly sensitive to the hit resolution model, but the 
momentum scale correction is not.  Changing the simulated COT hit resolution by 
$\pm 10~\mu$m, which corresponds to a $> 10 \sigma$ statistical variation, results 
in a $\pm 0.03 \times 10^{-3}$ change in $\Delta p/p$.  We include this in our 
systematic uncertainty estimate.  
\par
A $\pm 0.03 \times 10^{-3}$ uncertainty on $\Delta p/p$ from the background model 
is determined by changing its linear dependence on $m_{\mu\mu}$ to a constant.  
Finally, the world-average $J/\psi$ mass value used in this measurement contributes 
$\pm 0.01 \times 10^{-3}$ to the uncertainty on $\Delta p/p$.

\begin{table*}[htbp]
\begin{center}
\begin{tabular}{lccc}
\hline
\hline
Source                         &  $J/\psi$ ($\times 10^{-3}$) &  $\Upsilon$ ($\times 10^{-3}$)  & Common ($\times 10^{-3}$) \\
\hline
QED and energy loss model      & 0.20   &  0.13  &  0.13  \\
Magnetic field nonuniformities & 0.10   &  0.12  &  0.10  \\
Beam constraint bias	       & N/A    &  0.06  &  0     \\
Ionizing material scale        & 0.06   &  0.03  &  0.03  \\
COT alignment corrections      & 0.05   &  0.03  &  0.03  \\
Fit range	               & 0.05   &  0.02  &  0.02  \\
Trigger efficiency	       & 0.04   &  0.02  &  0.02  \\
Resolution model	       & 0.03   &  0.03  &  0.03  \\
Background model               & 0.03   &  0.02  &  0.02  \\
World-average mass value       & 0.01   &  0.03  &  0     \\
\hline
Statistical                    & 0.01   &  0.06  &  0     \\
\hline
Total                          & 0.25   & 0.21   &  0.17  \\
\hline
\hline
\end{tabular}
\caption{Uncertainties on the momentum scale correction derived from the 
$J/\psi$ and $\Upsilon$ mass measurements. }
\label{tbl:psiupssys}
\end{center}
\end{table*}

The final momentum scale correction derived from $J/\psi$ data is:
\begin{equation}
\Delta p/p = (-1.64 \pm 0.25) \times 10^{-3}.
\end{equation}

\subsection{$\Upsilon\rightarrow \mu\mu$ Calibration}

The $b\bar{b}$ resonance $\Upsilon$ provides a complementary 
momentum scale calibration tool to the $J/\psi$.  Its precisely 
measured mass $m_{\Upsilon} = (9460.30 \pm 0.26)$ MeV \cite{pdg} 
is three times larger than that of the $J/\psi$, so an $\Upsilon$
momentum scale calibration is less sensitive to the material and 
energy loss model than that of the $J/\psi$.  Because the $b\bar{b}$ 
resonances are the highest mass mesons, long-lived hadrons do not 
decay to the $\Upsilon$ and the muons from $\Upsilon$ decay 
effectively originate from the collision point.  We improve the 
accuracy of the muon measurements by constraining their tracks to 
the beam position, which is the same procedure applied to the $W$ 
and $Z$ decay lepton tracks.
\par
The $\Upsilon$ data sample is based on the same Level 1 trigger as 
the $J/\psi$ sample (Section \ref{sec:jpsidata}).  The Level 3 
requirements are:  one reconstructed track with $p_T > 4$ GeV and 
matching CMU and CMP track segments (CMUP); a second track with opposite 
charge to the first, $p_T > 3$ GeV, and a matching CMU or CMX track 
segment; and a reconstructed mass of the two tracks between 8 and 12 GeV.
Offline, the $p_T$ thresholds are increased to 4.2 (3.2) GeV for the 
track with a CMUP (CMU or CMX) track segment, and each track must have 
$|d_0| < 0.3$ cm and at least 5 hits in at least 3 axial and 3 stereo 
superlayers.  The two tracks are required to have a common vertex
($|\Delta z_0| < 3$ cm).
\par
\begin{figure}[!htbp]
\begin{center}
\includegraphics*[width=8.5cm]{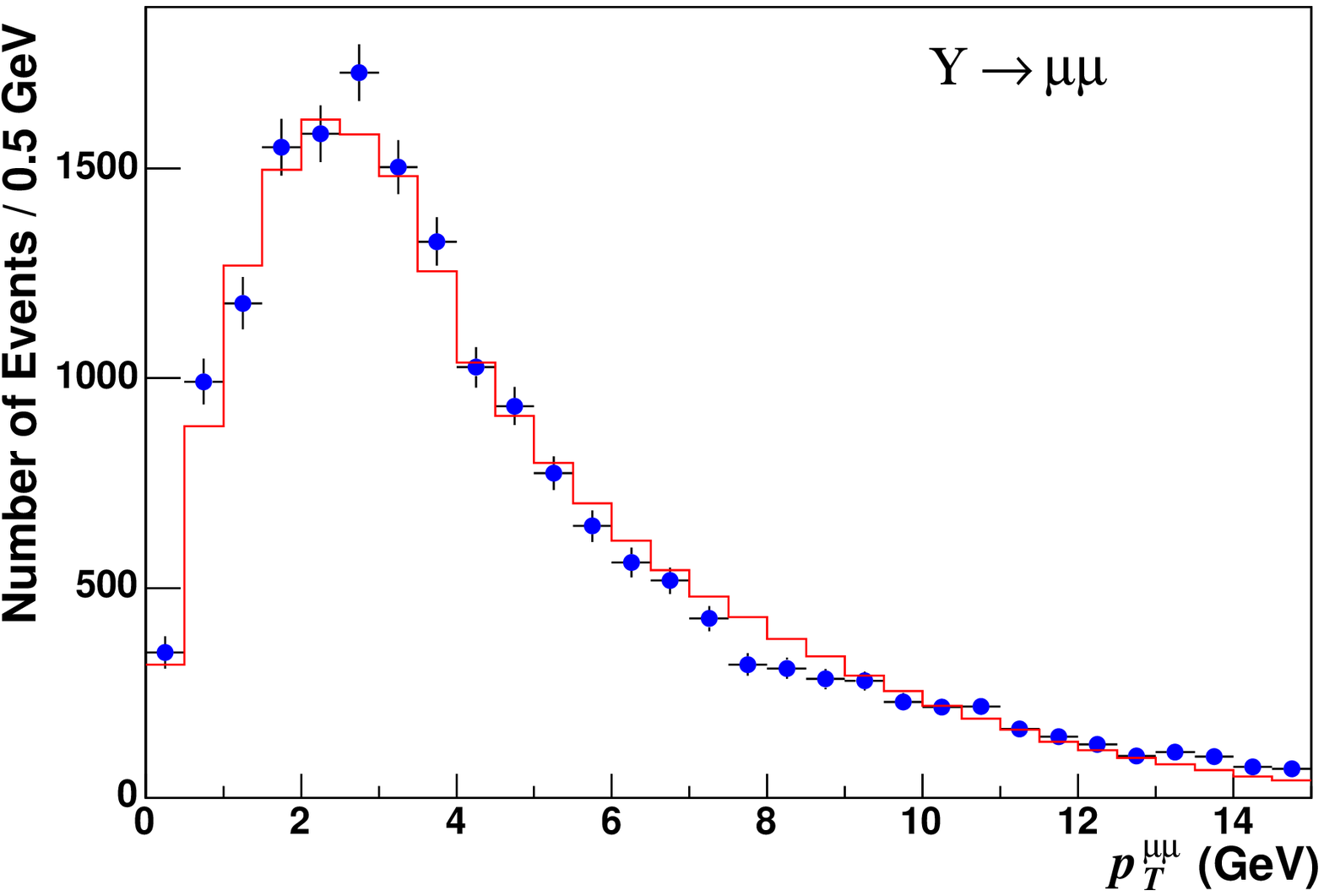}
\includegraphics*[width=8.5cm]{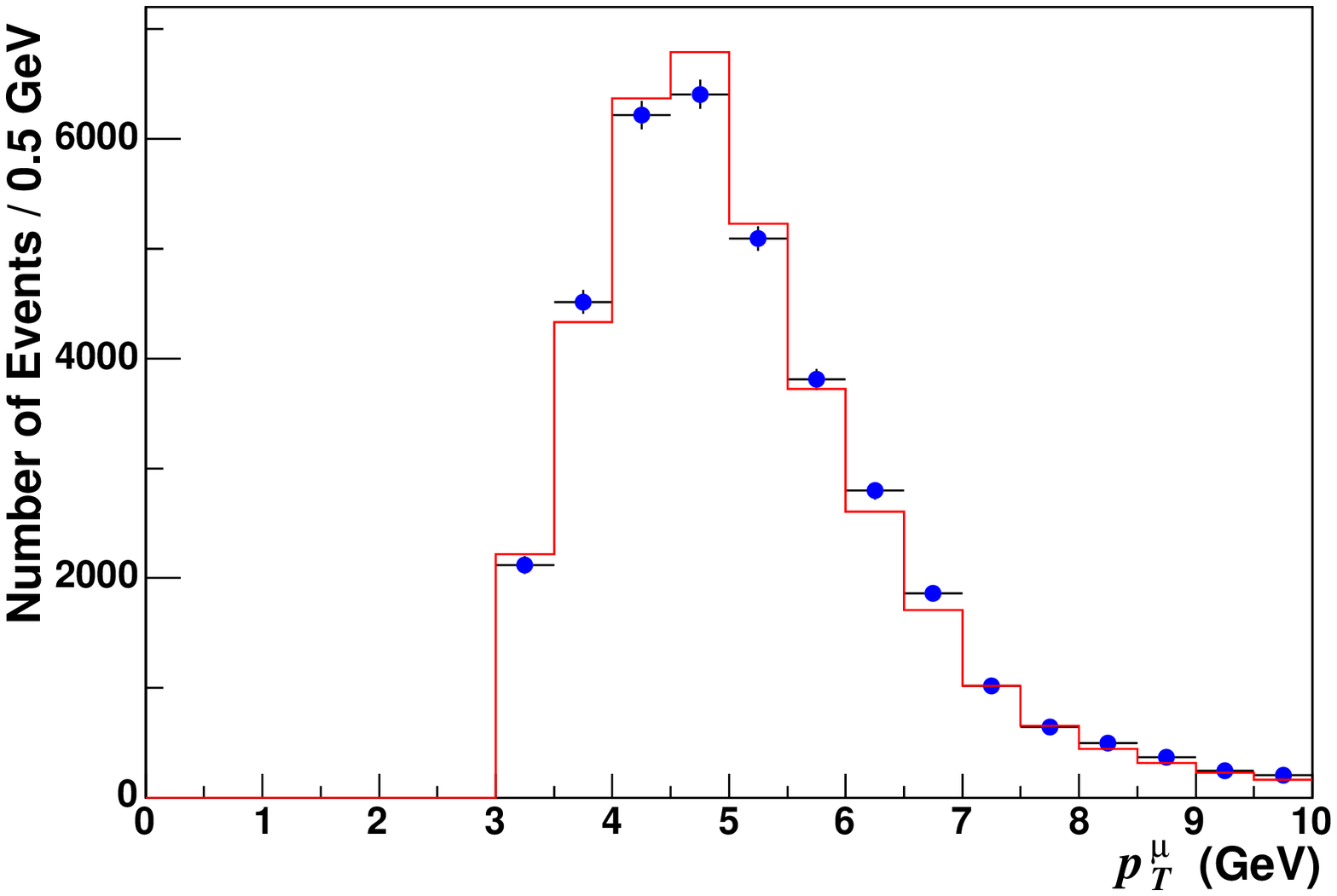}
\caption{The $p_T^{\mu\mu}$ (top) and $p_T^{\mu}$ (bottom) distributions from 
$\Upsilon \rightarrow \mu\mu$ decays for the data (points) and simulation 
(histogram). }
\label{fig:upsilonpt}
\end{center}
\end{figure}

We model $\Upsilon$ production and decay using {\sc pythia} \cite{pythia}, 
to which we apply the same tuning procedure as for $J/\psi$ generation.  
The data $p_T^{\Upsilon}$ distribution is matched in simulation by 
boosting the rapidity of each decay muon by $0.07 y_{\Upsilon}$ along 
$\hat{p}_{\Upsilon}$, where $y_{\Upsilon}$ is the $\Upsilon$ rapidity.  
Radiation of photons from the final state muons is simulated using the 
probability distribution of Eqns. \ref{eq:fsrpdf1} and \ref{eq:fsrpdf2}.  
The $p_T^{\mu\mu}$ distribution is shown in Fig.~\ref{fig:upsilonpt}, 
after subtracting background from the data.

\begin{figure}[!htbp]
\begin{center}
\includegraphics*[width=8.5cm]{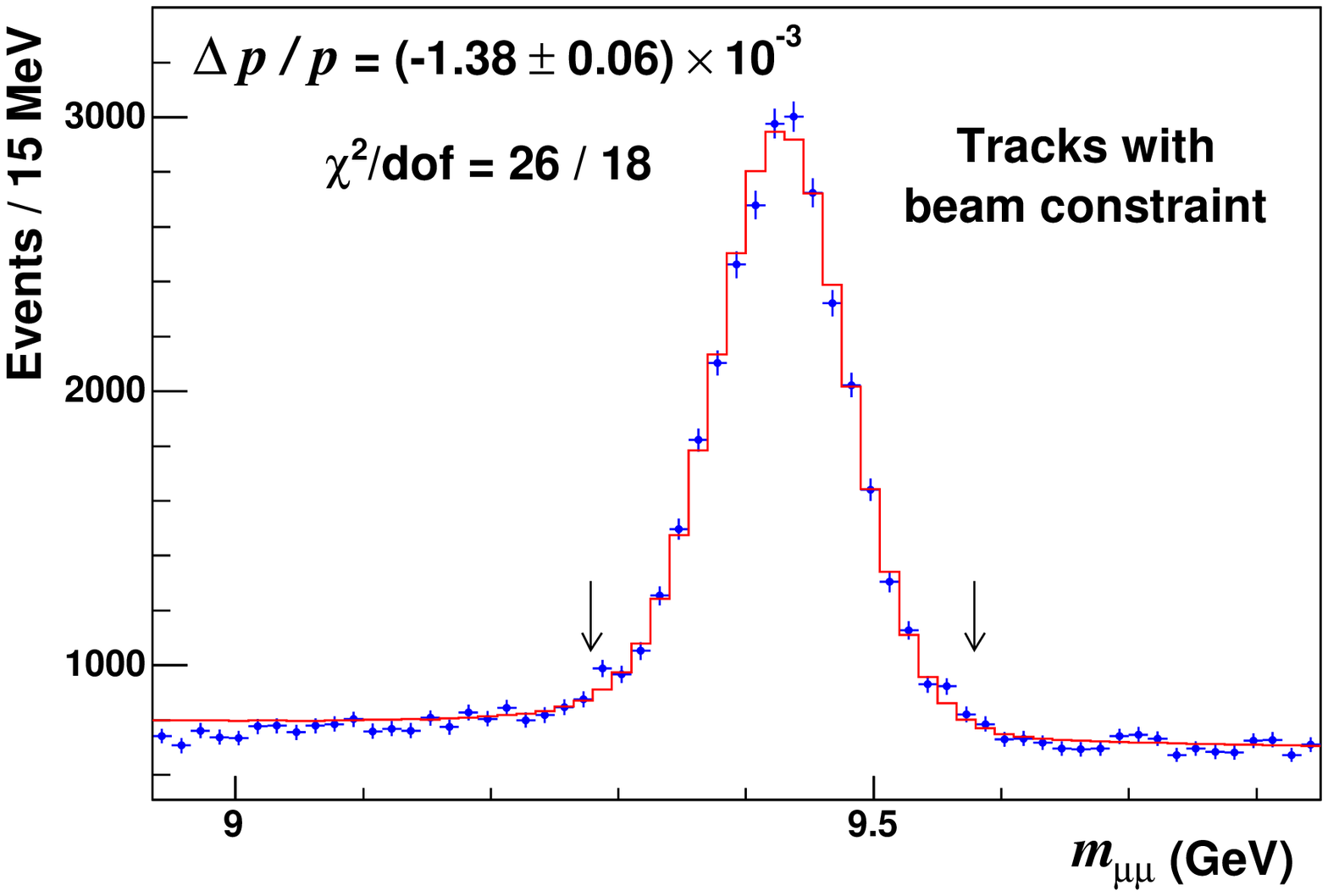}
\includegraphics*[width=8.5cm]{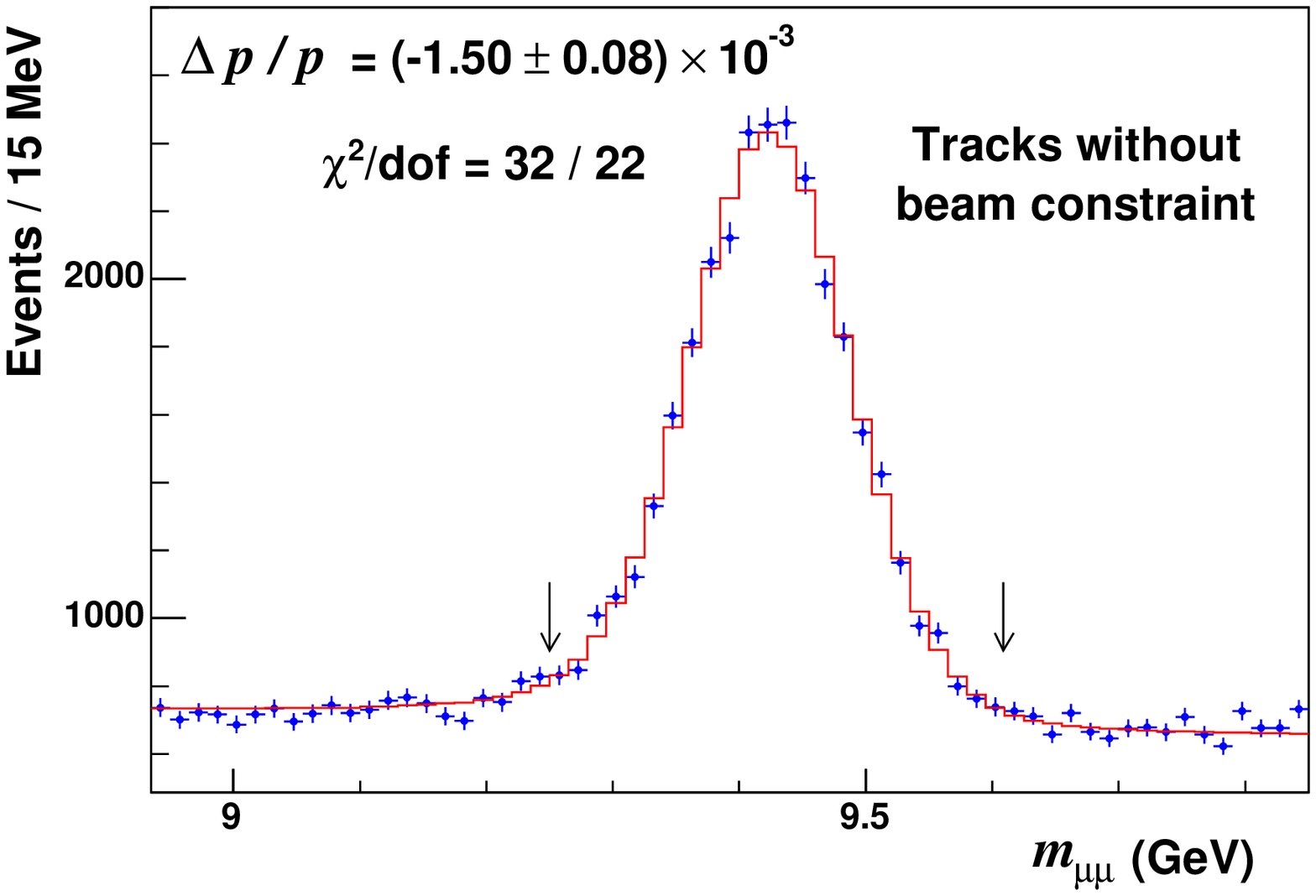}
\caption{The momentum scale correction $\Delta p/p$ derived from binned 
likelihood fits to the $m_{\mu\mu}$ data distribution (points) in the region 
dominated by $\Upsilon \rightarrow \mu\mu$ decays.  The small difference 
between fits using tracks with (top) and without (bottom) a beam constraint 
is incorporated into the systematic uncertainty.  The arrows indicate the fit 
region and the uncertainties are statistical only. }
\label{fig:upsilons}
\end{center}
\end{figure}

We test any possible beam-constraint bias by separately reconstructing 
charged muon tracks from $\Upsilon$ decays with and without incorporating 
the beam constraint.  For the sample with beam-constrained tracks we fit 
for $m_{\Upsilon}$ in the region $9.28$~GeV~$< m_{\mu\mu} < 9.58$~GeV, 
while for the sample with non-beam-constrained tracks we fit the region 
$9.25$~GeV~$< m_{\mu\mu} < 9.61$~GeV.  In $(190.8 \pm 11.1)$ pb$^{-1}$ of 
integrated luminosity, we have 34,618 $\Upsilon$ candidates with 
beam-constrained tracks and 35,622 candidates with non-beam-constrained 
tracks.  The two momentum scale measurements are shown in 
Fig.~\ref{fig:upsilons} and are consistent at the $2\sigma$ level when 
correlations are taken into account.  We define the $\Upsilon$ result to 
be the mean of the two values, and take half their difference 
($\Delta p/p = 0.06 \times 10^{-3}$) as a systematic uncertainty on the 
measurement.
\par
The remaining systematic uncertainties on the momentum scale measurement 
with $\Upsilon$ decays are common to those of the measurement with 
$J/\psi$ decays.  We use the same procedures as with the $J/\psi$ calibration
to estimate the sizes of the uncertainties, with one exception.  Since the 
$\Upsilon$ sample has $<10\%$ of the statistics of the $J/\psi$ sample, the 
QED and energy loss model cannot be tested with the $\chi^2$ of the 
$\Upsilon \rightarrow \mu \mu$ mass fit.  Instead, we change $Q$ in the photon 
radiation probability by the amount estimated for the $J/\psi$ systematic 
uncertainty (Section \ref{sec:jpsiuncertainty}).  We find that this variation 
affects $\Delta p/p$ by $\pm 0.13 \times 10^{-3}$ in the $\Upsilon$ calibration.

The final result of the $\Upsilon$ calibration is:
\begin{equation}
\Delta p/p = (-1.44 \pm 0.21) \times 10^{-3}.
\end{equation}

\noindent
We have verified that this result has no time dependence, at the level of 
the statistical precision of $\pm 0.13 \times 10^{-3}$.  When combined with 
the momentum scale correction from the $J/\psi$ calibration, we obtain:
\begin{equation}
\label{eq:pscale}
\Delta p/p = (-1.50 \pm 0.19) \times 10^{-3}. 
\end{equation}

\subsection{$Z\rightarrow \mu\mu$ Calibration}
\label{sec:zmm}

Given the precise momentum scale calibration from the $J/\psi$ 
and $\Upsilon$ decays, we measure the $Z$ boson mass and compare
it to the world-average value $m_Z = (91187.6 \pm 2.1)$ MeV \cite{pdg}.  
We then use the world-average $m_Z$ to derive an additional $\Delta p/p$
calibration and combine it with that of the $J/\psi$ and $\Upsilon$ 
decays.
\par
The systematic uncertainties of the $m_Z$ measurement are correlated
with those of the $m_W$ measurement, so a momentum scale calibration
with $Z$ bosons can reduce systematic uncertainties on the $m_W$ 
measurement.  However, the statistical uncertainty from the 
$Z \rightarrow \mu \mu$ sample is significantly larger than the 
calibration uncertainty from $J/\psi$ and $\Upsilon$ decays.  Thus, the 
main purpose of the $m_Z$ measurement is to confirm the momentum scale 
calibration and test our systematic uncertainty estimates.
\par
The $Z$ boson data sample is selected using the same single-muon trigger 
and offline muon selection as for the $W$ boson sample (Sections 
\ref{sec:trigger} and \ref{sec:wmusample}), with the exception that we 
remove the requirement of a track segment in a muon detector for one of 
the muons from the $Z$ boson decay.  Removing this requirement significantly
increases detector acceptance while negligibly affecting background.  $Z$ 
boson candidates are defined by $66~{\rm GeV} < m_{\mu\mu} < 116~{\rm GeV}$, 
$p_T^{\mu\mu} < 30$ GeV, $|\Delta t_0(\mu,\mu)| < 3$ ns, and oppositely 
charged muons.  A muon track's $t_0$ is defined as the time between the 
$p\bar{p}$ bunch crossing and the muon's production, and should be 
$(0 \pm 1)$ ns for $Z\rightarrow \mu\mu$ production and decay.  The 
track $t_0$ is measured using the time information from the track hits 
in the COT by incorporating $t_0$ into the helical fit.  The $|\Delta t_0| < 3$ 
ns requirement effectively removes cosmic ray muons passing through the 
detector.  An additional cosmic ray identification algorithm \cite{cosmic} 
reduces this background to a negligible size.  After applying all selection 
criteria, the $Z \rightarrow \mu \mu$ sample contains 4,960 events in 
$(190.8 \pm 11.1)$ pb$^{-1}$ of integrated luminosity.
\par
\begin{figure}[!tp]
\begin{center}
\epsfysize = 6.cm
\epsffile{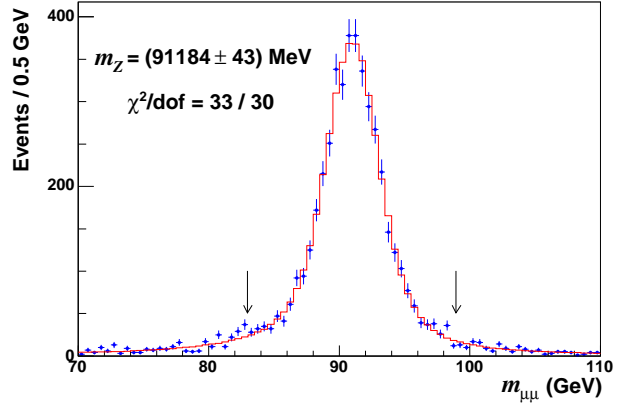}
\caption{The $m_Z$ fit to data (points) in the 83 GeV $< m_{\mu\mu} < 99$ GeV 
region (denoted by arrows).  The uncertainty is statistical only.}
\label{fig:zmm}
\end{center}
\end{figure}

We model $Z$ boson production and decay using the {\sc resbos} \cite{resbos}
event generator and a next-to-leading order QED calculation of photon radiation 
from the final-state muons \cite{wgrad} (Section \ref{sec:production}).  For 
$m_{\mu\mu}$ near the $Z$ boson resonance, the photon propagator and $Z/\gamma^*$ 
interference make small contributions to the shape of the $m_{\mu\mu}$ 
distribution.  We separately simulate these components and include them as fixed 
``background'' to the $Z$ lineshape.  We measure $m_Z$ using a binned likelihood 
template fit to the data in the range $83~{\rm GeV} < m_{\mu\mu} < 99~{\rm GeV}$ 
(Fig.~\ref{fig:zmm}).  Our measurement of $m_Z = [91.184 \pm 0.043 ({\rm stat})]$ 
GeV is in good agreement with the world-average value of 
$m_Z = (91.188 \pm 0.002)$ GeV \cite{pdg}.  

Systematic uncertainties on $m_Z$ are due to the momentum scale calibration 
(17 MeV), alignment correction uncertainties (7 MeV), and incomplete modeling of 
higher-order QED corrections (14 MeV).  The combined statistical and systematic 
uncertainty is 49 MeV.  
\par
Given the precise world-average measurement of $m_Z$, we use the $Z$ boson 
resonance as an additional calibration input to $\Delta p/p$.  We find that 
adding the $m_Z$ information reduces $\Delta p/p$ and its uncertainty by less 
than $0.01 \times 10^{-3}$ each.  
\par
Incorporating the alignment uncertainty (Section~\ref{sec:alignment}) into 
$\Delta p/p$ from Eqn. \ref{eq:pscale} gives the momentum scale correction
 applicable to the $W$ boson sample:
\begin{equation}
\Delta p/p = (-1.50 \pm 0.21) \times 10^{-3}.
\end{equation} 

\noindent
The corresponding uncertainty on the $m_W$ fits in the muon channel is 17 MeV.


\section{Electron Energy Measurement}
\label{sec:electrons}

An electron's energy is measured from its shower in the 
electromagnetic calorimeter.  We perform an initial data 
calibration by scaling the measured energy such that a 
Gaussian fit to the reconstructed dielectron mass in a 
region dominated by $Z$ decays (86-98 GeV) gives a mean of 
91 GeV.  This is slightly below the world-average $m_Z$ 
because the Gaussian fit is biased by the energy lost to 
final-state photon radiation (Section \ref{sec:fsr}).
This initial data calibration is accurate to $\approx 0.15$\%.
\par
To model the data, the simulated calorimeter energy is 
scaled to match the measured $E/p$ distribution of 
electrons in $W\rightarrow e\nu$ events.  A calibrated 
data measurement would result in an $E/p$ of unity for 
electrons that do not radiate before entering the 
calorimeter, and deposit all of their energy in the EM 
calorimeter.  We verify that the $E/p$ calibration is 
unbiased by using it to measure $m_Z$ in dielectron 
events.  Given consistency of the measured $m_Z$ with 
the world-average value, we incorporate the $m_Z$ fit 
into the calibration.  The final calibration has an 
accuracy of $0.037$\%.

\subsection{$E/p$ Calibration}
\label{sec:eop}

We transfer the precise tracker calibration to the calorimeter 
using the ratio of electron calorimeter energy to track momentum, 
$E/p$.  The material from the beam pipe to the inner COT wall 
causes bremsstrahlung that affects the measured position of the 
$E/p$ peak, and this material is scaled such that the simulation  
matches the data in the high $E/p$ region.  The non-linearity of 
the energy scale is removed by applying a correction to the 
simulation scale as a function of the calorimeter shower $p_T$
[Eq.~(\ref{eq:nonlinearity})].  Finally, corrections are applied 
to the data to improve uniformity in response as a function of 
detector tower and time.  After the complete set of corrections 
and simulation calibrations, the simulation energy scale $S_E$ 
is determined from a maximum likelihood template fit to the $E/p$ 
peak region.
\par
The shape of the $E/p$ distribution has a strong dependence on the 
material upstream of the COT.  Bremsstrahlung in this material 
reduces the measured electron momentum in the tracker while leaving
the measured calorimeter energy unchanged, since photons are radiated 
collinearly with the electron and deposit their energy in the same 
calorimeter tower as the electron.  Thus, the effect of bremsstrahlung 
is to shift the measured $E/p$ to values $> 1$.  If the material were 
not well modeled, the energy scale calibration would be biased to 
compensate for the mismodeling.
\par
A detailed accounting of the silicon and COT tracker material was 
performed at installation.  In the early data-taking period, the 
radial distribution of photon conversions was compared between data 
and a full {\sc geant} simulation.  The amount of copper cable was 
increased by a few percent of $X_0$ in the {\sc geant} simulation 
to correct observed discrepancies, and the three-dimensional lookup 
table of material properties (Section \ref{sec:ionization}) was 
produced from this corrected {\sc geant} simulation.  
\par
\begin{figure}[!tp]
\begin{center}
\epsfysize = 6.cm
\epsffile{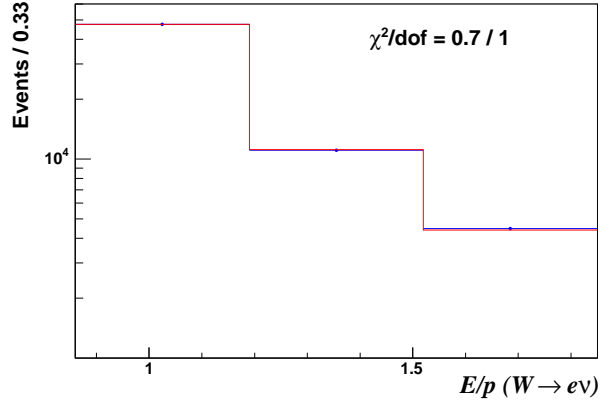}
\caption{The maximum likelihood fit to the tracker radiation length 
correction factor using the high $E/p$ region ($1.19 \leq E/p < 1.85$) 
in $W\rightarrow e\nu$ data (points). }
\label{fig:eopmat}
\end{center}
\end{figure}

For a final material tuning, we compare our parametrized simulation 
to the data in the high $E/p$ region ($1.19 \leq E/p < 1.85$) of 
electrons from $W$ boson decays.  Using the region 
$0.85 \leq E/p < 1.19$ for normalization, we perform a maximum 
likelihood fit to the $1.19 \leq E/p < 1.85$ region in two bins 
(Fig. \ref{fig:eopmat}) and measure a radiation length multiplicative 
correction factor of $S_{mat} = 1.004 \pm 0.009 ({\rm stat})$ 
\cite{matscale}.  As a further consistency check of the material 
lookup table, we determine $S_{mat}$ as a function of tower $|\eta|$, 
and find no statistically significant dependence on $|\eta|$.

\begin{figure}[!htbp]
\begin{center}
\includegraphics*[width=8.5cm]{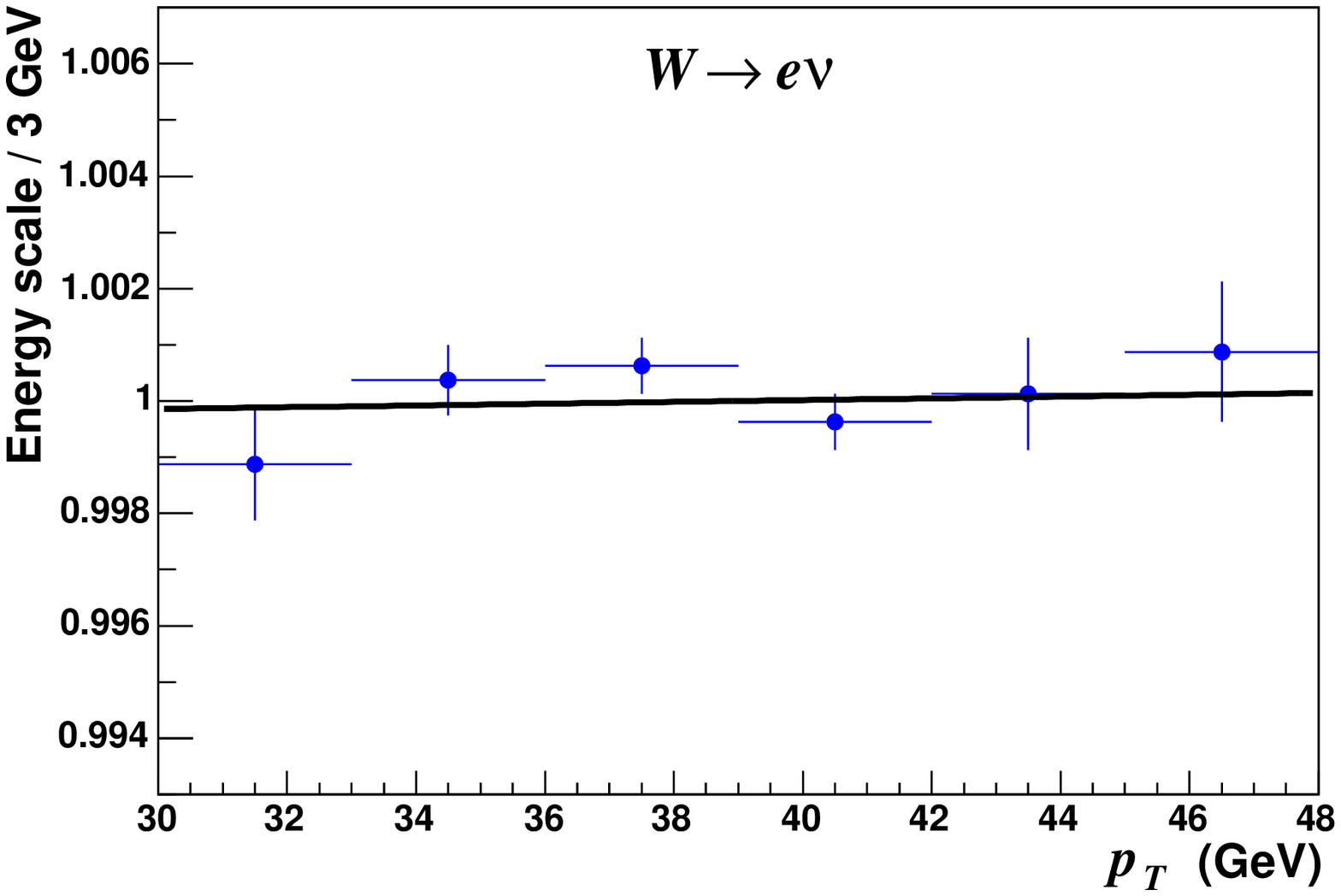}
\includegraphics*[width=8.5cm]{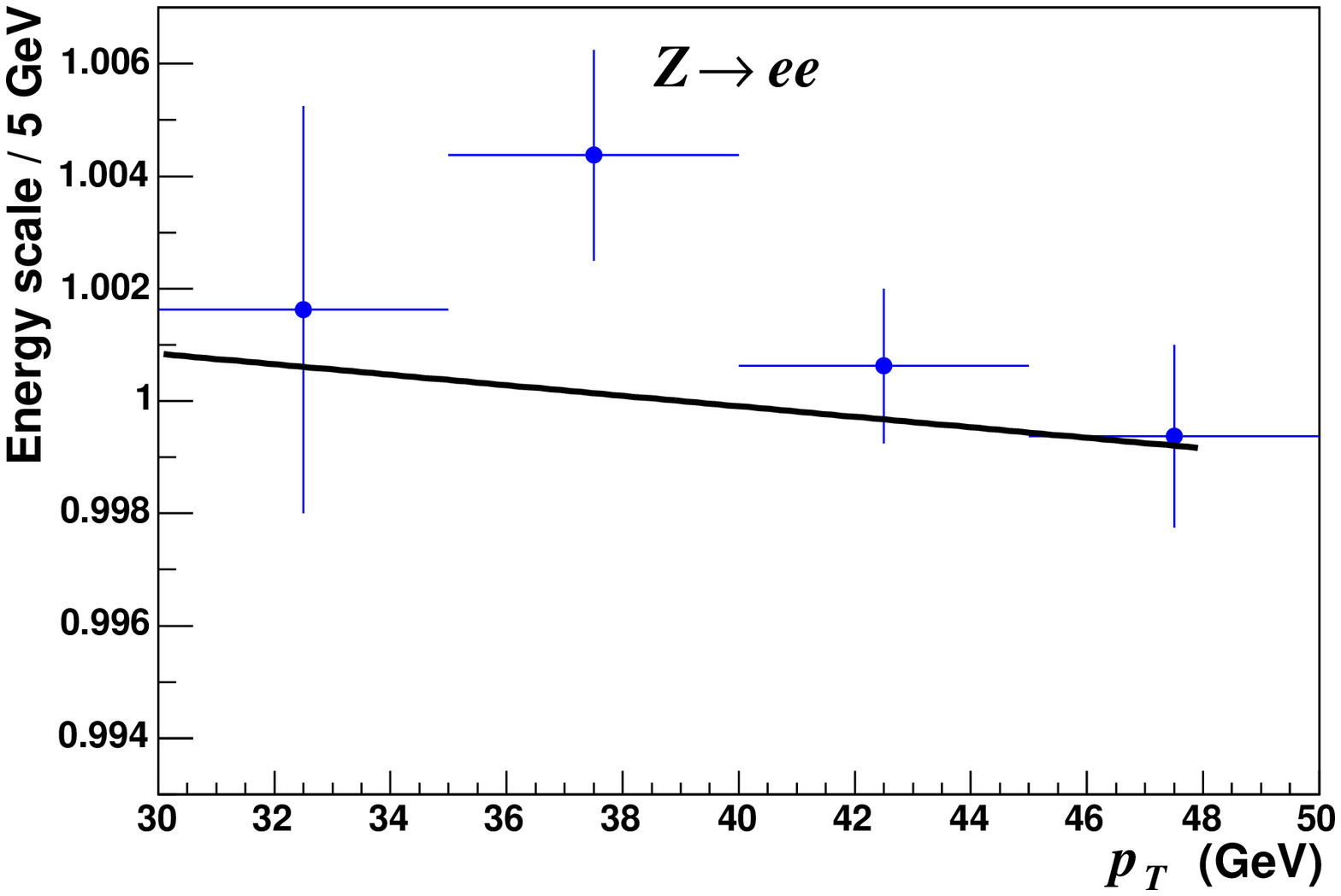}
\caption{The energy scale as a function of measured electron calorimeter 
$p_T$ for $W$ (top) and $Z$ (bottom) boson decays.  The fast simulation 
incorporates a per-particle non-linear response correction 
[Eq.~(\ref{eq:nonlinearity})] .  The combination of the linear fits to 
$\xi$ results in no energy dependence of the energy scale. }
\label{fig:eopet}
\end{center}
\end{figure}

Our simulation of electron interactions in the tracker and calorimeter
accounts for most of the energy dependence of the energy scale.  Any 
residual non-linearity is incorporated as a per-particle correction in the 
simulation (Section \ref{sec:ecal}).  To measure this non-linearity, we 
fit the $E/p$ peak region ($0.93 \leq E/p < 1.11$) for the energy scale 
in bins of measured electron calorimeter $p_T$ (Fig. \ref{fig:eopet}).  The 
resulting energy scale measurements are fit as a linear function of $p_T$, 
fixing the scale to 1 at the $W$ boson sample's 
$\left\langle p_T^e \right\rangle = 39$ GeV.   The error-weighted 
average, $\xi = [6 \pm 7 ({\rm stat})] \times 10^{-5}$, of the measurements 
of the non-linearity parameter from the $W$ and $Z$ boson samples is 
used in Eq.~(\ref{eq:nonlinearity}).  The linear fits in Fig. 
\ref{fig:eopet}, where the simulation includes this correction, show 
a constant energy scale \cite{zeop}.
  
To improve the energy resolution of the data, we apply time-dependent 
and tower-dependent calibrations derived from low-energy EM clusters.  At
level~3 the relevant trigger requires calorimeter and track $p_T$ greater than 
8 GeV each, as well as electron identification based on track-calorimeter 
matching and calorimeter shower shape properties.  Offline, candidates are 
required to have Had/EM$< 0.05$ and $E + p > 22$ GeV to remove any trigger 
bias.  Using the mean of the $E/p$ range $0.8-1.25$, we apply relative 
corrections of ${\cal{O}}(3\%)$ to remove variations as functions of tower and 
time.
\par
Because of bremsstrahlung radiation in the tracker, the mean $E/p$ 
correction has a small bias that depends on the electron path length.  
Since the path length increases as $|\eta|$ increases, we perform a 
final $|\eta|$-dependent calibration of the data.  Using template fits 
to the $E/p$ peak region of the $W\rightarrow e\nu$ sample in bins of 
$|\eta|$, we derive a relative correction for each bin.  This 
calibration removes $\approx 2\%$ residual variation in the calorimeter 
energy response.
\par
\begin{figure}[!tp]
\begin{center}
\epsfysize = 6.cm
\epsffile{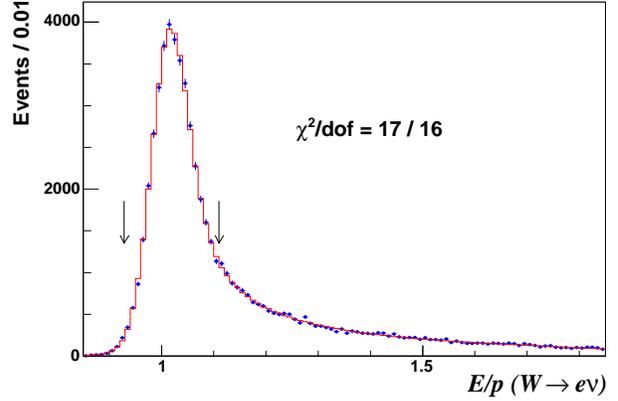}
\caption{The energy scale calibration using the peak $E/p$ region 
($0.93 \leq E/p < 1.11$, denoted by arrows) in $W\rightarrow e\nu$ data 
(points). }
\label{fig:eopfit}
\end{center}
\end{figure}

With the complete set of corrections applied to the data and simulation,
we calibrate the simulation energy scale using $W\rightarrow e\nu$ 
events.  The fit for $S_E$ [Eq.~(\ref{eq:nonlinearity})] to the $E/p$ peak 
region (Fig. \ref{fig:eopfit}) has a statistical uncertainty of 0.025\%.  
Including systematic uncertainties due to $S_{mat}$ ($\pm 0.011\%$) and 
the tracker momentum scale ($\pm 0.021\%$), we obtain a total uncertainty 
of 0.034\% on the $E/p$ calibration of the electron energy scale.

\begin{figure}[!tp]
\begin{center}
\epsfysize = 6.cm
\epsffile{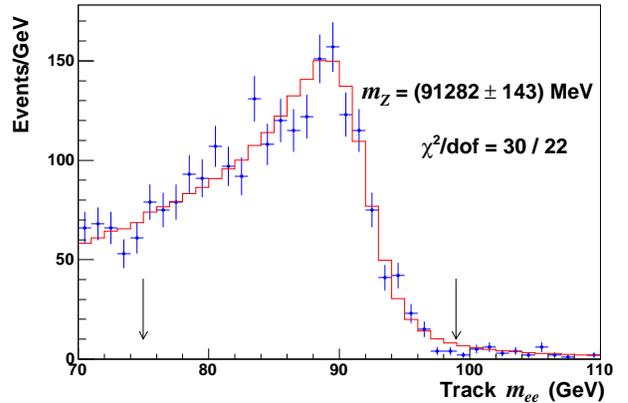}
\caption{The $m_Z$ fit to data (points) in the 75 GeV $< m_{ee} < 99$ GeV 
region (denoted by arrows), using reconstructed track information only.  The 
uncertainty is statistical only.}
\label{fig:zeetrack}
\end{center}
\end{figure}

The $E/p$ calibration requires an accurate simulation of electron
radiation in the tracker.  We test the track simulation by measuring 
$m_Z$ (Section \ref{sec:zee}) using electron track information only.  The 
measurement is a binned likelihood fit to the region 75~GeV~$< m_{ee} < 99$~GeV 
(Fig.~\ref{fig:zeetrack}), with $m_Z$ as the fit parameter.  Because of the 
significant radiated energy loss, the test is less precise than the measurement 
using the calorimeter (Fig.~\ref{fig:zee}).  Nevertheless, we obtain good 
consistency with the world-average $m_Z$, verifying that we do not have any 
significant mismodeling of electron radiation in the tracker.

\subsection{$Z\rightarrow ee$ Calibration}
\label{sec:zee}

Using the $E/p$-based calorimeter energy calibration, we measure 
the $Z$ boson mass from its decay to two electrons.  After 
confirming consistency of the result with the world-average mass, 
we fix $m_Z$ to this value and produce a combined calibration 
from the electron $E/p$-based method and $Z\rightarrow ee$ mass 
measurement.  
\par
We select $Z$ bosons using the same single-electron trigger and 
offline electron selection as for the $W$ boson sample (Sections 
\ref{sec:trigger} and \ref{sec:wesample}), and define candidates 
as oppositely-charged electrons with 66~GeV~$< m_{ee} < 116$~GeV 
and $p_T^{ee} < 30$ GeV.  The $Z$ boson sample contains 2,919 
events in $(218.1 \pm 12.4)$ pb$^{-1}$ of data.
\par
The sample includes a small component of multijet and $W +$ 
jet background.  From a comparison of the data with like-sign 
electrons to a prediction of the full {\sc geant} simulation, 
we estimate the background fraction to be $\lesssim 0.5$\%.  
Since $\left\langle m_{ee} \right\rangle$ of the background is 
$\approx 2$ GeV less than that of the $Z$ boson sample in the 
fit region, we estimate any corresponding bias on the measured 
$m_Z$ to be $\lesssim 10$ MeV.
\par
\begin{figure}[!tp]
\begin{center}
\epsfysize = 6.cm
\epsffile{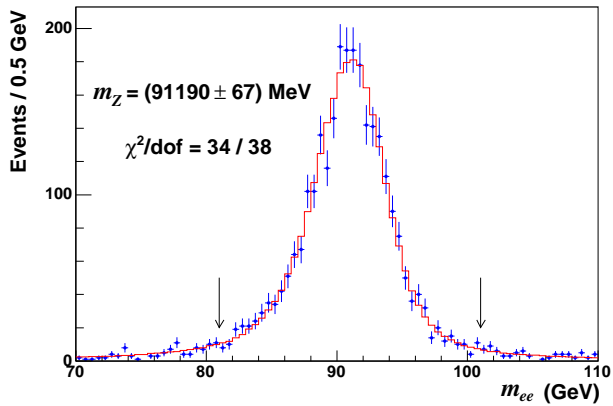}
\caption{The $m_Z$ fit to data (points) in the 81 GeV $< m_{ee} < 101$ GeV 
region (denoted by arrows).  The uncertainty is statistical only.}
\label{fig:zee}
\end{center}
\end{figure}

The model for $Z$ boson production and decay to electrons is the 
same as for the muon decay channel (Section~\ref{sec:zmm}).  We 
use the {\sc resbos} \cite{resbos} event generator and a 
next-to-leading order QED calculation of photon radiation from 
the final-state electrons \cite{wgrad} (Section~\ref{sec:production}).  
We include the virtual photon exchange and $Z/\gamma^*$ interference 
contributions as fixed ``backgrounds'' to the $Z$ boson lineshape, 
and determine $m_Z$ from a binned likelihood fit to the data in the 
range $81~{\rm GeV}~< m_{ee} < 101~{\rm GeV}$ (Fig.~\ref{fig:zee}).

Systematic uncertainties on the $m_Z$ measurement result from
the $E/p$ calibration (29 MeV), calorimeter non-linearity measurement 
(23 MeV), and higher-order QED radiation (14 MeV).  The measured 
$m_Z = [91.190 \pm 0.067 ({\rm stat})]$ GeV is consistent with the 
world-average value $m_Z = (91.188 \pm 0.002)$ GeV \cite{pdg}, given 
the total uncertainty of 78 MeV on the measurement.
\par
The uncorrelated uncertainties in the combination of the $m_Z$ 
and $E/p$ calibrations are the uncertainty on the non-linearity 
parameter $\xi$, the statistical uncertainty on the $m_Z$ 
measurement (0.073\%), and the uncertainty on the $E/p$ calibration 
(0.034\%).  Since the $m_W$ fit relies predominantly on the shape of 
the Jacobian edge of the $m_T$ distribution, the relevant electron 
transverse energies are in the $\approx 40-45$ GeV range.  The 
uncertainty on the energy dependence of the scale from the $Z$ boson 
mass is negligible, as the $\left\langle p_T^e \right\rangle$ is 
about 42 GeV in this sample.  The $E/p$-based calibration involves an 
extrapolation from $\left\langle p_T^e \right\rangle = 39$ GeV, so it 
receives an additional uncertainty contribution of 23 MeV to the $m_W$ 
measurement from the non-linearity parameter $\xi$.  Combining the two 
calibrations, we obtain a total electron energy measurement uncertainty 
of 30 MeV on the $m_W$ measurement in the electron channel.  Of this 
uncertainty, we take 17 MeV to be 100\% correlated with the muon channel 
through the momentum scale uncertainty.


\section{Recoil Measurement} 
\label{sec:recoil}

The recoil $\vec{u}_T$ (Fig. \ref{fig:recoilw}) in a $W$ boson
event results from quark or gluon radiation in the initial state,
and from photon radiation in the initial and final states.  A 
quark or gluon typically fragments into multiple hadrons, which 
are detected in the calorimeter.  Additional energy from the 
underlying event is also measured in the calorimeter and obscures 
the recoil measurement.  Rather than rely on detailed modeling of 
the underlying event, we develop an empirical model of the recoil 
$\vec{u}_T$ using $Z$ boson events, where the four-momentum of the 
$Z$ boson is measured precisely using its leptonic decays.  The 
model of the recoil energy measurement is tuned with these decays 
and applied to $W$ boson events.
\par
We measure the recoil energy using all calorimeter towers except 
those with ionization or shower energy from the charged leptons.  
To reduce potential bias and facilitate our model parametrization, 
we correct the measured energy in each tower for acceptance 
differences resulting from an uncentered beam.  In addition, we 
improve the measurement resolution by correcting for response 
differences between the central and plug calorimeters.

\subsection{Data Corrections}
\label{sec:recoildata}

The data used in this analysis have a relative offset of about 4~mm 
between the beam line and the center of the CDF II detector.  This results 
in a variation in calorimeter acceptance as a function of $\phi$ such 
that the calorimeter towers closest to the beam line have a larger 
acceptance, and thus a larger average measured energy per tower.  The 
variation is largest in the forward region, where the towers are in 
closest proximity to the beam line.  We suppress this azimuthal energy 
variation by applying a threshold on the combined EM and hadronic tower 
$p_T$ of 5 GeV for towers with detector $|\eta| > 2.6$.  The threshold 
strongly suppresses the forward tower energy variation in $W$ boson 
events, while retaining the energy from high-$p_T$ hadronic jets in 
multijet events.  All other towers have EM and hadronic energy 
thresholds of 20 MeV each.
\par
\begin{figure}[!tp]
\begin{center}
\epsfysize = 6.cm
\epsffile{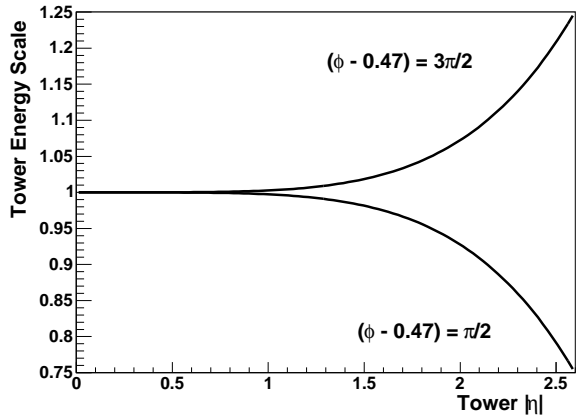}
\caption{The maximal energy correction factors applied to each tower 
as a function of $|\eta|$.  The decreasing curve corresponds to towers 
closest to the beam line, and the increasing curve corresponds to towers
on the opposite side.  The correction factors reduce the $\phi$ variation 
in acceptance due to the beam line displacement from the detector axis. }
\label{fig:beam}
\end{center}
\end{figure}

We reduce the residual azimuthal energy variation by applying 
a multiplicative correction factor to each measured tower energy 
according to the following empirical function (Fig.~\ref{fig:beam}):
\begin{equation}
S_{tower} = 1 - 0.6 (0.32|\eta|)^{4.74} \sin(\phi - 0.47).
\end{equation}

\noindent
This correction is determined using events collected by a 
minimum bias trigger, which requires evidence of an inelastic 
$p\bar{p}$ collision (Section \ref{sec:clc}).  
\par
The relative energy scale between the central and forward
calorimeters is initially determined from the calibration of 
high-$p_T$ hadronic jets.  The relative response has a 
significant energy dependence, however, and the initial 
calibration is not optimized for the low $p_T$ particles 
relevant to the $W$ boson recoil measurement.  Using the 
$E/p$ distribution of charged pions from minimum bias events, 
we find that a relative energy scale of $\approx 12\%$ 
between central and forward calorimeters is appropriate for 
particles with $p_T \lesssim 2$ GeV, the momentum region of a 
typical recoil particle.  To maintain the mean recoil energy 
scale, we scale the central (forward) calorimeter tower 
energies up (down) by 5\% (7\%).  This calibration improves the 
recoil resolution, and thus the statistical precision of the 
$m_W$ fits.  It also minimizes the sensitivity of the recoil 
model to differences in phase space sampled by the selected $W$ 
and $Z$ boson decays.

\subsection{Lepton Tower Removal}
\label{sec:leptonremoval}

The recoil $\vec{u}_T$ is measured as the sum of corrected $\vec{p}_T$ 
in all calorimeter towers (Sec.~\ref{sec:recoildata}), excluding the 
towers in which the lepton(s) deposit energy.  The exclusion of these 
towers also removes some recoil energy from the measurement, thus causing 
a bias in $u_{||}$.  We 
measure this bias from the data and incorporate it in the simulation.
\par
\begin{figure}[!tp]
\begin{center}
\includegraphics*[width=8.5cm]{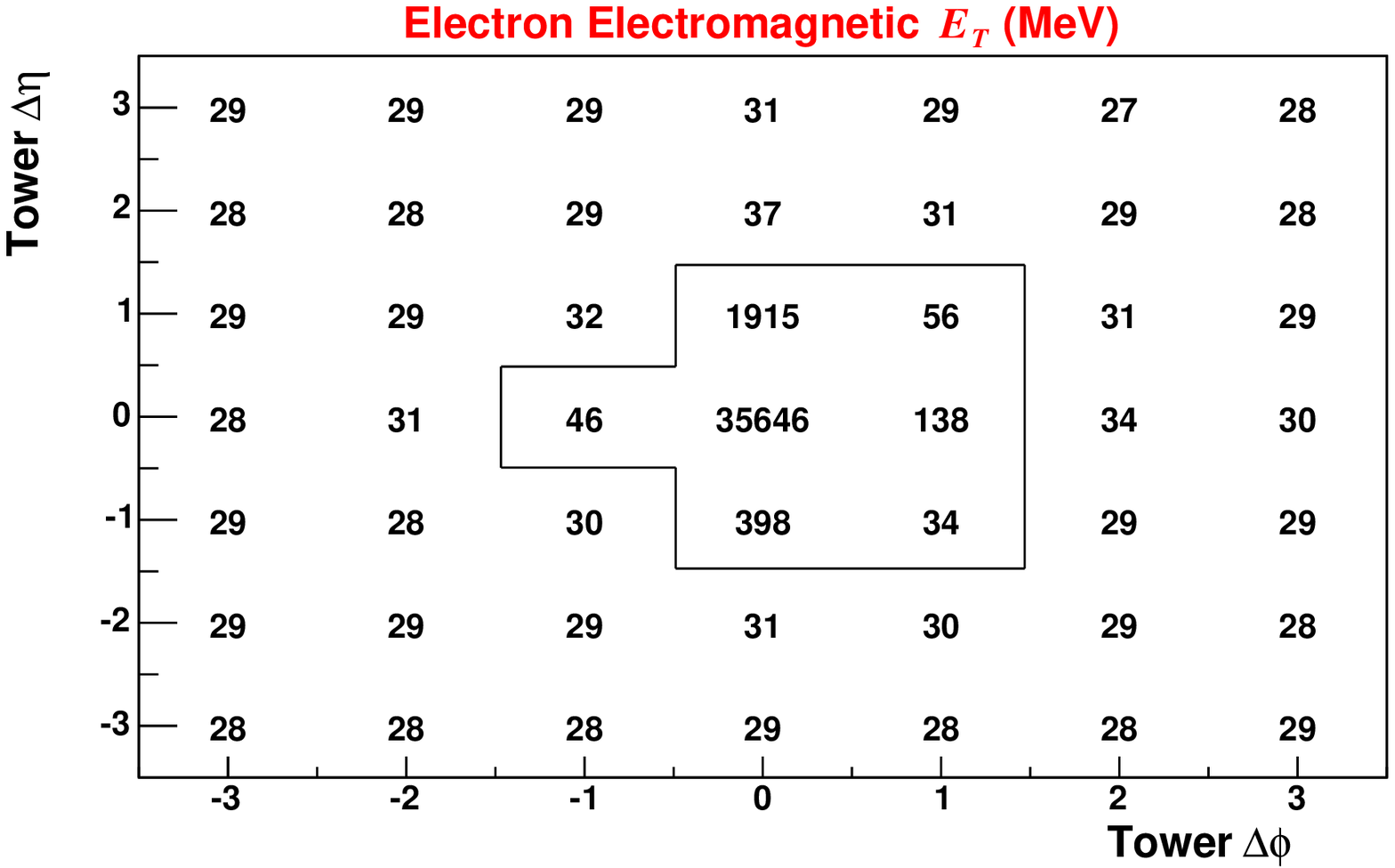}
\includegraphics*[width=8.5cm]{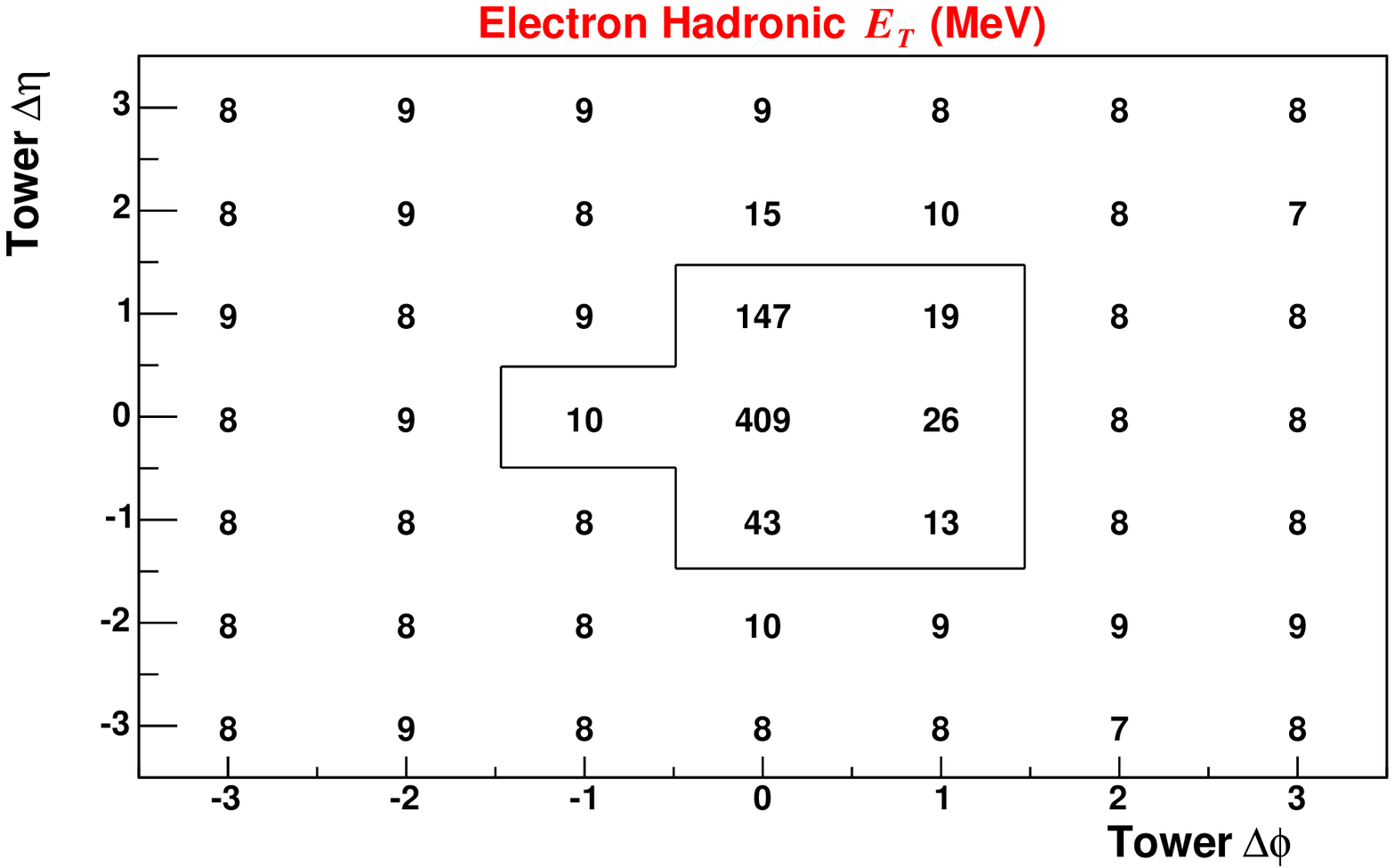}
\caption{The average energy collected in the electromagnetic (top) 
and hadronic (bottom) calorimeters in the vicinity of the electron 
shower in $W$ boson decays.  The differences $\Delta \phi$ and 
$\Delta \eta$ are signed such that positive differences correspond 
to towers closest to the electron position at shower maximum.  The 
central seven towers inside the box are removed from the recoil 
measurement.  Statistical uncertainties on the values outside the 
box are ${\cal{O}}(1$~MeV). }
\label{fig:eleremoval}
\end{center}
\end{figure}
\begin{figure}[!tp]
\begin{center}
\includegraphics*[width=8.5cm]{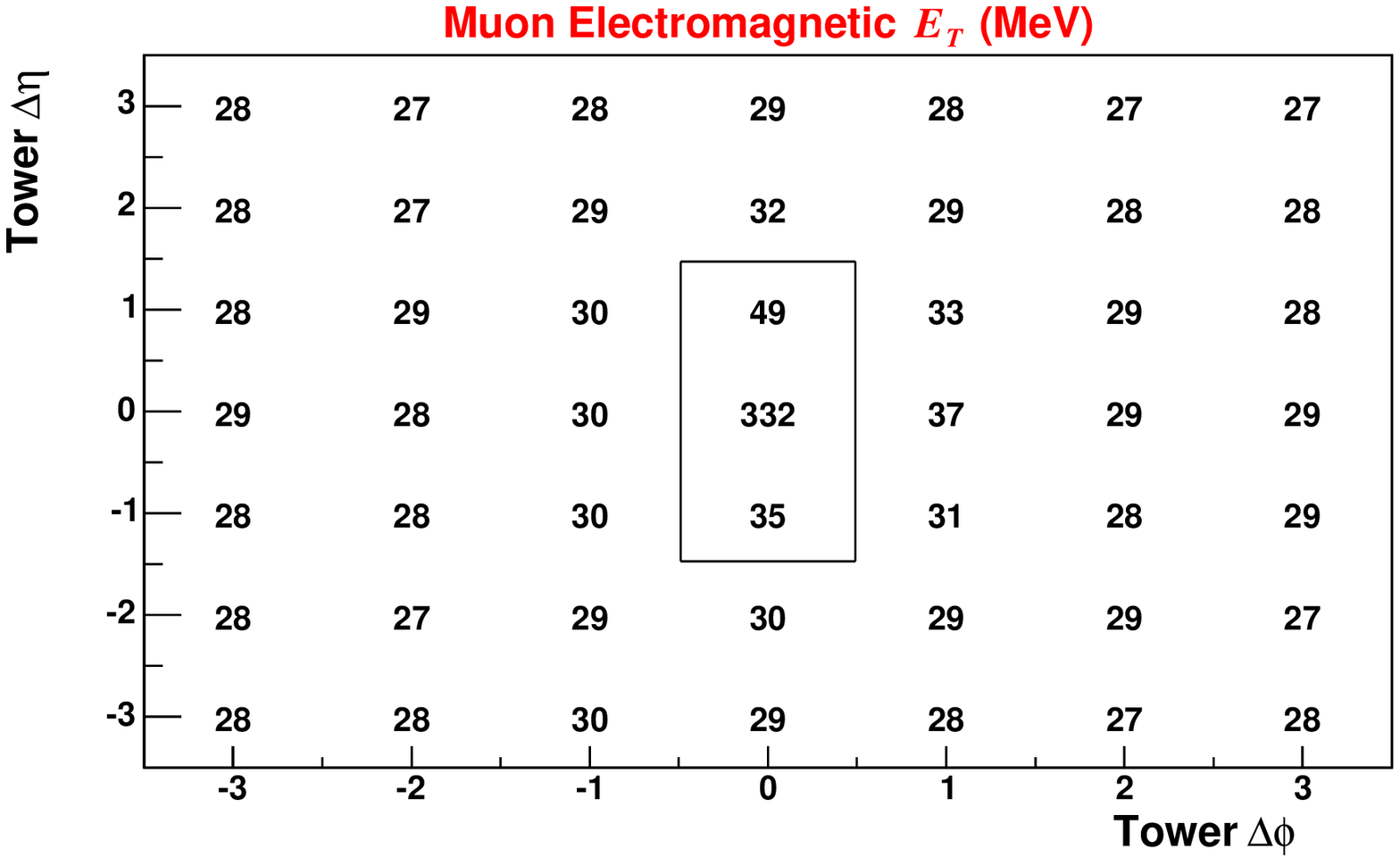}
\includegraphics*[width=8.5cm]{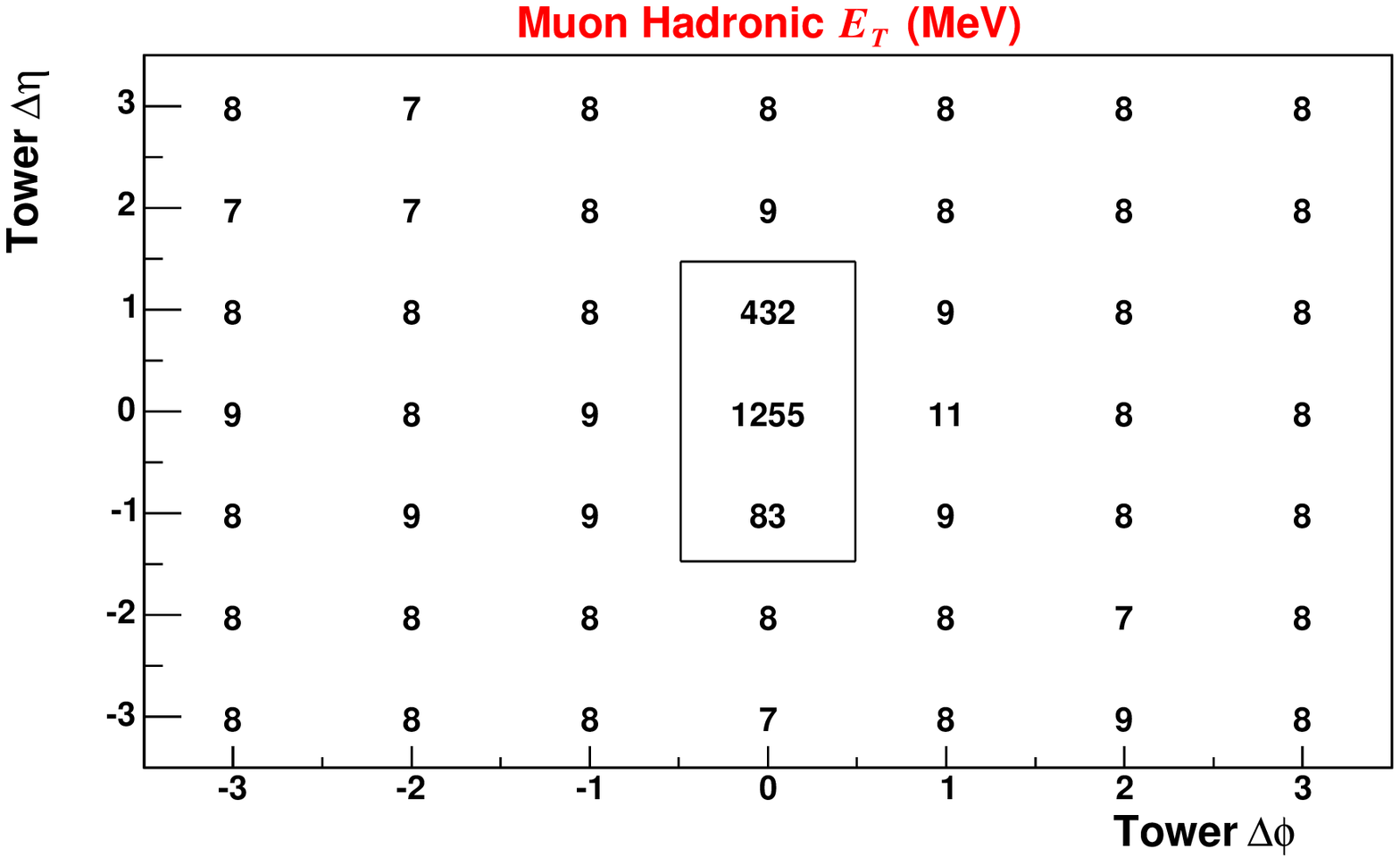}
\caption{The average energy collected in the electromagnetic (top) 
and hadronic (bottom) calorimeters in the vicinity of the muon in 
$W$ boson decays.  The differences $\Delta \phi$ and $\Delta \eta$ 
are signed such that positive differences correspond to towers 
closest to the muon position at the shower maximum detector.  The 
central three towers inside the box are removed from the recoil 
measurement.  Statistical uncertainties on the values outside the 
box are ${\cal{O}}(1$~MeV). }
\label{fig:muremoval}
\end{center}
\end{figure}

An electron shower typically distributes energy to two calorimeter towers, 
but can also contribute to a third tower if the electron is near a 
tower edge.  We remove each tower neighboring the electron's tower, as 
well as the corner towers closest to the electron's CES position 
(Fig.~\ref{fig:eleremoval}).  A muon near a tower edge can cross two 
towers, so we remove the two towers in $\eta$ neighboring the muon's tower 
(Fig.~\ref{fig:muremoval}).  The tower window definitions are motivated by 
the presence of excess energy in a given tower above the background energy 
from the underlying event.

\begin{figure}[!htbp]
\begin{center}
\includegraphics*[width=8.5cm]{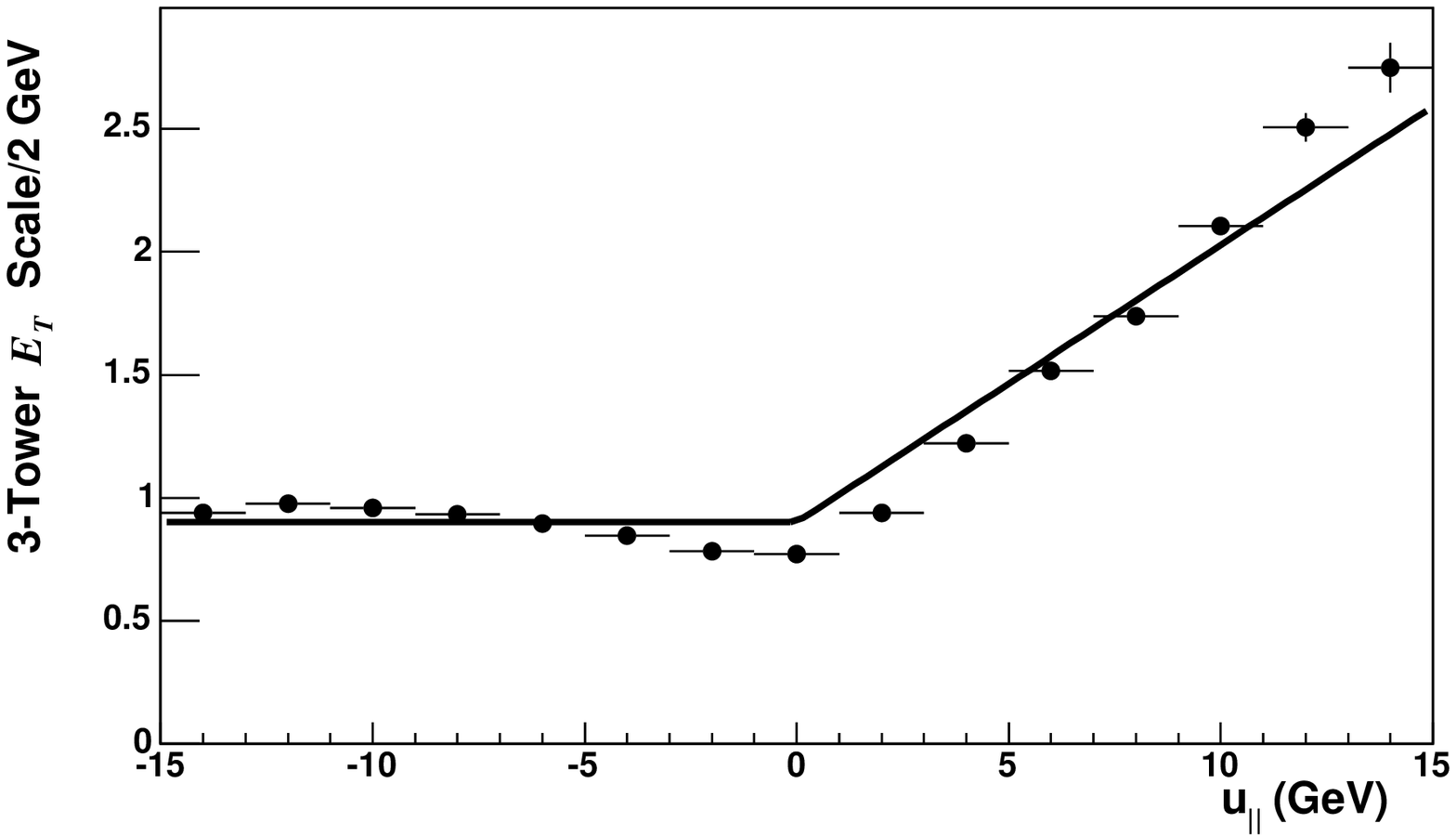}
\includegraphics*[width=8.5cm]{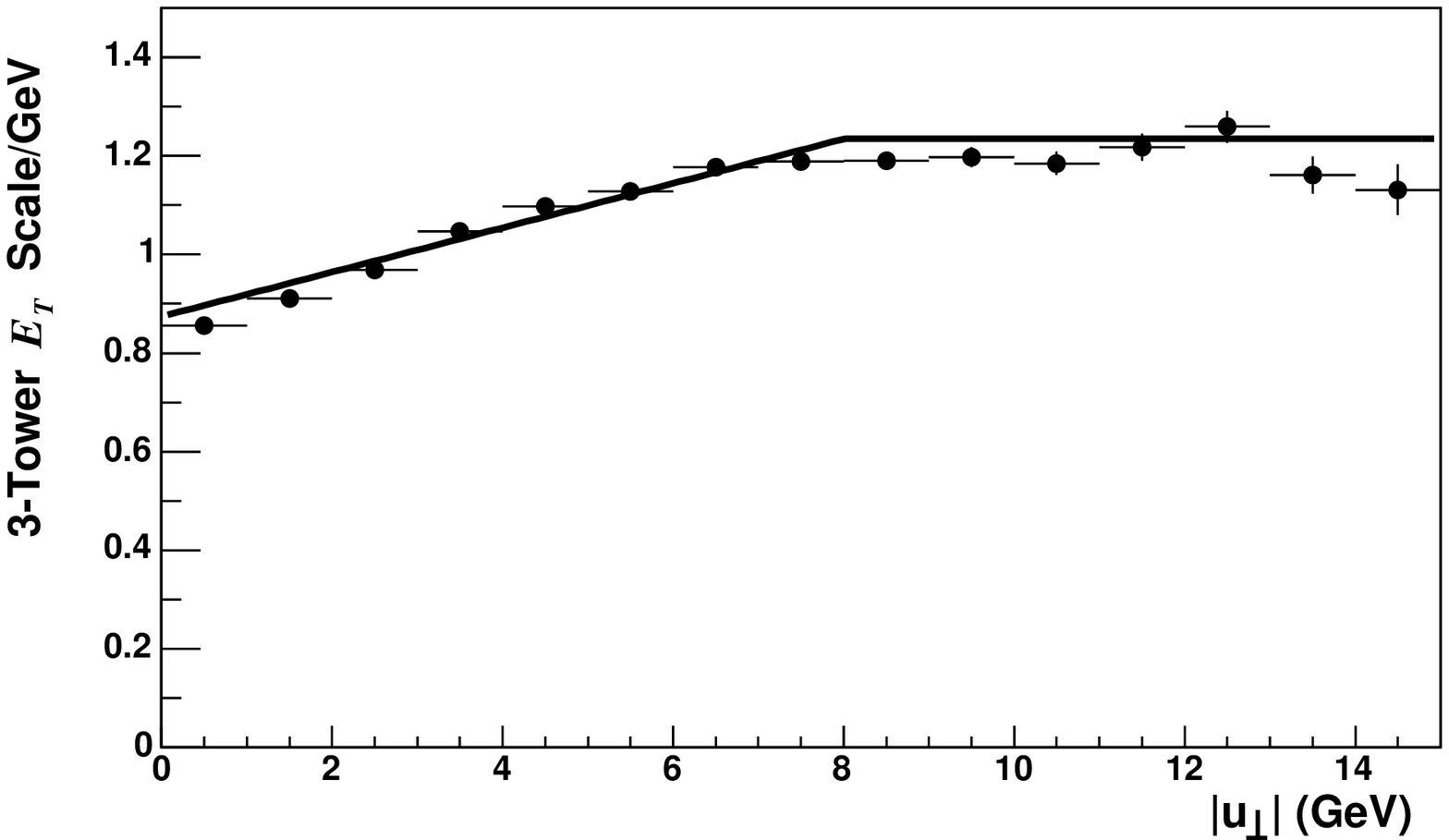}
\includegraphics*[width=8.5cm]{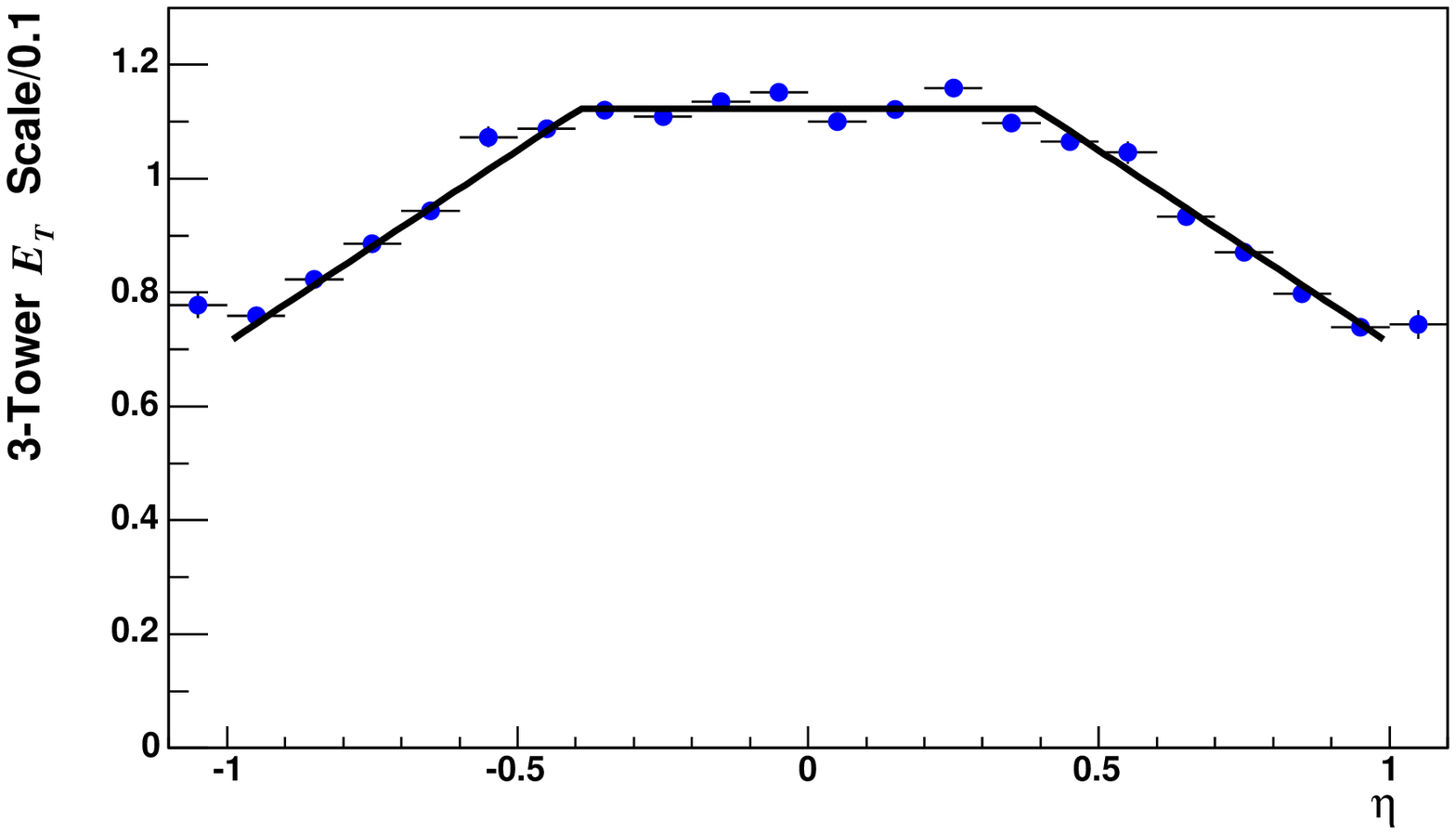}
\caption{The scales applied to the energy in the three removed muon 
towers in the simulation (solid lines), as functions of $u_{||}$ (top), 
$u_{\perp}$ (middle), and $\eta$ (bottom).  The points show the scales 
measured using towers separated in $\phi$ from the muon in $W\rightarrow 
\mu\nu$ data.  The scaling functions for removed electron towers have 
similar shapes. }
\label{fig:muscaling}
\end{center}
\end{figure}

We estimate the recoil energy flow into the excluded towers, denoted 
by $\Delta u_{||}$, using equivalent windows separated in $\phi$ from 
the lepton in $W\rightarrow l\nu$ events.  When simulating a $W$ or $Z$ 
boson event, we correct the simulated $\vec{u}_T$ by a $\Delta u_{||}$ 
taken from the measured distribution.  The simulated $\Delta u_{||}$ 
incorporates its measured dependence on $u_{||}$ and $u_{\perp}$, and 
lepton $|\eta|$.  These dependencies are shown for $W\rightarrow \mu\nu$ 
events in Fig.~\ref{fig:muscaling} and similar functions are defined for 
electrons.  The incorporation of these functions preserves 
$\left\langle \Delta u_{||} \right\rangle$, which is 269 MeV for 
electrons and 112 MeV for muons (with negligible statistical uncertainty). 

To estimate the systematic uncertainty associated with modeling the 
tower removal, we study the variation of $\Delta u_{||}$ in the data as 
a function of the $\phi$ separation from the lepton of the equivalent 
tower window.  We take half the variation as a systematic uncertainty:  
8 (5) MeV for removed electron (muon) towers.  To confirm our estimate 
of this uncertainty, we remove an additional window azimuthally 
opposite to the lepton ($\Delta\phi = \pi$), incorporate its model into 
the simulation, and compare the resulting simulation and data $u_{||}$ 
distributions.  We find the differences to be consistent within our 
quoted uncertainties.

\subsection{Recoil Model Parametrization}

The recoil consists of three separate components:  radiation in 
the $W$ or $Z$ boson production; radiation from the spectator 
partons; and energy from additional $p\bar{p}$ collisions in a 
given bunch crossing.  We use the {\sc resbos} \cite{resbos} 
generator to predict the net $p_T$ distribution of radiation in 
the $W$ or $Z$ boson production, and minimum bias data for the 
$p_T$ distribution from spectator partons and additional 
interactions.  The parameters for the detector response to the 
recoil are measured in $Z$ boson events.
\par
\begin{figure}[!tp]
\begin{center}
\epsfysize = 4.cm
\epsffile{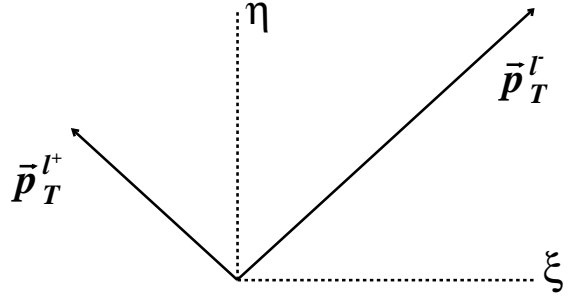}
\caption{The definitions of the $\eta$ and $\xi$ axes in 
$Z$ boson events.  The quark and gluon radiation from the boson 
production points predominantly in the $-\eta$ direction. }
\label{fig:axes}
\end{center}
\end{figure}

To facilitate tuning of the recoil model, we define axes such 
that quark and gluon radiation lies predominantly along one axis, 
denoted as the ``$\eta$'' axis (Fig. \ref{fig:axes}).  This axis 
is chosen to be the angular bisector of the two leptons, whose 
angles are precisely measured.  The orthogonal axis is denoted as 
the ``$\xi$'' axis.

\subsubsection{Recoil Energy Scale}

We tune the simulation to match the observed detector response
to the recoil radiation.  The recoil response is defined as 
$R \equiv \vec{u}_T^{~meas}\cdot\vec{u}_T^{~true}/u_T^{true}$, 
where $\vec{u}_T^{~true} = -\vec{p}_T^{~Z}$ is the generated net 
$\vec{p}_T$ of the initial-state-radiation, and $\vec{u}_T^{~meas}$ 
is the reconstructed vector of this transverse momentum.  
\par
\begin{figure}[!tp]
\begin{center}
\epsfysize = 6.cm
\epsffile{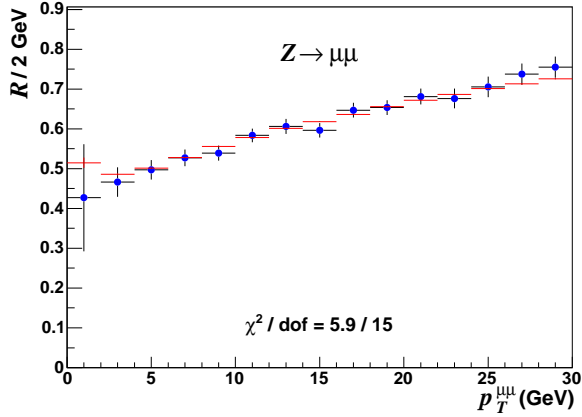}
\caption{The mean $-\vec{u}_T^{~meas}\cdot\vec{p}_T^{~\mu\mu}/p_T^{\mu\mu}$, 
which approximates the recoil response $R$, as a function of dimuon $p_T$ 
for $m_{\mu\mu}$ in the $Z$ boson mass region.  The simulation (solid) uses 
parameters fit from the electron and muon $p^{ll}_{\eta} + u_{\eta}$ 
distributions, and models the data (circles) well. }
\label{fig:recoilscale}
\end{center}
\end{figure}

To simulate the measured recoil, we parametrize the response as 
\begin{equation}
\label{eq:recoilscale}
R(A,B) = A \ln(u_T^{true} + B)/\ln(15 + B),
\end{equation}

\noindent
where $u_T^{true}$ is in units of GeV, and $A$ and $B$ are constants 
determined from the data.  Figure \ref{fig:recoilscale} shows 
$-\vec{u}_T^{~meas}\cdot\vec{p}_T^{~\mu\mu}/p_T^{\mu\mu}$, which 
approximates $R$, for $Z$ boson decays to muons.  The response $R$ 
is less than 1 due to calorimeter energy loss from particles curling 
in the tracker, particles passing through calorimeter cracks, and 
non-linearity of the hadronic calorimeter response. 
\par
\begin{figure}[!tp]
\begin{center}
\epsfysize = 6.cm
\includegraphics*[width=8.5cm]{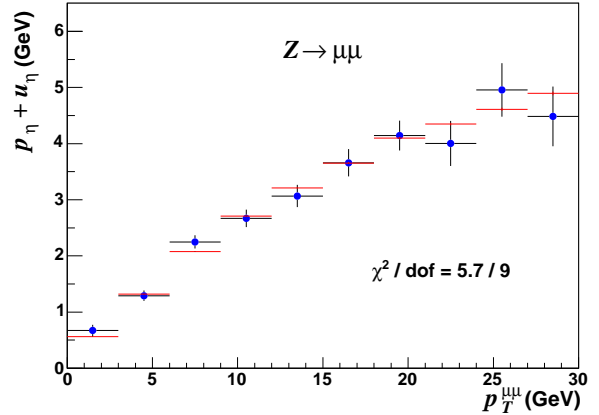}
\includegraphics*[width=8.5cm]{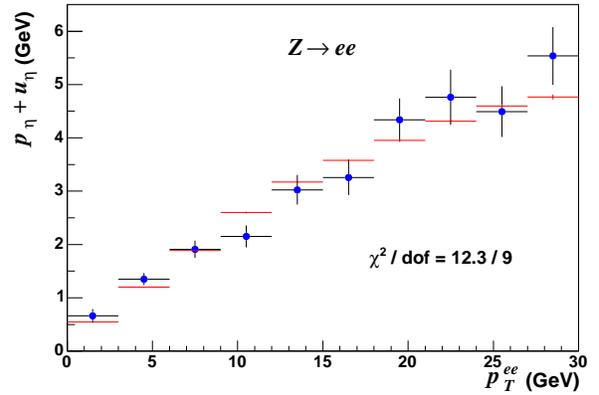}
\caption{The simulation (solid) and data (circles) distributions 
of $(p^{ll}_{\eta} + u_{\eta})$ for $Z$ boson decays to muons 
(top) and electrons (bottom).  The combined $\chi^2$ is minimized 
in the fit for the recoil detector response parameters. }
\label{fig:recoilscalefit}
\end{center}
\end{figure}

Projecting the lepton momenta and the recoil along the $\eta$ axis to 
obtain $p^{ll}_{\eta}$ and $u_{\eta}$, the sum $p^{ll}_{\eta} + u_{\eta}$ 
is sensitive to $R$.  This sum is zero for $R = 1$, and positive for 
$R < 1$.  We measure $A = 0.635 \pm 0.007 ({\rm stat})$ and 
$B = 6.68 \pm 1.04 ({\rm stat})$ by minimizing the combined $\chi^2$ of 
the electron and muon $(p^{ll}_{\eta} + u_{\eta})$ distributions 
as a function of $p_T^{ll}$ (Fig.~\ref{fig:recoilscalefit}).  We 
determine $A$ and $B$ with the $(p^{ll}_{\eta} + u_{\eta})$ distribution 
rather than the distribution of Fig.~\ref{fig:recoilscale} because 
$(p^{ll}_{\eta} + u_{\eta})$ is well-defined as $p_T^{Z} \rightarrow 0$~GeV, 
while $R$ is not.  The parameters $A$ and $B$ are statistically uncorrelated 
by construction.  We apply $R(A,B)$ to the generated recoil $\vec{u}_T$ in 
simulated $W$ and $Z$ boson events.  

\subsubsection{Spectator and Additional $p\bar{p}$ Interactions}

\begin{figure}[!tp]
\begin{center}
\epsfysize = 6.cm
\epsffile{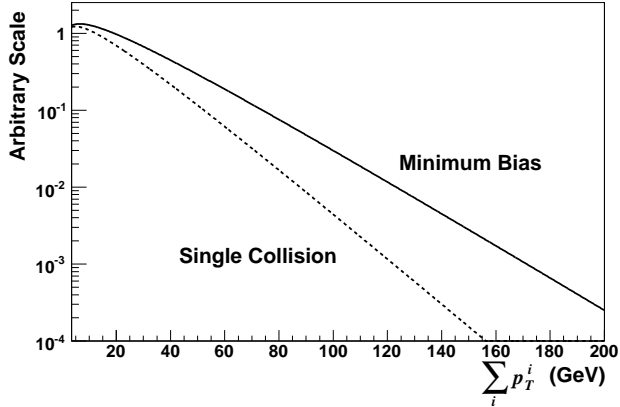}
\caption{The $\sum_i p_T^i$ distribution for minimum bias data (solid)
and a single $p\bar{p}$ collision (dashed), as derived from the minimum
bias distribution. }
\label{fig:sumetfunc}
\end{center}
\end{figure}

The net $\vec{p}_T$ flow from spectator quarks and additional interactions 
is negligible due to momentum conservation.  However, detector resolution
causes its measurement to generally be non-zero.  The resolution is
predominantly determined by the energy sampling in the calorimeter, 
and we expect it to increase as the square root of the scalar sum $\sum p_T$ 
of the calorimeter tower $p_T$.  We plot the width of the \met distribution, 
projected along the $x$ and $y$ axes, as a function of the $\sum p_T$ in 
minimum bias data.  We parametrize the dependence as a power law, with the 
fitted result:
\begin{equation}
\label{eqn:sigmaxy}
\sigma_{x,y} = 0.3842 \left(\sum p_T\right)^{0.5333}~{\rm GeV}, 
\end{equation}

\noindent
where $\sum p_T$ is defined in units of GeV.  The distribution of 
$\sum p_T$ from additional interactions, denoted $P_{MB}$, is parametrized 
as (Fig.~\ref{fig:sumetfunc}):
\begin{equation}
\label{eqn:pmb}
P_{MB}\left(\sum p_T\right) \propto \left(\sum p_T\right)^{0.325} e^{-\sum p_T/19.98},
\end{equation}

\noindent
with constants obtained from a fit to the minimum bias data.  In our 
simulation, we draw a value of $\sum p_T$ from this distribution, for 
the fraction of events containing at least one $p\bar{p}$ collision 
beyond that producing the $W$ or $Z$ boson.  This fraction is calculated 
from the average instantaneous luminosity of $2.137~(2.014) \times 10^{31}$ 
cm$^{-2}$ s$^{-1}$ for $W$ and $Z$ boson data in the muon (electron) 
channel, and the assumed instantaneous luminosity per additional collision 
($3.3 \times 10^{31}$ cm$^{-2}$ s$^{-1}$).  

The observed $\sum p_T$ from spectator partons in the $p\bar{p}\rightarrow W$ 
or $Z$ boson interaction is modeled from the minimum bias data, which correspond 
to one or more $p\bar{p}$ collisions.  We deconvolute the $\sum p_T$ spectrum of 
Eqn.~\ref{eqn:pmb} with the distribution of the number of collisions in minimum 
bias data to derive the following single-collision $\sum p_T$ distribution 
$P_{1-col.}$ applicable to $W$ or $Z$ production (Fig.~\ref{fig:sumetfunc}): 
\begin{equation}
\label{eqn:p1}
P_{1-col.}\left(\sum p_T\right) \propto \left(\sum p_T\right)^{0.345} e^{-\sum p_T/14.27}.
\end{equation}

The $\sum p_T$ produced in a single minimum bias collision can be different 
from that produced by the spectator partons in $W$ or $Z$ boson production.  In 
order to allow for a difference, we scale the $\sum p_T$ drawn from the 
single-collision spectrum by a parameter $N_{W,Z}$, which we tune on the $Z$ 
boson data.
\par
With this model, the $\sum p_T$ in a simulated event is obtained by adding
the contributions from the spectator partons and the additional interactions.  
The corresponding recoil resolution is generated according to Eqn. \ref{eqn:sigmaxy}, 
with a single tunable parameter $N_{W,Z}$.

\subsubsection{Recoil Energy Resolution}

\begin{figure}[!tp]
\begin{center}
\epsfysize = 6.cm
\includegraphics*[width=8.5cm]{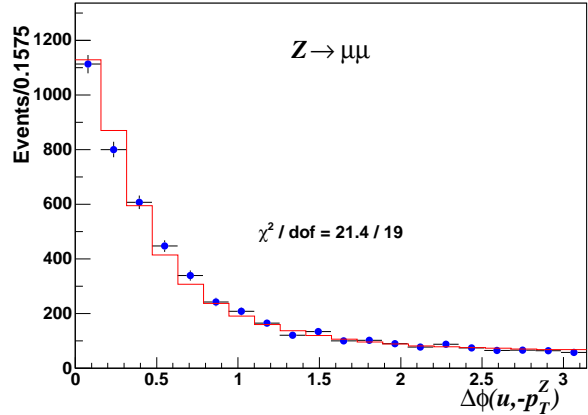}
\includegraphics*[width=8.5cm]{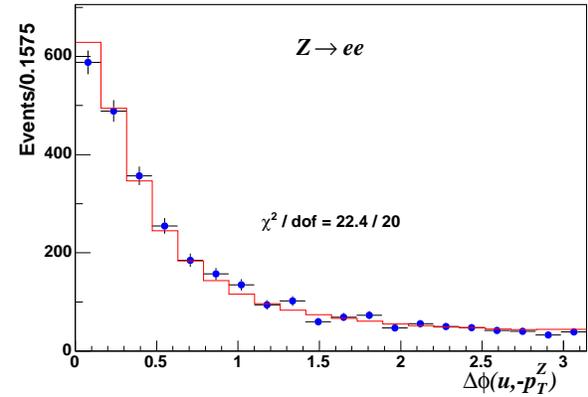}
\caption{The angle between the measured recoil and the direction 
opposite $p_T^Z$, for simulation (histogram) and data (circles) 
events where the $Z$ boson decays to muons (top) or electrons 
(bottom).  The $\chi^2$ from the $Z\rightarrow \mu \mu$ sample is 
minimized in the fit to the recoil angular resolution.  The 
corresponding uncertainty on $m_W$ is negligible. }
\label{fig:recoilangle}
\end{center}
\end{figure}

The measurement of the quark and gluon radiation is affected by 
detector energy resolution, which in turn affects the measured
recoil direction.  We model the recoil angular resolution as a 
Gaussian distribution with $\sigma_{\phi} = 0.14 \pm 0.01 ({\rm stat})$, 
determined from fits to the $\Delta \phi(\vec{u}_T,-\vec{p}_T^{~ll})$ 
distribution in $Z$ boson events (Fig. \ref{fig:recoilangle}).  Since 
the lepton directions are precisely measured, the width of the peak 
at $\Delta \phi = 0$ is dominated by the recoil angular resolution.

The energy resolution of the quark and gluon radiation is predominantly 
determined by stochastic fluctuations in the hadronic calorimeter, which
motivate the functional form $\sigma_{u_T} \propto \sqrt{u_T^{true}}$.  We measure 
the proportionality constant $s_{hard}$ using $Z$ boson data.
\par
\begin{figure}[!tp]
\begin{center}
\epsfysize = 6.cm
\includegraphics*[width=8.5cm]{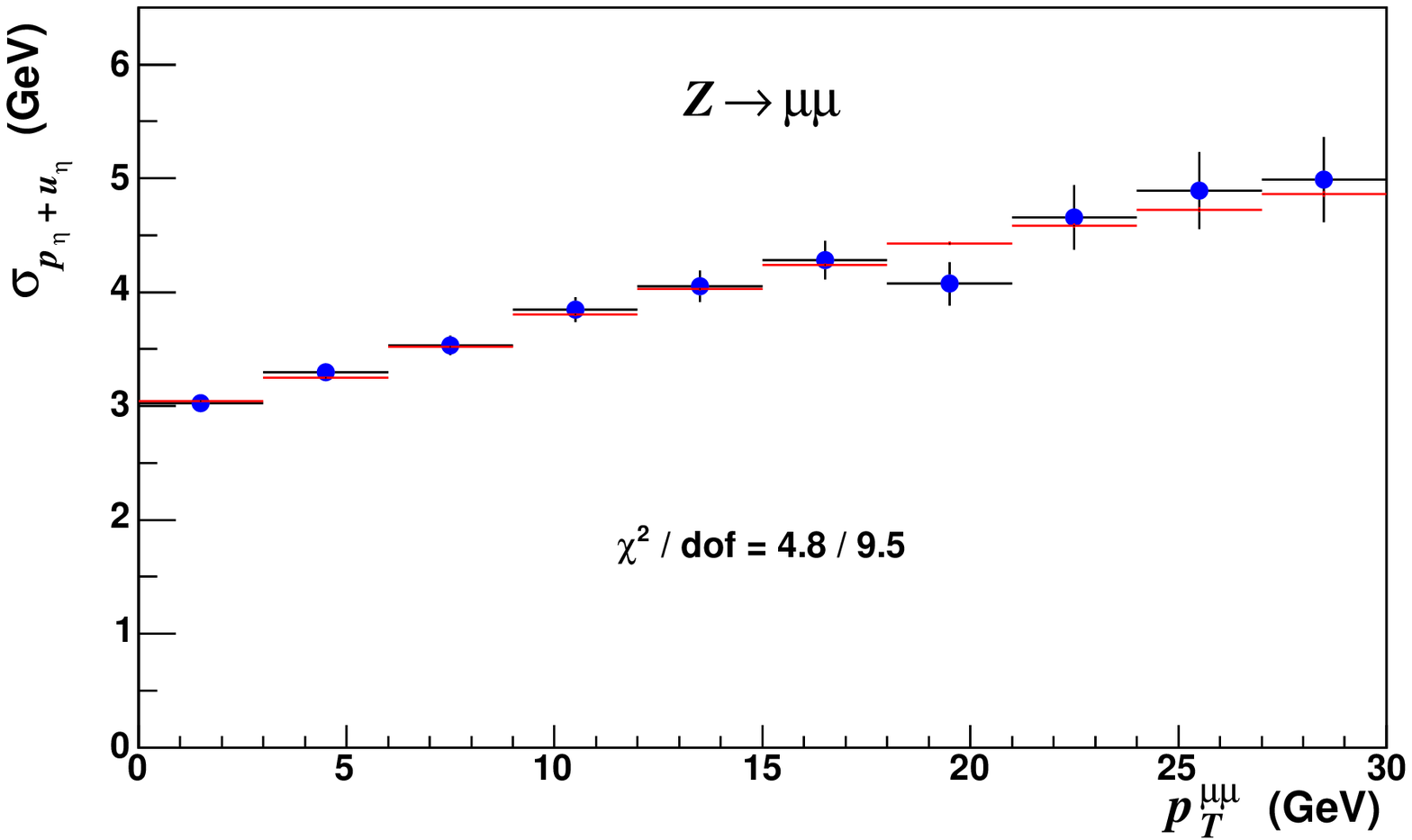}
\includegraphics*[width=8.5cm]{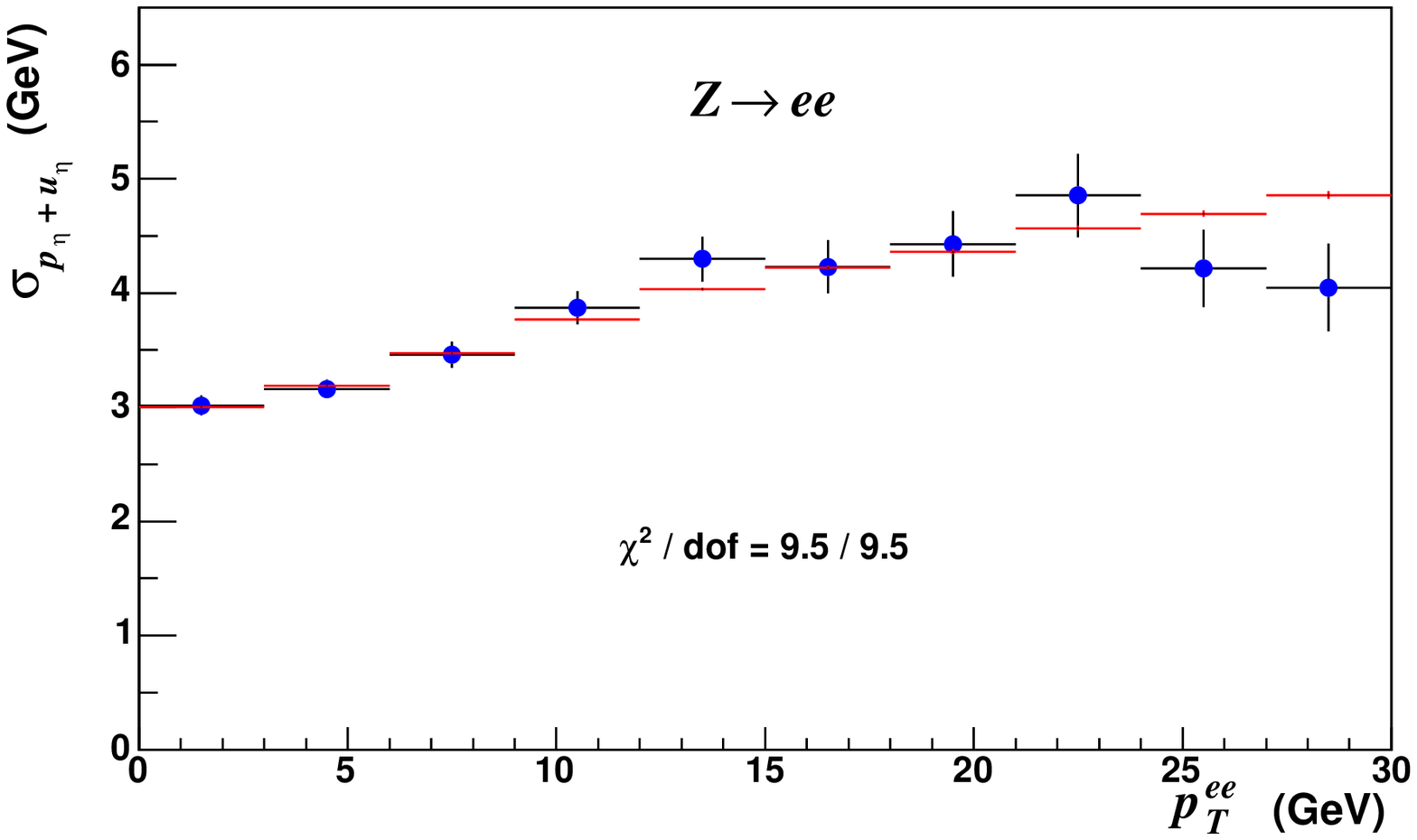}
\includegraphics*[width=8.5cm]{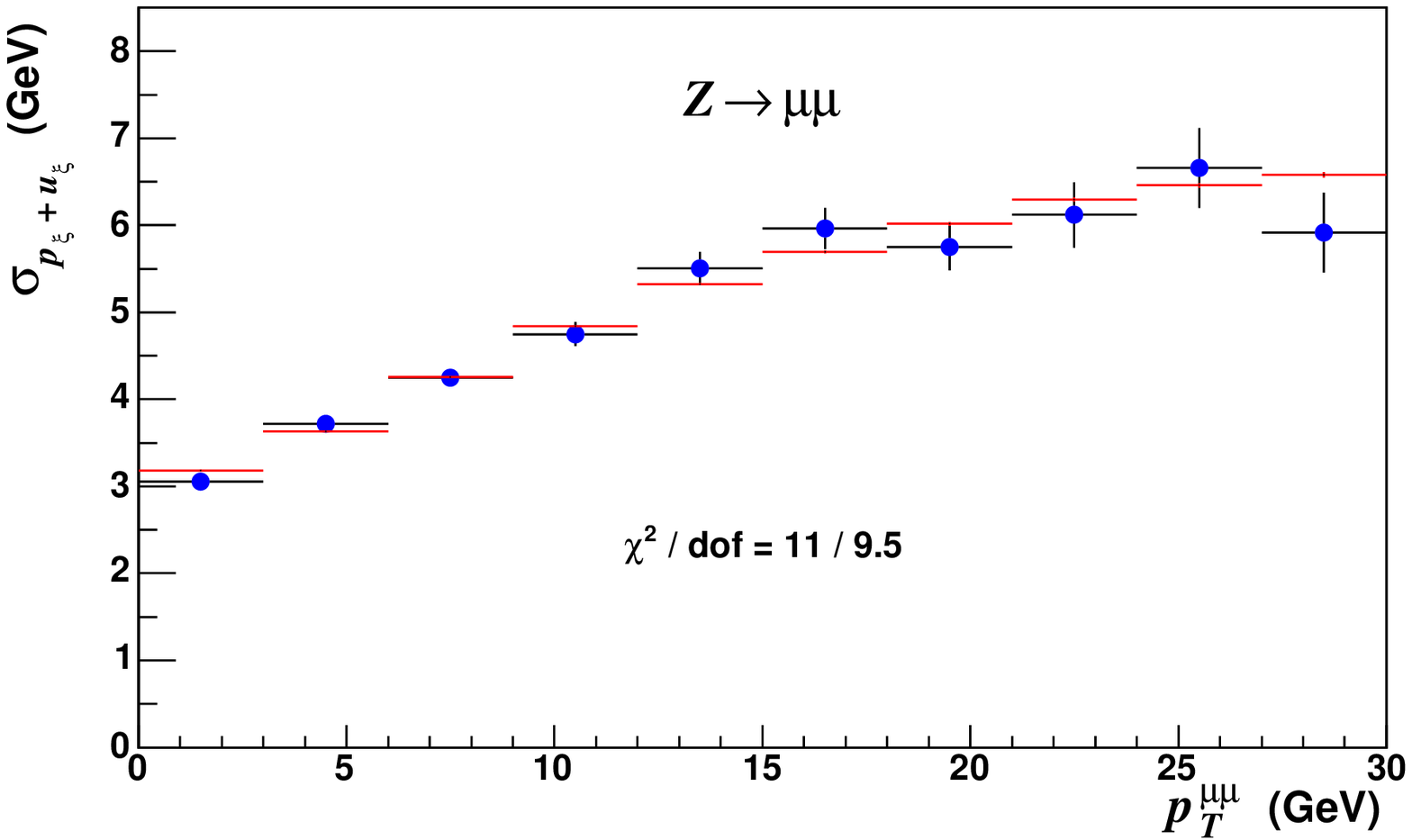}
\includegraphics*[width=8.5cm]{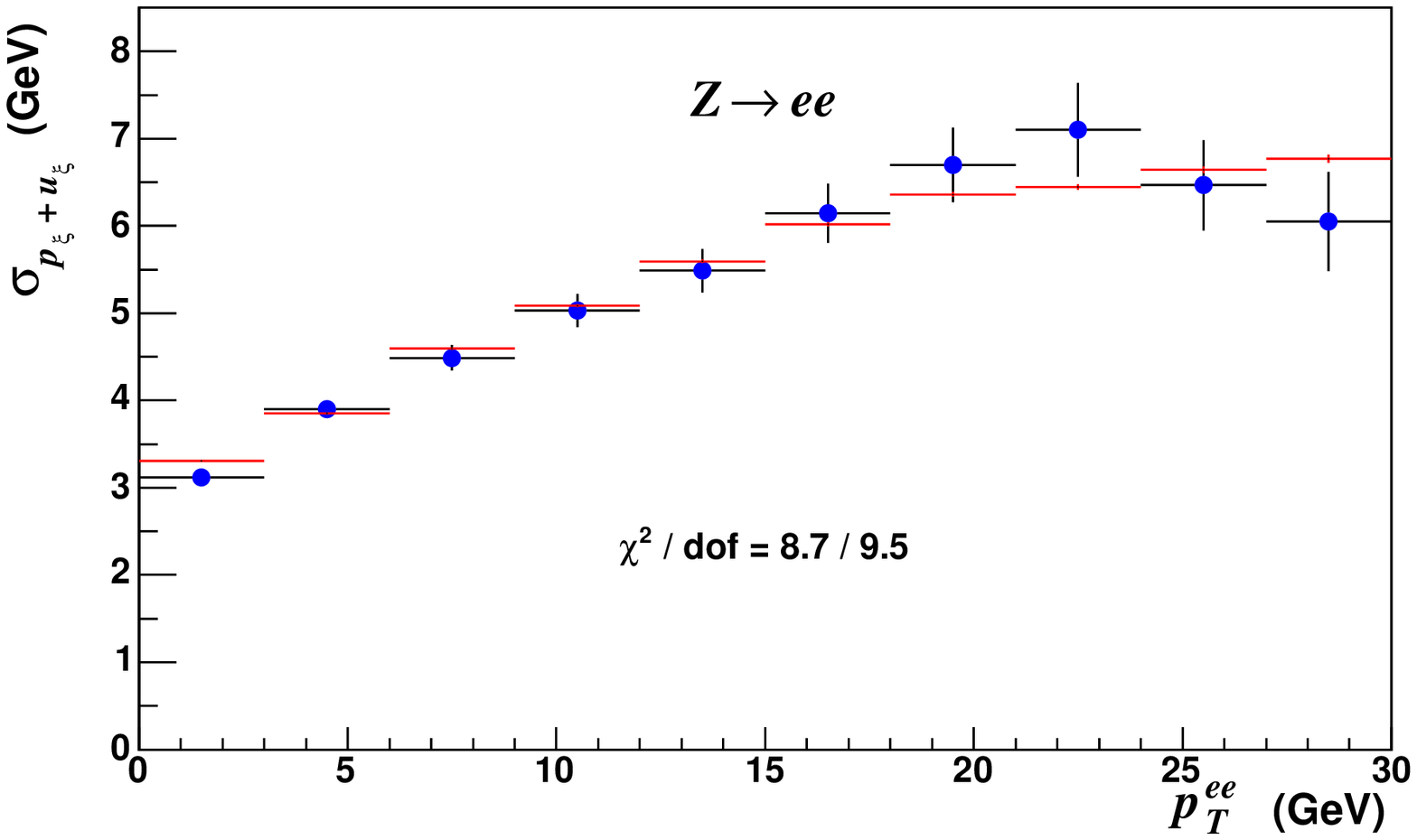}
\caption{The simulation (solid) and data (circles) $p^{ll}_{\eta} + u_{\eta}$
(top) and $p^{ll}_{\xi} + u_{\xi}$ (bottom) resolutions for $Z$ boson decays
to muons and electrons.  The sum of the four $\chi^2$ values is minimized in 
the fit for the recoil resolution parameters $N_{W,Z}$ and $s_{hard}$.  Since 
there are four distributions and two fit parameters, each distribution contributes 
half a degree of freedom to the fit. }
\label{fig:recoilres}
\end{center}
\end{figure}

To tune $s_{hard}$ and $N_{W,Z}$, we project the momentum imbalance
$\vec{p}_T^{~ll} + \vec{u}_T$ along the $\eta$ and $\xi$ axes in $Z$ 
boson decays (Fig. \ref{fig:recoilres}).  The width of these projections 
as a function of $p_T^{ll}$ provides information on $N_{W,Z}$ and 
$s_{hard}$.  At low $p_T^Z$ the resolution is dominantly affected by 
$N_{W,Z}$, with the $s_{hard}$ contribution increasing as the boson 
$p_T$ increases.  We compare the widths of the data and simulation 
projections as a function of $p_T^{ll}$ and compute the $\chi^2$.  
Minimizing this $\chi^2$, we obtain $N_{W,Z} = 1.167 \pm 0.026 ({\rm stat})$ 
and \mbox{$s_{hard} = [0.828 \pm 0.028 ({\rm stat})]$ GeV$^{1/2}$}. The tuning 
is performed such that the statistical uncertainties on these parameters are 
uncorrelated.

\subsection{Recoil Model Cross-Checks}
\label{sec:recoilcheck}

The full recoil model, with parameters tuned from $Z$ boson events, is 
applied to the simulated $W$ boson sample.  We compare the data to the 
predictions of distributions that can affect the final mass measurement:  
the projections of the recoil along ($u_{||}$) and perpendicular to 
($u_{\perp}$) the charged lepton; and the total recoil $u_T$. 
\par
The $u_{||}$ distribution is directly affected by the measurements of 
lepton efficiency as a function of $u_{||}$ (Figs. \ref{fig:uparmu} and
\ref{fig:uparele}) and the modeling of lepton tower removal (Figs. 
\ref{fig:eleremoval} and \ref{fig:muremoval}).  The $u_{||}$ is 
also sensitive to the boson $p_T$ (Sec.~\ref{sec:bosonpt}) and decay 
angular distributions, and to the recoil response and resolutions.  
\par
Since $u_T$ is much less than the charged lepton $p_T$ for our event 
selection, \met~$\approx |p_T + u_{||}|$.  Thus, $m_T$ can be written as:
\begin{equation}
m_T \approx 2 p_T \sqrt{1 + u_{||}/p_T} \approx 2 p_T + u_{||}.
\end{equation}

\begin{figure}[!tp]
\begin{center}
\epsfysize = 6.cm
\includegraphics*[width=8.5cm]{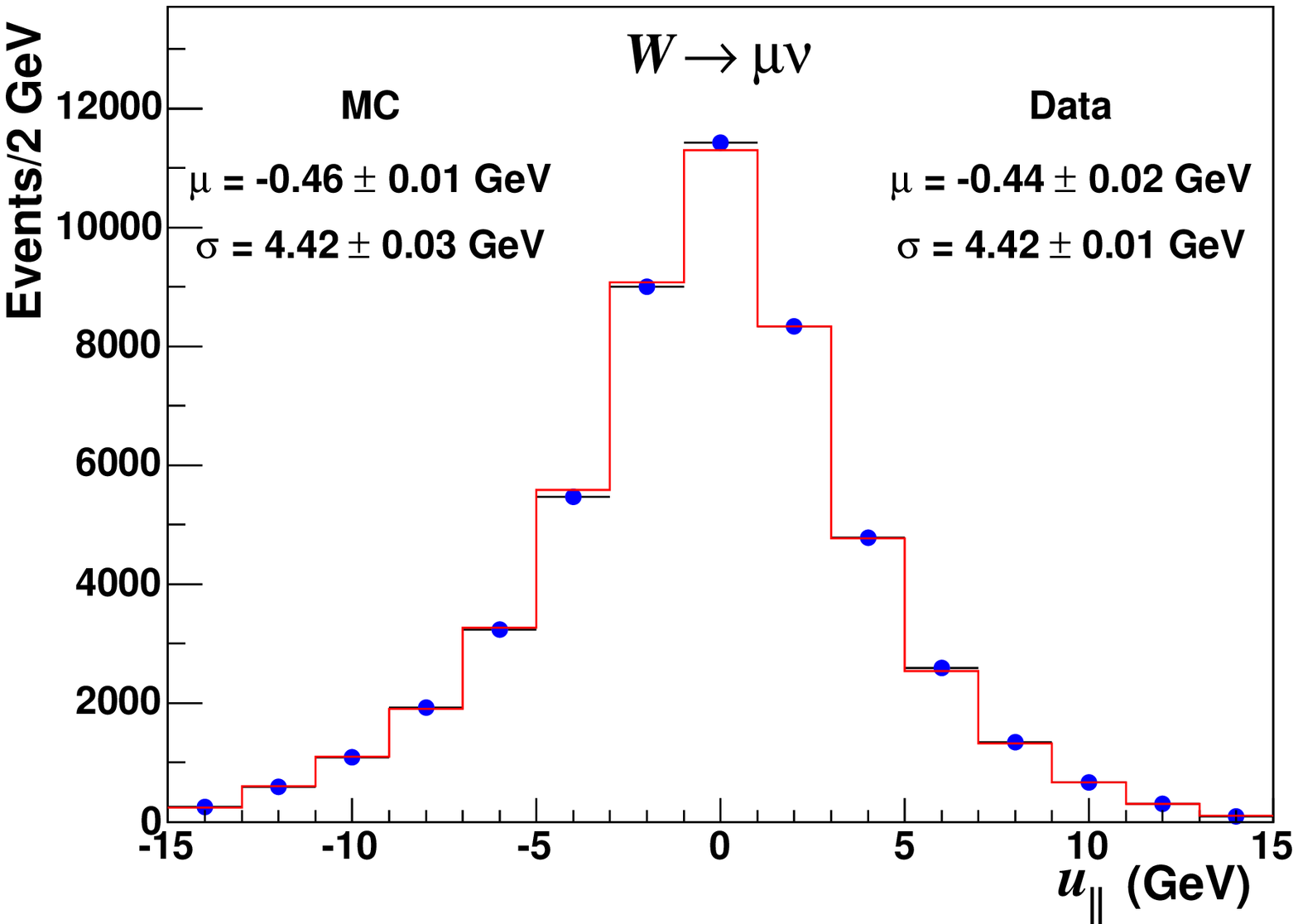}
\includegraphics*[width=8.5cm]{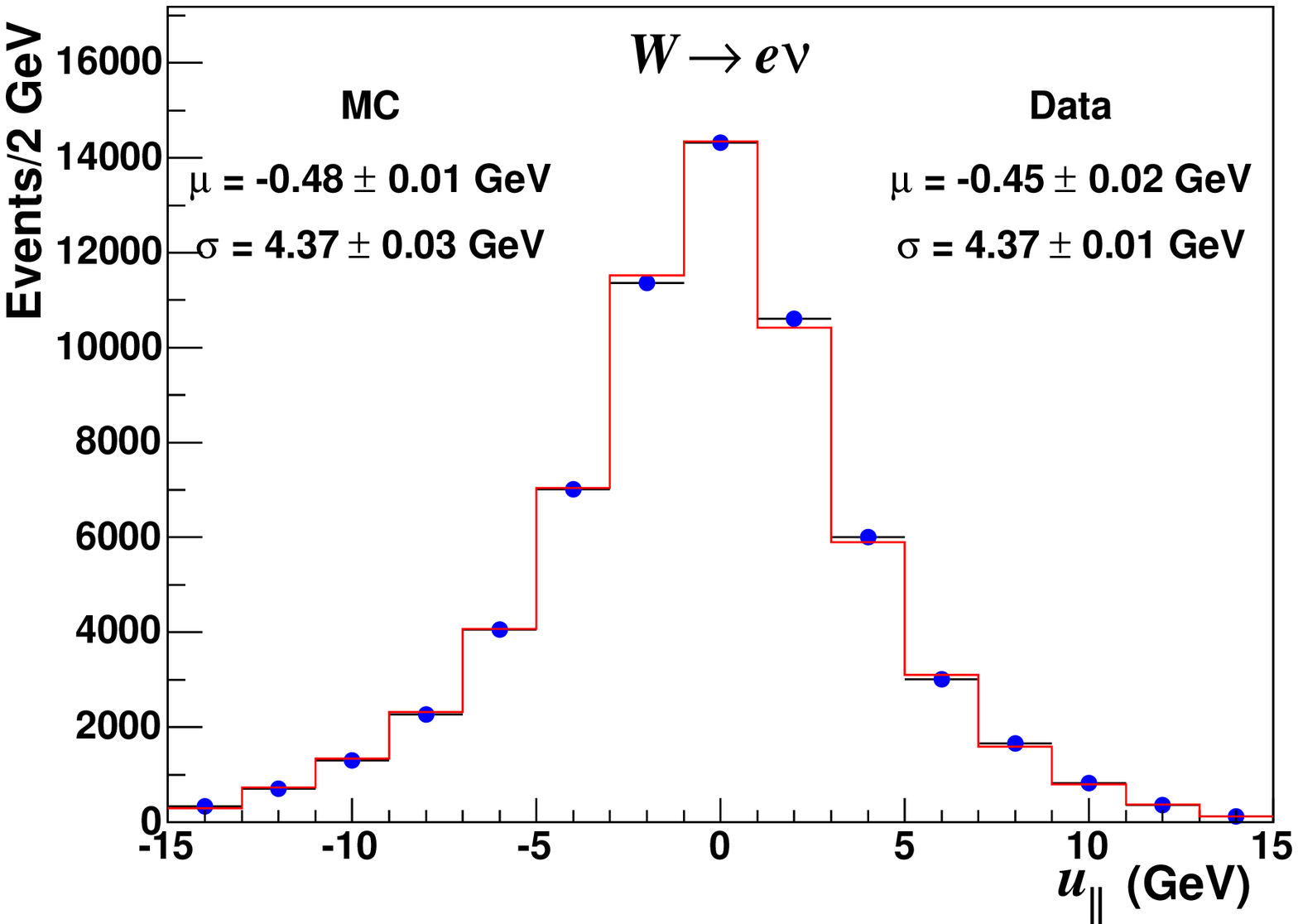}
\caption{The simulation (solid) and data (circles) $u_{||}$ 
distributions for $W$ boson decays to $\mu\nu$ (top) and 
$e\nu$ (bottom).  The simulation uses parameters fit from 
$Z$ boson data, and the uncertainty on the simulation is 
due to the statistical uncertainty on these parameters.  The 
data mean ($\mu$) and RMS ($\sigma$) are well-modeled by 
the simulation.}
\label{fig:upar}
\end{center}
\end{figure}

\noindent
To a good approximation, any bias in $u_{||}$ directly enters as a bias in 
the $m_T$ fit.  We compare the $u_{||}$ distributions in data and simulation 
$u_{||}$ in Fig. \ref{fig:upar}, and observe no evidence of a bias at the 
level of the data statistics and simulation systematics derived from the 
recoil model parameters.   All backgrounds (Section \ref{sec:background}) 
are included in the comparison, except $W\rightarrow \tau\nu$, which has 
similar distributions to the other $W$ leptonic decays.

\begin{figure}[!tp]
\begin{center}
\epsfysize = 6.cm
\includegraphics*[width=8.5cm]{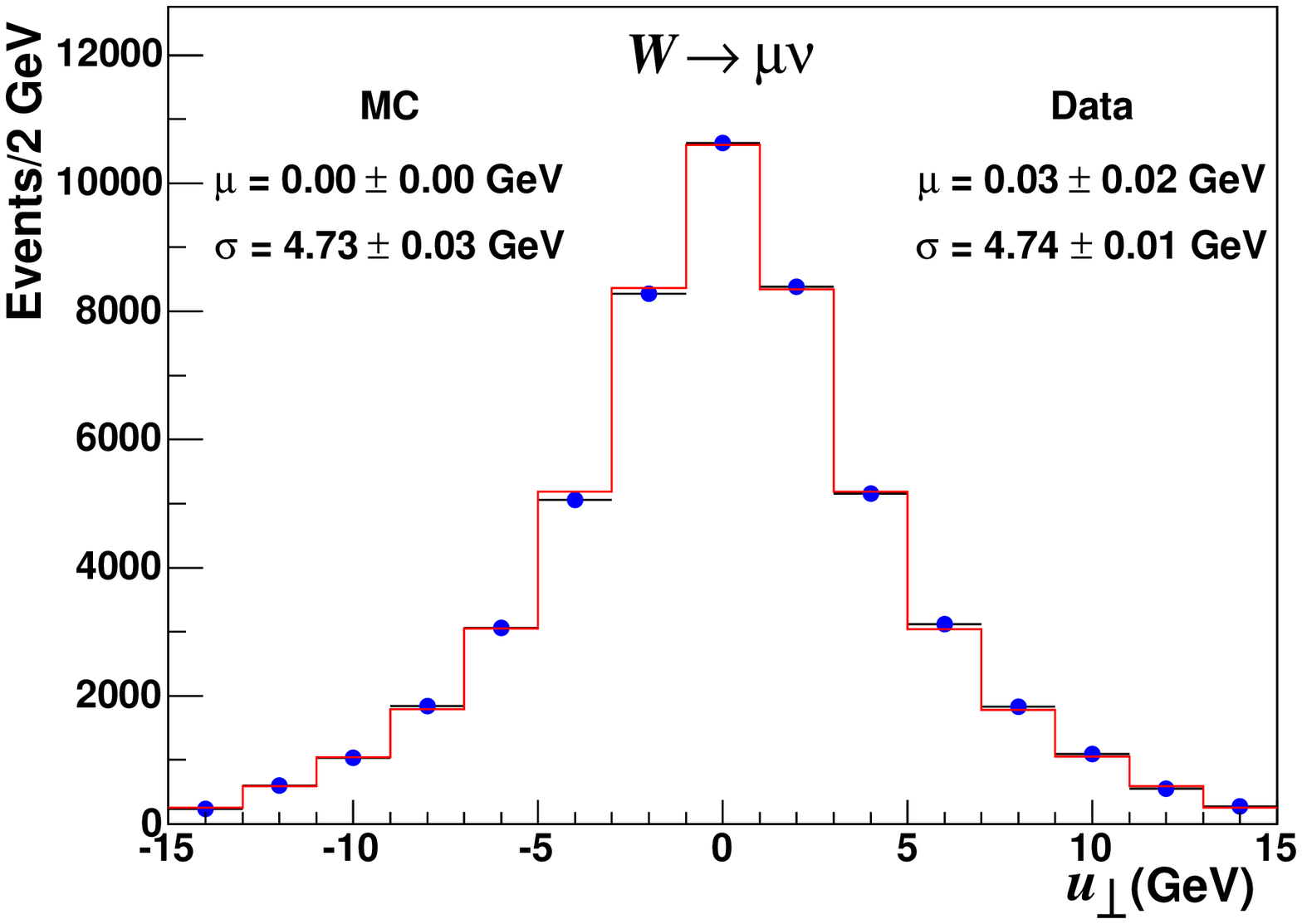}
\includegraphics*[width=8.5cm]{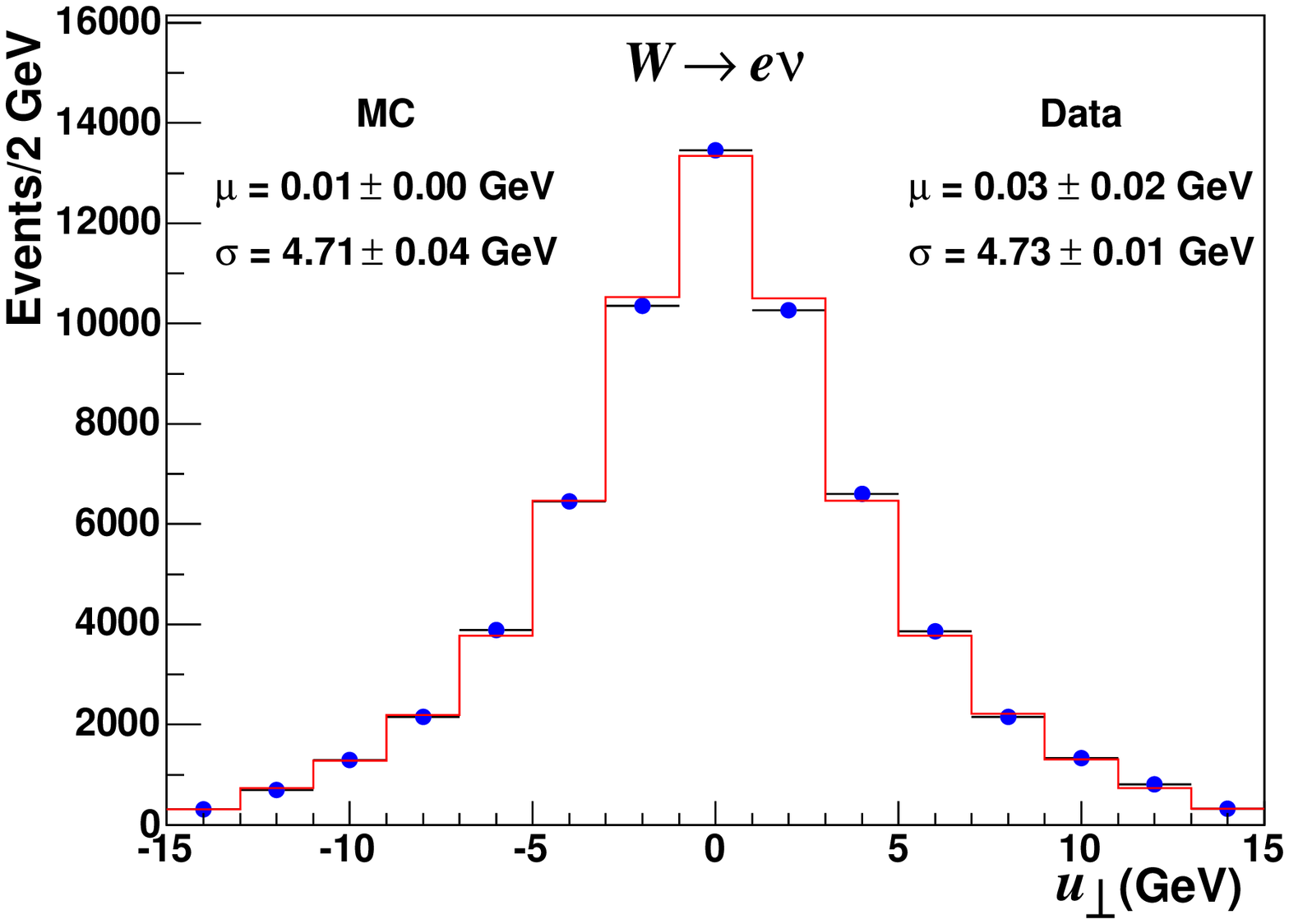}
\caption{The simulation (solid) and data (circles) $u_{\perp}$ 
distributions for $W$ boson decays to $\mu\nu$ (top) and 
$e\nu$ (bottom).  The simulation uses parameters fit from 
$Z$ boson data, and the uncertainty on the simulation is 
due to the statistical uncertainty on these parameters.  The 
data mean ($\mu$) and RMS ($\sigma$) are well-modeled by 
the simulation.}
\label{fig:uper}
\end{center}
\end{figure}

The $u_{\perp}$ distribution is dominantly affected by the recoil resolution,
with a smaller contribution from the recoil response.  The simulation models 
this distribution well for both $W\rightarrow e\nu$ and $W\rightarrow \mu\nu$
samples (Fig. \ref{fig:uper}).

\begin{figure}[!tp]
\begin{center}
\epsfysize = 6.cm
\includegraphics*[width=8.5cm]{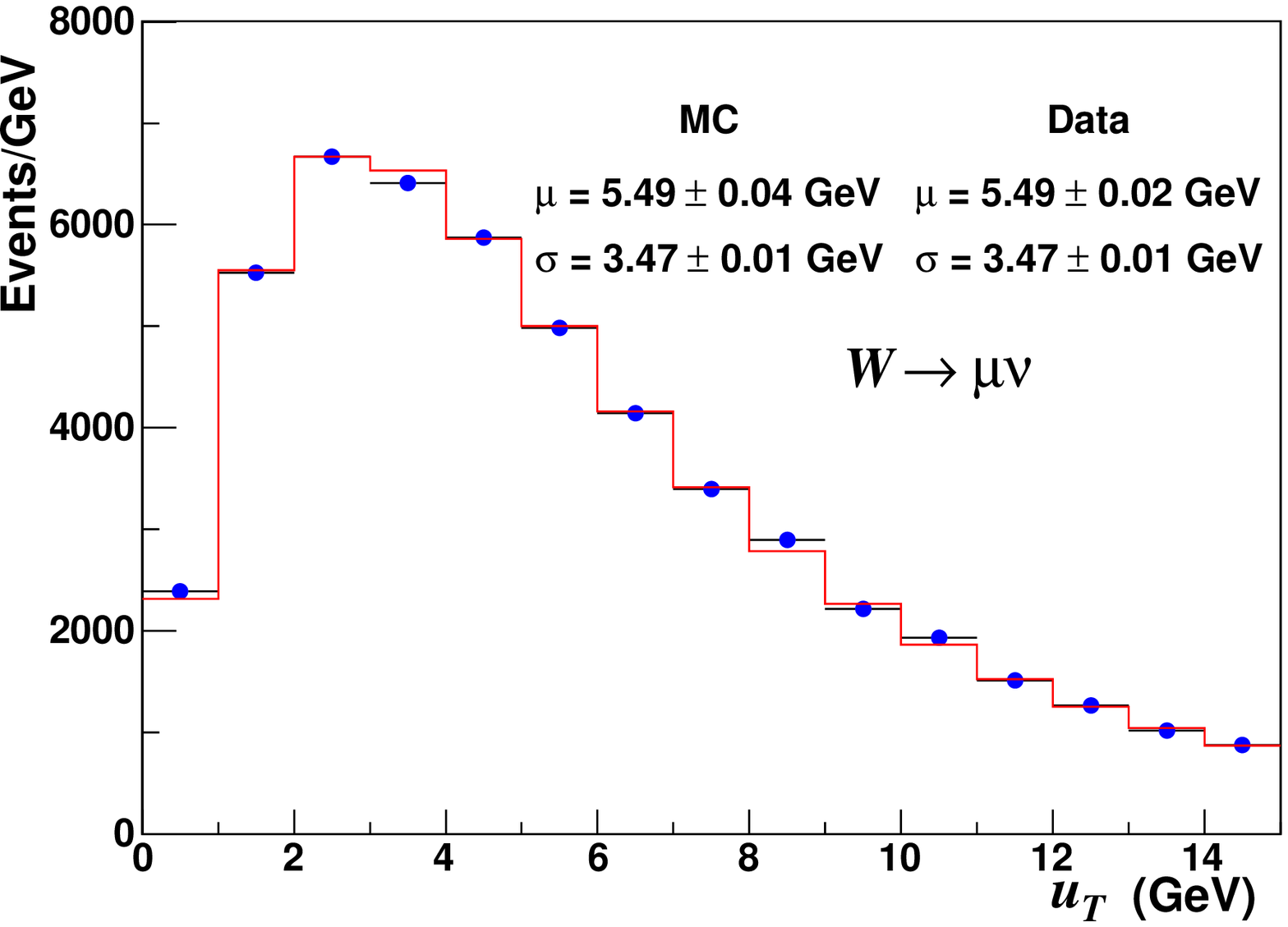}
\includegraphics*[width=8.5cm]{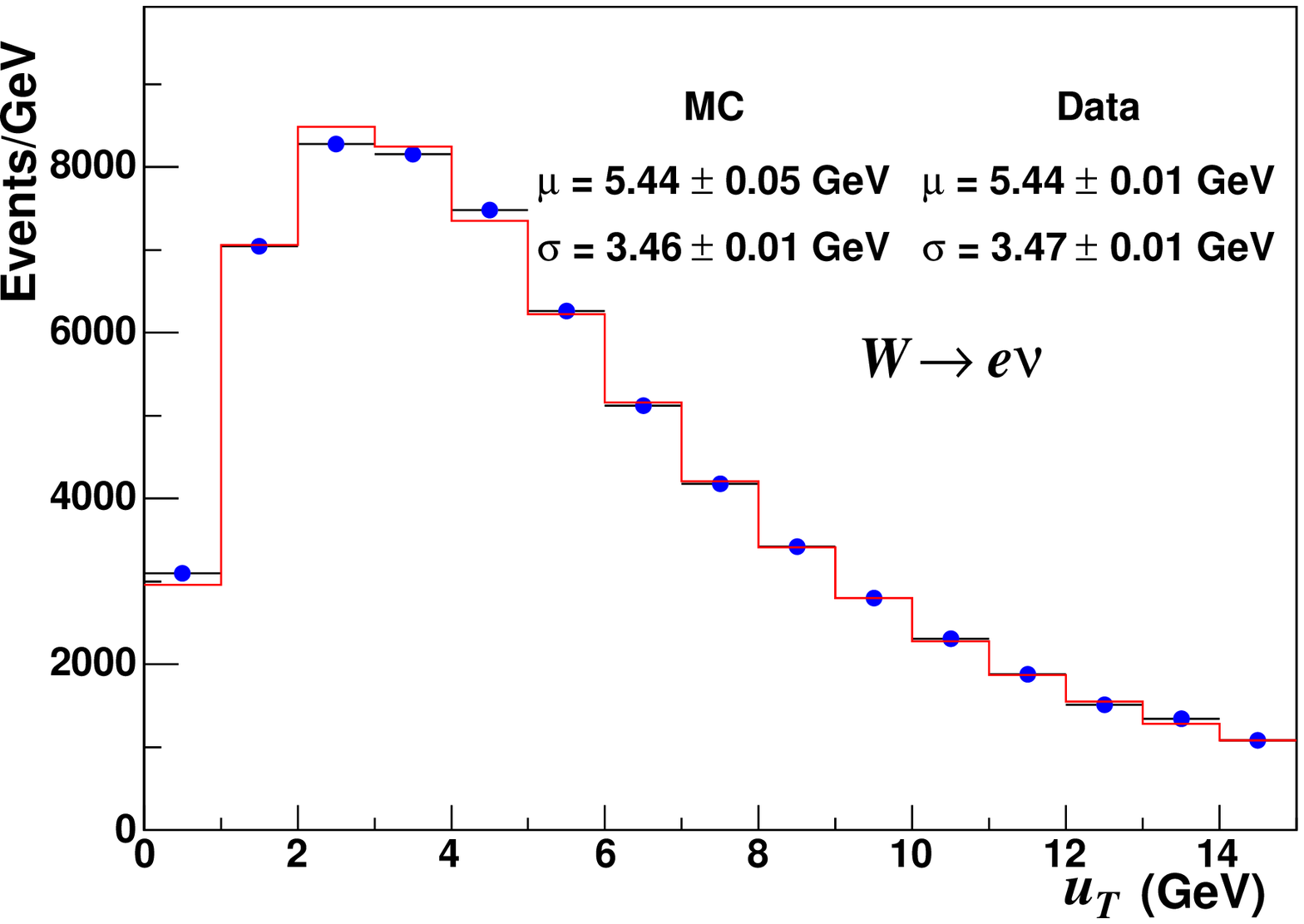}
\caption{The simulation (solid) and data (circles) $u_T$ 
distributions for $W$ boson decays to $\mu\nu$ (top) and 
$e\nu$ (bottom).  The simulation uses parameters fit from 
$Z$ boson data, and the uncertainty on the simulation is 
due to the statistical uncertainty on these parameters.  The 
data mean ($\mu$) and RMS ($\sigma$) are well-modeled by 
the simulation.}
\label{fig:ut}
\end{center}
\end{figure}

The mean of the $u_T$ distribution is sensitive to the recoil response and 
the boson $p_T$, and is affected to a lesser extent by the resolution.  The 
reverse is the case for the RMS of the $u_T$ distribution.  Both are modeled well 
by the simulation for both $W\rightarrow e\nu$ and $W\rightarrow \mu\nu$ samples 
(Fig. \ref{fig:ut}).

The uncertainties on the $m_W$ fits from the recoil parameters 
(Table~\ref{tbl:recoilsys}) are determined by varying each parameter 
by $\pm 3\sigma$ and assuming linear variation of the fit $m_W$ with 
the parameter.  Since all uncertainties are uncorrelated, we add them 
in quadrature to obtain total recoil model uncertainties of 12, 17, and 
34 MeV on $m_W$ from the $m_T$, $p_T$, and \met fits, respectively.  
The uncertainties are the same and 100\% correlated for the electron 
and muon channels, since the recoil parameters are obtained from combined
fits to $Z\rightarrow ee$ and $Z\rightarrow \mu\mu$ data.  The uncertainty 
on the $p_T$ fit arises predominantly from the modeling of the $u_T < 15$ 
GeV threshold used to select $W$ boson events (Section~\ref{sec:wsample}). 

\begin{table}[!htbp]
\begin{center}
\begin{tabular}{cccc}
\hline
\hline
  Input & \multicolumn{3}{c}{Shift (MeV)} \\
 parameter      & ~~$m_T$~~      & ~~$p_T$~~     & ~~\met~~   \\
\hline
$A$             & $-9$       & $-8$     &  $2$ \\
$B$             & $-2$       & $15$     &  $15$ \\
\hline
$N_{W,Z}$       & $5$        & $0$      &  $22$ \\
$s_{hard}$	& $5$        & $-3$     &  $21$ \\
\hline
\hline
\end{tabular}
\caption{Signed shifts in the $m_W$ fits due to $1\sigma$ increases in 
the recoil model parameters. }
\label{tbl:recoilsys}
\end{center}
\end{table}


\section{Backgrounds}
\label{sec:background}

The event selection criteria (Section~\ref{sec:wsample}) result 
in $W$ boson samples with high purity.  However, the small 
residual backgrounds affect the distributions used for the $m_W$ 
fits.  Both the $W\rightarrow e\nu$ and \mbox{$W\rightarrow \mu\nu$} 
samples receive contributions from: $Z/\gamma^*\rightarrow ll$, 
where one lepton is not detected; $W\rightarrow \tau\nu$, where 
the $\tau$ decay products are reconstructed as a charged lepton; 
and multijet production, where at least one jet is misreconstructed.  
The $W\rightarrow \mu\nu$ sample also contains backgrounds from 
cosmic rays, where a muon passing through the COT is reconstructed 
on only one side of the COT, and long-lived hadrons decaying to 
$\mu \nu X$, where the muon momentum is misreconstructed.

\subsection{$W \rightarrow e\nu$ Backgrounds}
\label{sec:webd}

We model the $W\rightarrow \tau\nu$ and $Z/\gamma^*\rightarrow ee$ backgrounds 
using events generated with {\sc pythia} \cite{pythia} and simulated 
with a full {\sc geant}-based detector simulation \cite{GEANT, CDFSIM}.  The 
full simulation models global detector inefficiencies and is thus more 
appropriate for predicting background normalizations than the custom fast 
simulation.  The multijet background is estimated using a data-based 
approach.
\par
In the standard model the branching ratio for $W\rightarrow e\nu$
is the same as for $W\rightarrow \tau\nu$, neglecting lepton masses.  
Measurements from LEP \cite{LEP} test this prediction with a precision 
of 2.9\%, and a slight discrepancy from the standard model is observed 
with a significance of $2.6\sigma$.  In estimating the $W\rightarrow \tau\nu$ 
background, we assume the standard model prediction and determine the 
ratio of $W\rightarrow \tau\nu$ to $W\rightarrow e\nu$ events from the 
ratio of acceptances of these two processes, as determined by the full 
{\sc geant}-based detector simulation.  We include an uncertainty of 
2.9\%, corresponding to the statistical precision of the tests of this 
assumption.  We estimate the $W\rightarrow \tau\nu$ background to be
$(0.93 \pm 0.03)$\% of the $W\rightarrow e\nu$ sample.
\par
The $Z/\gamma^*$ background is determined from the ratio of 
$Z/\gamma^*\rightarrow ee$ to $W\rightarrow e\nu$ acceptances determined 
from the {\sc geant}-based detector simulation, multiplied by the 
corresponding ratio of cross sections times branching ratios.  The ratio 
$\sigma \cdot BR (Z \rightarrow ee) / \sigma \cdot BR (W\rightarrow e \nu)$ 
has been calculated in the standard model to be $10.69 \pm 0.08$ \cite{WZxsec} 
\cite{wzprd}, and measurements are consistent with this value \cite{wzprd} 
\cite{wzrun1} \cite{wzd0}.  We take an uncertainty of $\pm 0.43$ on this 
value from the CDF Run~I measurement, and estimate the 
$Z/\gamma^*\rightarrow ee$ background in the $W\rightarrow e\nu$ candidate 
sample to be $(0.24 \pm 0.01)$\%.
\par
\begin{figure}[!tp]
\begin{center}
\epsfysize = 6.cm
\epsffile{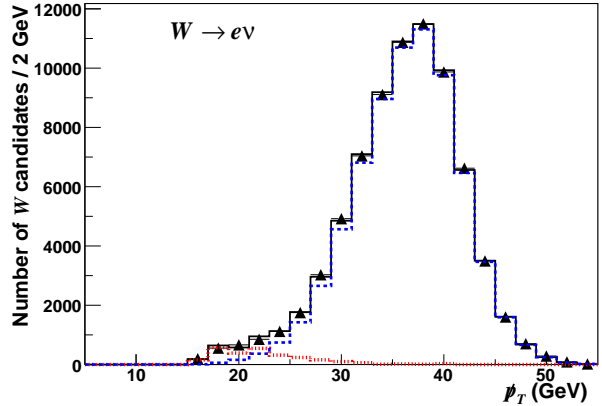}
\caption{The \met distribution of the $W\rightarrow e\nu$ candidate sample
(triangles) and prediction (solid), with the \met and $m_T$ selection cuts 
removed.  We fit for the normalization of the hadronic jet background (dotted) 
after fixing the normalization of the $W \rightarrow e\nu$ distribution (dashed) 
in the peak.  Not shown are the $Z/\gamma^*\rightarrow ee$ and 
$W\rightarrow \tau \nu$ backgrounds, whose relative normalizations are fixed 
from the simulation. }
\label{fig:qcdfit}
\end{center}
\end{figure}

Multijet background enters the signal data sample when a hadronic jet is 
misreconstructed as an electron and a second jet results in large \met
through energy misreconstruction or the semi-leptonic decay of a hadron.  
To estimate this background, we remove the \met threshold in our signal 
event selection to include the background-dominated kinematic region of 
low \met \!.  We then fit the observed \met spectrum to the combination of 
the hadronic jet, $W\rightarrow e\nu$, $Z/\gamma^*\rightarrow ee$, and 
$W\rightarrow \tau\nu$ components, floating only the hadronic jet shape 
normalization (Fig.~\ref{fig:qcdfit}).  

In this fit, the shapes and normalizations for the $W\rightarrow e\nu$, 
$Z/\gamma^*\rightarrow ee$, and $W\rightarrow \tau\nu$ components are determined 
from the {\sc geant}-based simulation.  The shape of the \met spectrum of the 
hadronic jet background is determined from the single-electron events that pass an 
anti-electron identification requirement based on a neural network discriminant 
$NN$.  The discriminant is determined by combining the electron quality variables 
(Section~\ref{sec:wesample}) into a neural network \cite{jetnet} trained with 
single-electron data events, using \met to separate signal from background.  
\par
Electron candidates in the $W\rightarrow e\nu$ sample with low $NN$ values 
have a high probability to be jets misreconstructed as electrons.  Events 
with such candidates provide a \met distribution characteristic of hadronic 
jet production.  We apply a small correction to this distribution to account 
for the expected contribution from $W\rightarrow e\nu$ decay electrons with 
low $NN$ values. 
\par
This method relies on the assumption that the hadronic jet background has a 
\met distribution that is independent of the electron identification variables.  
As a test of this assumption, we perform the same fit for the jet background 
normalization, using only the isolation variable (Section~\ref{sec:wmusample}) 
instead of the $NN$ to select a hadronic jet subsample.  We take a weighted 
average of the two fitted background normalizations, and assign an uncertainty 
that covers the range of the two results.  The resulting background estimate is 
$(0.25 \pm 0.15)$\% of the $W\rightarrow e\nu$ sample.
\par
\begin{figure}[!tp]
\begin{center}
\epsfysize = 6.5cm
\includegraphics*[width=8.75cm]{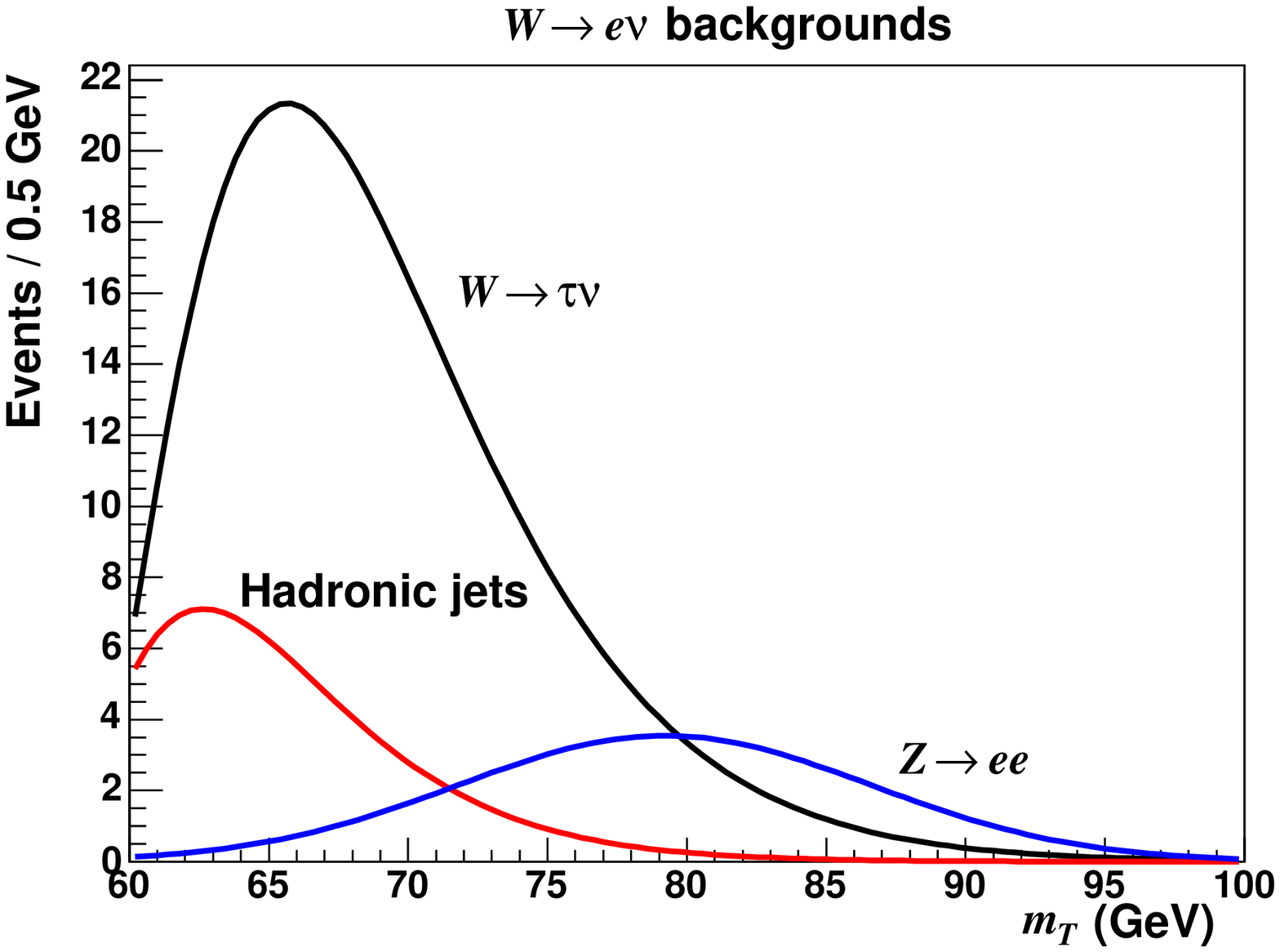}
\includegraphics*[width=8.75cm]{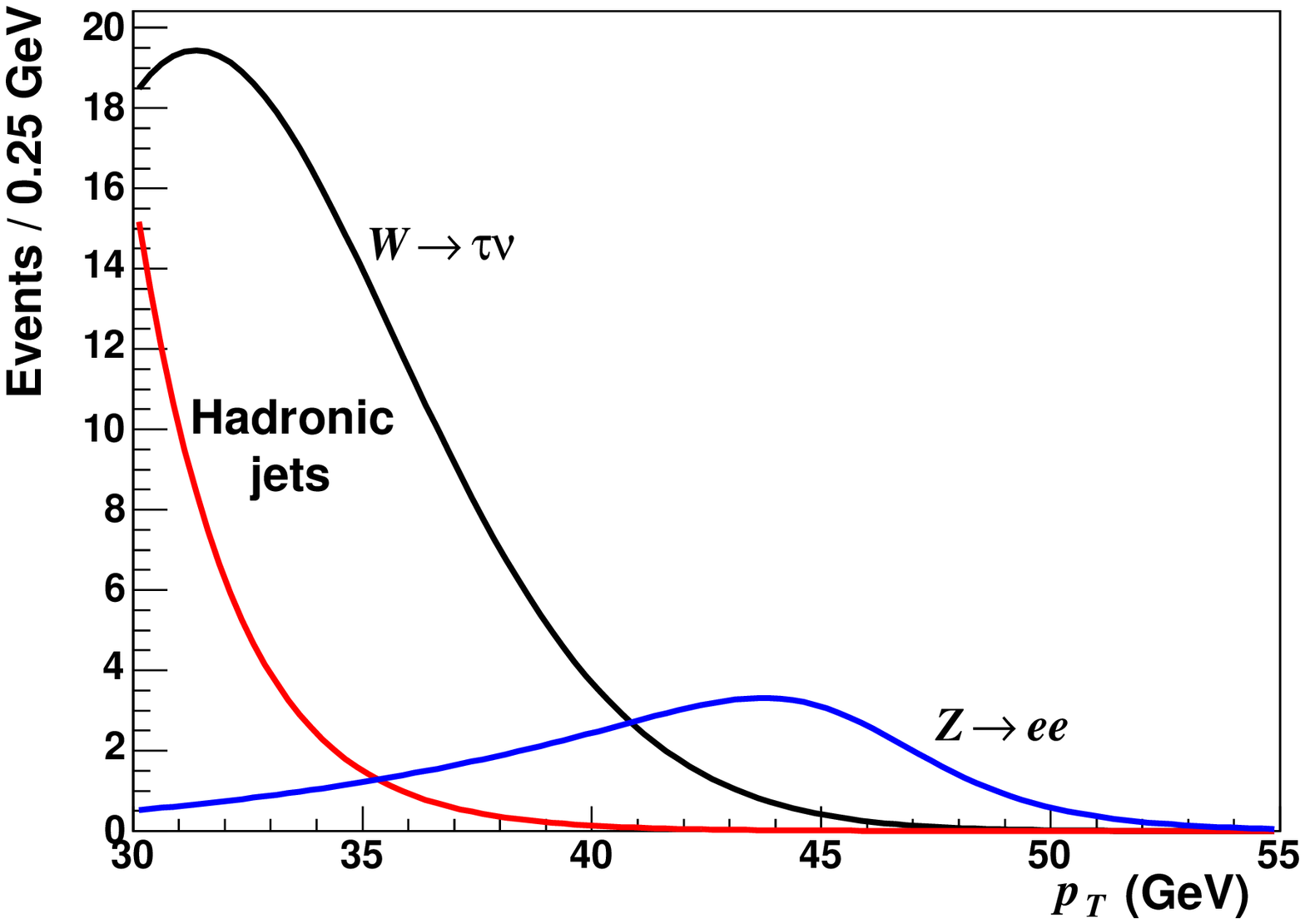}
\includegraphics*[width=8.75cm]{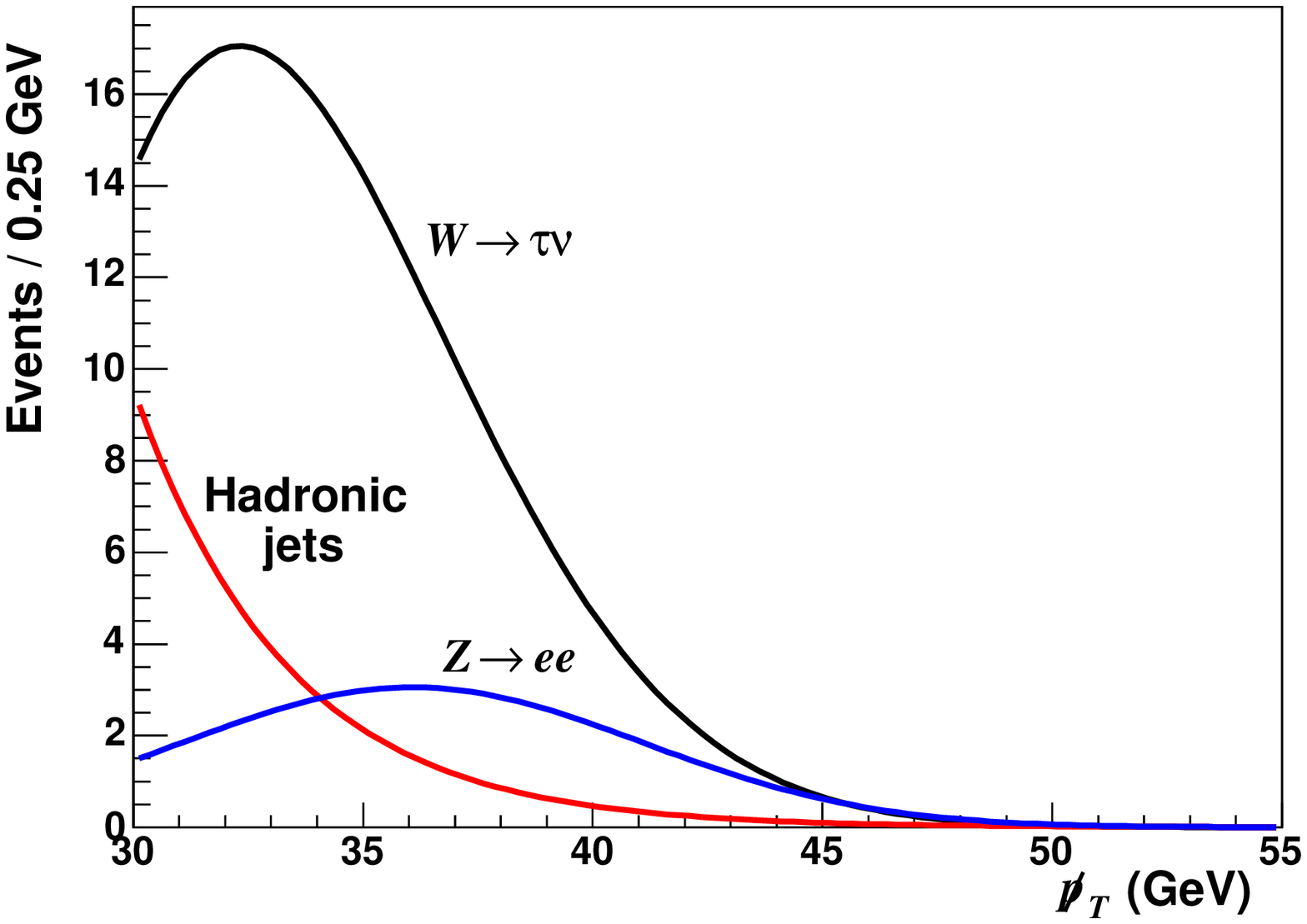}
\caption{The parametrizations of the backgrounds to the $W\rightarrow e\nu$ 
data sample.  The backgrounds to the $m_T$ (top), $p_T$ (middle), and \met 
(bottom) distributions are included in the $m_W$ fits.}
\label{fig:elebd}
\end{center}
\end{figure}

The $m_T$, $p_T$, and \met distributions are obtained from the {\sc geant}-based
simulation for $W$ and $Z$ boson backgrounds, and from events in the 
$W\rightarrow e\nu$ sample with low-$NN$ electron candidates for the hadronic jet 
background.  We fit these distributions (Fig.~\ref{fig:elebd}) and include their 
shapes and relative normalizations in the $m_W$ template fits.  The uncertainties 
on the background estimates result in uncertainties of 8, 9, and 7 MeV on $m_W$ 
from the $m_T$, $p_T$, and \met fits, respectively (Table~\ref{tbl:elebd}). 

\begin{table}[!ht]
\begin{center}
\begin{tabular}{ccccc}
\hline
\hline 
  & \% of & \multicolumn{3}{c}{$\delta m_W$ (MeV)} \\
Background          & $W\rightarrow e\nu$ data & \, $m_T$ fit \, 
                    & \, $p_T$ fit \, & \, \met fit \, \\
\hline
$W\rightarrow\tau\nu$      & 0.93 $\pm$ 0.03 & 2 & 2 & 2 \\
Hadronic jets              & 0.25 $\pm$ 0.15 & 8 & 9 & 7 \\
$Z/\gamma^*\rightarrow ee$ & 0.24 $\pm$ 0.01 & 1 & 1 & 0 \\
\hline
Total                      & 1.42 $\pm$ 0.15 & 8 & 9 & 7 \\
\hline
\hline
\end{tabular}
\end{center}
\vskip -0.1in
\caption{The percentages of the various backgrounds in the
$W\rightarrow e\nu$ data set, and the corresponding uncertainties
on the $m_T$, $p_T$ and \met fits for $m_W$. }
\label{tbl:elebd}
\end{table}

\subsection{$W \rightarrow \mu\nu$ Backgrounds}

The $W\rightarrow \tau\nu$ and $Z/\gamma^*\rightarrow \mu\mu$ backgrounds are 
modeled using events generated with {\sc pythia} \cite{pythia} and simulated 
with the {\sc geant} \cite{GEANT}-based detector simulation.  We use the data 
to estimate backgrounds from cosmic rays, multijets, and hadrons decaying in 
flight to $\mu\nu X$.
\par
Backgrounds from $W \rightarrow \tau\nu$ and $Z/\gamma^*\rightarrow \mu\mu$ to 
the $W\rightarrow \mu\nu$ sample are modeled in the same manner as for the 
$W\rightarrow e\nu$ sample (Section \ref{sec:webd}).  We determine the ratio 
of the acceptance for $W\rightarrow \tau\nu$ or $Z/\gamma^*\rightarrow \mu\mu$ 
events to the acceptance for $W\rightarrow \mu\nu$ events using the 
{\sc geant}-based detector simulation.  We assume equal branching ratios for the 
two $W$ boson decay modes, and use the ratio $\sigma \cdot BR (Z \rightarrow \mu\mu) 
/ \sigma \cdot BR (W\rightarrow \mu \nu) = 10.69 \pm 0.43$ (Section~\ref{sec:webd}).  
We estimate the fraction of $W\rightarrow \tau\nu$ ($Z/\gamma^*\rightarrow \mu\mu$) 
events in the $W \rightarrow \mu\nu$ candidate sample to be $(0.89 \pm 0.02)$\% 
[$(6.6 \pm 0.3)$\%].  The $Z/\gamma^*\rightarrow \mu\mu$ background is large 
because our event selection does not identify muons with $|\eta| \gtrsim 1.2$.  The 
tracker and muon detectors have incomplete or no coverage in the forward rapidity 
region, and the muons deposit only a few GeV of energy in the calorimeter.  Thus, a 
$Z/\gamma^* \rightarrow \mu \mu$ event with one central and one forward muon is 
measured as a single-muon event with large \met \!.
\par
\begin{figure}[!tp]
\begin{center}
\epsfysize = 6.cm
\epsffile{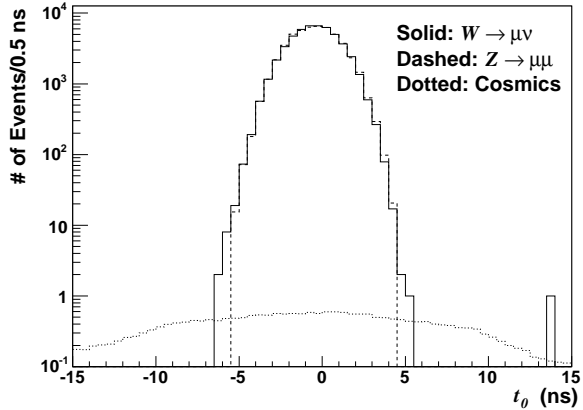}
\caption{The $t_0$ distributions for $W \rightarrow \mu\nu$ candidates 
(solid), $Z \rightarrow \mu\mu$ candidates (dashed), and identified cosmic
rays (dotted).  For comparison, the distribution from $Z$ candidates has 
been normalized to that of the $W$ candidates, while the cosmic distribution 
has been normalized to 0.05\% of the $W$ boson sample. }
\label{fig:cosmics}
\end{center}
\end{figure}

Cosmic-ray muons passing close to the beam line are a source of background to 
the $W\rightarrow \mu\nu$ sample when the muon track is reconstructed on only 
one side of the COT.  The cosmic-ray identification algorithm \cite{cosmic} 
searches for unreconstructed tracks and removes cosmic rays with high 
efficiency.  The residual cosmic-ray background is estimated using the 
reconstructed interaction time $t_0$ and impact parameter $d_0$ from the COT track 
fit.  Figure~\ref{fig:cosmics} compares the $t_0$ distributions of the 
$W \rightarrow \mu\nu$ candidate sample, $Z/\gamma^*\rightarrow \mu\mu$ candidates, 
and identified cosmic rays.  The cosmic ray fraction is fit by minimizing the 
$\chi^2$ of the sum of the $Z/\gamma^*\rightarrow\mu\mu$ and cosmic ray distributions
with respect to the $W\rightarrow \mu\nu$ distribution.  We obtain an alternative 
background estimate by comparing the $d_0$ distribution of identified cosmic rays to 
the $d_0$ distribution of $W \rightarrow \mu\nu$ candidates with the $d_0$ selection 
cut removed.  The high impact parameter region of the $W$ boson sample is enriched with 
cosmic rays, and is used to estimate the cosmic ray background within the selection 
region $|d_0| < 1$~mm.  We take the cosmic-ray background to be $(0.05 \pm 0.05)$\%, 
where the uncertainty covers the range of results from the two estimates.

Decay of a long-lived meson to a muon can result in a reconstructed track with high 
momentum and large event \met \!.  A low-momentum pion or kaon ($\lesssim 10$~GeV) 
that decays in the tracking chamber can be reconstructed as a high-momentum muon if 
the decay is in an azimuthal direction opposite the meson's curvature (i.e., a kink
in the trajectory).  Such misreconstruction typically results in a poor COT track 
$\chi^2$ and a large impact parameter.  For each of these quantities we obtain a 
prompt muon distribution from $Z$ boson decays and a meson decay-in-flight 
distribution from the $W$ boson sample by requiring either high COT track $\chi^2$ or 
high impact parameter.  We fit for the background fraction by summing the prompt 
muon distribution with the decay-in-flight distribution, and minimizing the $\chi^2$
with respect to the muon distribution from the $W$ boson sample.  We obtain a background
fraction of $(0.3 \pm 0.2)$\%, where the uncertainty covers the range of the estimates
obtained using the COT track $\chi^2$ and impact parameter distributions.  
\par
A separate class of hadronic background results from high-momentum muons from short-lived
hadronic decays, or energetic hadrons penetrating the calorimeter to the muon detectors.  
These background muon candidates are typically accompanied by significant hadronic energy
due to an associated hadronic jet, and can be separated using a muon isolation variable.
Two such variables are determined by using either calorimeter energy or track momenta in 
an $\eta-\phi$ cone of size 0.4 surrounding the muon candidate.  Using the low \met 
region to select a jet-dominated sample, we fit the track and calorimeter isolation 
distributions of the $W$ boson candidate sample to the sum of the expected distributions 
from $Z \rightarrow \mu \mu$ events and jet-dominated events.  As a third method, we fit 
the \met distribution, using muon candidates with high-isolation values to provide the 
\met distribution of the hadronic-jet background.  From the range of results of the 
three methods, we obtain a jet background estimate of $(0.1 \pm 0.1)$\%.  
\par
The distributions for the $m_W$ fit variables are obtained from the {\sc geant}-based
simulation for $W$ and $Z$ boson backgrounds, from identified cosmic ray events for the 
cosmic ray background, and from events in the $W\rightarrow \mu\nu$ sample with 
high-$\chi^2$ (isolation) muons for the decay-in-flight (hadronic jet) background.  
Including uncertainties on the shapes of the distributions, the total uncertainties on 
the background estimates result in uncertainties of 9, 19, and 11 MeV on $m_W$ from the 
$m_T$, $p_T$, and \met fits, respectively (Table~\ref{tbl:mubd}).

\begin{figure}[!tp]
\begin{center}
\epsfysize = 6.5cm
\includegraphics*[width=8.75cm]{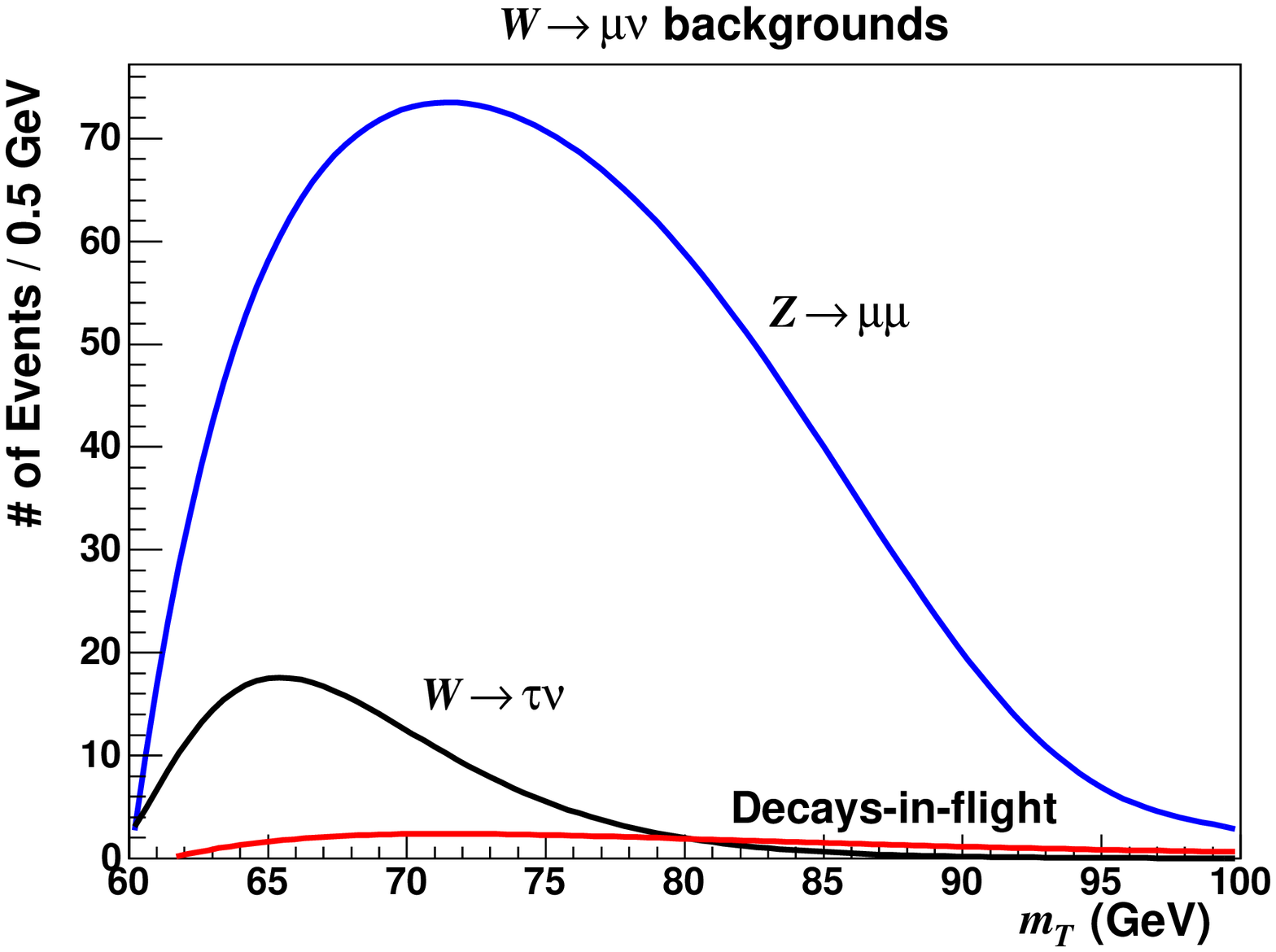}
\includegraphics*[width=8.75cm]{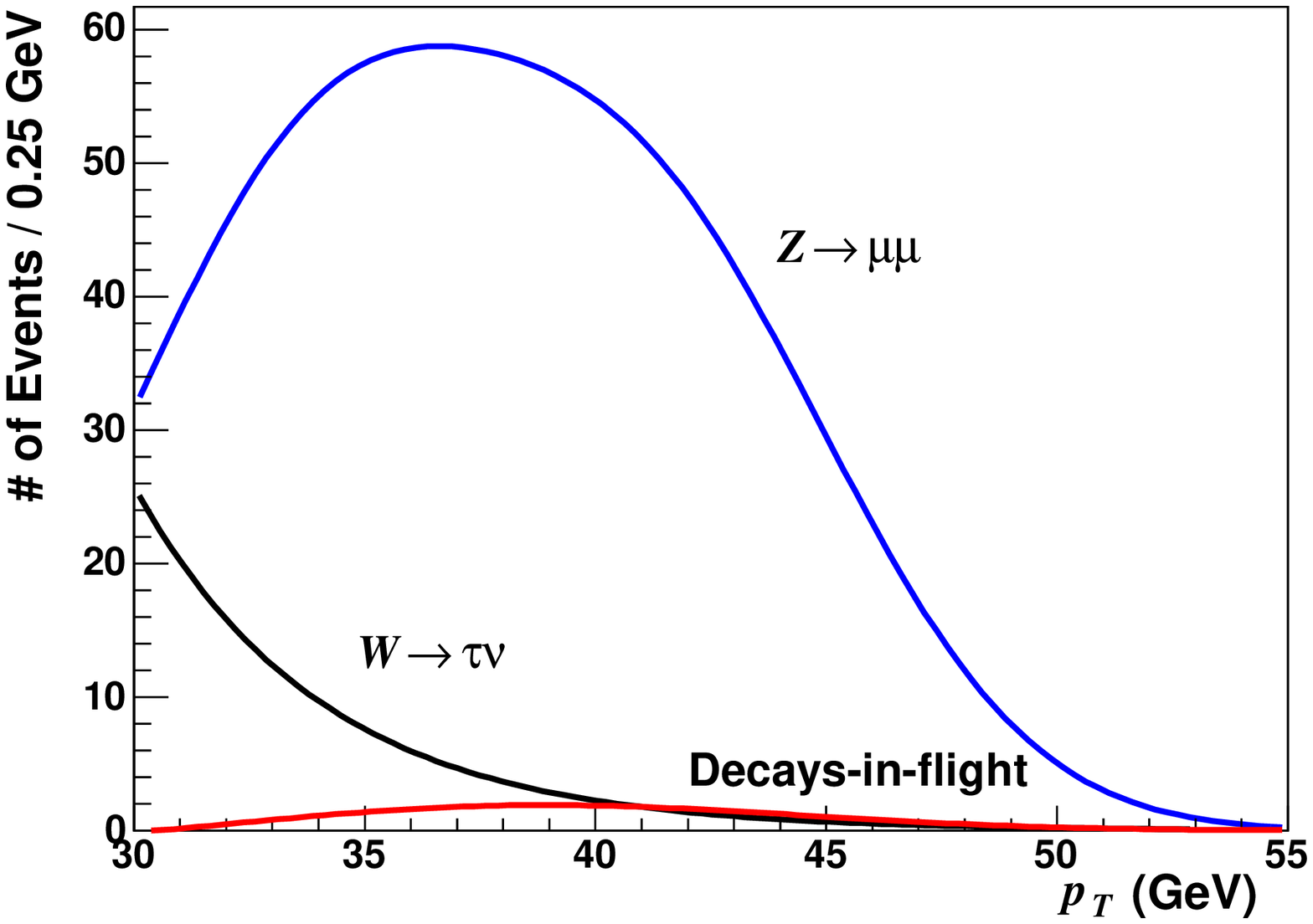}
\includegraphics*[width=8.75cm]{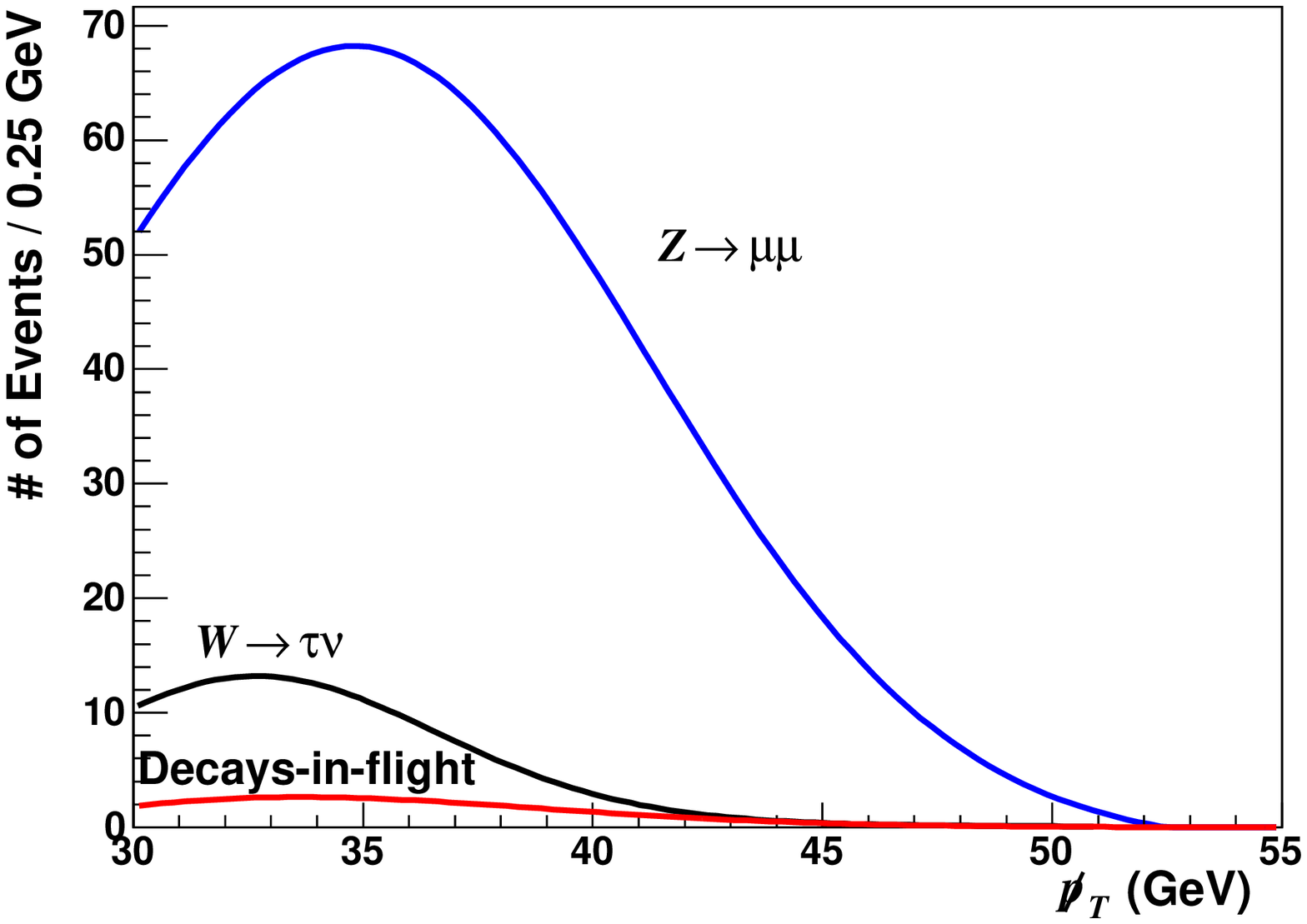}
\caption{The parametrizations of the backgrounds to the $W\rightarrow \mu\nu$ 
data sample.  The backgrounds to the $m_T$ (top), $p_T$ (middle), and \met (bottom) 
distributions are included in the $m_W$ fits.  Not shown are the small hadronic-jet 
and cosmic-ray background distributions.}
\label{fig:mubd}
\end{center}
\end{figure}

\begin{table}[!ht]
\begin{center}
\begin{tabular}{ccccc}
\hline
\hline 
  & \% of & \multicolumn{3}{c}{$\delta m_W$ (MeV)} \\
Background        & $W\rightarrow \mu\nu$ data & ~$m_T$ fit~
                  & ~$p_T$ fit~ & ~\met fit~ \\
\hline
$Z/\gamma^*\rightarrow \mu\mu$ & 6.6 $\pm$ 0.3   & 6 & 11 & 5 \\
$W\rightarrow\tau\nu$          & 0.89 $\pm$ 0.02 & 1 & 7  & 8 \\
Decays in flight               & 0.3 $\pm$ 0.2   & 5 & 13 & 3 \\
Hadronic jets                  & 0.1 $\pm$ 0.1   & 2 & 3  & 4 \\
Cosmic rays                    & 0.05 $\pm$ 0.05 & 2 & 2  & 1 \\
\hline
Total                          & 7.9 $\pm$ 0.4   & 9 & 19 & 11 \\
\hline
\hline
\end{tabular}
\end{center}
\vskip -0.1in
\caption{The percentages of the various backgrounds in the
$W\rightarrow \mu\nu$ data set, and the corresponding uncertainties
on the $m_T$, $p_T$ and \met fits for $m_W$. }
\label{tbl:mubd}
\end{table}


\section{Production and Decay Models}
\label{sec:production}

The measurement of the $W$ boson mass relies on a complete model of $W$ 
and $Z$ boson production and decay.  The production process is described 
by perturbative QCD and a parametrization of non-perturbative QCD effects, 
with parameters determined from global fits to hadron-hadron and 
lepton-hadron collision data.  $W$ and $Z$ boson decay are modeled using a 
next-to-leading-order electroweak calculation and includes QCD corrections 
for the lepton angular distributions, as a function of boson $p_T$.  The 
most important process in the decay is photon radiation off the final-state 
charged lepton, which has been calculated at next-to-leading order \cite{wgrad}.

\subsection{Parton Distribution Functions}

The longitudinal momentum of the produced $W$ or $Z$ boson depends on the 
momenta of the interacting partons.  These momenta, generally expressed in 
terms of the fractions $x_i$ of the colliding (anti-)proton energies, are 
not known on an event-by-event basis.  The $x_i$ parton distribution 
functions (PDFs) are however well constrained by hadron-hadron and 
lepton-hadron collision data.  The distributions have been parametrized as 
simple functional forms for the quarks, antiquarks, and gluons inside a 
proton.  Two independent fits to the global data, performed by the MRST 
\cite{MRST} and CTEQ \cite{CTEQ} collaborations, constrain the parameters 
in these PDFs.  
\par
We model the quark momentum fractions using the next-to-leading-order 
CTEQ6M parton distribution functions.  The CTEQ parametrization \cite{CTEQ} 
for most of the distribution functions inside the proton is:
\begin{equation}
x_pf_a(x_p, Q_0) = A_0 x_p^{A_1} (1-x_p)^{A_2} e^{A_3 x_p} (1 + A_4 x_p)^{A_5},
\end{equation}
\noindent
where $f_a$ are the distributions of a particular quark or gluon combination 
$a$, $A_i$ are the fit parameters, and $Q_0$ is the energy scale at which 
the parameters are defined.  The functions at a particular energy scale $Q$ 
are determined by a perturbative evolution calculation known as the DGLAP 
equation \cite{DGLAP}.
\par
The uncertainty on the $m_W$ measurement arising from uncertainties on the 
PDF parameters is determined using a set of 40 PDFs provided by the CTEQ 
collaboration.  The set covers the $\pm 1.6\sigma$ (90\% C.L.) uncertainties 
\cite{pdfuncertainty} for the eigenvectors of the parametrization.  The mass 
shift of a particular $+1.6\sigma$ PDF, relative to the corresponding 
$-1.6\sigma$ PDF, determines the uncertainty due to that eigenvector.  We 
calculate the total PDF uncertainty using the quadrature sum of all eigenvector 
contributions 
\cite{CTEQ}:
\begin{equation}
\delta m_W^{PDF} = \frac{1}{1.6}\left[\frac{1}{2}\sqrt{\sum_i(m_W^{i+} - m_W^{i-})^2}\right],
\end{equation}
\noindent
where $m_W^{i\pm}$ represents the mass fits for the $\pm 1.6 \sigma$ shifts 
in eigenvector $i$.  These fits are performed using templates and simulated 
pseudoexperiments both generated with {\sc pythia} \cite{pythia}.  The 
resulting $\delta m_W^{PDF}$ are 11, 9, and 13 MeV, for the $m_T$, $p_T$, and 
\met fits, respectively.  A fit to pseudodata using the MRST PDF set results 
in $m_W$ shifts smaller than these uncertainties.

\subsection{$W$ and $Z$ Boson $p_T$}
\label {sec:bosonpt}

Because mass is a Lorentz invariant, the $W$ boson transverse mass 
is only weakly sensitive to the $W$ boson transverse momentum $p_T^W$.  
However, the decay lepton $p_T$ spectra are more significantly affected 
by the $p_T^W$ distribution.
\par
\begin{figure}[!tp]
\begin{center}
\epsfysize = 6.cm
\includegraphics*[width=8.5cm]{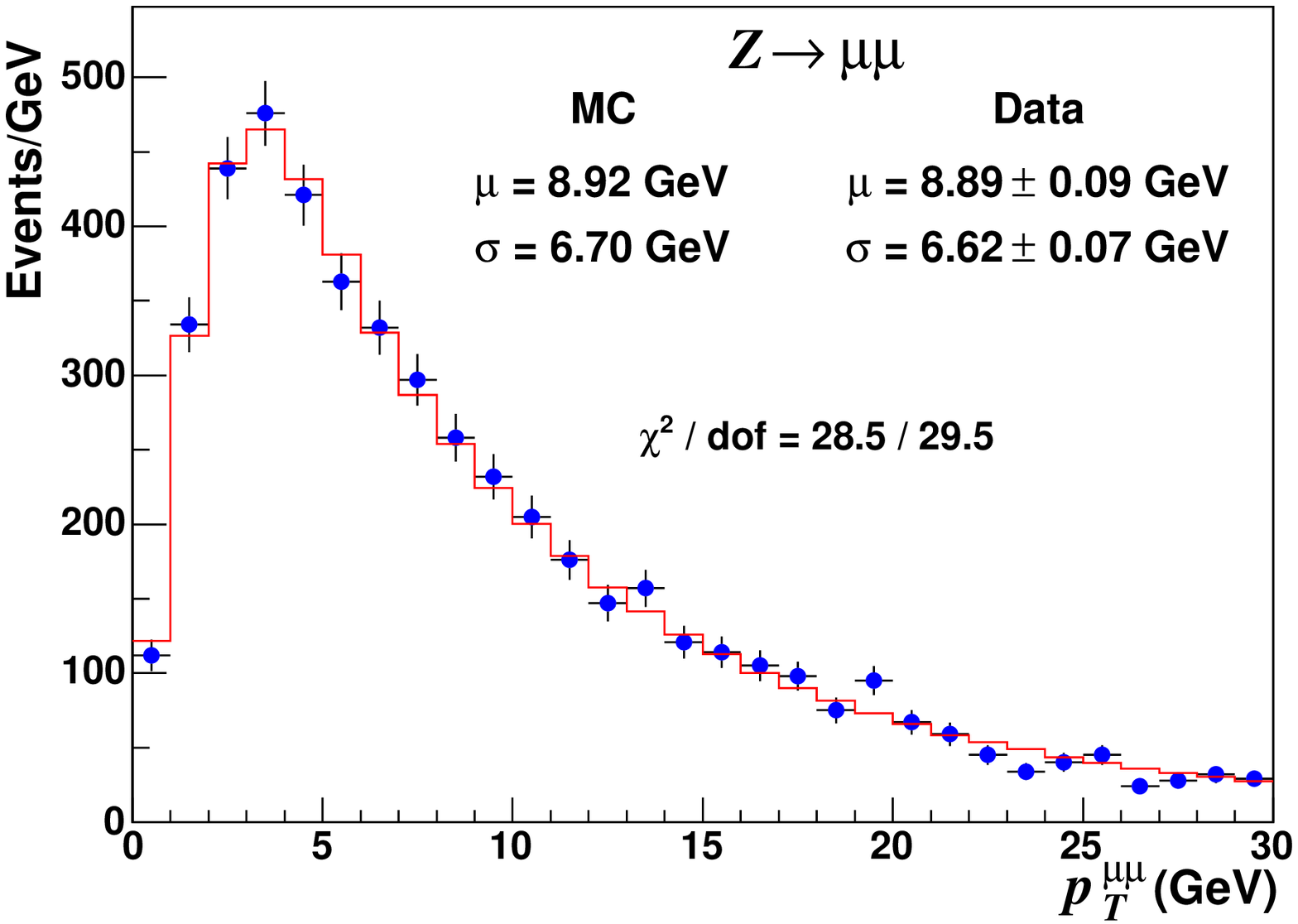}
\includegraphics*[width=8.5cm]{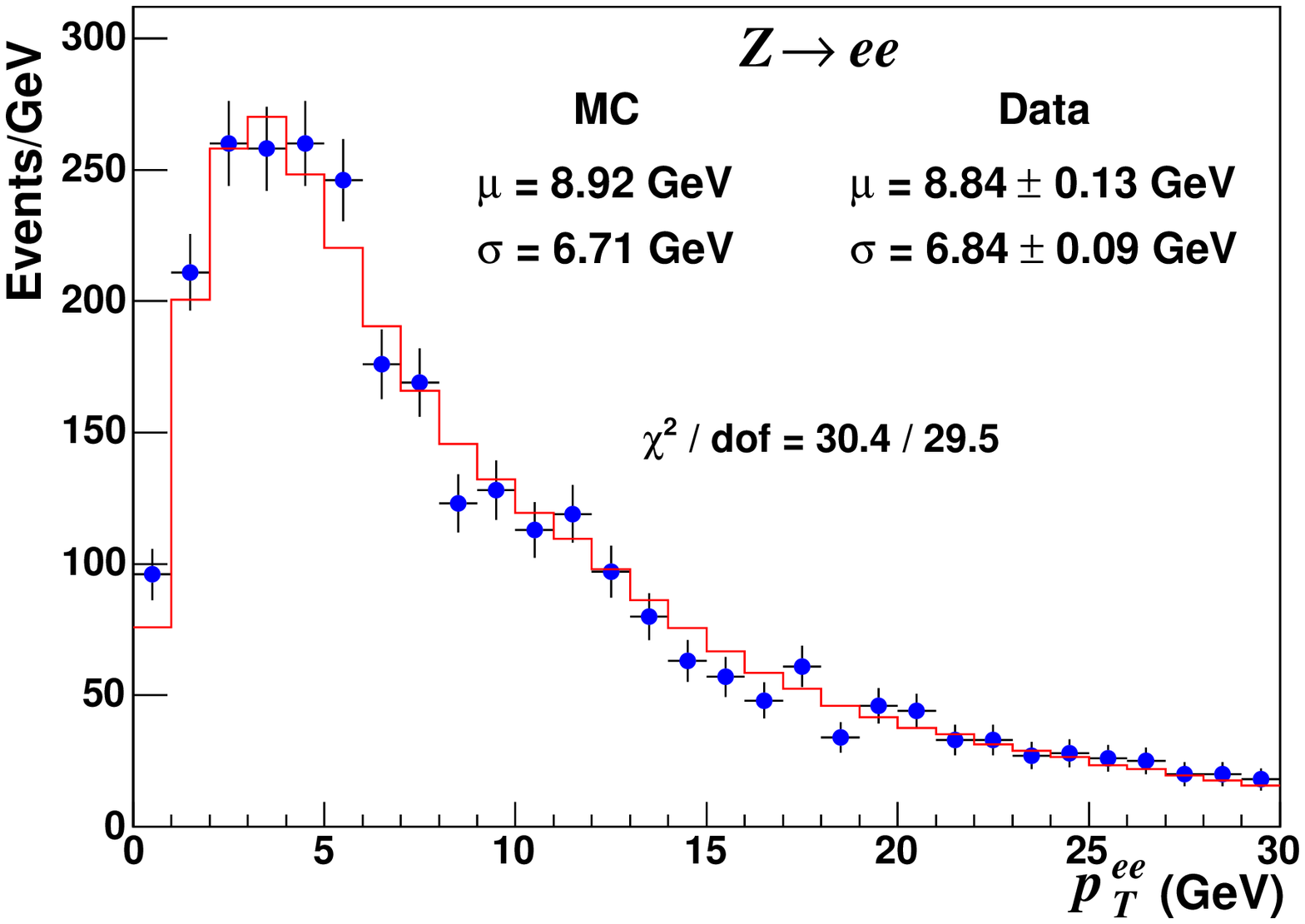}
\caption{The simulation (solid) and data (circles) $p_T^Z$ 
distributions for $Z$ boson decays to $\mu\mu$ (top) and 
$ee$ (bottom).  The distributions are used to fit for the 
non-perturbative parameter $g_2$, which determines the most 
probable value of $p_T^Z$.  Since there are two distributions 
for one fit parameter, each distribution contributes half of 
a degree of freedom.  The mean ($\mu$) and RMS ($\sigma$) 
are consistent between data and simulation. }
\label{fig:zpt}
\end{center}
\end{figure}

At the Tevatron, the $p_T$ spectra of $W$ and $Z$ bosons peak at 
a few GeV (Fig.~\ref{fig:zpt}), where the shapes are predominantly determined 
by non-perturbative QCD interactions.  We model the distribution with the 
{\sc resbos} generator \cite{resbos}, which uses the Collins-Soper-Sterman 
(CSS) \cite{css} resummation formalism and a parametrized non-perturbative 
form factor.  In this formalism, the cross section for $W$ boson production 
is written as:
\begin{align}
\frac{d\sigma(p\bar{p}\rightarrow W + X)}{d\hat{s} d^2\vec{p}_T^{~W} dy} & =  
\frac{1}{2\pi^2} \delta(\hat{s} - m_W^2) \int d^2b e^{i\vec{p}_T^{~W}\cdot\vec{b}} 
\notag \\
& \times \tilde{W}_{j\bar{k}}(\vec{b},\hat{s},x_i) + Y(p_T^W,\hat{s},x_i),
\end{align}

\noindent
where $x_i$ are the parton energy fractions of the (anti-)proton, 
$y = 0.5\ln(x_p/x_{\bar{p}})$ is the boson rapidity, $\vec{b}$ is 
the relative impact parameter of the partons in the collision, $Y$ is a 
function calculable at fixed order, and $\tilde{W}$ can be separated into 
its perturbative and non-perturbative components.  We use the 
Brock-Landry-Nadolsky-Yuan (BLNY) form for the non-perturbative component:
\begin{equation}
\tilde{W}_{j\bar{k}}^{NP} = e^{[g_1 - g_2 \ln(\frac{Q}{2Q_0}) - g_1g_3 \ln
(100 x_p x_{\bar{p}})]b^2},
\end{equation}

\noindent
where $Q_0 = 1.6$ GeV and $g_i$ are parameters suggested by the CSS formalism 
to be universal to processes with initial state quarks and colorless objects in 
the final state \cite{resbos}.  
\par
The $g_2$ parameter affects the position of the 
most probable $p_T^W$ and is the most relevant for the $m_W$ measurement.  We use 
$g_1 = 0.21$ GeV$^2$, $g_2 = 0.68$ GeV$^2$, and $g_3 = -0.60$, which are determined 
from fits to global Drell-Yan data \cite{resbos}.  We verify the applicability of 
these values to our data by fitting the dilepton $p_T$ distribution 
(Fig.~\ref{fig:zpt}) for $g_2$.  We find $g_2 = [0.685 \pm 0.048({\rm stat})]$~GeV$^2$, 
consistent with the global fits.  Varying $g_2$ by $\pm 3\sigma$ in pseudoexperiments 
and taking the fit $m_W$ to be linearly dependent on $g_2$, we find that the 
uncertainty of $\delta g_2 = 0.048$~GeV$^2$ results in uncertainties of 3, 9, and 5 
MeV, on $m_W$ for the $m_T$, $p_T$, and \met fits, respectively.  These uncertainties 
are the same and 100\% correlated between the electron and muon channels, since $g_2$
is fit using $Z\rightarrow ee$ and $Z\rightarrow \mu\mu$.  Neglecting correlations 
between PDFs and the $g_i$ parameters, we find that uncertainties on the other $g_i$ 
do not significantly affect the $m_W$ measurement.

\subsection{$W$ Boson Decay}

The $m_W$ measurement is sensitive to the charged lepton decay angle 
relative to the boson $p_T$.  The mismodeling of this angle can bias the 
projection of the recoil along the lepton ($u_{||}$), which in turn 
affects $m_W$ measured from the $m_T$ fit (Section \ref{sec:recoilcheck}).  
\par
The lepton decay angle is predicted by the matrix element calculation in 
the {\sc resbos} generator, which computes the differential cross section 
$\frac{d^4\sigma}{dp_T^W dy_W dQ^2 d\Omega}$.  The angular distributions 
are defined in the Collins-Soper rest frame of the $W$ boson \cite{collinssoper}.  
In this frame, the $z$-axis is defined to bisect the angle between the proton 
momentum and the opposite of the antiproton momentum.
\par
The angular component of the differential cross section can be written as 
\cite{mirkes},
\begin{align}
\frac{d\sigma}{d\Omega} \propto & (1 + \cos^2\theta) + 
   \frac{1}{2}A_0(1-3\cos^2\theta) \notag \\
 & + A_1\sin2\theta \cos\phi + \frac{1}{2}A_2\sin^2\theta \cos2\phi \notag \\
 & + A_3 \sin\theta \cos\phi + A_4 \cos\theta + A_5 \sin^2\theta\sin2\phi \notag \\
 & + A_6 \sin2\theta \sin\phi + A_7 \sin\theta\sin\phi,
\end{align}
\noindent
where the $A_i(p_T^W, y_W)$ have been calculated to NLO in $\alpha_s$.  
Because of the $V - A$ structure of the electroweak interaction, for leading-order 
valence quark interactions all $A_i$ are zero except $A_4 = 2$.  The $A_i$ can 
be determined experimentally through a moments analysis \cite{strologas} of the 
lepton angle in the Collins-Soper frame.
\par
We have performed a moments analysis to extract the $A_i$ from the {\sc resbos}
generator, and compared the results to those obtained \cite{strologas} from 
the {\sc dyrad} event generator \cite{dyrad}, which produces $W +$ jet events to 
order $\alpha_s^2$.  The two generators give consistent results in the overlapping 
region 15~GeV~$< p_T^W < 100$~GeV.

\subsection{Photon Radiation}
\label{sec:fsr}

The quarks, the $W$ boson, and charged lepton have non-zero electromagnetic 
charge and can radiate photons in the $W$ boson production process.  
Radiation off the initial-state quarks and the $W$ boson propagator have a 
negligible effect on the invariant mass distribution of the $W$ boson.
Radiation off the final-state charged lepton reduces the measured transverse 
mass (relative to the $W$ boson mass) and must be accurately modelled.
\par
We study photon radiation using the {\sc wgrad} event generator \cite{wgrad}, 
which models the full next-to-leading-order (NLO) electroweak physics.  The 
generator allows an independent study of photon radiation from the 
initial-state quarks (ISR), the $W$ boson propagator, and the final-state 
charged lepton (FSR).  Interference between the contributing diagrams can 
also be studied independently.  We verify that the initial-state, propagator, 
and interference effects do not affect the measured $W$ boson mass, within 
the 5 MeV statistical uncertainty of the simulation.
\par
We simulate final-state photon radiation in our {\sc resbos}-generated $W$ and 
$Z$ boson events by generating a photon for each charged lepton.  The energy 
and angular spectra are taken from the {\sc wgrad} generator using the appropriate 
boson mass.  To avoid the infrared divergence that arises when the photon momentum 
goes to zero, we require $E_{\gamma} > \delta_s \sqrt{\hat{s}}/2$, where 
$\delta_s = 10^{-4}$.  We find that increasing $\delta_s$ to $10^{-3}$ does not 
affect the $m_W$ measurement, at the level of the 5 MeV statistical precision of 
the tests.  
\par
\begin{figure}[!t]
\begin{center}
\epsfysize = 6.cm
\includegraphics*[width=8.5cm]{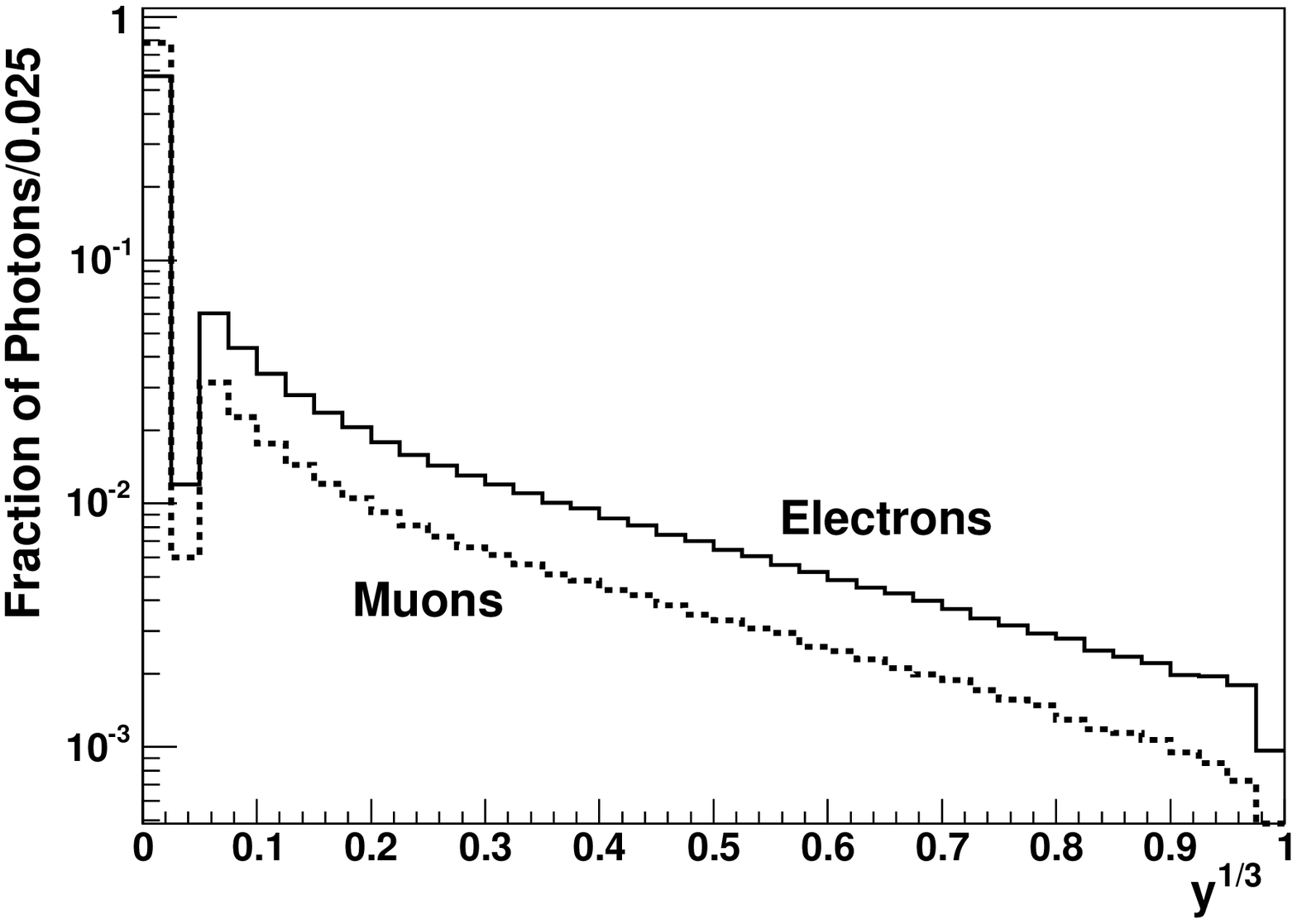}
\includegraphics*[width=8.5cm]{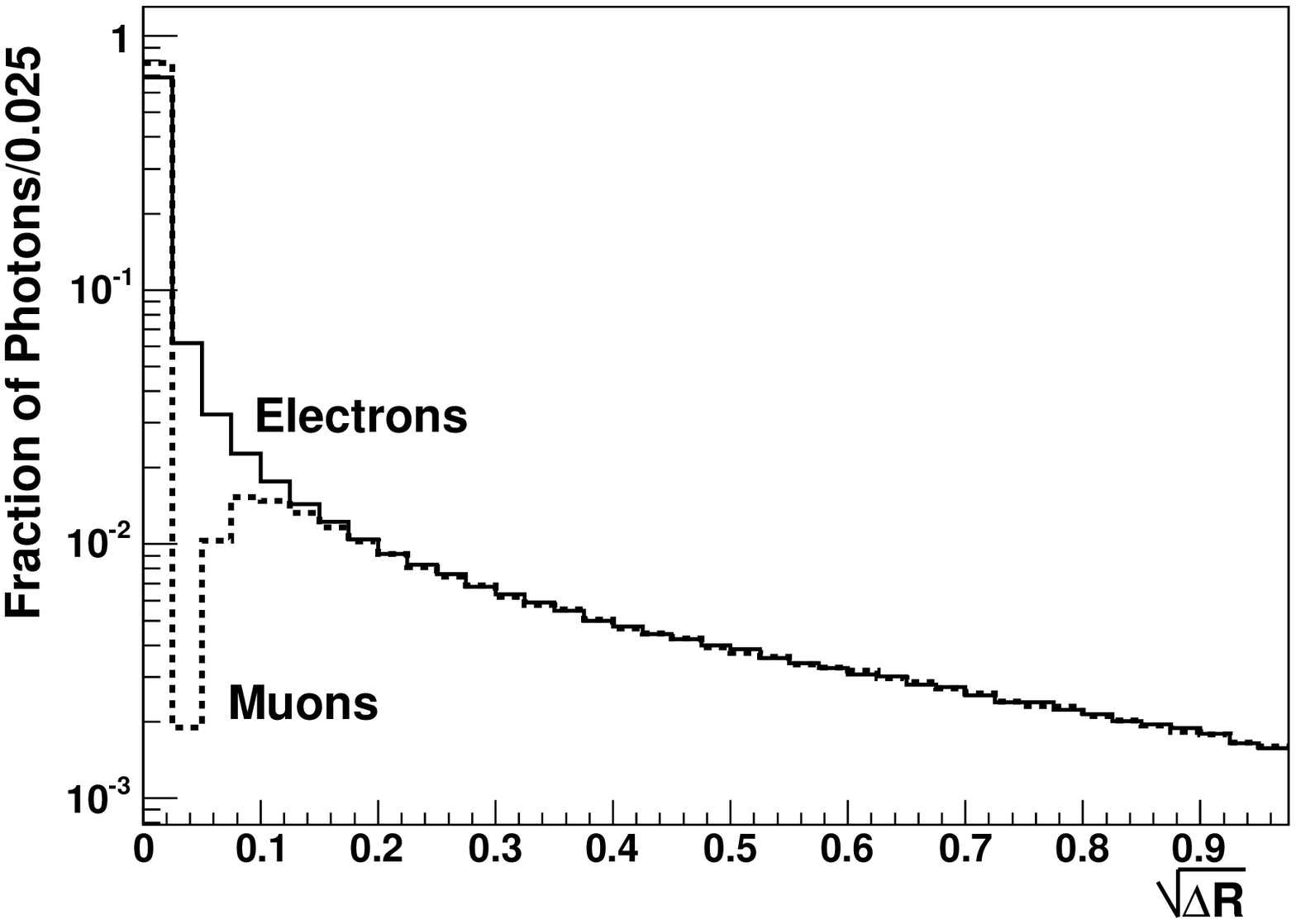}
\caption{The cube root of the fraction of electron (solid) or muon (dashed) 
momentum contained in the radiated photon (top), and the square root of the 
angle $\Delta R$ (bottom) between the radiated photon and the electron (solid)
or muon (dashed). }
\label{fig:fsr}
\end{center}
\end{figure}

The energy of a photon in a given event is calculated from the fraction 
$y = E_{\gamma}/E_l$.  The photon angle $\Delta R = \sqrt{(\Delta \eta)^2 
+ (\Delta \phi)^2}$ is taken with respect to the charged lepton.  To improve the 
phase space sampling, we sample from a two-dimensional distribution of the 
variables $\sqrt{\Delta R}$ and $y^{1/3}$ when selecting a photon.  The individual 
distributions of these variables are shown in Fig.~\ref{fig:fsr}.

We validate our photon simulation by fitting a sample of events generated with FSR 
using {\sc wgrad} to templates generated with leading-order {\sc wgrad} and photons 
simulated according to our model.  We find our FSR model to be consistent with that 
of {\sc wgrad} at the level of the 5 MeV statistical precision of the test.  The total 
effect of including FSR is shown in Table \ref{tbl:fsr}.  Since FSR reduces the charged 
lepton momentum, the shift is largest for the $p_T$ fit and smallest for the \met fit.  
The effects are smaller for electrons than for muons because the electron calorimeter 
energy measurement recovers much of the energy of FSR photons.

\begin{table}[!tbp]
\begin{center}
\begin{tabular}{ccc}
\hline
\hline
  Fit~    & $~\delta m_{W,Z}$($\mu$)~ (MeV) & $~\delta m_{W,Z}$($e$) (MeV)~ \\
\hline
 $m_T$   & -158   & -138  \\
 $p_T$   & -206   & -186  \\
\met     & -77    & -59  \\
$m_{ll}$ & -196   & -215  \\
\hline
\hline
\end{tabular}
\caption{The mass shifts obtained by fitting events generated with our 
simulation of single-photon radiation to templates generated without 
final-state photon radiation.  The shifts are for the $W$ boson $m_T$, 
$p_T$, and \met fits, and for the $Z$ boson $m_{ll}$ fit.  The shifts 
have statistical uncertainties of 7 MeV each. }
\label{tbl:fsr}
\end{center}
\end{table}

We approximate the effect of next-to-next-to-leading-order FSR by 
increasing the photon's momentum fraction ($y$) by 10\%, consistent 
with the results of a study of higher-order photon radiation \cite{nnloqed}.
We take half the correction as a systematic uncertainty to account for 
higher-order QED effects.
\par
The total uncertainty due to photon radiation is the quadrature sum of:  
uncertainties on ISR, interference between ISR and FSR, and radiation off 
the propagator (5 MeV); uncertainty due to the infrared cutoff of the FSR 
photon (5 MeV); the FSR model (5 MeV); and uncertainties on higher-order 
FSR corrections (7 MeV for the electron and 8 MeV for the muon $m_T$ fits).  
The total uncertainties are 12 (11), 13 (13), and 10 (9) MeV, for the muon 
(electron) $m_T$, $p_T$, and \met fits, respectively.


\section{$W$ Boson Mass Fits}
\label{sec:fits}

We fit the $W$ boson data distributions to a sum of background and simulated 
signal templates of the $m_T$, $p_T$, and \met distributions, fixing the 
normalization of the sum to the number of data events.  The fit minimizes the 
negative log likelihood (Section~\ref{sec:templates}) as a function of the 
template parameter $m_W$, which is defined by the relativistic Breit-Wigner 
mass distribution \cite{pdg}:
\begin{equation}
\frac{d\sigma}{dm} \propto \frac{m^2}{(m^2 - m_W^2)^2 + m^4 \Gamma_W^2/m_W^2},
\end{equation}

\noindent
where $m$ is the invariant mass of the propagator.  The likelihood is 
calculated in $m_W$ steps of 1 MeV.  We use the standard model $W$ boson width 
$\Gamma_W = 2.094$ GeV, which has an accuracy of 2 MeV and is calculated for 
$m_W = 80.393$ GeV.  Using pseudoexperiments, we find the input $\Gamma_W$ 
affects the fit $m_W$ according to the relation $dm_W/d\Gamma_W = 0.14 \pm 0.04$.

\subsection{Fit Results}
 
\begin{figure}[!tp]
\begin{center}
\epsfysize = 6.cm
\includegraphics*[width=8.5cm]{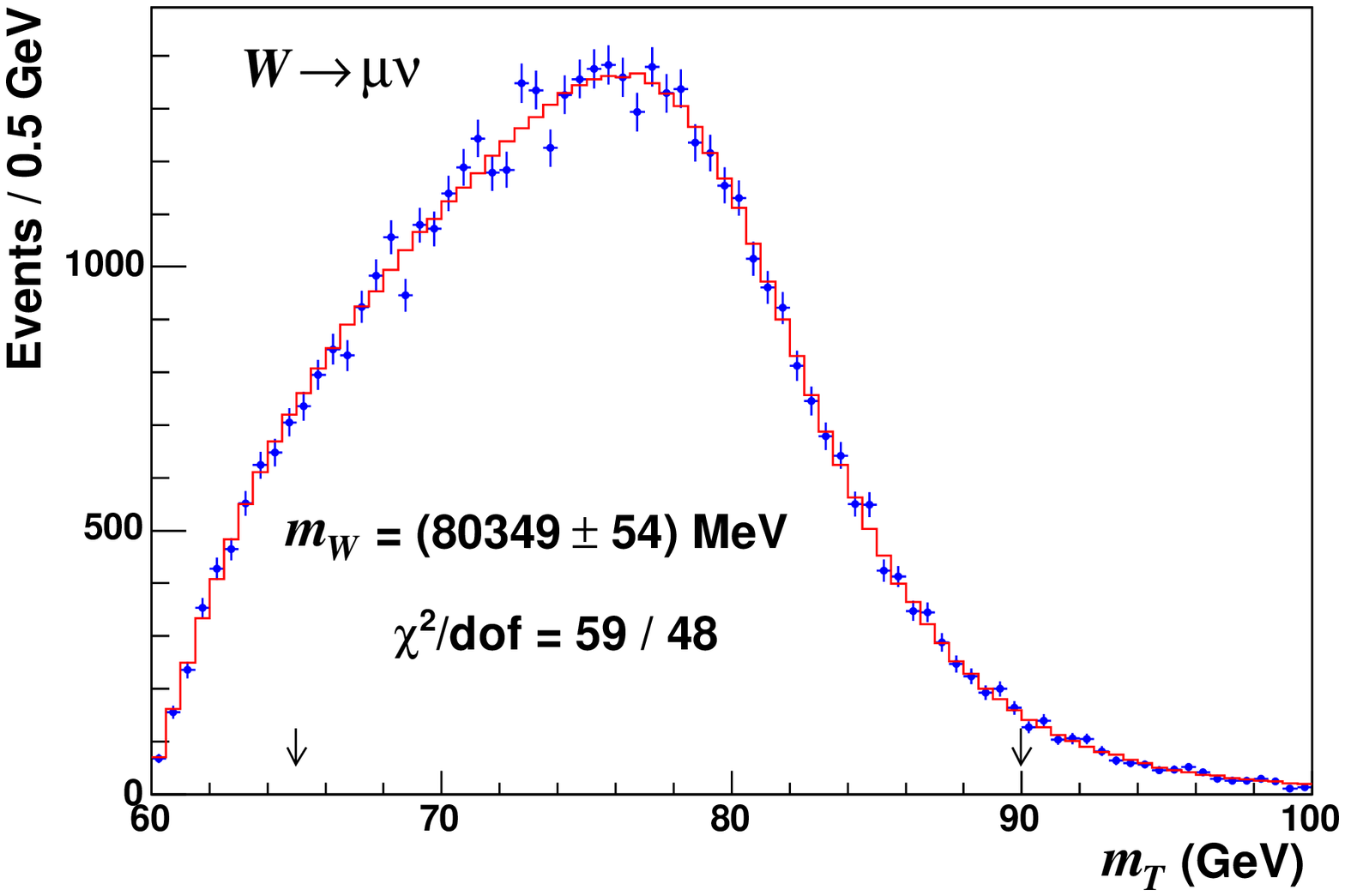}
\includegraphics*[width=8.5cm]{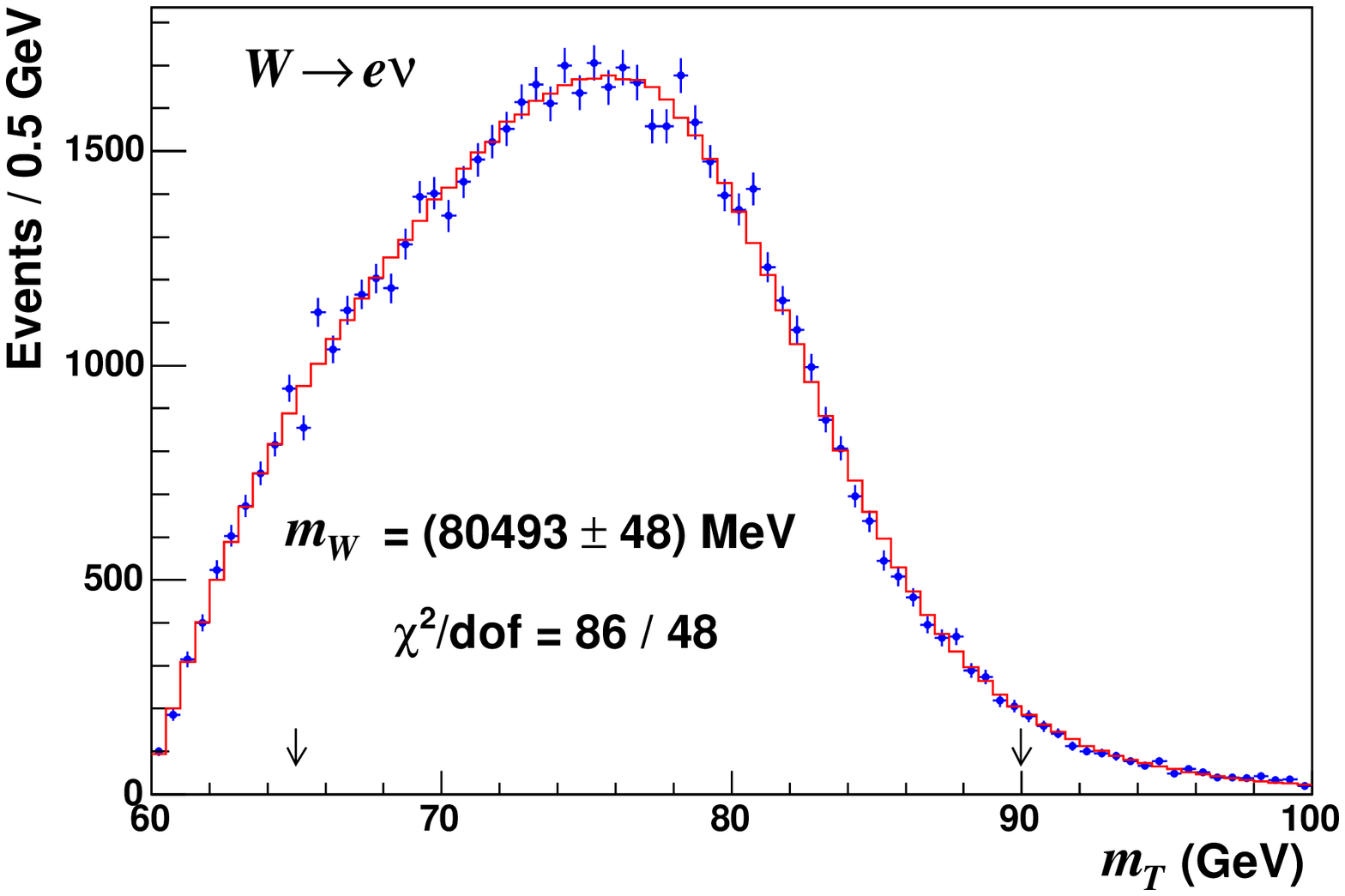}
\caption{The simulation (solid) and data (points) $m_T$ distributions 
for $W$ boson decays to $\mu\nu$ (top) and $e\nu$ (bottom).  The 
simulation corresponds to the best-fit $m_W$, determined using events 
between the two arrows.  The uncertainty is statistical only.  The 
large $\chi^2$ for the electron fit is due to individual bin fluctuations 
(Fig.~\ref{fig:signedchimt}) and does not bias the fit result, as 
evidenced by the small change in the fit $m_W$ when the fit window is 
varied (Fig.~\ref{mass_vs_end_mt}). }
\label{fig:mt}
\end{center}
\end{figure}

\begin{figure}[!tp]
\begin{center}
\epsfysize = 6.cm
\includegraphics*[width=8.5cm]{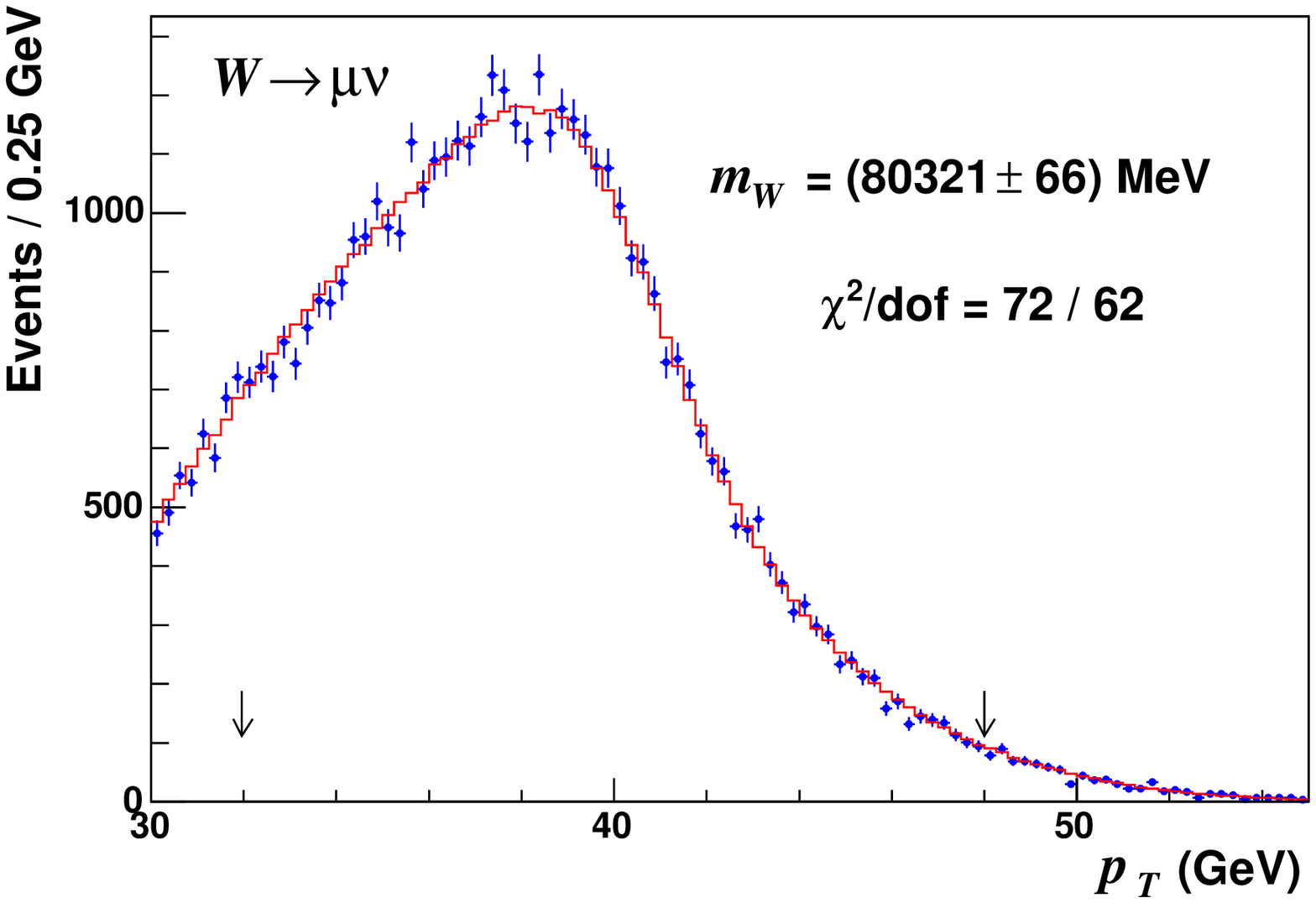}
\includegraphics*[width=8.5cm]{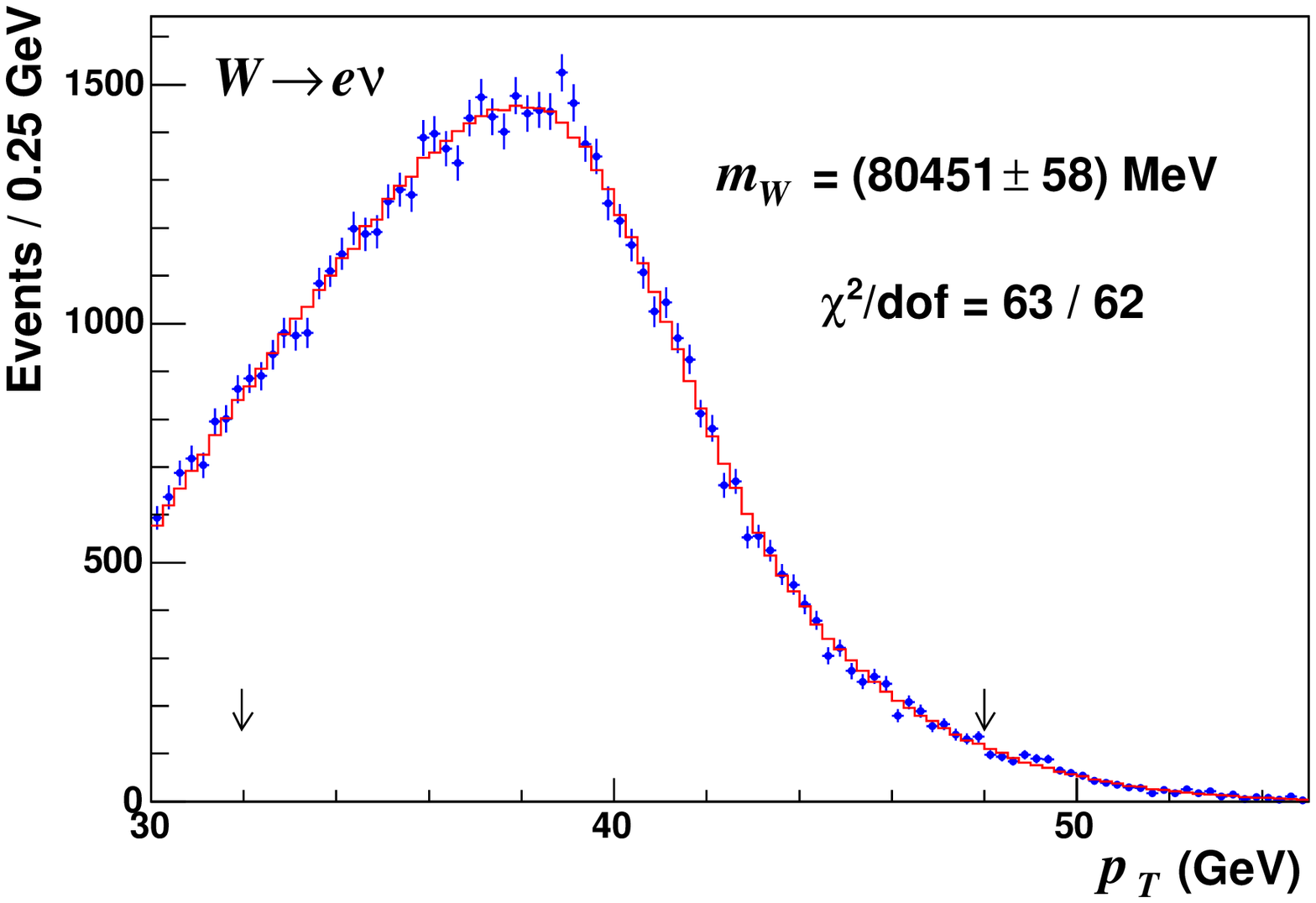}
\caption{The simulation (solid) and data (points) $p_T$ 
distributions for $W$ boson decays to $\mu\nu$ (top) and 
$e\nu$ (bottom).  The simulation corresponds to the best-fit 
$m_W$, determined using events between the two arrows.  The 
uncertainty is statistical only.  }
\label{fig:pt}
\end{center}
\end{figure}

\begin{figure}[!tp]
\begin{center}
\epsfysize = 6.cm
\includegraphics*[width=8.5cm]{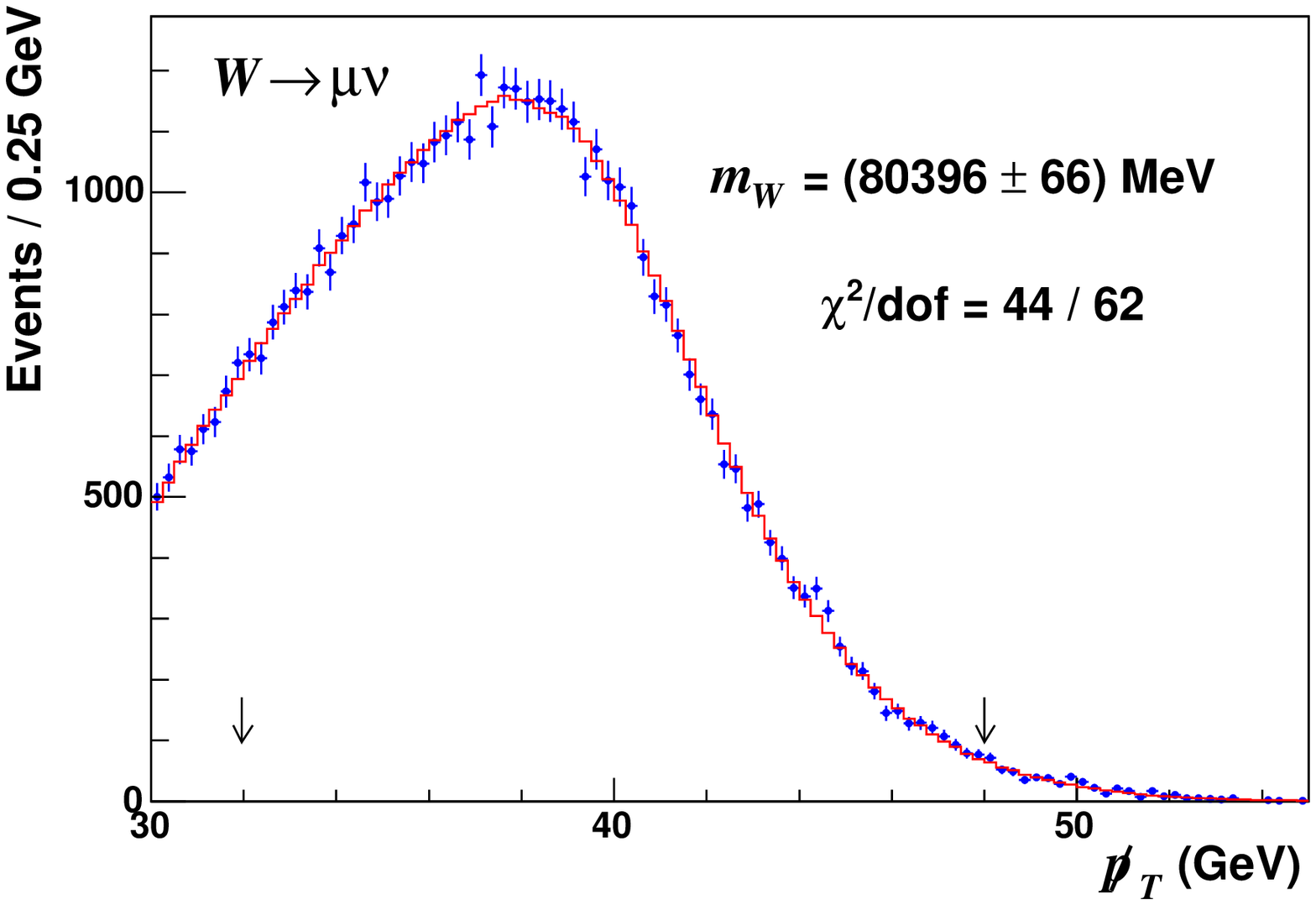}
\includegraphics*[width=8.5cm]{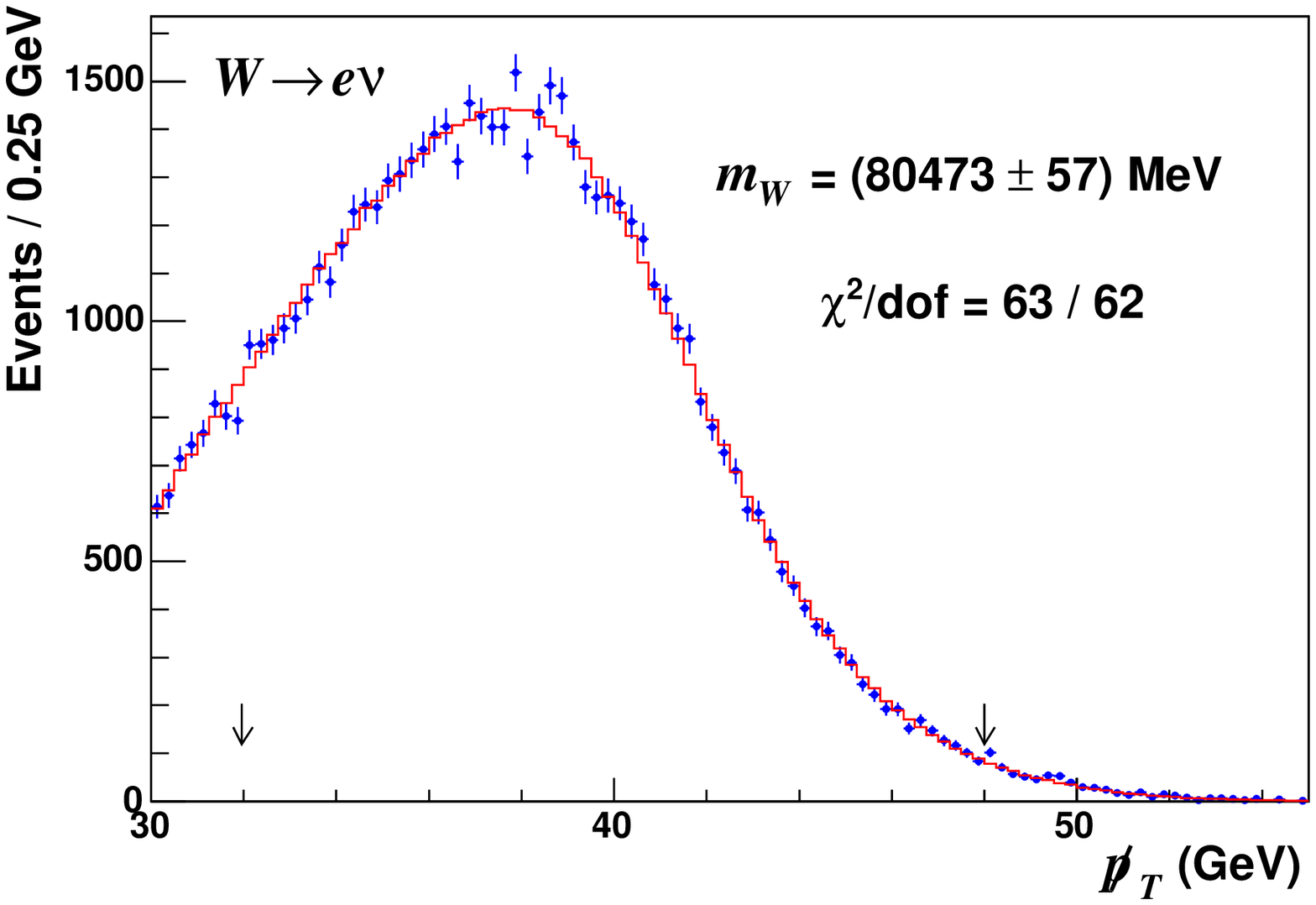}
\caption{The simulation (solid) and data (points) \met 
distributions for $W$ boson decays to $\mu\nu$ (top) and 
$e\nu$ (bottom).  The simulation corresponds to the best-fit 
$m_W$, determined using events between the two arrows.  The 
uncertainty is statistical only.  }
\label{fig:met}
\end{center}
\end{figure}

\begin{figure}[!tp]
\begin{center}
\epsfysize = 6.cm
\includegraphics*[width=8.5cm]{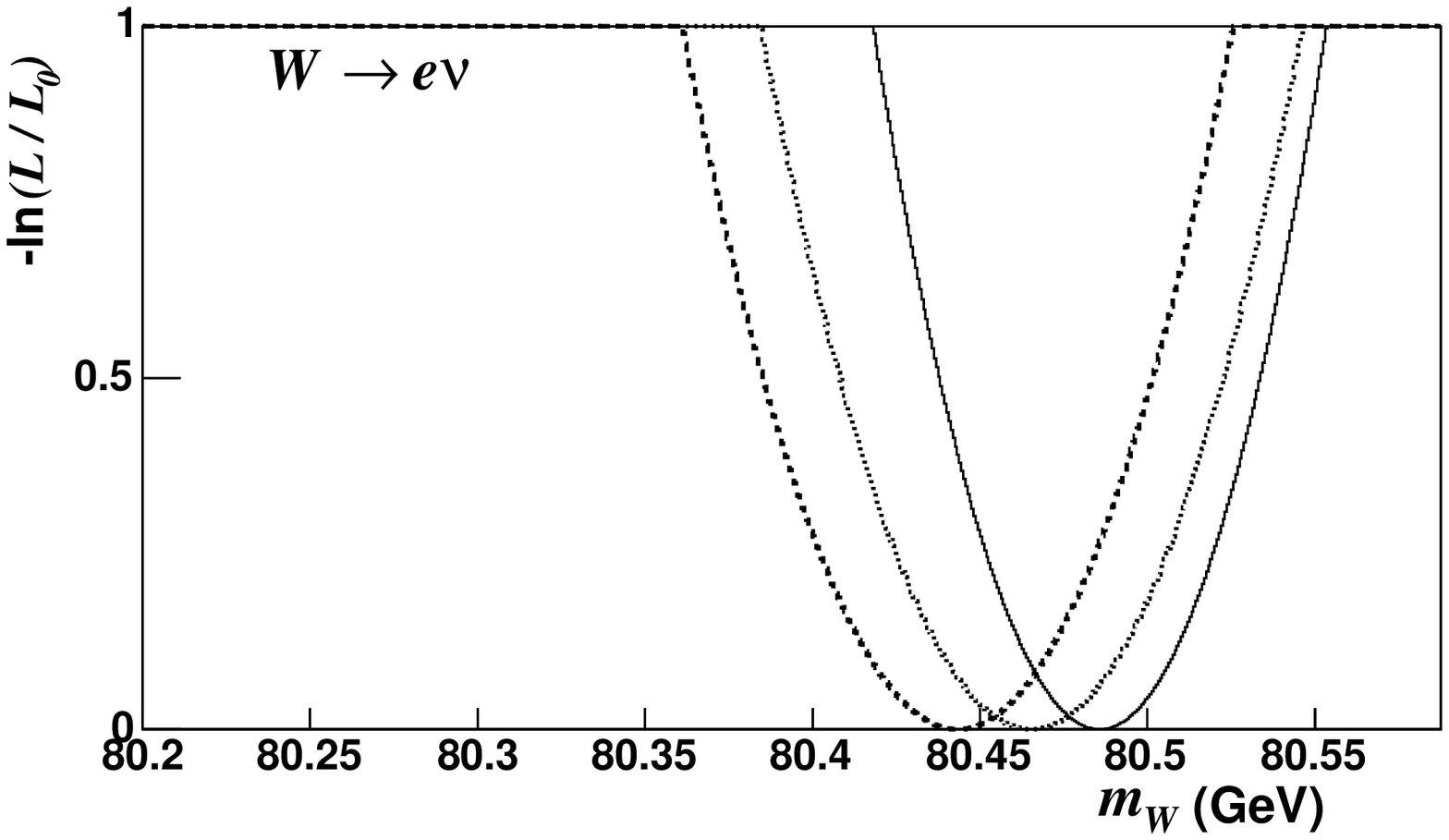}
\includegraphics*[width=8.5cm]{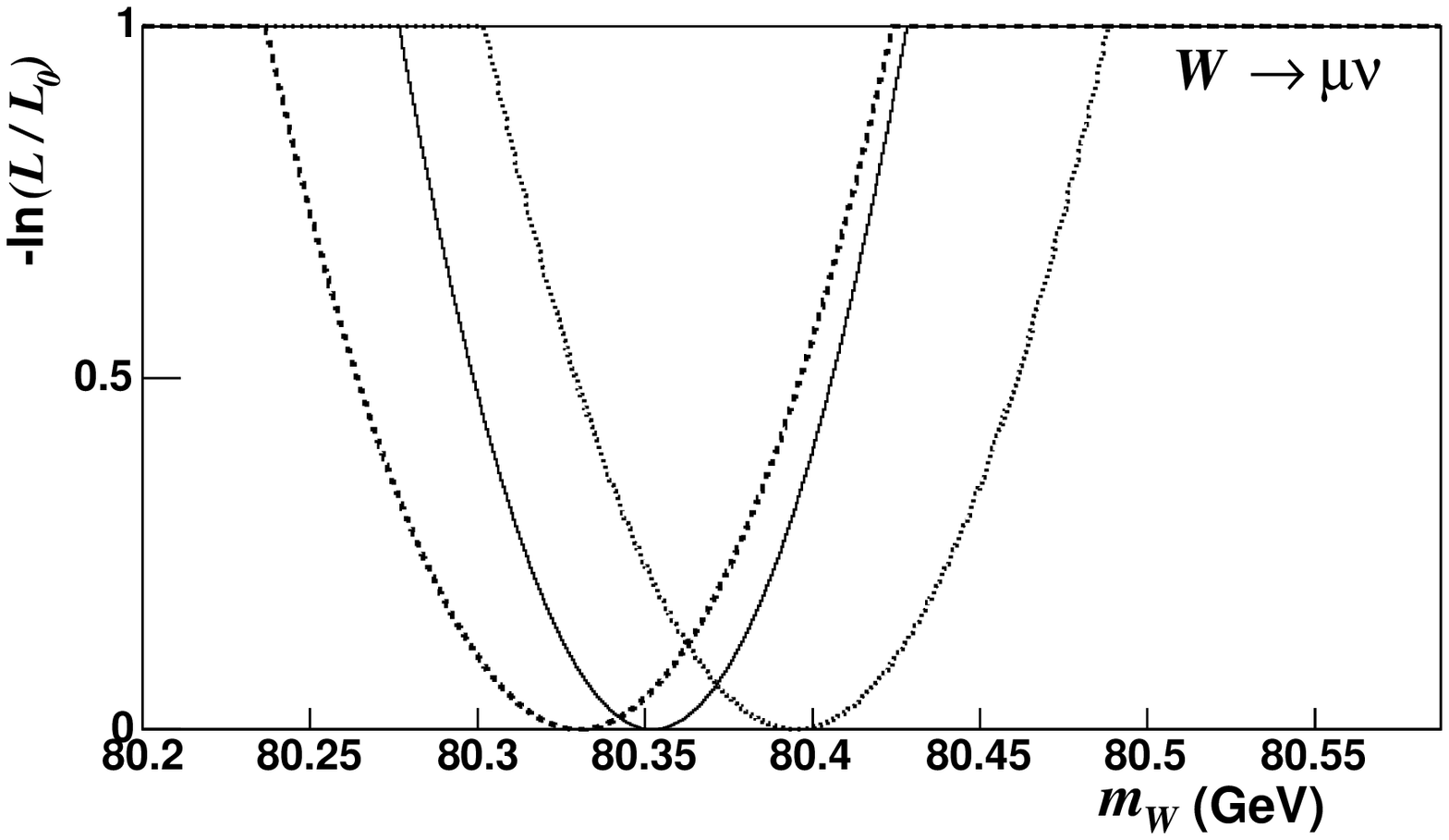}
\caption{The negative log of the likelihood ratio ${\cal{L}}/{\cal{L}}_0$, where 
${\cal{L}}_0$ is the maximum likelihood, as a function of $m_W$ for the $m_T$ (solid), 
$p_T$ (dashed), and \met (dotted) fits in the electron (top) and muon (bottom) 
channels. }
\label{fig:likelihood}
\end{center}
\end{figure}

The results of the $m_T$ fits are shown in Fig. \ref{fig:mt}, and 
Table \ref{tbl:mt} gives a summary of the 68\% confidence level 
uncertainties associated with the fits.  We fit for $m_W$ in the 
range 65 GeV $< m_T < 90$ GeV, where the fit range has been chosen 
to minimize the total uncertainty on $m_W$.  The $p_T$ and \met 
distributions are fit in the range 32 GeV $< p_T < 48$ GeV 
(Figs.~\ref{fig:pt} and \ref{fig:met}, respectively) and have 
uncertainties shown in Tables~\ref{tbl:pt} and \ref{tbl:met}, 
respectively.  We show the individual fit results in 
Table~\ref{tbl:allfits}, and the negative log-likelihoods of all 
fits in Fig.~\ref{fig:likelihood}.

\begin{table}[htbp]
\begin{center}
\begin{tabular}{lccc}
\hline
\hline
\multicolumn{4}{c}{$m_T$ Fit Uncertainties} \\
Source                   & $W\rightarrow \mu\nu$ & $W\rightarrow e\nu$ & Correlation \\
\hline
Tracker Momentum Scale   & 17  &  17  &  100\% \\
Calorimeter Energy Scale & 0   &  25  &  0\% \\
Lepton Resolution        & 3   &  9   &  0\%  \\
Lepton Efficiency	 & 1   &  3   &  0\%  \\
Lepton Tower Removal     & 5   &  8   &  100\%  \\
Recoil Scale	         & 9   &  9   &  100\%  \\
Recoil Resolution	 & 7   &  7   &  100\%  \\
Backgrounds 	         & 9   &  8   &  0\%  \\
PDFs                     & 11  &  11  &  100\% \\
$W$ Boson $p_T$          & 3   &  3   &  100\%  \\
Photon Radiation         & 12  &  11  &  100\% \\
\hline
Statistical              & 54   & 48  &  0\%  \\
\hline
Total                    & 60   & 62  &  - \\
\hline
\hline
\end{tabular}
\caption{Uncertainties in units of MeV on the transverse mass fit for 
$m_W$ in the $W\rightarrow \mu\nu$ and $W\rightarrow e\nu$ samples. }
\label{tbl:mt}
\end{center}
\end{table}

\begin{table}[htbp]
\begin{center}
\begin{tabular}{lccc}
\hline
\hline
\multicolumn{4}{c}{$p_T$ Fit Uncertainties} \\
Source                   & $W\rightarrow \mu\nu$ & $W\rightarrow e\nu$ & Correlation \\
\hline
Tracker Momentum Scale   & 17  &  17  &  100\% \\
Calorimeter Energy Scale & 0   &  25  &  0\% \\
Lepton Resolution        & 3   &  9   &  0\%  \\
Lepton Efficiency	 & 6   &  5   &  0\%  \\
Lepton Tower Removal     & 0   &  0   &  0\%  \\
Recoil Scale		 & 17  &  17  &  100\% \\
Recoil Resolution	 & 3   &  3   &  100\% \\
Backgrounds 	         & 19  &  9   &  0\%  \\
PDFs                     & 20  &  20  &  100\% \\
$W$ Boson $p_T$          & 9   &  9   &  100\%  \\
Photon Radiation         & 13  &  13  &  100\% \\
\hline
Statistical              & 66   & 58  &  0\%  \\
\hline
Total                    & 77   & 73  &  - \\
\hline
\hline
\end{tabular}
\caption{Uncertainties in units of MeV on the charged lepton transverse momentum 
fit for $m_W$ in the $W\rightarrow \mu\nu$ and $W\rightarrow e\nu$ samples. }
\label{tbl:pt}
\end{center}
\end{table}

\begin{table}[htbp]
\begin{center}
\begin{tabular}{lccc}
\hline
\hline
\multicolumn{4}{c}{\met Fit Uncertainties} \\
Source                   & $W\rightarrow \mu\nu$ & $W\rightarrow e\nu$ & Correlation \\
\hline
Tracker Momentum Scale   & 17  &  17  &  100\% \\
Calorimeter Energy Scale & 0   &  25  &  0\% \\
Lepton Resolution        & 5   &  9   &  0\%  \\
Lepton Efficiency	 & 13  &  16  &  0\%  \\
Lepton Tower Removal     & 10  &  16  &  100\%  \\
Recoil Scale		 & 15  &  15  &  100\% \\
Recoil Resolution	 & 30  &  30  &  100\% \\
Backgrounds 	         & 11  &  7   &  0\%  \\
PDFs                     & 13  &  13  &  100\% \\
$W$ Boson $p_T$          & 5   &  5   &  100\% \\
Photon Radiation         & 10  &  9   &  100\% \\
\hline
Statistical              & 66   & 57  &  0\%  \\
\hline
Total                    & 80   & 79  &  - \\
\hline
\hline
\end{tabular}
\caption{Uncertainties in units of MeV on the missing transverse momentum 
fit for $m_W$ in the $W\rightarrow \mu\nu$ and $W\rightarrow e\nu$ samples. } 
\label{tbl:met}
\end{center}
\end{table}

\begin{table}[htbp]
\begin{center}
\begin{tabular}{lcc}
\hline
\hline
Distribution   & $m_W$ (GeV)                               & $\chi^2$/dof \\
\hline
$m_T(e,\nu)$   & $80.493 \pm 0.048 \pm 0.039$ & 86/48 \\
$p_T(e)$       & $80.451 \pm 0.058 \pm 0.045$ & 63/62 \\
\met(e)        & $80.473 \pm 0.057 \pm 0.054$ & 63/62 \\ 
\hline
$m_T(\mu,\nu)$ & $80.349 \pm 0.054 \pm 0.027$ & 59/48 \\
$p_T(\mu)$     & $80.321 \pm 0.066 \pm 0.040$ & 72/62 \\
\met($\mu$)    & $80.396 \pm 0.066 \pm 0.046$ & 44/62 \\ 
\hline
\hline
\end{tabular}
\caption{The results of the fits for $m_W$ to the $m_T$, $p_T$, and \met 
distributions in the electron and muon decay channels.  The first uncertainty
is statistical and the second is systematic.}
\label{tbl:allfits}
\end{center}
\end{table}

We combine results from the $W\rightarrow \mu\nu$ and $W\rightarrow e\nu$
fits using the Best Linear Unbiased Estimator (BLUE) \cite{blue}.  The 
BLUE algorithm defines a procedure for constructing a complete covariance
matrix using the derivative of $m_W$ with respect to each model 
parameter \cite{DZEROEC}.  We construct this matrix assuming each source 
of systematic uncertainty is independent of any other source of uncertainty.  
The resulting covariance matrix (Table~\ref{tbl:covariance}) is then used 
to combine all six $m_W$ fits.  When combining any subset of fits, the 
appropriate smaller covariance matrix is used.

\begin{table}[htbp]
\begin{center}
\begin{tabular}{lcccccc}
\hline
\hline
      & $m_T(e,\nu)$ & $m_T(\mu,\nu)$ & $p_T(e)$ & $p_T(\mu)$ & \met$(e)$ & \met$(\mu)$ \\
\hline
$m_T(e,\nu)$   & $64^2$ & $27^2$ & $61^2$ & $27^2$ & $61^2$ & $28^2$ \\
$m_T(\mu,\nu)$ &        & $61^2$ & $27^2$ & $59^2$ & $28^2$ & $59^2$ \\
$p_T(e)$       &        &        & $75^2$ & $35^2$ & $51^2$ & $32^2$ \\
$p_T(\mu)$     &        &        &        & $77^2$ & $32^2$ & $53^2$ \\
\met$(e)$      &        &        &        &        & $81^2$ & $43^2$ \\
\met$(\mu)$    &        &        &        &        &        & $81^2$ \\
\hline
\hline
\end{tabular}
\caption{The complete covariance matrix for the $m_T$, $p_T$, and \met fits 
in the electron and muon decay channels, in units of MeV$^2$.  The matrix is 
symmetric. }
\label{tbl:covariance}
\end{center}
\end{table}

The result of combining the $m_W$ fits to the $m_T$ distribution in the 
$W\rightarrow \mu\nu$ and $W\rightarrow e\nu$ channels is
\begin{equation}
m_W = 80.417 \pm 0.048~{\rm GeV}.
\end{equation}

\noindent
The $\chi^2$/dof of the combination is 3.2/1 and the probability that 
two measurements of the same quantity would have a $\chi^2$/dof at least 
as large as this is 7\%.
\par
The combination of the fits to the $p_T$ distribution yields 
\begin{equation}
m_W = 80.388 \pm 0.059~{\rm GeV,}
\end{equation}

\noindent
with a $\chi^2$/dof of 1.8/1 and an 18\% probability for the two 
measurements to obtain a $\chi^2$/dof~$\geq$~1.8.
\par
The results of the fits to the \met distribution gives
\begin{equation}
m_W = 80.434 \pm 0.065~{\rm GeV,}
\end{equation}

\noindent
with a 43\% probability of obtaining a $\chi^2$/dof at least as large as observed
(0.6/1).
\par
Combining the $m_T$, $p_T$, and \met fits within the individual decay channels 
gives \mbox{$m_W = (80.352 \pm 0.060)$ GeV} with a $\chi^2$/dof of 1.4/2 for the 
$W\rightarrow \mu\nu$ channel and $m_W = (80.477 \pm 0.062)$ GeV with a $\chi^2$/dof 
of 0.8/2 for the $W\rightarrow e\nu$ channel.   
\par
We combine the six fits with the BLUE procedure to obtain our final result of
\begin{equation}
m_W = 80.413 \pm 0.048~{\rm GeV,}
\end{equation}

\noindent
which has statistical and systematic uncertainties of 34 MeV each.  The 
statistical correlations between the fits, determined from simulation 
pseudoexperiments, are shown in Table \ref{tbl:correlation}.  The relative
weights of the fits are 47.7\% (32.3\%), 3.4\% (8.9\%), 0.9\% (6.8\%) for the 
$m_T$, $p_T$ and \met fit distributions, respectively, in the muon (electron) 
channel.  The combination establishes an {\it a priori} procedure to incorporate
all the information from individual fits, and yields a $\chi^2$/dof of 4.8/5.  
The probability to obtain a $\chi^2$ at least as large as this is 44\%. 

\begin{table}[htbp]
\begin{center}
\begin{tabular}{lcc}
\hline
\hline
Correlation      & $W\rightarrow \mu\nu$ & $W\rightarrow e\nu$ \\
\hline
$m_T-p_T$        & 70\%  & 68\% \\
$m_T-$\met       & 72\%  & 63\% \\
$p_T-$\met	 & 38\%  & 17\% \\
\hline
\hline
\end{tabular}
\caption{The statistical correlations between the $m_T$, $p_T$, and \met fits 
in the electron and muon decay channels. }
\label{tbl:correlation}
\end{center}
\end{table}

\subsection{Cross-Checks}

\begin{figure}[!tp]
\begin{center}
\epsfysize = 6.cm
\includegraphics*[width=8.5cm]{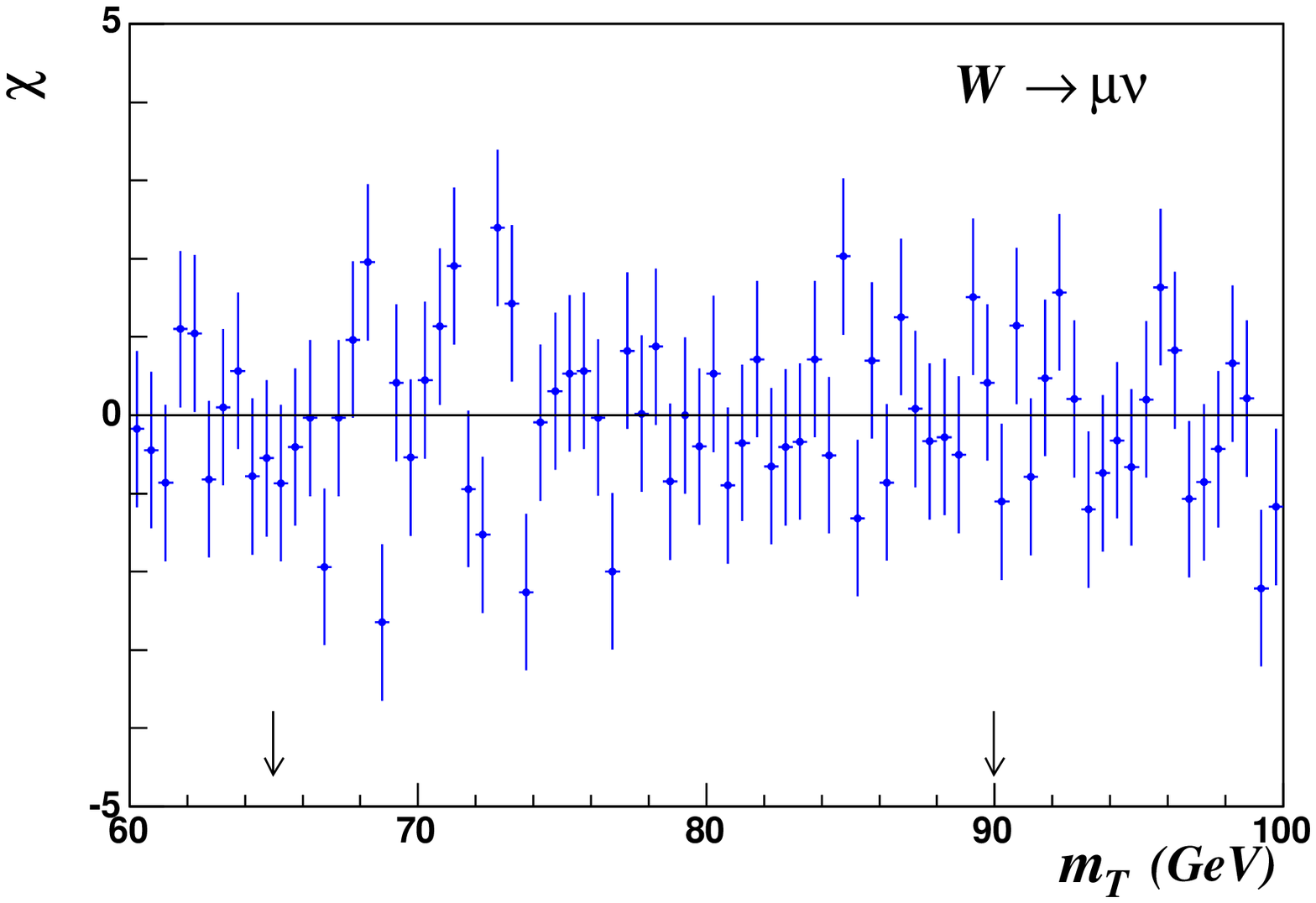}
\includegraphics*[width=8.5cm]{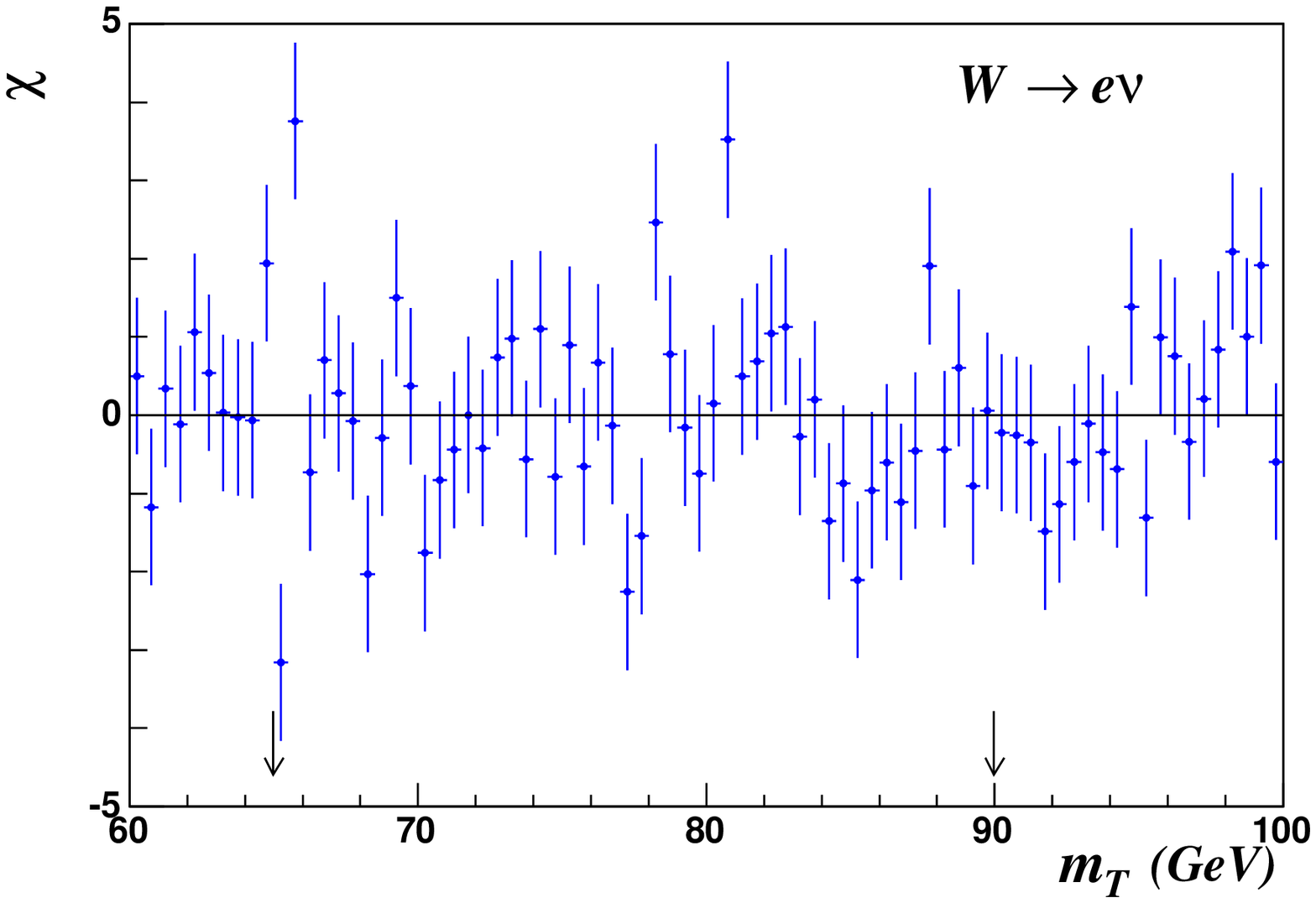}
\caption{The difference between the data and simulation, divided by the 
statistical uncertainty on the prediction, for the $m_T$ distributions in the 
muon (top) and electron (bottom) channels.  The arrows indicate the fit region. }
\label{fig:signedchimt}
\end{center}
\end{figure}

\begin{figure}[!tp]
\begin{center}
\epsfysize = 6.cm
\includegraphics*[width=8.5cm]{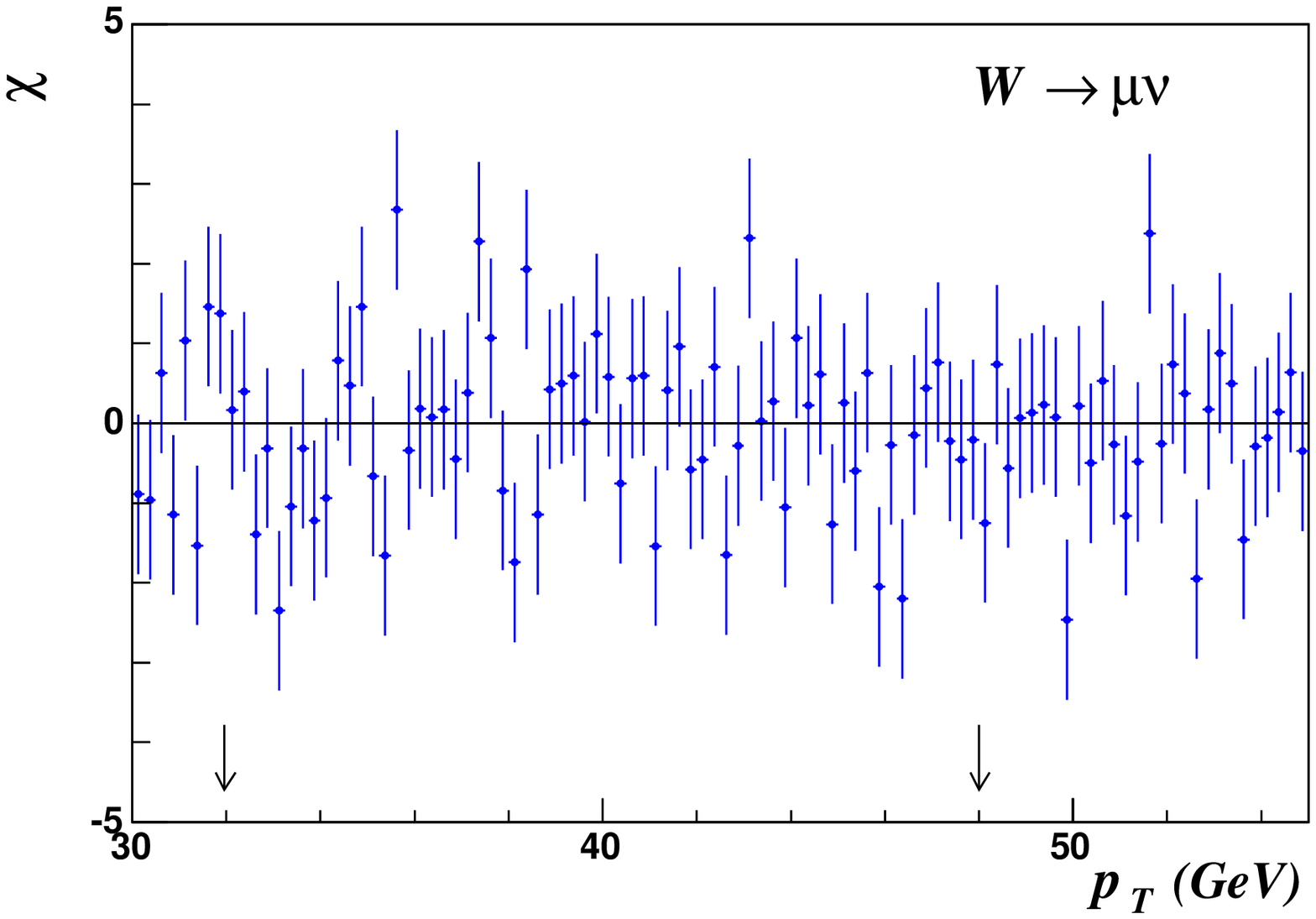}
\includegraphics*[width=8.5cm]{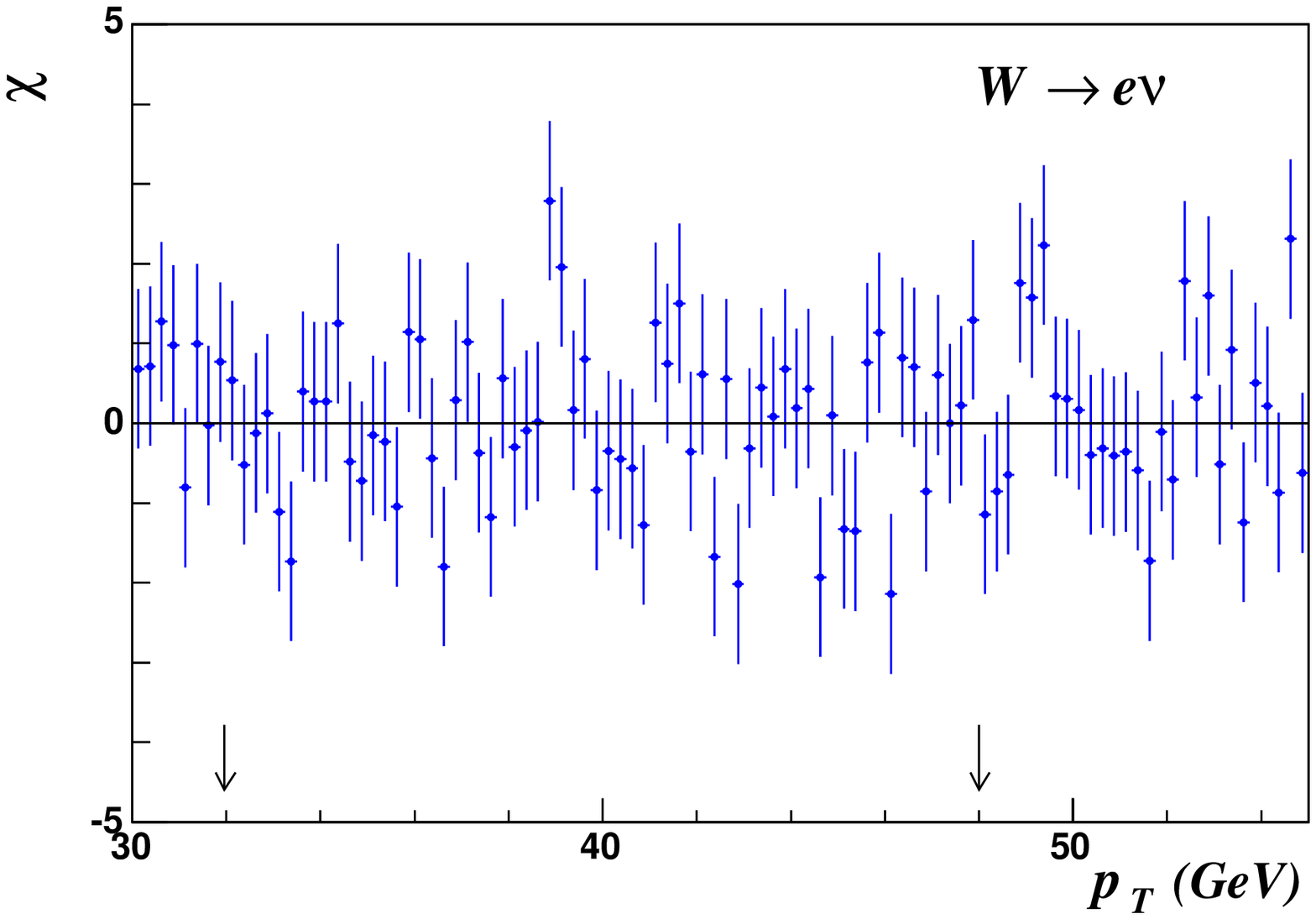}
\caption{The difference between the data and simulation, divided by the 
statistical uncertainty on the prediction, for the $p_T$ distributions in the 
muon (top) and electron (bottom) channels.  The arrows indicate the fit region. }
\label{fig:signedchipt}
\end{center}
\end{figure}

\begin{figure}[!tp]
\begin{center}
\epsfysize = 6.cm
\includegraphics*[width=8.5cm]{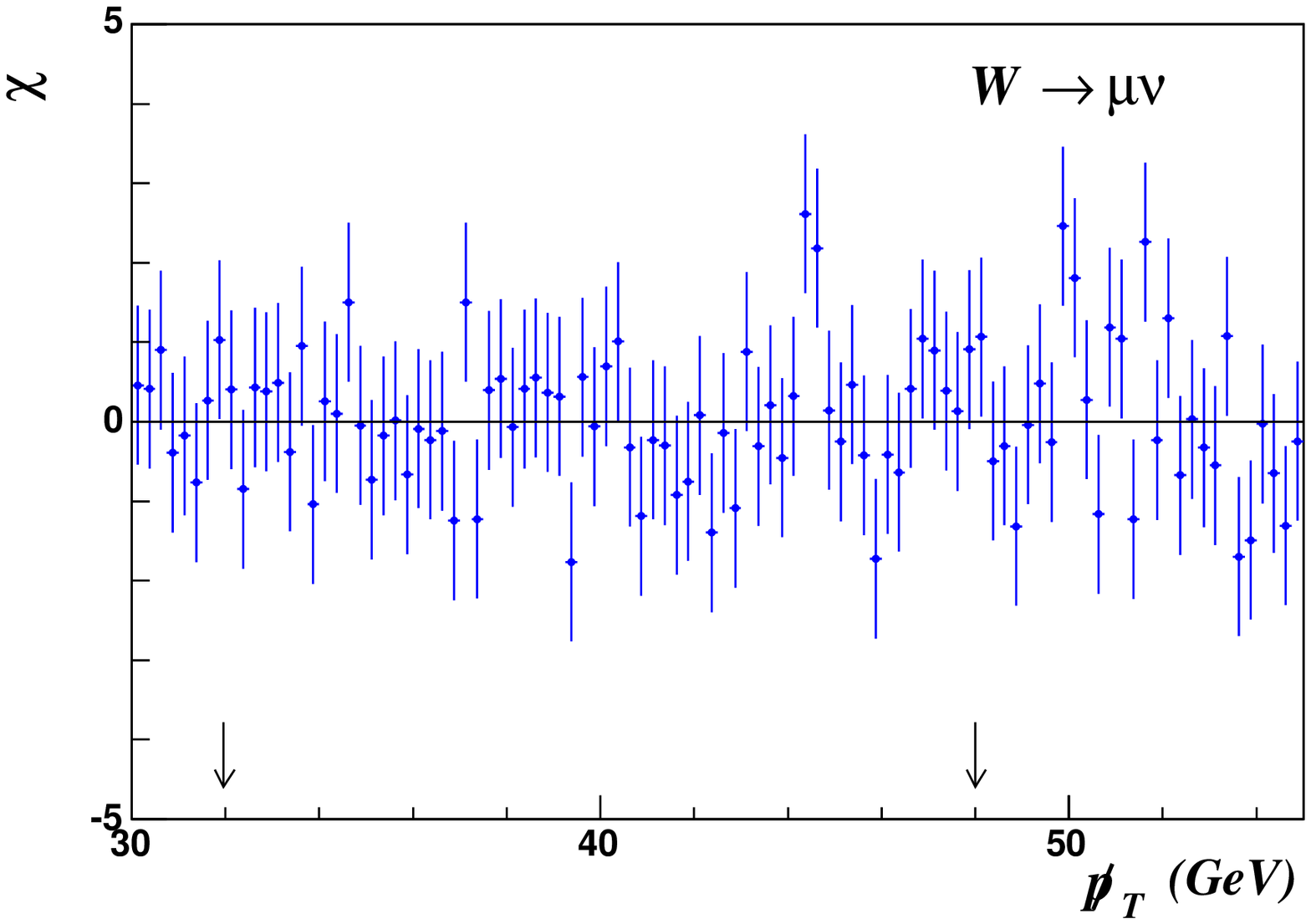}
\includegraphics*[width=8.5cm]{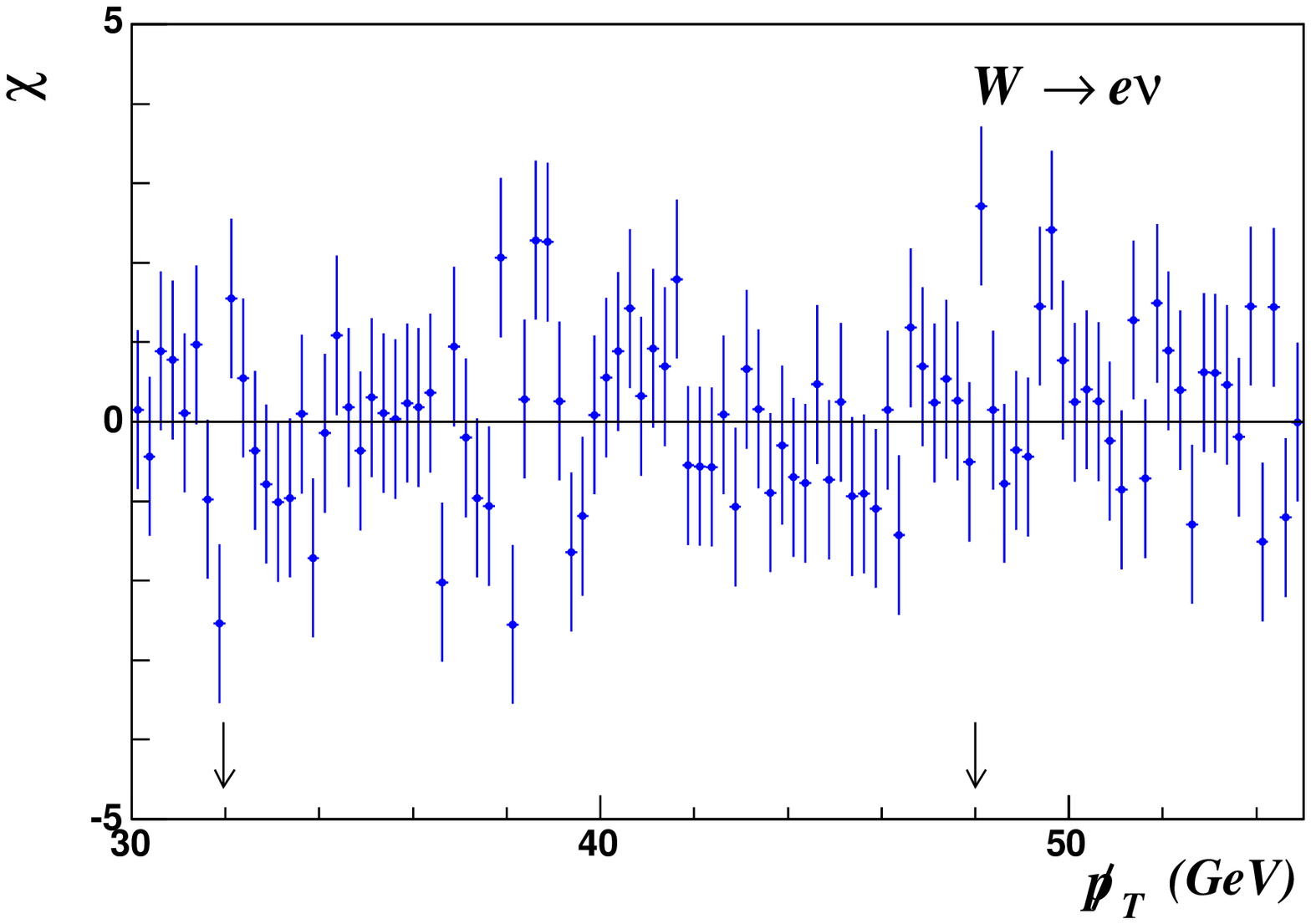}
\caption{The difference between the data and simulation, divided by the 
statistical uncertainty on the data points, for the \met distributions in the 
muon (top) and electron (bottom) channels.  The arrows indicate the fit region. }
\label{fig:signedchimet}
\end{center}
\end{figure}

\begin{figure}
\begin{center}
\epsfysize = 6.cm
\includegraphics*[width=8.5cm]{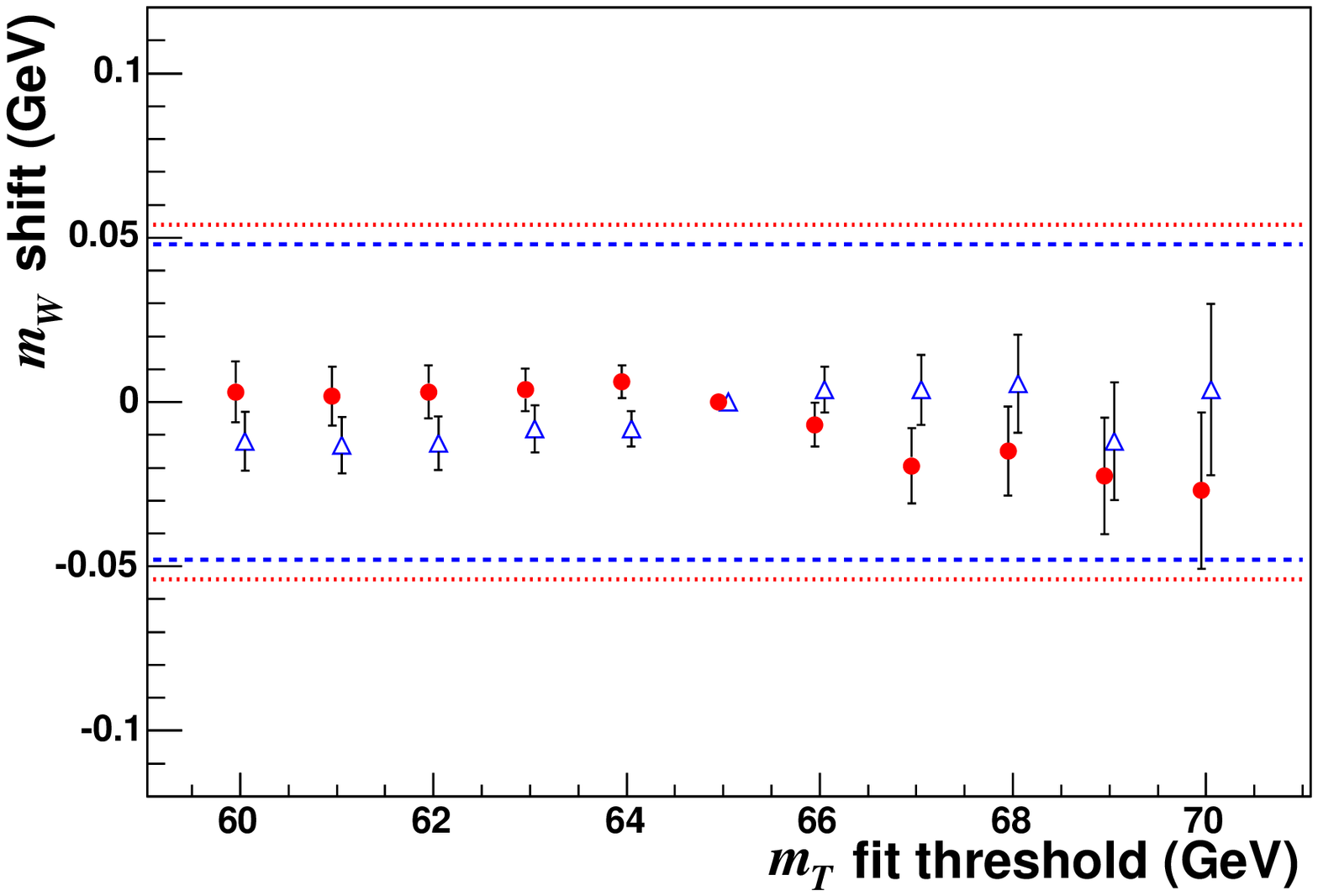}
\includegraphics*[width=8.5cm]{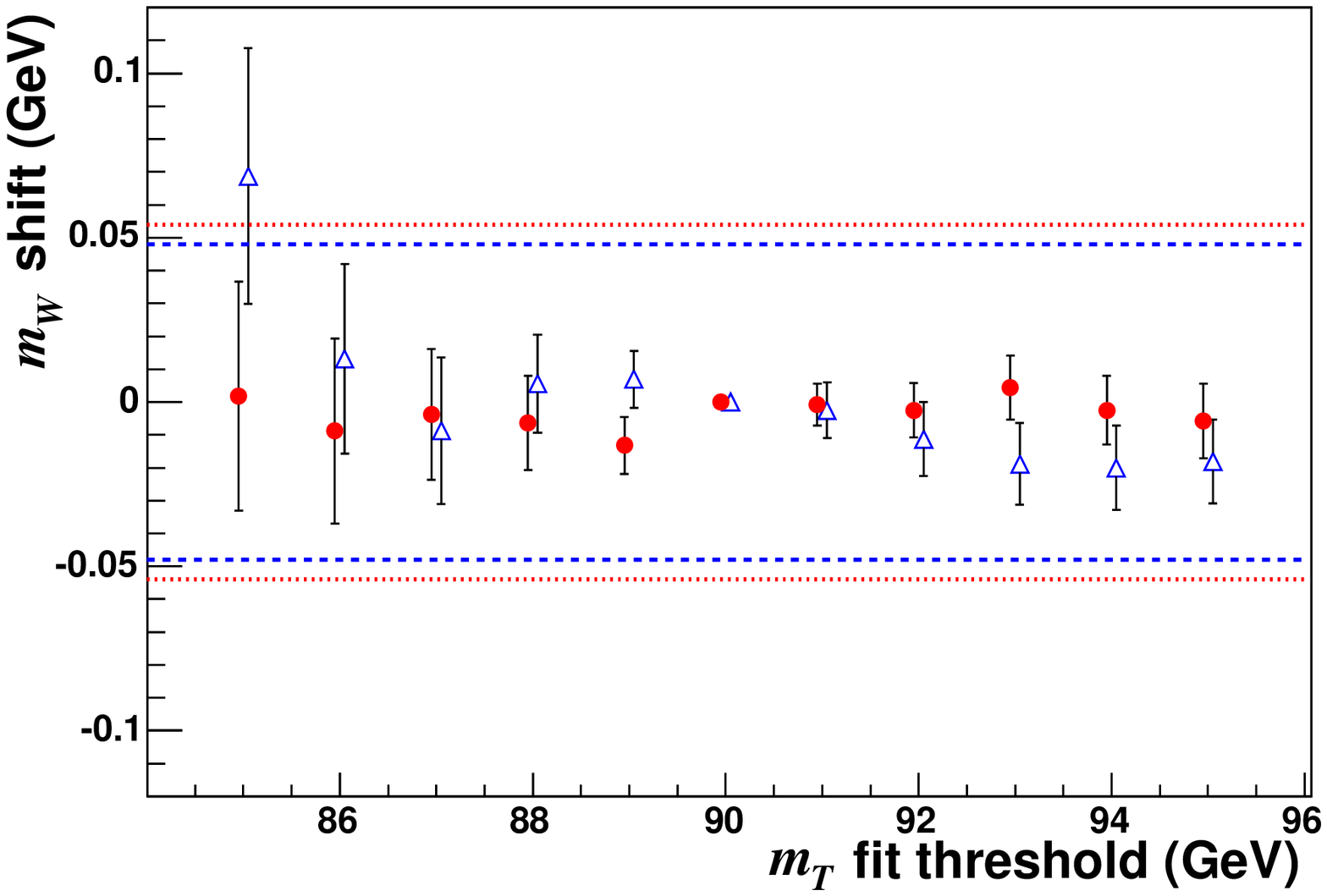}
\caption{The shifts in measured $m_W$ for variations in the lower (top) and upper 
(bottom) edges of the $m_T$ fit range.  The electron channel is denoted by 
open triangles and the muon channel by solid circles.  The error bars indicate the expected 
statistical variations from simulation pseudoexperiments.  The dashed (dotted) lines 
indicate the statistical uncertainty on the $m_W$ fit using the default fit range in the 
electron (muon) channel.  }
\label{mass_vs_end_mt}
\end{center}
\end{figure}

\begin{figure}
\begin{center}
\epsfysize = 6.cm
\includegraphics*[width=8.5cm]{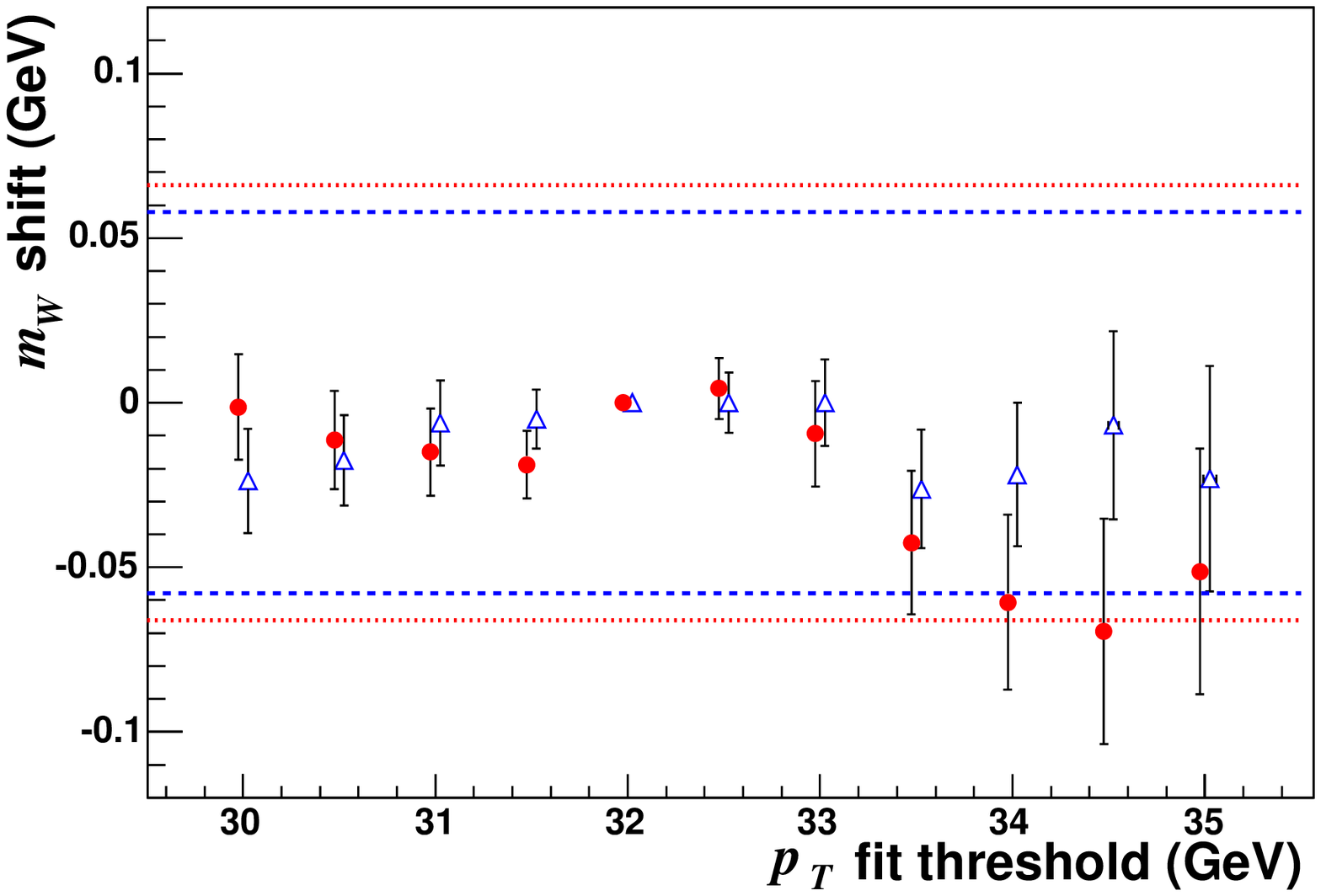}
\includegraphics*[width=8.5cm]{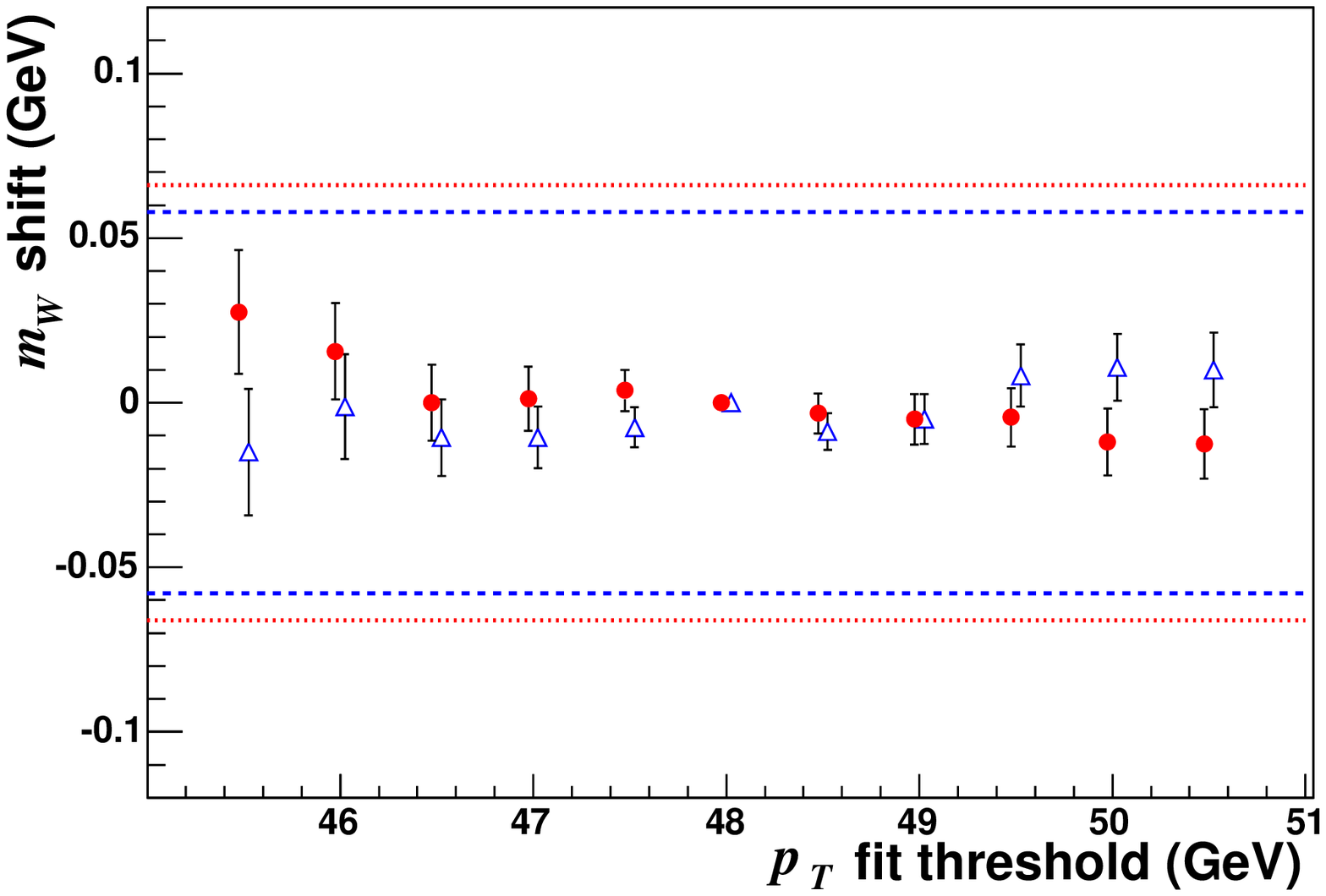}
\caption{The shifts in measured $m_W$ for variations in the lower (top) and upper 
(bottom) edges of the $p_T$ fit range.  The electron channel is denoted by open
triangles and the muon channel by solid circles.  The error bars indicate the expected 
statistical variations from simulation pseudoexperiments.  The dashed (dotted) lines indicate
the statistical uncertainty on the $m_W$ fit using the default fit range in the electron
(muon) channel.  }
\label{mass_vs_end_pt}
\end{center}
\end{figure}

\begin{figure}
\begin{center}
\epsfysize = 6.cm
\includegraphics*[width=8.5cm]{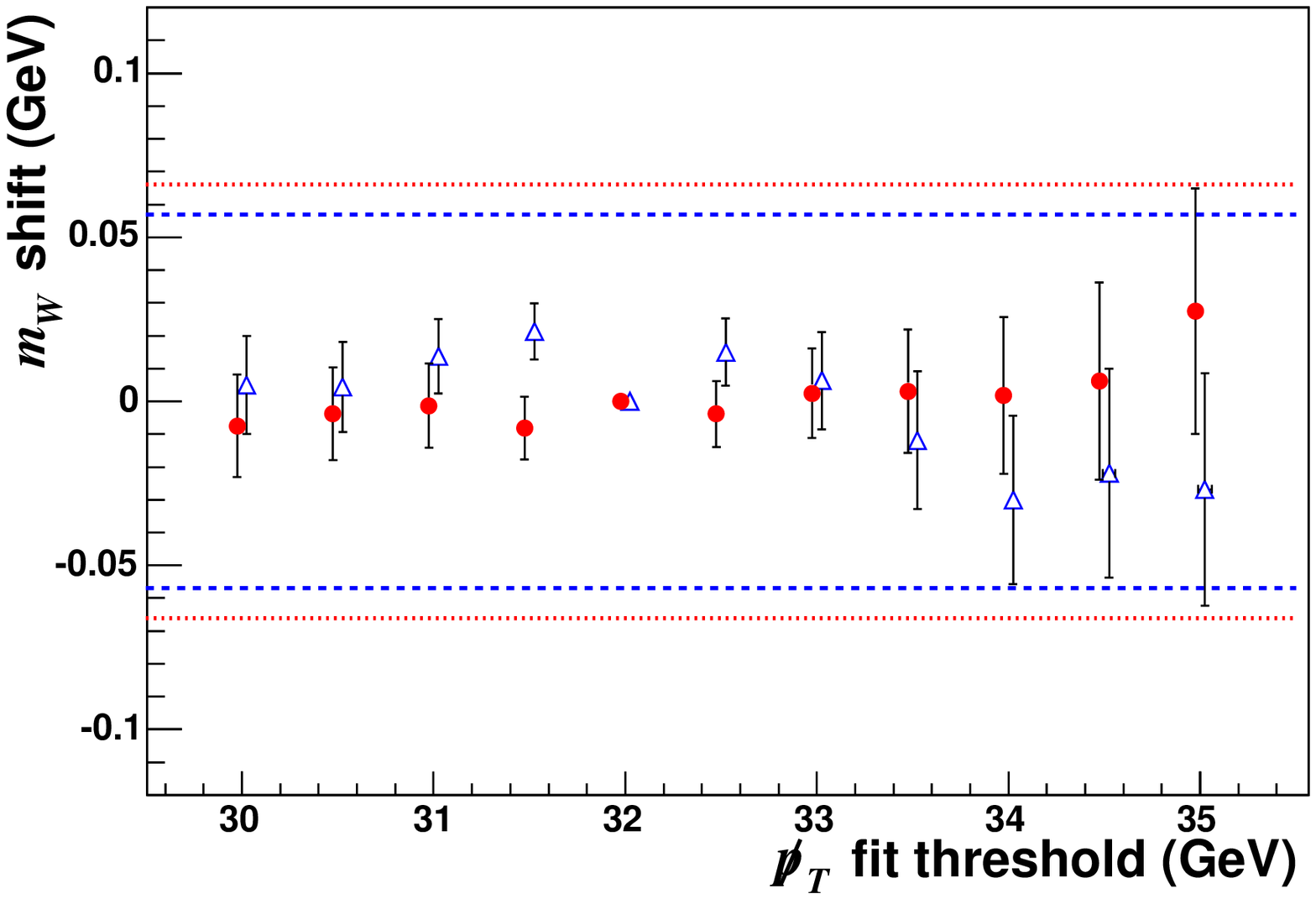}
\includegraphics*[width=8.5cm]{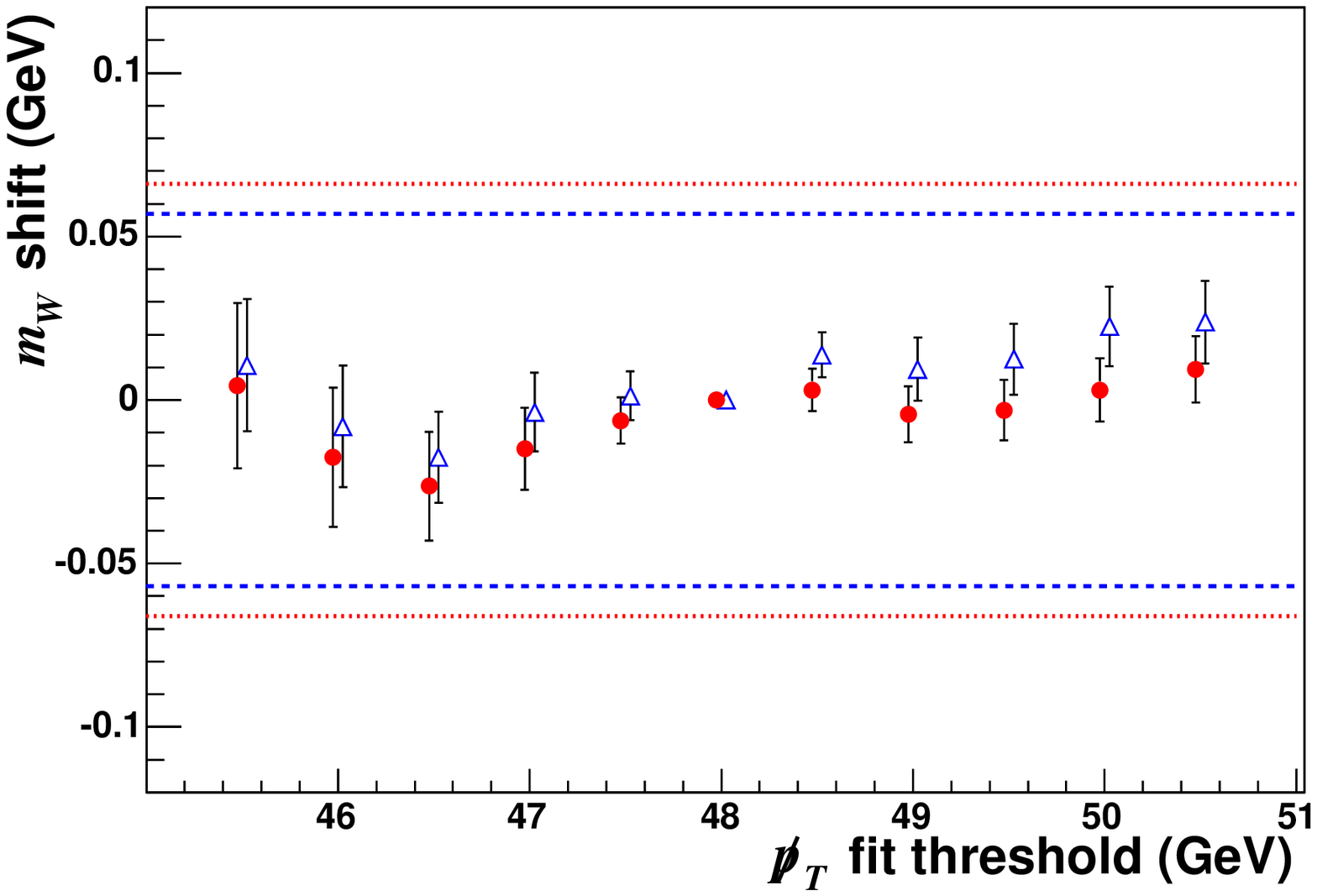}
\caption{The shifts in measured $m_W$ for variations in the lower (top) and upper 
(bottom) edges of the \met fit range.  The electron channel is denoted by open
triangles and the muon channel by solid circles.  The error bars indicate the expected 
statistical variations from simulation pseudoexperiments.  The dashed (dotted) lines 
indicate the statistical uncertainty on the $m_W$ fit using the default fit range in the 
electron (muon) channel.  }
\label{mass_vs_end_met}
\end{center}
\end{figure}

Figures~\ref{fig:signedchimt}-\ref{fig:signedchimet} show the differences between 
data and simulation, divided by the statistical uncertainties on the predictions, for 
the $m_T$, $p_T$ and \met distributions.  Figures~\ref{mass_vs_end_mt}-\ref{mass_vs_end_met}
show the variations of the fitted mass values, relative to the nominal results, as 
the fit regions are varied.  These plots show variations consistent with statistical 
fluctuations.


\par
The variation of the $p_T$ fits with time, detector region, and lepton charge 
(Table \ref{tbl:crosschecks}) show no evidence of dependence on time or detector 
region.  There is a difference between positive and negative lepton mass fits at 
the level of $\approx 2\sigma$ in each decay channel.  The largest systematic 
uncertainty in this difference arises in the muon channel from the uncertainty 
on the alignment parameters $a_0$ and $a_2$ (Table \ref{tbl:eoppars}).  The 
uncertainties on the mass difference due to these parameters are 49 MeV and 
56 MeV, respectively, for a total uncertainty of 75 MeV.  Any bias in 
these parameters affects the positive and negative lepton fits in opposite
directions, and thus has a negligible net effect when the two are combined.  

\begin{table}[htbp]
\begin{center}
\begin{tabular}{lcc}
\hline
\hline
Fit difference      & $W\rightarrow \mu\nu$  & $W\rightarrow e\nu$ \\
\hline
$m_W$($l^+$)$-m_W$($l^-$)                 & $286 \pm 152$   & $257 \pm 117$ \\
$m_W$($\phi_l > 0$)$ - m_W$($\phi_l < 0$) & $0 \pm 133$     & $116 \pm 117$ \\ 
$m_W$(Mar, 2002-Apr, 2003) - & & \\ 
$m_W$(Apr, 2003-Sep, 2003) & $75 \pm 135$ & $-107 \pm 117$ \\
\hline
\hline
\end{tabular}
\caption{Differences of $m_W$ in the $p_T$ fits between positively and negatively 
charged leptons, leptons in the upper and lower halves of the detector, and early 
and late data.  The units are MeV. }
\label{tbl:crosschecks}
\end{center}
\end{table}

\section{Summary}
\label{sec:summary}

We have performed a measurement of the $W$ boson mass using 
$200$ pb$^{-1}$ of data collected by the CDF II detector at 
$\sqrt{s} = 1.96$ TeV.  From fits to $m_T$, $p_T$, and \met 
distributions of the $W\rightarrow \mu\nu$ and $W\rightarrow e\nu$ 
data samples, we obtain
\begin{equation}
m_W = 80.413 \pm 0.048~{\rm GeV},
\end{equation}

\noindent
which is the single most precise determination of $m_W$ to date.  The 
uncertainty includes statistical and systematic contributions of 34 MeV 
each.
\par
Combining this result with the Run I Tevatron measurements using the 
method in \cite{run1combo}, we obtain a CDF Run I/II combined result of 
\begin{equation}
m_W = 80.418 \pm 0.042~{\rm GeV},
\end{equation}

\noindent
and a combined Tevatron result of
\begin{equation}
m_W = 80.429 \pm 0.039~{\rm GeV.}
\end{equation}

\noindent
In these combinations, we take the uncertainties due to PDFs and
photon radiation to be fully correlated between our measurement 
and the previous Tevatron measurements.  In the BLUE combination 
method \cite{blue}, each uncertainty source contributes its 
covariance matrix, and all covariance matrices are summed to obtain 
the total covariance matrix.  We evaluate an individual contribution 
to the uncertainty on our result by ignoring its respective covariance 
matrix and repeating the six-fold combination of our individual 
electron and muon channel $m_T$, $p_T$ and \met fits.  The difference 
in quadrature between the total uncertainty, including and excluding 
a given covariance matrix contribution, is taken to be the uncertainty 
due to that source.  Following this procedure, we obtain the systematic
uncertainty contributions due to PDFs and QED radiative corrections 
to be 12.6 MeV and 11.6 MeV respectively (Table \ref{tbl:totalsys}), 
for a combined uncertainty of 17.2 MeV.  

\begin{table}[htbp]
\begin{center}
\begin{tabular}{lc}
\hline
\hline
Source                   & Uncertainty (MeV) \\
\hline
Lepton Scale             & 23.1 \\
Lepton Resolution        & 4.4 \\
Lepton Efficiency	 & 1.7 \\
Lepton Tower Removal     & 6.3 \\
Recoil Energy Scale	 & 8.3 \\
Recoil Energy Resolution & 9.6 \\
Backgrounds 	         & 6.4 \\
PDFs                     & 12.6 \\
$W$ Boson $p_T$          & 3.9 \\
Photon Radiation         & 11.6 \\
\hline
\hline
\end{tabular}
\caption{Systematic uncertainties in units of MeV on the combination of the 
six fits in the electron and muon channels.  Each uncertainty has been 
estimated by removing its covariance and repeating the six-fold combination. } 
\label{tbl:totalsys}
\end{center}
\end{table}

Assuming no correlation between the Tevatron and LEP 
measurements, we obtain a new world average of 
\begin{equation}
\label{worldaverage}
m_W = 80.398 \pm 0.025~{\rm GeV.}
\end{equation}

\noindent
Our measurement reduces the world uncertainty to 31 parts in $10^5$, 
and further constrains the properties of the Higgs boson and other 
new particles coupling to the $W$ and $Z$ bosons.  Within the context 
of the standard model, fits made to high energy precision electroweak 
data in 2006 gave $m_H = 85^{+39}_{-28}$~GeV, with $m_H < 166$~GeV at 
the 95\% confidence level \cite{LEP}.  The values used for the 
top quark and $W$ boson masses in these fits were 
$m_t = (171.4 \pm 2.1)$~GeV and $m_W = (80.392 \pm 0.029)$~GeV, 
respectively.  Updating these fits with the most recent world average 
values of $m_t = (170.9 \pm 1.8)$~GeV and $m_W = (80.398 \pm 0.025)$~GeV 
[Eq.~(\ref{worldaverage})], and using the methods and data described in 
\cite{LEP} and \cite{renton}, gives $m_H = 76^{+33}_{-24}$~GeV, with 
$m_H < 144$~GeV at the 95\% confidence level.  The effect of the new $m_W$ 
value alone is to reduce the predicted value of the standard model Higgs 
boson mass by 6 GeV.
\par
We anticipate a significant reduction in the uncertainty of future CDF $m_W$ 
measurements using larger available data sets.  The dominant uncertainties 
on this measurement are due to $W$ boson statistics and to the lepton energy 
scale calibration (Table~\ref{tbl:totalsys}), and will be reduced with 
increased statistics in the $W$ boson and calibration data samples.

\appendix
\section{Electron and Photon Interactions}
\label{app:egammaint}

The simulation of electrons and photons (Section~\ref{sec:elesim}) uses the 
Bethe-Heitler differential cross sections for electron bremsstrahlung and 
photon conversion \cite{tsai}.  Defining $y$ as the final state energy divided 
by the initial state energy, the bremsstrahlung cross section is:
\begin{widetext}
\begin{equation}
\frac{d\sigma}{dy} = 4\alpha_{EM} r_e^2 \left[\left(\frac{4}{3y} - 
\frac{4}{3} + y\right)\psi_1(Z)  \\ 
  + \left(\frac{1}{y} - 1 \right) \frac{\psi_2(Z)}{9}\right], 
\end{equation}
\end{widetext}

\noindent
where: 
\begin{eqnarray}
\begin{array}{cll}
\psi_1(Z) & = & Z^2 [\ln(184.15Z^{-1/3})-f] + \\ 
          &   & ~~~~Z \ln(1194Z^{-2/3}), \\
\psi_2(Z) & = & Z^2 + Z, \\
f & = & a^2[(1+a^2)^{-1} + 0.20206 - 0.0369a^2 + \\
  &   & ~~~~0.0083a^4 - 0.002a^6],
\end{array}
\end{eqnarray}

\noindent
and $a = \alpha_{EM} Z$.  We define a material's radiation length $X_0$ 
according to \cite{tsai}:
\begin{eqnarray}
\begin{array}{llc}
X_0^{-1} & \equiv & 4\alpha_{EM} r_e^2 N_A\rho \psi_1(Z) / A ,
\end{array}
\end{eqnarray}

\noindent
where $\rho$ is the density of the material.  In terms of the radiation 
length, the cross section is
\begin{equation}
\label{eq:pdgbremapp}
\frac{d\sigma }{dy} = \frac{A}{N_A X_0\rho} \left[\left(\frac{4}{3} + C\right)
\left(\frac{1}{y}-1\right) + y\right], 
\end{equation}

\noindent
where
\begin{equation}
C \equiv \frac{\psi_2(Z)/9}{ \psi_1(Z) }.
\end{equation}

\noindent
The conversion cross section takes a similar form, since the relevant Feynman 
diagram is a rotation of the bremsstrahlung process \cite{tsai}:
\begin{equation}
\frac{d\sigma}{dy} = \frac{A}{N_A X_0 \rho}\left[1 - (4/3 + C)y(1-y)\right].
\label{eq:ppapp}
\end{equation}

The Compton scattering cross section as a function of scattering angle is 
given by the Klein-Nishina formula \cite{hubble}:
\begin{equation}
\label{eq:compton}
\frac{d\sigma}{d\Omega} = \frac{r_e^2}{2}\biggl[\frac{1 + \cos^2\theta}
       {[1 + k(1-\cos\theta)]^2} + \frac{k^2(1-\cos\theta)^2}
       {[1 + k(1-\cos\theta)]^3}\biggr],
\end{equation}

\noindent
where $k \equiv E_{\gamma}/m_e$.  The scattering angle is kinematically 
related to the energy loss by \cite{hubble} 
\begin{equation}
y = k'/k = [1 + k(1-\cos\theta)]^{-1},
\end{equation}

\noindent
where $k'$ is the energy of the photon after scattering, in
units of $m_e$.  Using this equation, the differential
cross section with respect to $y$ can be written as \cite{GEANT}:
\begin{equation}
\frac{d\sigma}{dy} \propto 1/y + y, \\
\end{equation}

\noindent
ignoring terms containing $1/k$.

\begin{acknowledgments}
We thank the Fermilab staff and the technical staffs of the participating institutions 
for their vital contributions. This work was supported by the U.S. Department of Energy 
and National Science Foundation; the Italian Istituto Nazionale di Fisica Nucleare; the 
Ministry of Education, Culture, Sports, Science and Technology of Japan; the Natural 
Sciences and Engineering Research Council of Canada; the National Science Council of the 
Republic of China; the Swiss National Science Foundation; the A.P. Sloan Foundation; the 
Bundesministerium f\"ur Bildung und Forschung, Germany; the Korean Science and Engineering 
Foundation and the Korean Research Foundation; the Science and Technology Facilities Council 
and the Royal Society, UK; the Institut National de Physique Nucleaire et Physique des 
Particules/CNRS; the Russian Foundation for Basic Research; the Comisi\'on Interministerial 
de Ciencia y Tecnolog\'{\i}a, Spain; the European Community's Human Potential Programme; 
the European Commission under the Marie Curie Programme; the Slovak R\&D Agency; and the 
Academy of Finland. 
\end{acknowledgments}



\begin{thebibliography}{75}

\bibitem{WZdiscovery}
G. Arnison {\it et al.} (UA1 Collaboration), Phys. Lett. B {\bf 122}, 103 (1983);
M. Banner {\it et al.} (UA2 Collaboration), Phys. Lett. B {\bf 122}, 476 (1983);
G. Arnison {\it et al.} (UA1 Collaboration), Phys. Lett. B {\bf 126}, 398 (1983);
P. Bagnaia {\it et al.} (UA2 Collaboration), Phys. Lett. B {\bf 129}, 130 (1983).
     
\bibitem{GWS}
S.~Glashow, Nucl. Phys. {\bf 22}, 579 (1961); S. Weinberg, Phys. Rev. Lett.
  {\bf 19}, 1264 (1967); A. Salam, {\it Elementary Particle Theory: Relativistic Groups 
    and Analyticity (Nobel Symposium No. 8)}, edited by N. Svartholm (Almqvist and
    Wiksell, Stockholm), p. 367 (1968).

\bibitem{corrections}
W. J. Marciano, Phys. Rev. D {\bf 20}, 274 (1979); F. Antonelli, M. Consoli, and 
G. Corbo, Phys. Lett. B {\bf 91}, 90 (1980); M. Veltman, {\it ibid.}, 95 (1980).


\bibitem{sirlin}
A. Sirlin, Phys. Rev. D {\bf 22}, 971 (1980).  

\bibitem{prlreference}
D. Acosta {\it et al.} (CDF Collaboration), submitted to Phys. Rev. Lett.

\bibitem{wtb}
D. C. Kennedy and B. W. Lynn, Nucl. Phys. {\bf B322}, 1 (1989);
M. B. Einhorn, D. R. T. Jones, and M. Veltman, Nucl. Phys. {\bf B191},
146 (1981).

\bibitem{susyloop}
S. Heinemeyer {\it et al.}, J. High Energy Phys. {\bf 08}, 052 (2006). 

\bibitem{dmdpar}
M. Awramik {\it et al.}, Phys. Rev. D {\bf 69}, 053006 (2004).

\bibitem{top}
E. Brubaker {\it et al.} (CDF and D\O\ Collaborations), hep-ex/0703034 (2007).

\bibitem{alpha}
H. Burkhardt and B. Pietrzyk, Phys. Rev. D {\bf 72}, 057501 (2005).

\bibitem{pdg}
W.-M. Yao {\it et al.}, J. Phys. G {\bf 33}, 1 (2006).

\bibitem{ALEPH}
S. Schael {\it et al.} (ALEPH Collaboration), Eur. Phys. J. C {\bf 47}, 309 (2006). 

\bibitem{OPAL}
G. Abbiendi {\it et al.} (OPAL Collaboration), Eur. Phys. J. C {\bf 45}, 307 (2006).

\bibitem{L3}
P. Achard {\it et al.} (L3 Collaboration), Eur. Phys. J. C {\bf 45}, 569 (2006).

\bibitem{DELPHI}
J. Abdallah {\it et al.} (DELPHI Collaboration), submitted to Z. Phys. C.

\bibitem{CDF}
T. Affolder {\it et al.} (CDF Collaboration), Phys. Rev. D {\bf 64}, 052001 (2001).

\bibitem{DZERO}
V. M. Abazov {\it et al.} (D\O\ Collaboration), Phys. Rev. D {\bf 66}, 012001 (2002);
B. Abbott {\it et al.} (D\O\ Collaboration), Phys. Rev. D {\bf 58}, 092003 (1998).

\bibitem{DZEROEC}
B. Abbott {\it et al.} (D\O\ Collaboration), Phys. Rev. D {\bf 62}, 092006 (2000).

\bibitem{LEP}
J. Alcarez {\it et al.} (LEP Collaborations), hep-ex/0612034 (2006).

\bibitem{TEV}
V. M. Abazov {\it et al.}, Phys. Rev. D {\bf 70}, 092008 (2004).

\bibitem{theses}
O. Stelzer-Chilton, Ph. D. thesis, University of Toronto, Fermilab-Thesis-2005-71 
(2005); I. Vollrath, Ph. D. thesis, University of Toronto, Fermilab-Thesis-2007-07 
(2007).

\bibitem{correlations}
There is in principle a correlation due to final-state photon radiation and the $W$ boson 
width, but these are sufficiently small that they are ignored when combining $m_W$ 
measurements.

\bibitem{MRST}
A. D. Martin, R. G. Roberts, W. J. Stirling, and R. S. Thorne, Eur. Phys. J. C 
{\bf 28}, 455 (2003).

\bibitem{CTEQ}
J. Pumplin {\it et al.}, J. High Energy Phys. 0207, 012 (2002).

\bibitem{WZxsec}
P. Sutton, A. Martin, R. Roberts, and W. Stirling, Phys. Rev. D {\bf 45}, 2349 (1992);
R. Rijken and W. van Neerven, Phys. Rev. D {\bf 51} 44 (1995); R. Harlander and 
W. Kilgore, Phys. Rev. Lett. {\bf 88}, 201801 (2002).

\bibitem{wzprd}
A. Abulencia {\it et al.} (CDF Collaboration), hep-ex/0508029, submitted to J. Phys. G;
D. Acosta {\it et al.} (CDF Collaboration), Phys. Rev. Lett. {\bf {94}}, 091803 (2005).

%
\bibitem{blinding} In order to obtain thesis results, two of the authors removed 
the blinding offset at two different points in the analysis.  Access to these theses 
was denied to all authors outside of the University of Toronto.

\bibitem{jpsi}
D. Acosta {\it et al.} (CDF Collaboration), Phys. Rev. D {\bf 71}, 032001 (2005);

\bibitem{COT}
T. Affolder {\it et al.}, Nucl. Instrum. Methods Phys. Res. A {\bf 526}, 249 (2004).

\bibitem{CEM}
L. Balka {\it et al.}, Nucl. Instrum. Methods Phys. Res. A {\bf 267}, 272 (1988);
K. Yasuoka {\it et al.}, Nucl. Instrum. Methods Phys. Res. A {\bf 267}, 315 (1988).

\bibitem{HAD}
S. Bertolucci {\it et al.}, Nucl. Instrum. Methods Phys. Res. A {\bf 267}, 301 (1988).

\bibitem{plug}
M. Albrow {\it et al.}, Nucl. Instrum. Meth. Phys. Res. A {\bf 480}, 524 (2002);
G. Apollinari {\it et al.}, Nucl. Instrum. Methods Phys. Res. A {\bf 412}, 515 (1998).

\bibitem{CMU}
G. Ascoli {\it et al.}, Nucl. Instrum. Methods Phys. Res. A {\bf 268}, 33 (1988).

\bibitem{cdftdr}
CDF Collaboration, Fermilab-Pub-96/390-E (1996).

\bibitem{XFT}
E. J. Thomson {\it et al.}, IEEE Trans. Nucl. Sc. {\bf 49}, 1063 (2002).

\bibitem{XFTcaveat} 
For a small fraction of the data, only 10 hits out of 12 sense wires were required 
to create XFT segments.

\bibitem{XTRP}
R. Downing {\it et al.}, Nucl. Instr. Methods Phys. Res. A {\bf 570}, 36 (2007).

\bibitem{CLC} 
J. Elias {\it et al.}, Nucl. Instr. Methods Phys. Res. A {\bf 441}, 366 (2000).

\bibitem{ue}
The underlying event refers to the spectator parton and additional inelastic 
$p\bar{p}$ interactions that produce low $p_T$ particles roughly uniform in 
phase space.

\bibitem{GEANT}
R. Brun and F. Carminati, CERN Program Library Long Writeup, W5013, 1993
(unpublished), version 3.15.

\bibitem{CDFSIM}
E. Gerchtein and M. Paulini, physics/0306031 (2003).

\bibitem{silimap}
The map is divided into 333 longitudinal and 120-1000 azimuthal sections, 
with the number of azimuthal sections increasing as radius increases.  

\bibitem{ms}
G. R. Lynch and O. I. Dahl, Nucl. Instrum. Methods Phys. Res. B {\bf 58}, 6 (1991).

\bibitem{mstail}
D. Attwood {\it et al.} (MuScat Collaboration), Nucl. Instrum. Methods Phys. Res. B 
{\bf 251}, 41 (2006).

\bibitem{trackresolution}
The radial distribution of the hits has a small impact on parameter resolution, 
with the importance depending on the parameter.  We do not attempt to model the 
radial hit distribution.

\bibitem{ionim}
C. Hays {\it et al.}, Nucl. Instrum. Methods Phys. Res. A {\bf 538}, 249 (2005). 

\bibitem{ttxsec}
D. Acosta {\it et al.} (CDF Collaboration), Phys. Rev. D {\bf 71}, 052003 
(2005).

\bibitem{tsai}
Y.-S. Tsai, Rev. Mod. Phys. {\bf 46}, 815 (1974).

\bibitem{ymin}
Increasing $y_0$ from $10^{-4}$ to $10^{-3}$ has about a 50 MeV effect 
on the $E/p$ calibration for electrons from $W$ boson decays; reducing it to 
$5 \times 10^{-5}$ has less than a 5 MeV effect.

\bibitem{expsuppression}
P. L. Anthony {\it et al.}, Phys. Rev. Lett. {\bf 76}, 3550 (1996).

\bibitem{migdal}
A. B. Migdal, Phys. Rev. {\bf 103}, 1811 (1956); L. D. Landau and I. J. Pomeranchuk,
Dokl. Akad. Nauk. SSSR {\bf 92}, 535 (1953); {\bf 92}, 735 (1953).

\bibitem{klein}
S. Klein, Rev. Mod. Phys. {\bf 71}, 1501 (1999).

\bibitem{dielectric}
M. L. Ter-Mikaelian, Dokl. Akad. Nauk. SSSR {\bf 94}, 1033 (1954).

\bibitem{hubble}
J. S. Hubbell, H. Gimm, and I. \O\ $\! \! \!$verb\o\ $\! \!$, J. Phys. Chem. Ref.
Data {\bf 9}, 1023 (1980).

\bibitem{eop}
Because the model of the width of the $E/p$ peak depends on the model of the 
momentum resolution, there is a $0.13$\% systematic uncertainty on $\kappa$ 
associated with the momentum resolution model.

\bibitem{wzrun1}
F. Abe {\it et al.} (CDF Collaboration), Phys. Rev. D {\bf 52}, 2624 (1995).

\bibitem{cosmic}
\rm A. V. Kotwal, H. K. Gerberich, and C. Hays,
Nucl. Instrum. Methods Phys. Res. A {\bf 506}, 110 (2003).

\bibitem{pdg2002}
K. Hagiwara {\it et al.}, Phys. Rev. D {\bf 66}, 010001 (2002).

\bibitem{pythia}
T. Sj$\rm\ddot{o}$strand, Comput. Phys. Commun. {\bf 82}, 74 (1994).  We 
used version 6.129 for $W$ and $Z$ production, version 6.136 for $\Upsilon$ 
production, and version 6.157 for $J/\psi$ production. 

\bibitem{fsrapprox}
R. Kleiss {\it et al.}, CERN 89-08, vol. 3, 1989 (unpublished).


\bibitem{resbos}
F. Landry, R. Brock, P.M. Nadolsky, and C.-P. Yuan, Phys.Rev. D {\bf 67}, 073016 (2003);
C. Balazs and C.-P. Yuan, Phys.Rev. D {\bf 56}, 5558 (1997);
G.A. Ladinsky and C.-P. Yuan, Phys.Rev. D {\bf 50}, 4239 (1994).

\bibitem{wgrad}
U. Baur, S. Keller, and D. Wackeroth, Phys. Rev. D {\bf 59}, 013002 (1998).

\bibitem{matscale}
We do not require the correction factor for the number of radiation lengths 
(0.4 $\pm$ 0.9)\% to be the same as that for the ionization energy loss 
(-6 $\pm$ 2)\% because of our assumption that copper cable needs to be added 
to the {\sc geant} simulation to model the radial distribution of photon 
conversions.  Since the bremsstrahlung and conversion cross sections scale as 
$X_0^{-1} \propto Z^2$ (Appendix~\ref{app:egammaint}) and the ionization energy 
loss scales as $Z$ (Section~\ref{sec:muonsim}), the correction factors can be 
different if the type of material is incompletely known.

\bibitem{zeop}
The statistical significance of the slope in the $Z$ boson sample is $0.5\sigma$.
Averaging with the $W$ boson sample results in zero slope.

\bibitem{wzd0}
B. Abbott {\it et al.} (D\O\ Collaboration), Phys. Rev. D {\bf 61}, 072001
(2000).

\bibitem{jetnet}
C. Peterson, T. R$\rm\ddot{o}$gnvaldsson, and L. L$\rm\ddot{o}$nnblad, Comput.
Phys. Commun. {\bf 81}, 185 (1994).

\bibitem{DGLAP}
V. N. Gribov and L. N. Lipatov, Sov. J. Nucl. Phys. {\bf 15}, 438 (1972);
G. Altarelli and G. Parisi, Nucl. Phys. {\bf B126}, 298 (1977);
Y. L. Dokshitzer, Sov. Phys. JETP {\bf 46}, 641 (1977).

\bibitem{pdfuncertainty}
J. M. Campbell, J. W. Huston, and W. J. Stirling, Rep. Prog. Phys. {\bf 70}, 89 (2007).

\bibitem{css}
J. C. Collins, D. Soper, and G. Sterman, Nucl. Phys. {\bf B250}, 199 (1985).

\bibitem{collinssoper} 
J. C. Collins and D. E. Soper, Phys. Rev. D {\bf 16}, 2219 (1977).

\bibitem{mirkes} E. Mirkes, Nucl. Phys. {\bf B387}, 3 (1992).

\bibitem{strologas} J. Strologas and S. Errede, Phys. Rev. D {\bf 73}, 052001 (2006).

\bibitem{dyrad} W. T. Giele, E. W. N. Glover, and D. A. Kosower, Nucl. Phys. {\bf B403},
633 (1993).

\bibitem{nnloqed} C. M. Carloni Calame, G. Montagna, O. Nicrosini, and 
M. Treccani, Phys. Rev. D {\bf 69}, 037301 (2004).

\bibitem{blue}
L. Lyons, D. Gibaut, and P. Clifford, Nucl. Instrum. Methods Phys. Res. 
A {\bf 270}, 110 (1988).

\bibitem{run1combo}
V. M. Abazov {\it et al.} (CDF and D\O\ Collaborations), Phys. Rev. D 
{\bf 70}, 092008 (2004).

\bibitem{renton}
P.B. Renton, Rep. Prog. Phys. {\bf 65}, 1271 (2002).

\end{thebibliography}
\end{document}